\documentclass[longauth]{aa}

\usepackage{graphicx}
\usepackage{natbib}
\usepackage{scalerel}

\usepackage[table]{xcolor}

\usepackage{txfonts}

\usepackage[pdfencoding=auto,psdextra]{hyperref}
\hypersetup{
    colorlinks=true,
    linkcolor=blue,
    filecolor=magenta,      
    urlcolor=blue,
    citecolor=blue
}
\makeatletter
\renewcommand*\aa@pageof{, page \thepage{} of \pageref*{LastPage}}
\makeatother

\usepackage[utf8]{inputenc}

\usepackage[switch, modulo]{lineno}

\usepackage{euclid}

\newcommand{\av}{$A_V$}

\newcommand{\ebv}{$E(B-V)$}

\newcommand{\msun}{$M_\odot$}

\newcommand{\Io}{\ensuremath{I_\sfont{E,0}}\xspace}
\newcommand{\Yo}{\ensuremath{Y_\sfont{E,0}}\xspace}
\newcommand{\Jo}{\ensuremath{J_\sfont{E,0}}\xspace}
\newcommand{\Ho}{\ensuremath{H_\sfont{E,0}}\xspace}
\newcommand{\IoHo}{\ensuremath{I_\sfont{E,0}-H_\sfont{E,0}}\xspace}
\newcommand{\IoYo}{\ensuremath{I_\sfont{E,0}-Y_\sfont{E,0}}\xspace}
\newcommand{\YoHo}{\ensuremath{Y_\sfont{E,0}-H_\sfont{E,0}}\xspace}
\newcommand{\IH}{\ensuremath{I_\sfont{E}-H_\sfont{E}}\xspace}

\defcitealias{ERONearbyGals}{H25}
\defcitealias{EROData}{C25}
\defcitealias{bp24}{BP24}

\begin{document}

%
%
\title{\Euclid: Early Release Observations -- The extended stellar component of the IC10 dwarf galaxy\thanks{This paper is published on behalf of the Euclid Consortium.}}


\newcommand{\orcid}[1]{} 
\author{F.~Annibali\thanks{\email{francesca.annibali@inaf.it}}\inst{\ref{aff1}}
\and A.~M.~N.~Ferguson\inst{\ref{aff2}}
\and P.~M.~Sanchez-Alarcon\orcid{0000-0002-6278-9233}\inst{\ref{aff3},\ref{aff4}}
\and P.~Dimauro\orcid{0000-0001-7399-2854}\inst{\ref{aff5},\ref{aff6}}
\and L.~K.~Hunt\orcid{0000-0001-9162-2371}\inst{\ref{aff7}}
\and R.~Pascale\orcid{0000-0002-6389-6268}\inst{\ref{aff1}}
\and M.~Bellazzini\orcid{0000-0001-8200-810X}\inst{\ref{aff1}}
\and A.~Lan\c{c}on\orcid{0000-0002-7214-8296}\inst{\ref{aff8}}
\and P.~Jablonka\orcid{0000-0002-9655-1063}\inst{\ref{aff9}}
\and J.~M.~Howell\orcid{0009-0002-2242-6515}\inst{\ref{aff2}}
\and K.~Voggel\orcid{0000-0001-6215-0950}\inst{\ref{aff8}}
\and J.-C.~Cuillandre\orcid{0000-0002-3263-8645}\inst{\ref{aff10}}
\and Abdurro'uf\orcid{0000-0002-5258-8761}\inst{\ref{aff11}}
\and G.~Battaglia\orcid{0000-0002-6551-4294}\inst{\ref{aff12}}
\and L.~R.~BEDIN\orcid{0000-0003-4080-6466}\inst{\ref{aff13}}
\and Michele~Cantiello\orcid{0000-0003-2072-384X}\inst{\ref{aff14}}
\and D.~Carollo\orcid{0000-0002-0005-5787}\inst{\ref{aff15}}
\and P.-A.~Duc\orcid{0000-0003-3343-6284}\inst{\ref{aff8}}
\and S.~S.~Larsen\orcid{0000-0003-0069-1203}\inst{\ref{aff16}}
\and M.~Libralato\orcid{0000-0001-9673-7397}\inst{\ref{aff13}}
\and F.~R.~Marleau\orcid{0000-0002-1442-2947}\inst{\ref{aff17}}
\and D.~Massari\orcid{0000-0001-8892-4301}\inst{\ref{aff1}}
\and T.~Saifollahi\orcid{0000-0002-9554-7660}\inst{\ref{aff8}}
\and C.~Tortora\orcid{0000-0001-7958-6531}\inst{\ref{aff18}}
\and M.~Urbano\orcid{0000-0001-5640-0650}\inst{\ref{aff8}}
\and M.~Gatto\orcid{0000-0003-4636-6457}\inst{\ref{aff18}}
\and I.~McDonald\orcid{0000-0003-0356-0655}\inst{\ref{aff19}}
\and M.~Baes\orcid{0000-0002-3930-2757}\inst{\ref{aff20}}
\and J.~Rom\'an\orcid{0000-0002-3849-3467}\inst{\ref{aff21}}
\and E.~Dalessandro\orcid{0000-0003-4237-4601}\inst{\ref{aff1}}
\and E.~Iodice\orcid{0000-0003-4291-0005}\inst{\ref{aff18}}
\and R.~Ragusa\inst{\ref{aff18}}
\and S.~Pearson\orcid{0000-0003-0256-5446}\inst{\ref{aff22}}
\and S.~Andreon\orcid{0000-0002-2041-8784}\inst{\ref{aff23}}
\and N.~Auricchio\orcid{0000-0003-4444-8651}\inst{\ref{aff1}}
\and C.~Baccigalupi\orcid{0000-0002-8211-1630}\inst{\ref{aff24},\ref{aff15},\ref{aff25},\ref{aff26}}
\and M.~Baldi\orcid{0000-0003-4145-1943}\inst{\ref{aff27},\ref{aff1},\ref{aff28}}
\and A.~Balestra\orcid{0000-0002-6967-261X}\inst{\ref{aff13}}
\and S.~Bardelli\orcid{0000-0002-8900-0298}\inst{\ref{aff1}}
\and P.~Battaglia\orcid{0000-0002-7337-5909}\inst{\ref{aff1}}
\and A.~Biviano\orcid{0000-0002-0857-0732}\inst{\ref{aff15},\ref{aff24}}
\and E.~Branchini\orcid{0000-0002-0808-6908}\inst{\ref{aff29},\ref{aff30},\ref{aff23}}
\and M.~Brescia\orcid{0000-0001-9506-5680}\inst{\ref{aff31},\ref{aff18}}
\and S.~Camera\orcid{0000-0003-3399-3574}\inst{\ref{aff32},\ref{aff33},\ref{aff34}}
\and G.~Ca\~nas-Herrera\orcid{0000-0003-2796-2149}\inst{\ref{aff2},\ref{aff35}}
\and V.~Capobianco\orcid{0000-0002-3309-7692}\inst{\ref{aff34}}
\and C.~Carbone\orcid{0000-0003-0125-3563}\inst{\ref{aff36}}
\and J.~Carretero\orcid{0000-0002-3130-0204}\inst{\ref{aff37},\ref{aff38}}
\and S.~Casas\orcid{0000-0002-4751-5138}\inst{\ref{aff39},\ref{aff40}}
\and M.~Castellano\orcid{0000-0001-9875-8263}\inst{\ref{aff6}}
\and G.~Castignani\orcid{0000-0001-6831-0687}\inst{\ref{aff1}}
\and S.~Cavuoti\orcid{0000-0002-3787-4196}\inst{\ref{aff18},\ref{aff41}}
\and A.~Cimatti\inst{\ref{aff42}}
\and C.~Colodro-Conde\inst{\ref{aff3}}
\and G.~Congedo\orcid{0000-0003-2508-0046}\inst{\ref{aff2}}
\and C.~J.~Conselice\orcid{0000-0003-1949-7638}\inst{\ref{aff19}}
\and L.~Conversi\orcid{0000-0002-6710-8476}\inst{\ref{aff43},\ref{aff44}}
\and Y.~Copin\orcid{0000-0002-5317-7518}\inst{\ref{aff45}}
\and F.~Courbin\orcid{0000-0003-0758-6510}\inst{\ref{aff46},\ref{aff47},\ref{aff48}}
\and H.~M.~Courtois\orcid{0000-0003-0509-1776}\inst{\ref{aff49}}
\and M.~Cropper\orcid{0000-0003-4571-9468}\inst{\ref{aff50}}
\and H.~Degaudenzi\orcid{0000-0002-5887-6799}\inst{\ref{aff51}}
\and G.~De~Lucia\orcid{0000-0002-6220-9104}\inst{\ref{aff15}}
\and H.~Dole\orcid{0000-0002-9767-3839}\inst{\ref{aff52}}
\and F.~Dubath\orcid{0000-0002-6533-2810}\inst{\ref{aff51}}
\and C.~A.~J.~Duncan\orcid{0009-0003-3573-0791}\inst{\ref{aff2}}
\and X.~Dupac\inst{\ref{aff44}}
\and S.~Escoffier\orcid{0000-0002-2847-7498}\inst{\ref{aff53}}
\and M.~Farina\orcid{0000-0002-3089-7846}\inst{\ref{aff54}}
\and R.~Farinelli\inst{\ref{aff1}}
\and S.~Ferriol\inst{\ref{aff45}}
\and F.~Finelli\orcid{0000-0002-6694-3269}\inst{\ref{aff1},\ref{aff55}}
\and M.~Frailis\orcid{0000-0002-7400-2135}\inst{\ref{aff15}}
\and E.~Franceschi\orcid{0000-0002-0585-6591}\inst{\ref{aff1}}
\and M.~Fumana\orcid{0000-0001-6787-5950}\inst{\ref{aff36}}
\and S.~Galeotta\orcid{0000-0002-3748-5115}\inst{\ref{aff15}}
\and K.~George\orcid{0000-0002-1734-8455}\inst{\ref{aff56}}
\and B.~Gillis\orcid{0000-0002-4478-1270}\inst{\ref{aff2}}
\and C.~Giocoli\orcid{0000-0002-9590-7961}\inst{\ref{aff1},\ref{aff28}}
\and J.~Gracia-Carpio\inst{\ref{aff57}}
\and A.~Grazian\orcid{0000-0002-5688-0663}\inst{\ref{aff13}}
\and F.~Grupp\inst{\ref{aff57},\ref{aff58}}
\and S.~V.~H.~Haugan\orcid{0000-0001-9648-7260}\inst{\ref{aff59}}
\and H.~Hoekstra\orcid{0000-0002-0641-3231}\inst{\ref{aff35}}
\and W.~Holmes\inst{\ref{aff60}}
\and I.~M.~Hook\orcid{0000-0002-2960-978X}\inst{\ref{aff61}}
\and F.~Hormuth\inst{\ref{aff62}}
\and A.~Hornstrup\orcid{0000-0002-3363-0936}\inst{\ref{aff63},\ref{aff64}}
\and K.~Jahnke\orcid{0000-0003-3804-2137}\inst{\ref{aff65}}
\and M.~Jhabvala\inst{\ref{aff66}}
\and E.~Keih\"anen\orcid{0000-0003-1804-7715}\inst{\ref{aff67}}
\and S.~Kermiche\orcid{0000-0002-0302-5735}\inst{\ref{aff53}}
\and A.~Kiessling\orcid{0000-0002-2590-1273}\inst{\ref{aff60}}
\and B.~Kubik\orcid{0009-0006-5823-4880}\inst{\ref{aff45}}
\and M.~K\"ummel\orcid{0000-0003-2791-2117}\inst{\ref{aff58}}
\and M.~Kunz\orcid{0000-0002-3052-7394}\inst{\ref{aff68}}
\and H.~Kurki-Suonio\orcid{0000-0002-4618-3063}\inst{\ref{aff69},\ref{aff70}}
\and R.~Laureijs\inst{\ref{aff71}}
\and A.~M.~C.~Le~Brun\orcid{0000-0002-0936-4594}\inst{\ref{aff72}}
\and S.~Ligori\orcid{0000-0003-4172-4606}\inst{\ref{aff34}}
\and P.~B.~Lilje\orcid{0000-0003-4324-7794}\inst{\ref{aff59}}
\and V.~Lindholm\orcid{0000-0003-2317-5471}\inst{\ref{aff69},\ref{aff70}}
\and I.~Lloro\orcid{0000-0001-5966-1434}\inst{\ref{aff73}}
\and G.~Mainetti\orcid{0000-0003-2384-2377}\inst{\ref{aff74}}
\and D.~Maino\inst{\ref{aff75},\ref{aff36},\ref{aff76}}
\and E.~Maiorano\orcid{0000-0003-2593-4355}\inst{\ref{aff1}}
\and O.~Mansutti\orcid{0000-0001-5758-4658}\inst{\ref{aff15}}
\and S.~Marcin\inst{\ref{aff77}}
\and O.~Marggraf\orcid{0000-0001-7242-3852}\inst{\ref{aff78}}
\and M.~Martinelli\orcid{0000-0002-6943-7732}\inst{\ref{aff6},\ref{aff79}}
\and N.~Martinet\orcid{0000-0003-2786-7790}\inst{\ref{aff80}}
\and F.~Marulli\orcid{0000-0002-8850-0303}\inst{\ref{aff81},\ref{aff1},\ref{aff28}}
\and R.~J.~Massey\orcid{0000-0002-6085-3780}\inst{\ref{aff82}}
\and E.~Medinaceli\orcid{0000-0002-4040-7783}\inst{\ref{aff1}}
\and S.~Mei\orcid{0000-0002-2849-559X}\inst{\ref{aff83},\ref{aff84}}
\and M.~Melchior\inst{\ref{aff85}}
\and Y.~Mellier\thanks{Deceased}\inst{\ref{aff86},\ref{aff87}}
\and M.~Meneghetti\orcid{0000-0003-1225-7084}\inst{\ref{aff1},\ref{aff28}}
\and E.~Merlin\orcid{0000-0001-6870-8900}\inst{\ref{aff6}}
\and G.~Meylan\inst{\ref{aff9}}
\and A.~Mora\orcid{0000-0002-1922-8529}\inst{\ref{aff88}}
\and M.~Moresco\orcid{0000-0002-7616-7136}\inst{\ref{aff81},\ref{aff1}}
\and L.~Moscardini\orcid{0000-0002-3473-6716}\inst{\ref{aff81},\ref{aff1},\ref{aff28}}
\and R.~Nakajima\orcid{0009-0009-1213-7040}\inst{\ref{aff78}}
\and C.~Neissner\orcid{0000-0001-8524-4968}\inst{\ref{aff89},\ref{aff38}}
\and S.-M.~Niemi\orcid{0009-0005-0247-0086}\inst{\ref{aff90}}
\and C.~Padilla\orcid{0000-0001-7951-0166}\inst{\ref{aff89}}
\and S.~Paltani\orcid{0000-0002-8108-9179}\inst{\ref{aff51}}
\and F.~Pasian\orcid{0000-0002-4869-3227}\inst{\ref{aff15}}
\and K.~Pedersen\inst{\ref{aff22}}
\and W.~J.~Percival\orcid{0000-0002-0644-5727}\inst{\ref{aff91},\ref{aff92},\ref{aff93}}
\and V.~Pettorino\orcid{0000-0002-4203-9320}\inst{\ref{aff90}}
\and S.~Pires\orcid{0000-0002-0249-2104}\inst{\ref{aff10}}
\and G.~Polenta\orcid{0000-0003-4067-9196}\inst{\ref{aff94}}
\and M.~Poncet\inst{\ref{aff95}}
\and L.~A.~Popa\inst{\ref{aff96}}
\and L.~Pozzetti\orcid{0000-0001-7085-0412}\inst{\ref{aff1}}
\and F.~Raison\orcid{0000-0002-7819-6918}\inst{\ref{aff57}}
\and R.~Rebolo\orcid{0000-0003-3767-7085}\inst{\ref{aff3},\ref{aff97},\ref{aff4}}
\and A.~Renzi\orcid{0000-0001-9856-1970}\inst{\ref{aff98},\ref{aff99}}
\and J.~Rhodes\orcid{0000-0002-4485-8549}\inst{\ref{aff60}}
\and G.~Riccio\inst{\ref{aff18}}
\and E.~Romelli\orcid{0000-0003-3069-9222}\inst{\ref{aff15}}
\and M.~Roncarelli\orcid{0000-0001-9587-7822}\inst{\ref{aff1}}
\and R.~Saglia\orcid{0000-0003-0378-7032}\inst{\ref{aff58},\ref{aff57}}
\and Z.~Sakr\orcid{0000-0002-4823-3757}\inst{\ref{aff100},\ref{aff101},\ref{aff102}}
\and D.~Sapone\orcid{0000-0001-7089-4503}\inst{\ref{aff103}}
\and B.~Sartoris\orcid{0000-0003-1337-5269}\inst{\ref{aff58},\ref{aff15}}
\and M.~Schirmer\orcid{0000-0003-2568-9994}\inst{\ref{aff65}}
\and P.~Schneider\orcid{0000-0001-8561-2679}\inst{\ref{aff78}}
\and A.~Secroun\orcid{0000-0003-0505-3710}\inst{\ref{aff53}}
\and G.~Seidel\orcid{0000-0003-2907-353X}\inst{\ref{aff65}}
\and S.~Serrano\orcid{0000-0002-0211-2861}\inst{\ref{aff104},\ref{aff105},\ref{aff106}}
\and P.~Simon\inst{\ref{aff78}}
\and C.~Sirignano\orcid{0000-0002-0995-7146}\inst{\ref{aff98},\ref{aff99}}
\and G.~Sirri\orcid{0000-0003-2626-2853}\inst{\ref{aff28}}
\and L.~Stanco\orcid{0000-0002-9706-5104}\inst{\ref{aff99}}
\and J.~Steinwagner\orcid{0000-0001-7443-1047}\inst{\ref{aff57}}
\and P.~Tallada-Cresp\'{i}\orcid{0000-0002-1336-8328}\inst{\ref{aff37},\ref{aff38}}
\and A.~N.~Taylor\inst{\ref{aff2}}
\and I.~Tereno\orcid{0000-0002-4537-6218}\inst{\ref{aff107},\ref{aff108}}
\and N.~Tessore\orcid{0000-0002-9696-7931}\inst{\ref{aff50}}
\and S.~Toft\orcid{0000-0003-3631-7176}\inst{\ref{aff109},\ref{aff110}}
\and R.~Toledo-Moreo\orcid{0000-0002-2997-4859}\inst{\ref{aff111}}
\and F.~Torradeflot\orcid{0000-0003-1160-1517}\inst{\ref{aff38},\ref{aff37}}
\and I.~Tutusaus\orcid{0000-0002-3199-0399}\inst{\ref{aff106},\ref{aff104},\ref{aff101}}
\and L.~Valenziano\orcid{0000-0002-1170-0104}\inst{\ref{aff1},\ref{aff55}}
\and J.~Valiviita\orcid{0000-0001-6225-3693}\inst{\ref{aff69},\ref{aff70}}
\and T.~Vassallo\orcid{0000-0001-6512-6358}\inst{\ref{aff15}}
\and A.~Veropalumbo\orcid{0000-0003-2387-1194}\inst{\ref{aff23},\ref{aff30},\ref{aff29}}
\and Y.~Wang\orcid{0000-0002-4749-2984}\inst{\ref{aff112}}
\and J.~Weller\orcid{0000-0002-8282-2010}\inst{\ref{aff58},\ref{aff57}}
\and G.~Zamorani\orcid{0000-0002-2318-301X}\inst{\ref{aff1}}
\and I.~A.~Zinchenko\orcid{0000-0002-2944-2449}\inst{\ref{aff113}}
\and E.~Zucca\orcid{0000-0002-5845-8132}\inst{\ref{aff1}}
\and J.~Garc\'ia-Bellido\orcid{0000-0002-9370-8360}\inst{\ref{aff114}}
\and J.~Mart\'{i}n-Fleitas\orcid{0000-0002-8594-569X}\inst{\ref{aff115}}
\and V.~Scottez\orcid{0009-0008-3864-940X}\inst{\ref{aff86},\ref{aff116}}}
										   
\institute{INAF-Osservatorio di Astrofisica e Scienza dello Spazio di Bologna, Via Piero Gobetti 93/3, 40129 Bologna, Italy\label{aff1}
\and
Institute for Astronomy, University of Edinburgh, Royal Observatory, Blackford Hill, Edinburgh EH9 3HJ, UK\label{aff2}
\and
Instituto de Astrof\'{\i}sica de Canarias, E-38205 La Laguna, Tenerife, Spain\label{aff3}
\and
Universidad de La Laguna, Dpto. Astrof\'\i sica, E-38206 La Laguna, Tenerife, Spain\label{aff4}
\and
Observatorio Nacional, Rua General Jose Cristino, 77-Bairro Imperial de Sao Cristovao, Rio de Janeiro, 20921-400, Brazil\label{aff5}
\and
INAF-Osservatorio Astronomico di Roma, Via Frascati 33, 00078 Monteporzio Catone, Italy\label{aff6}
\and
INAF-Osservatorio Astrofisico di Arcetri, Largo E. Fermi 5, 50125, Firenze, Italy\label{aff7}
\and
Universit\'e de Strasbourg, CNRS, Observatoire astronomique de Strasbourg, UMR 7550, 67000 Strasbourg, France\label{aff8}
\and
Institute of Physics, Laboratory of Astrophysics, Ecole Polytechnique F\'ed\'erale de Lausanne (EPFL), Observatoire de Sauverny, 1290 Versoix, Switzerland\label{aff9}
\and
Universit\'e Paris-Saclay, Universit\'e Paris Cit\'e, CEA, CNRS, AIM, 91191, Gif-sur-Yvette, France\label{aff10}
\and
Department of Astronomy, Indiana University, 727 East Third Street, Bloomington, IN 47405, USA\label{aff11}
\and
 Instituto de Astrof\'{\i}sica de Canarias, E-38205 La Laguna; Universidad de La Laguna, Dpto. Astrof\'\i sica, E-38206 La Laguna, Tenerife, Spain\label{aff12}
\and
INAF-Osservatorio Astronomico di Padova, Via dell'Osservatorio 5, 35122 Padova, Italy\label{aff13}
\and
INAF - Osservatorio Astronomico d'Abruzzo, Via Maggini, 64100, Teramo, Italy\label{aff14}
\and
INAF-Osservatorio Astronomico di Trieste, Via G. B. Tiepolo 11, 34143 Trieste, Italy\label{aff15}
\and
Department of Astrophysics/IMAPP, Radboud University, PO Box 9010, 6500 GL Nijmegen, The Netherlands\label{aff16}
\and
Universit\"at Innsbruck, Institut f\"ur Astro- und Teilchenphysik, Technikerstr. 25/8, 6020 Innsbruck, Austria\label{aff17}
\and
INAF-Osservatorio Astronomico di Capodimonte, Via Moiariello 16, 80131 Napoli, Italy\label{aff18}
\and
Jodrell Bank Centre for Astrophysics, Department of Physics and Astronomy, University of Manchester, Oxford Road, Manchester M13 9PL, UK\label{aff19}
\and
Sterrenkundig Observatorium, Universiteit Gent, Krijgslaan 281 S9, 9000 Gent, Belgium\label{aff20}
\and
Departamento de F{\'\i}sica de la Tierra y Astrof{\'\i}sica, Universidad Complutense de Madrid, Plaza de las Ciencias 2, E-28040 Madrid, Spain\label{aff21}
\and
DARK, Niels Bohr Institute, University of Copenhagen, Jagtvej 155, 2200 Copenhagen, Denmark\label{aff22}
\and
INAF-Osservatorio Astronomico di Brera, Via Brera 28, 20122 Milano, Italy\label{aff23}
\and
IFPU, Institute for Fundamental Physics of the Universe, via Beirut 2, 34151 Trieste, Italy\label{aff24}
\and
INFN, Sezione di Trieste, Via Valerio 2, 34127 Trieste TS, Italy\label{aff25}
\and
SISSA, International School for Advanced Studies, Via Bonomea 265, 34136 Trieste TS, Italy\label{aff26}
\and
Dipartimento di Fisica e Astronomia, Universit\`a di Bologna, Via Gobetti 93/2, 40129 Bologna, Italy\label{aff27}
\and
INFN-Sezione di Bologna, Viale Berti Pichat 6/2, 40127 Bologna, Italy\label{aff28}
\and
Dipartimento di Fisica, Universit\`a di Genova, Via Dodecaneso 33, 16146, Genova, Italy\label{aff29}
\and
INFN-Sezione di Genova, Via Dodecaneso 33, 16146, Genova, Italy\label{aff30}
\and
Department of Physics "E. Pancini", University Federico II, Via Cinthia 6, 80126, Napoli, Italy\label{aff31}
\and
Dipartimento di Fisica, Universit\`a degli Studi di Torino, Via P. Giuria 1, 10125 Torino, Italy\label{aff32}
\and
INFN-Sezione di Torino, Via P. Giuria 1, 10125 Torino, Italy\label{aff33}
\and
INAF-Osservatorio Astrofisico di Torino, Via Osservatorio 20, 10025 Pino Torinese (TO), Italy\label{aff34}
\and
Leiden Observatory, Leiden University, Einsteinweg 55, 2333 CC Leiden, The Netherlands\label{aff35}
\and
INAF-IASF Milano, Via Alfonso Corti 12, 20133 Milano, Italy\label{aff36}
\and
Centro de Investigaciones Energ\'eticas, Medioambientales y Tecnol\'ogicas (CIEMAT), Avenida Complutense 40, 28040 Madrid, Spain\label{aff37}
\and
Port d'Informaci\'{o} Cient\'{i}fica, Campus UAB, C. Albareda s/n, 08193 Bellaterra (Barcelona), Spain\label{aff38}
\and
Institute for Theoretical Particle Physics and Cosmology (TTK), RWTH Aachen University, 52056 Aachen, Germany\label{aff39}
\and
Deutsches Zentrum f\"ur Luft- und Raumfahrt e. V. (DLR), Linder H\"ohe, 51147 K\"oln, Germany\label{aff40}
\and
INFN section of Naples, Via Cinthia 6, 80126, Napoli, Italy\label{aff41}
\and
Dipartimento di Fisica e Astronomia "Augusto Righi" - Alma Mater Studiorum Universit\`a di Bologna, Viale Berti Pichat 6/2, 40127 Bologna, Italy\label{aff42}
\and
European Space Agency/ESRIN, Largo Galileo Galilei 1, 00044 Frascati, Roma, Italy\label{aff43}
\and
ESAC/ESA, Camino Bajo del Castillo, s/n., Urb. Villafranca del Castillo, 28692 Villanueva de la Ca\~nada, Madrid, Spain\label{aff44}
\and
Universit\'e Claude Bernard Lyon 1, CNRS/IN2P3, IP2I Lyon, UMR 5822, Villeurbanne, F-69100, France\label{aff45}
\and
Institut de Ci\`{e}ncies del Cosmos (ICCUB), Universitat de Barcelona (IEEC-UB), Mart\'{i} i Franqu\`{e}s 1, 08028 Barcelona, Spain\label{aff46}
\and
Instituci\'o Catalana de Recerca i Estudis Avan\c{c}ats (ICREA), Passeig de Llu\'{\i}s Companys 23, 08010 Barcelona, Spain\label{aff47}
\and
Institut de Ciencies de l'Espai (IEEC-CSIC), Campus UAB, Carrer de Can Magrans, s/n Cerdanyola del Vall\'es, 08193 Barcelona, Spain\label{aff48}
\and
UCB Lyon 1, CNRS/IN2P3, IUF, IP2I Lyon, 4 rue Enrico Fermi, 69622 Villeurbanne, France\label{aff49}
\and
Mullard Space Science Laboratory, University College London, Holmbury St Mary, Dorking, Surrey RH5 6NT, UK\label{aff50}
\and
Department of Astronomy, University of Geneva, ch. d'Ecogia 16, 1290 Versoix, Switzerland\label{aff51}
\and
Universit\'e Paris-Saclay, CNRS, Institut d'astrophysique spatiale, 91405, Orsay, France\label{aff52}
\and
Aix-Marseille Universit\'e, CNRS/IN2P3, CPPM, Marseille, France\label{aff53}
\and
INAF-Istituto di Astrofisica e Planetologia Spaziali, via del Fosso del Cavaliere, 100, 00100 Roma, Italy\label{aff54}
\and
INFN-Bologna, Via Irnerio 46, 40126 Bologna, Italy\label{aff55}
\and
University Observatory, LMU Faculty of Physics, Scheinerstr.~1, 81679 Munich, Germany\label{aff56}
\and
Max Planck Institute for Extraterrestrial Physics, Giessenbachstr. 1, 85748 Garching, Germany\label{aff57}
\and
Universit\"ats-Sternwarte M\"unchen, Fakult\"at f\"ur Physik, Ludwig-Maximilians-Universit\"at M\"unchen, Scheinerstr.~1, 81679 M\"unchen, Germany\label{aff58}
\and
Institute of Theoretical Astrophysics, University of Oslo, P.O. Box 1029 Blindern, 0315 Oslo, Norway\label{aff59}
\and
Jet Propulsion Laboratory, California Institute of Technology, 4800 Oak Grove Drive, Pasadena, CA, 91109, USA\label{aff60}
\and
Department of Physics, Lancaster University, Lancaster, LA1 4YB, UK\label{aff61}
\and
Felix Hormuth Engineering, Goethestr. 17, 69181 Leimen, Germany\label{aff62}
\and
Technical University of Denmark, Elektrovej 327, 2800 Kgs. Lyngby, Denmark\label{aff63}
\and
Cosmic Dawn Center (DAWN), Denmark\label{aff64}
\and
Max-Planck-Institut f\"ur Astronomie, K\"onigstuhl 17, 69117 Heidelberg, Germany\label{aff65}
\and
NASA Goddard Space Flight Center, Greenbelt, MD 20771, USA\label{aff66}
\and
Department of Physics and Helsinki Institute of Physics, Gustaf H\"allstr\"omin katu 2, University of Helsinki, 00014 Helsinki, Finland\label{aff67}
\and
Universit\'e de Gen\`eve, D\'epartement de Physique Th\'eorique and Centre for Astroparticle Physics, 24 quai Ernest-Ansermet, CH-1211 Gen\`eve 4, Switzerland\label{aff68}
\and
Department of Physics, P.O. Box 64, University of Helsinki, 00014 Helsinki, Finland\label{aff69}
\and
Helsinki Institute of Physics, Gustaf H{\"a}llstr{\"o}min katu 2, University of Helsinki, 00014 Helsinki, Finland\label{aff70}
\and
Kapteyn Astronomical Institute, University of Groningen, PO Box 800, 9700 AV Groningen, The Netherlands\label{aff71}
\and
Laboratoire d'etude de l'Univers et des phenomenes eXtremes, Observatoire de Paris, Universit\'e PSL, Sorbonne Universit\'e, CNRS, 92190 Meudon, France\label{aff72}
\and
SKAO, Jodrell Bank, Lower Withington, Macclesfield SK11 9FT, UK\label{aff73}
\and
Centre de Calcul de l'IN2P3/CNRS, 21 avenue Pierre de Coubertin 69627 Villeurbanne Cedex, France\label{aff74}
\and
Dipartimento di Fisica "Aldo Pontremoli", Universit\`a degli Studi di Milano, Via Celoria 16, 20133 Milano, Italy\label{aff75}
\and
INFN-Sezione di Milano, Via Celoria 16, 20133 Milano, Italy\label{aff76}
\and
University of Applied Sciences and Arts of Northwestern Switzerland, School of Computer Science, 5210 Windisch, Switzerland\label{aff77}
\and
Universit\"at Bonn, Argelander-Institut f\"ur Astronomie, Auf dem H\"ugel 71, 53121 Bonn, Germany\label{aff78}
\and
INFN-Sezione di Roma, Piazzale Aldo Moro, 2 - c/o Dipartimento di Fisica, Edificio G. Marconi, 00185 Roma, Italy\label{aff79}
\and
Aix-Marseille Universit\'e, CNRS, CNES, LAM, Marseille, France\label{aff80}
\and
Dipartimento di Fisica e Astronomia "Augusto Righi" - Alma Mater Studiorum Universit\`a di Bologna, via Piero Gobetti 93/2, 40129 Bologna, Italy\label{aff81}
\and
Department of Physics, Institute for Computational Cosmology, Durham University, South Road, Durham, DH1 3LE, UK\label{aff82}
\and
Universit\'e Paris Cit\'e, CNRS, Astroparticule et Cosmologie, 75013 Paris, France\label{aff83}
\and
CNRS-UCB International Research Laboratory, Centre Pierre Bin\'etruy, IRL2007, CPB-IN2P3, Berkeley, USA\label{aff84}
\and
University of Applied Sciences and Arts of Northwestern Switzerland, School of Engineering, 5210 Windisch, Switzerland\label{aff85}
\and
Institut d'Astrophysique de Paris, 98bis Boulevard Arago, 75014, Paris, France\label{aff86}
\and
Institut d'Astrophysique de Paris, UMR 7095, CNRS, and Sorbonne Universit\'e, 98 bis boulevard Arago, 75014 Paris, France\label{aff87}
\and
Telespazio UK S.L. for European Space Agency (ESA), Camino bajo del Castillo, s/n, Urbanizacion Villafranca del Castillo, Villanueva de la Ca\~nada, 28692 Madrid, Spain\label{aff88}
\and
Institut de F\'{i}sica d'Altes Energies (IFAE), The Barcelona Institute of Science and Technology, Campus UAB, 08193 Bellaterra (Barcelona), Spain\label{aff89}
\and
European Space Agency/ESTEC, Keplerlaan 1, 2201 AZ Noordwijk, The Netherlands\label{aff90}
\and
Waterloo Centre for Astrophysics, University of Waterloo, Waterloo, Ontario N2L 3G1, Canada\label{aff91}
\and
Department of Physics and Astronomy, University of Waterloo, Waterloo, Ontario N2L 3G1, Canada\label{aff92}
\and
Perimeter Institute for Theoretical Physics, Waterloo, Ontario N2L 2Y5, Canada\label{aff93}
\and
Space Science Data Center, Italian Space Agency, via del Politecnico snc, 00133 Roma, Italy\label{aff94}
\and
Centre National d'Etudes Spatiales -- Centre spatial de Toulouse, 18 avenue Edouard Belin, 31401 Toulouse Cedex 9, France\label{aff95}
\and
Institute of Space Science, Str. Atomistilor, nr. 409 M\u{a}gurele, Ilfov, 077125, Romania\label{aff96}
\and
Consejo Superior de Investigaciones Cientificas, Calle Serrano 117, 28006 Madrid, Spain\label{aff97}
\and
Dipartimento di Fisica e Astronomia "G. Galilei", Universit\`a di Padova, Via Marzolo 8, 35131 Padova, Italy\label{aff98}
\and
INFN-Padova, Via Marzolo 8, 35131 Padova, Italy\label{aff99}
\and
Institut f\"ur Theoretische Physik, University of Heidelberg, Philosophenweg 16, 69120 Heidelberg, Germany\label{aff100}
\and
Institut de Recherche en Astrophysique et Plan\'etologie (IRAP), Universit\'e de Toulouse, CNRS, UPS, CNES, 14 Av. Edouard Belin, 31400 Toulouse, France\label{aff101}
\and
Universit\'e St Joseph; Faculty of Sciences, Beirut, Lebanon\label{aff102}
\and
Departamento de F\'isica, FCFM, Universidad de Chile, Blanco Encalada 2008, Santiago, Chile\label{aff103}
\and
Institut d'Estudis Espacials de Catalunya (IEEC),  Edifici RDIT, Campus UPC, 08860 Castelldefels, Barcelona, Spain\label{aff104}
\and
Satlantis, University Science Park, Sede Bld 48940, Leioa-Bilbao, Spain\label{aff105}
\and
Institute of Space Sciences (ICE, CSIC), Campus UAB, Carrer de Can Magrans, s/n, 08193 Barcelona, Spain\label{aff106}
\and
Departamento de F\'isica, Faculdade de Ci\^encias, Universidade de Lisboa, Edif\'icio C8, Campo Grande, PT1749-016 Lisboa, Portugal\label{aff107}
\and
Instituto de Astrof\'isica e Ci\^encias do Espa\c{c}o, Faculdade de Ci\^encias, Universidade de Lisboa, Tapada da Ajuda, 1349-018 Lisboa, Portugal\label{aff108}
\and
Cosmic Dawn Center (DAWN)\label{aff109}
\and
Niels Bohr Institute, University of Copenhagen, Jagtvej 128, 2200 Copenhagen, Denmark\label{aff110}
\and
Universidad Polit\'ecnica de Cartagena, Departamento de Electr\'onica y Tecnolog\'ia de Computadoras,  Plaza del Hospital 1, 30202 Cartagena, Spain\label{aff111}
\and
Caltech/IPAC, 1200 E. California Blvd., Pasadena, CA 91125, USA\label{aff112}
\and
Astronomisches Rechen-Institut, Zentrum f\"ur Astronomie der Universit\"at Heidelberg, M\"onchhofstr. 12-14, 69120 Heidelberg, Germany\label{aff113}
\and
Instituto de F\'isica Te\'orica UAM-CSIC, Campus de Cantoblanco, 28049 Madrid, Spain\label{aff114}
\and
Aurora Technology for European Space Agency (ESA), Camino bajo del Castillo, s/n, Urbanizacion Villafranca del Castillo, Villanueva de la Ca\~nada, 28692 Madrid, Spain\label{aff115}
\and
ICL, Junia, Universit\'e Catholique de Lille, LITL, 59000 Lille, France\label{aff116}}    

 \date{\bf }

%
%
   \abstract{
We present a detailed analysis of the old, extended stellar component of the Local Group dwarf galaxy IC\,10 using deep resolved-star photometry in the VIS and NISP bands of the \Euclid Early Release Observations. Leveraging \Euclid's unique combination of a wide field of view and high spatial resolution, we traced red giant branch (RGB) stars out to $\sim$\,8\,kpc from the galaxy centre, reaching azimuthally averaged surface brightness levels as faint as $\mu_{\HE}\sim$\,29\,mag\,arcsec$^{-2}$. Our analysis reveals that IC\,10’s stellar distribution is significantly more extended than previously assumed. After correcting for foreground extinction and subtracting contamination from Milky Way stars and background galaxies, we derived a radial stellar density profile from the RGB star counts. The profile shows a marked flattening beyond $\sim$\,5\,kpc and it is best fit by a two-component (Sérsic + exponential) model, yielding a total stellar mass in old (age $\gtrsim$\,1\,Gyr) stars of $M_{\star}=(6.7$--8.1)$\times10^8$\,\msun.  The origin of the outer stellar component is unclear.  It might have been accreted or even possibly associated with the counter-rotating \ion{H}{i} gas in the outer regions of IC\,10; alternatively, it might represent an ancient `in situ' stellar halo. We tentatively detected two symmetric stellar overdensities at the edge of our imagery, which are roughly aligned with the direction of IC\,10's orbit around M31, suggesting that they could be signatures of tidal stripping. As part of our analysis, we derived a new distance to IC\,10 based on the tip of the RGB, finding $D=(762\pm 20)$\,kpc with a distance modulus of $(m-M)_0=24.41\pm 0.05$.}
%
%
%
%
   \titlerunning{\Euclid\/: The extended stellar component of IC\,10}
   \authorrunning{F. Annibali et al.}
   
   \maketitle
\nolinenumbers
   
%
%
%
%
   
\section{\label{sc:Intro}Introduction}

In the $\Lambda$ cold dark matter ($\Lambda$CDM) cosmological framework \citep{Peebles1982}, structure formation proceeds hierarchically through the successive merging and accretion of smaller dark-matter haloes \citep{White1978,Springel2006}, a process that begins at early cosmic times and continues down to the present day \citep{Fakhouri2010,DeLucia2008}. As a result, galaxies today are expected to be embedded in extended stellar haloes formed from the disrupted remnants of accreted satellites and to be surrounded and orbited by numerous smaller companions and stellar streams \citep{Bekki2001,Bullock2005,Johnston2008,Cooper2010}.

This picture has been confirmed extensively for massive galaxies, both spirals and ellipticals, which show abundant observational evidence of past accretion events \citep[e.g.][]
{Ibata2001a,Ibata2001b,Ferguson2002,Belokurov2006,Martinez10,Crnojevic2016}. Dwarf galaxies, traditionally seen as the fundamental building blocks of larger systems, are likewise expected to grow via the accretion of even smaller structures \citep{Diemand2008,Wheeler2015}.
However, direct evidence for accretion events around dwarfs remains much more elusive than for massive galaxies \citep{Martinez08,Koposov2018,Paudel2018,carlin19,Kado2020,Correnti2025}. It is still unclear whether dwarf galaxies host stellar haloes at all \citep{Chiti2021,Tarumi2021,Deason2022,Kado-Fong2022}. While $\Lambda$CDM simulations predict similar relative amounts of substructure across halo mass scales \citep{Diemand2008}, the amount of stellar mass these systems can accrete over time is poorly constrained
\citep{Kang2019,Martin2021,Deason2022,Cooper2025}. This uncertainty largely stems from the poorly known stellar mass–halo mass (SMHM) relation and the low halo occupation fraction at the smallest masses \citep[][and references therein]{Sales2022}. As a matter of fact, all simulations consistently indicate that the stellar haloes of individual dwarf galaxies, if present,  are expected to be extremely faint and challenging to detect, having typical surface brightness values fainter than $\mu\sim35$\,mag\,arcsec$^{-2}$ \citep[see e.g.][]{Deason2022}.  
On the other hand, cold tidal features, such as streams and shells, 
tend to stand out above the average surface brightness profile and can be as bright as $\mu\sim27$--29\,mag\,arcsec$^{-2}$, making them comparatively easier to observe \citep{martinez12,Annibali2016,Belokurov2016,Roman2023,Sacchi2024}.
An additional challenge in detecting stellar haloes around dwarf galaxies is that much of the accreted, ex situ stellar mass is expected to project onto the central regions of the galaxy, where it is blended with the typically dominant in situ stellar population \citep[e.g.][]{Pillepich2015,Cooper2025}. This overlap complicates efforts to disentangle the two components observationally. As a result, robust identification of the accreted stellar halo requires tracing the stellar distribution to large galactocentric radii and reaching extremely low surface brightness levels \citep{Chiti2021,Sestito2023,Jensen2024}.

In this paper, we search for and study the extended stellar component around the Local Group dwarf galaxy IC\,10
 \citep[$M_V\sim-15$,][]{McConnachie2012}. Argued to be the closest example of a starburst galaxy \citep{Richer01}, IC\,10 is a fascinating object  whose stellar component has been poorly studied due to its low Galactic latitude ($b\sim \ang{-3.3;;}$). This sightline leads to significant foreground contamination from the stars and dust of the Milky Way (MW), with the result being that many of IC\,10's properties are still rather uncertain. In particular, estimates of its current star-formation rate range from  0.02--0.5\,$M_\odot \,\rm {yr}^{-1}$ \citep{Koch2025TheLGLBS, Binder2025The10}, and distance estimates range from $\sim$\,500\,kpc to $\sim$\,1\,Mpc \citep[e.g.][]{Jacobs2009, Sanna2008, Hunter2001, Gholami2025}.  
 Also, its stellar mass is highly uncertain, with quoted values ranging from ${\sim}\,0.9\times10^8$\,\msun\ to ${\sim}\,4\times10^8$\,\msun, a spread that cannot be attributed solely to distance uncertainties \citep{Lee2003,McConnachie2012,Pace2025,nersesian19}.

As a distant satellite of M31, IC\,10 may have experienced one or more interactions with other systems.  
Several studies have shown that IC\,10 hosts an extended and disturbed \ion{H}{i} disc, including counter-rotating gas in its outskirts \citep{Shostak1989,wilcots98,namumba19}. \citet{nidever13} also identified a \ang{1.3;;}-long north-west  \ion{H}{i} feature with a strong velocity gradient reaching to ${\sim}\,65$ km\,s$^{-1}$ lower than the IC\,10 systemic velocity, later confirmed by \citet{namumba19}. Using the proper motions  from \citet{Brunthaler2007}, they argued that this structure is unlikely to result from tidal forces from M31 or ram-pressure stripping from the group environment, and instead may reflect an interaction or merger with another dwarf. Deep imaging has not revealed clear stellar tidal features \citep{Demers2004,Sanna2010}, and an apparent offset between young and old populations \citep{gerbrandt15} was not confirmed by \citet{Howell2026}. Orbital modelling by \citet{Bennet2024} suggests a ${\sim}\,130$ kpc pericentre passage around M31 about 1--2\,Gyr ago. Altogether, these findings point to a past interaction, making the structure of IC\,10’s extended stellar component a valuable  piece of information in understanding its evolutionary history.

In Sect.~\ref{sc:data} of this paper, we describe the data reduction and the photometric analysis, while 
in Sect.~\ref{sc:ebv_fg}, we present our procedure for reddening correction and removal of 
foreground and background contaminants from the catalogues.  
In Sect.~\ref{sc:cmd} we present the final cleaned, reddening-corrected colour-magnitude diagrams (CMDs) 
and derive the galaxy distance from the tip of the red giant branch (TRGB).
Section~\ref{sc:spatial} describes the spatial distribution of IC\,10's stellar population and the inferred radial profile. We discuss our results in a larger context in Sect.~\ref{sc:conclusions}, and give our summary in Sect.~\ref{sc:summary}.

\section{\label{sc:data}Data reduction and photometry}

\subsection{\label{sc:obs_image} Observations and image reduction}

IC\,10 was observed in 2023 during the \Euclid performance verification (PV) phase as part of the Showcase Galaxies Early Release Observation (ERO) program \citep[][hereafter \citetalias{EROData,ERONearbyGals}]{EROData,ERONearbyGals}. While the other galaxies from this programme were acquired with one standard Reference Observation Sequence (ROS) consisting of four dithered exposures of 560\,s for the Visible Camera \citep[VIS; ][]{EuclidSkyVIS} and 87\,s for each of the  Near-Infrared Spectrometer and Photometer \citep[NISP;][]{EuclidSkyNISP} bands\footnote{The quoted values correspond to the effective exposure times, while the total durations reported in \cite{Scaramella-EP1} also account for observational overheads.}
(\YE, \JE, and \HE), typical of the Euclid Wide Survey \citep[EWS,][]{Scaramella-EP1}, IC\,10 was observed for a longer exposure time consisting of 2 ROS, totalling about 2 hours with overheads.

As with the other ERO galaxies, IC\,10 was not reduced with the standard Science Ground Segment pipeline
\citep{Q1-TP004}, but rather with a procedure optimised for the study of the low surface brightness (LSB) emission, developed by \citetalias{EROData}.
In brief, this pipeline processes the calibrated level 1 raw VIS and NISP images removing cosmic rays, correcting for geometric 
distortions, masking persistence effects due to preceding spectroscopic observations and applying a `super flat field' that includes the illumination pattern and low-level flux non-linearity.

\begin{figure}[h!]
\includegraphics[width=\columnwidth]{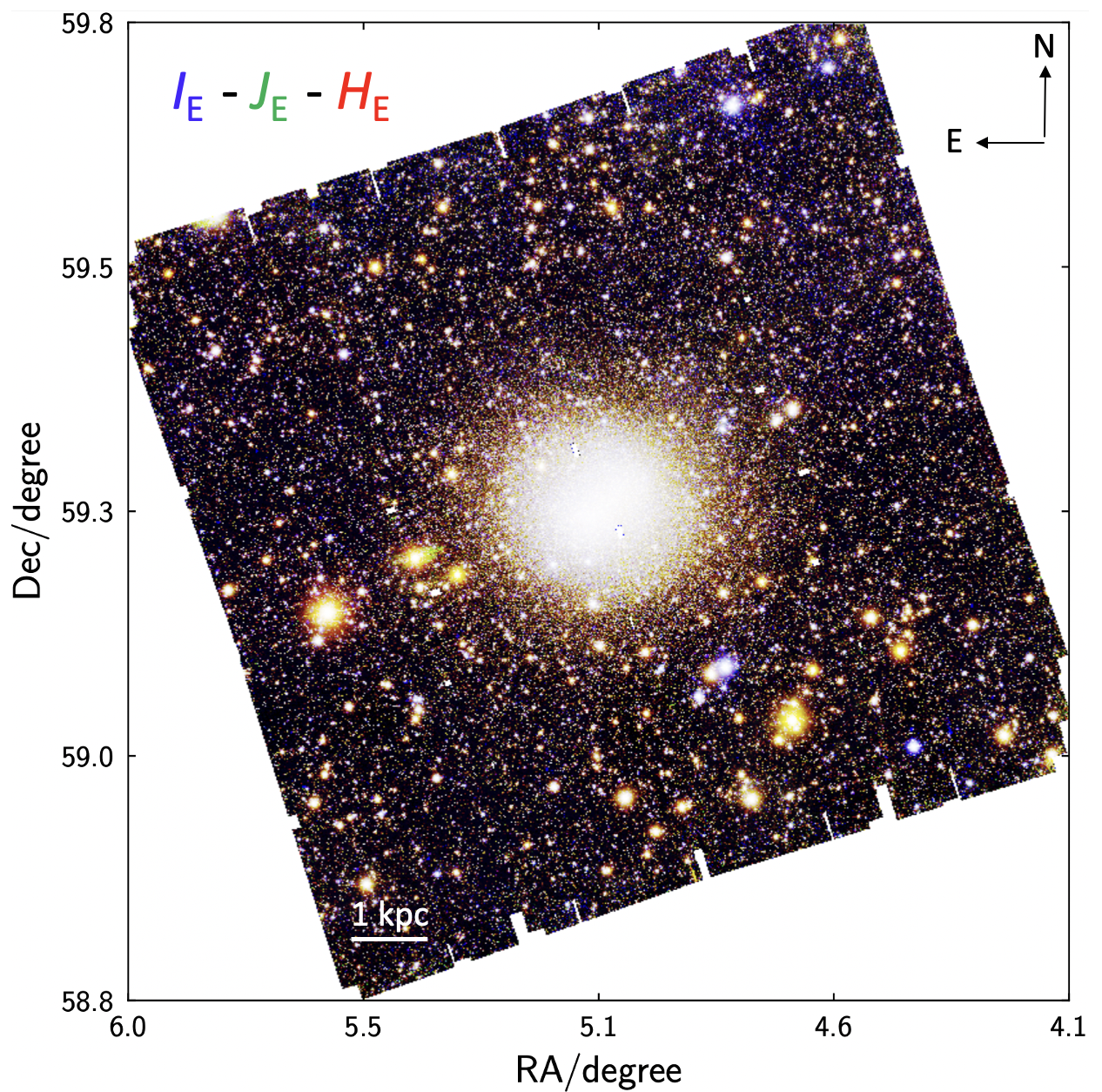}
\caption{\Euclid colour-combined image of IC\,10  with the stretch optimised to highlight the morphology of the old stellar component. The horizontal scale bar corresponds to 1\,kpc. North is up and east is left. 
} 
\label{fig:ic10_image}
\end{figure}

The final astrometrically aligned, stacked frames have a median point spread function (PSF) full width at half maximum  (FWHM) of \mbox{\ang{;;0.16}}, \mbox{\ang{;;0.47}}, \mbox{\ang{;;0.47}}, and  \mbox{\ang{;;0.49}} (1.57, 1.57, 1.58, 1.65\,pixels) in \IE, \YE, \JE, and \HE, respectively. 
For both instruments, the PSF is slightly undersampled, given the pixel sizes of 
\ang{;;0.1} and \ang{;;0.3} of the VIS and NISP images, respectively.  
The ERO data were rescaled to have a nominal zero point of ZP\,=30.13\,AB mag for \IE and ZP\,=\,30.0\,AB mag for \YE, \JE, and \HE \citepalias[][]{EROData,ERONearbyGals}. Figure~\ref{fig:ic10_image} shows a colour-combined image of the galaxy (\IE=blue, \JE=green, and \HE=red) with the stretch optimised to highlight the morphology of the old stellar component.

\subsection{\label{sc:photometry} Point source photometry}

Photometry was performed with the \texttt{DAOPHOT} package \citep{Stetson1987} in the 
\texttt{IRAF} environment.\footnote{\texttt{IRAF} is distributed by the National Optical Astronomy Observatory, which
is operated by the Association of Universities for Research in Astronomy, Inc.,
under cooperative agreement with the National Science Foundation.} Given our focus on the extended low stellar density  component of IC\,10, aperture photometry is well suited to our goals, offering a level of performance comparable to PSF-fitting methods in this regime.

We detected sources by running the 
\texttt{DAOFIND} routine independently in each stack of the four \Euclid photometric bands,  adopting a threshold of three times the average background standard deviation, $\sigma_{\mathrm{bg}}$, derived from several 20\,pixel$\times$20\,pixel external regions devoid of stars. 
For each band, aperture photometry was performed at the position of the detected sources with the \texttt{PHOT} task, adopting an aperture of 1.5 pixels in radius, 
corresponding to \ang{;;0.15} and \ang{;;0.45} in VIS and NISP, respectively  (i.e. corresponding roughly to the PSF FWHM). The local background was computed within an annulus of 15--20 pixel radius centred on each source. For VIS, we also computed photometry in a larger aperture of 2 pixel radius to allow for the removal of saturated stars and extended sources, as detailed below. 
The aperture corrections involved raising a 1.5 pixel radius to larger apertures  of 3\arcsecond\ for VIS and 9\arcsecond\
for the NISP images, totalling about 20 times the PSF FWHM and capturing $\gtrsim$\,98\% of the total PSF flux \citepalias{EROData}. These corrections were computed by selecting the brightest unsaturated, most isolated stars and then subtracting nearby neighbours through a PSF model derived for each band with the  \texttt{DAOPHOT PSF} task  (see Appendix~\ref{sc:artificial} for more details on the PSF model assumptions). The derived aperture corrections are $-0.56\pm0.02$, $-0.32\pm0.05$, $-0.34\pm0.05$, and $-0.46\pm0.05$\,mag in \IE, \YE, \JE, and \HE, respectively. Finally, the magnitudes were calibrated by applying the zero points of \citetalias{EROData}.

The four photometric catalogues in \IE, \YE, \JE, and \HE were cross-matched, adopting a \ang{;;0.5}
maximum tolerance in position  (about the size of the NISP PSF FWHM) and resulting in
\num{418484} sources that have photometry in all bands. 
The cross-match removes the majority of spurious detections, such as cosmic rays, peaks on bright star spikes, and residual artifacts from the image reduction pipeline. We applied selection cuts based on the \texttt{DAOPHOT} photometric errors in all four bands, retaining only sources within three times the standard deviation of the error distribution as a function of magnitude.  

\begin{figure}
\includegraphics[width=\columnwidth]{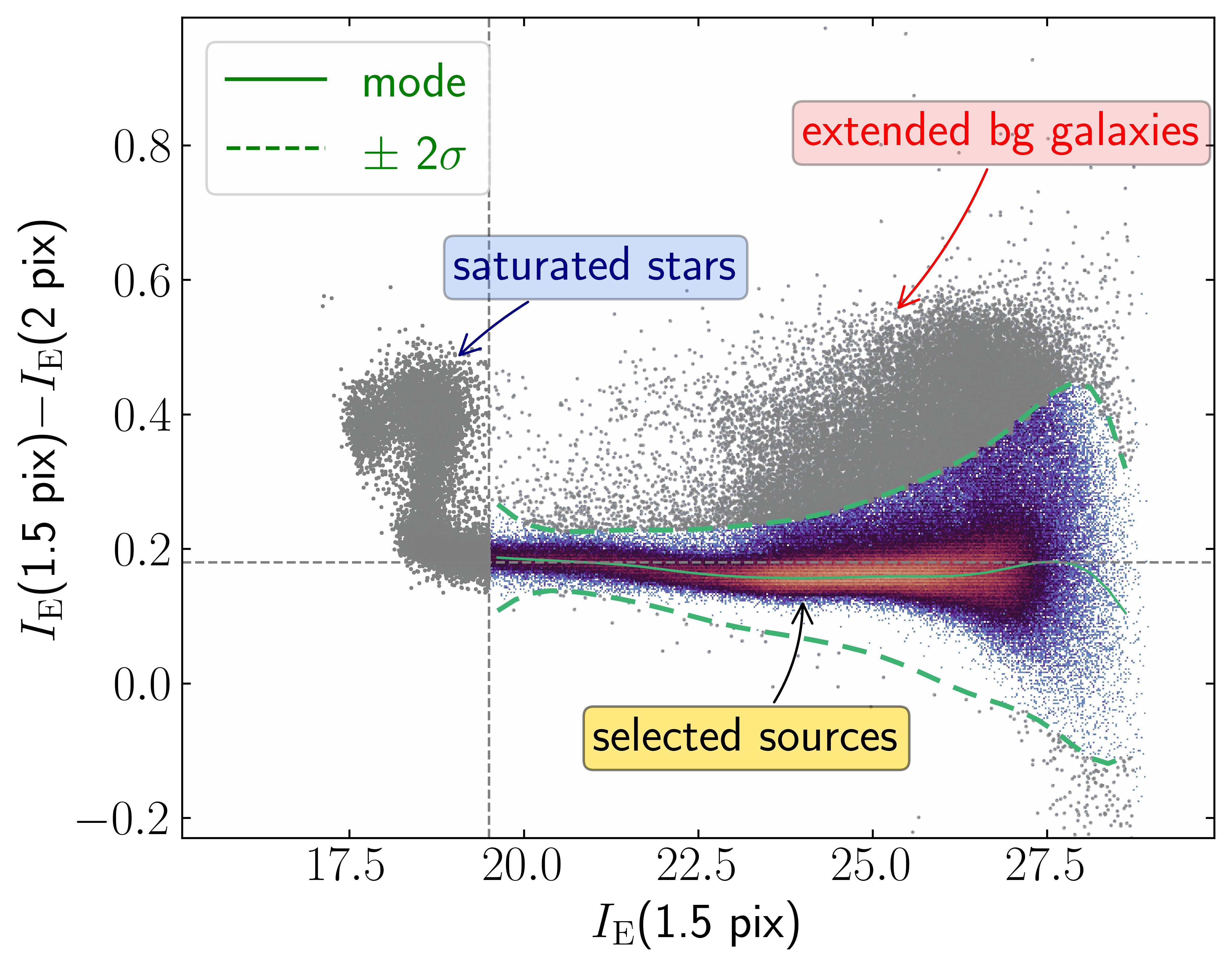}
\caption{Our source selection based on the distribution in the 
${\IE({\rm 1.5\,pix})}-{\IE({\rm 2\,pix})}$ versus $\IE{\rm (1.5\,pix)}$ plane. 
Point sources populate a narrow region around $\Delta\IE\sim0.18$, while bright saturated stars and extended background galaxies tend to have larger values. We retained all sources within $2\,\sigma$ of the mode of the distribution, as indicated by the dashed green line, and discarded bright stars with $\IE{\rm(1.5\,pix)}\le19.5$.} 
\label{fig:ap_selection}
\end{figure}

Bright, saturated stars and extended objects were then removed based on diagnostics that quantify the deviation from a stellar profile.
More specifically, we identified objects that deviate from the PSF based on their `concentration index'; namely, by comparing aperture photometry within two different apertures \citep[see, e.g.][]{Peng2011}. 
Our approach is illustrated in Fig.~\ref{fig:ap_selection}, where we show the VIS photometry computed within a 
1.5\,pixel radius aperture, 
$\IE{\rm (1.5\,pix)}$, versus ${\Delta\IE}={\IE({\rm 1.5\,pix})}-{\IE({\rm 2\,pix})}$; this is effectively the difference in magnitude between photometry at 1.5 and at 2 pixels, respectively. 
Relatively bright unsaturated stars populate a narrow region around $\Delta\IE\sim0.18$.
On the other hand, bright saturated stars with $\IE{\rm(1.5\,pix)}\le19.5$ tend to have larger 
$\Delta\IE$ values, due to the loss of flux in the most central pixels. 
At fainter magnitudes, large positive $\Delta\IE$ values are typically due to extended background galaxies, with flatter profiles than the PSF, while negative values are associated with spurious detections and residual bad pixels. We selected sources within two times the standard deviation from the mode of the distribution and conservatively discarded all sources brighter than $\IE{\rm(1.5\,pix)}=19.5$,  independently of their 
$\Delta\IE$ value, as potentially saturated (see Fig.~\ref{fig:ap_selection}). In the end, we were left with \num{336957} sources. 
However, these selections removed only the most obvious contaminants, while compact background galaxies, as well as compact stellar clusters, were retained in the photometric catalogue. 

\section{\label{sc:ebv_fg} Reddening and source contamination}

\subsection{\label{sc:ebv}Reddening correction}

\begin{figure}[h!]
\includegraphics[width=\columnwidth]{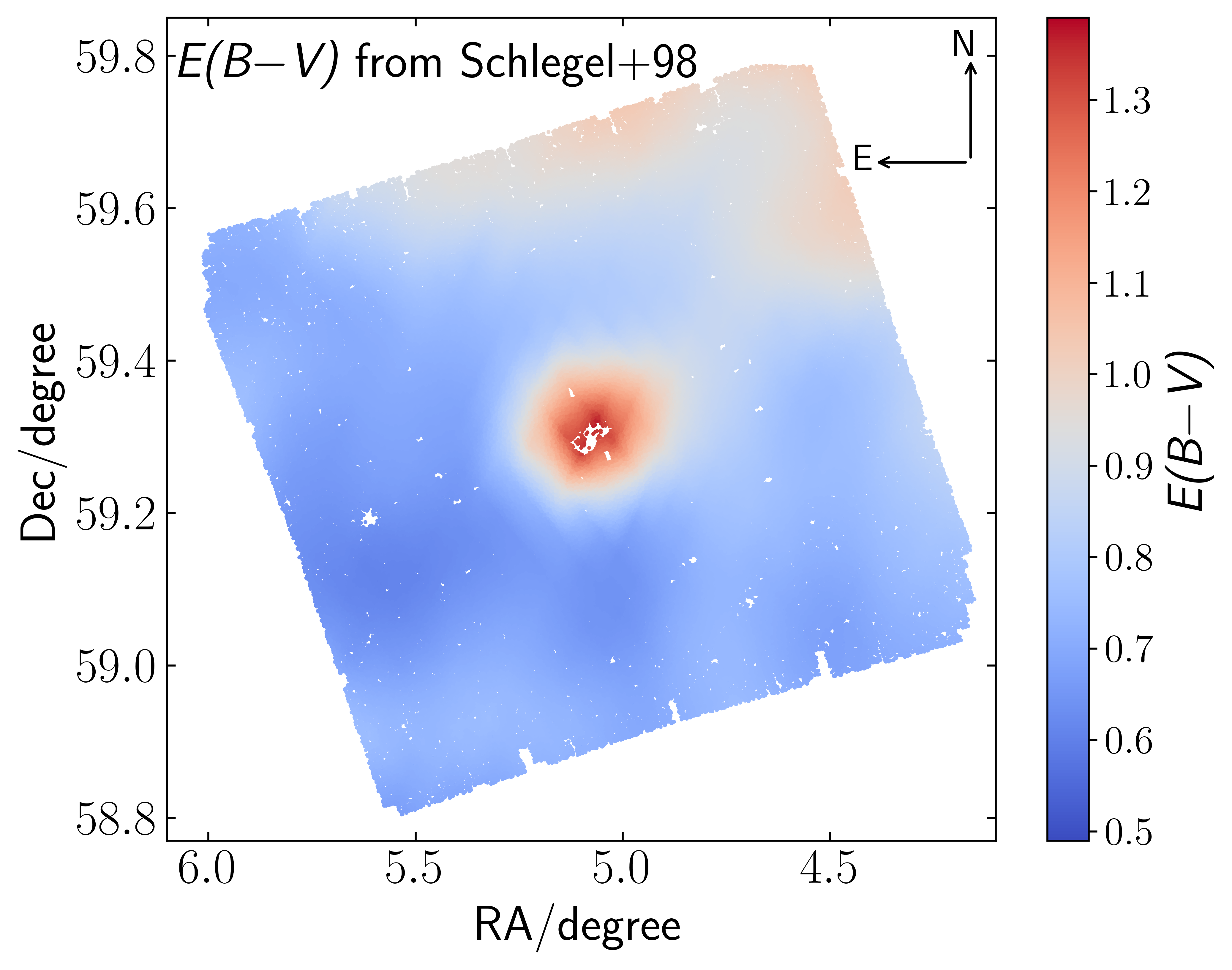}
\includegraphics[width=\columnwidth]{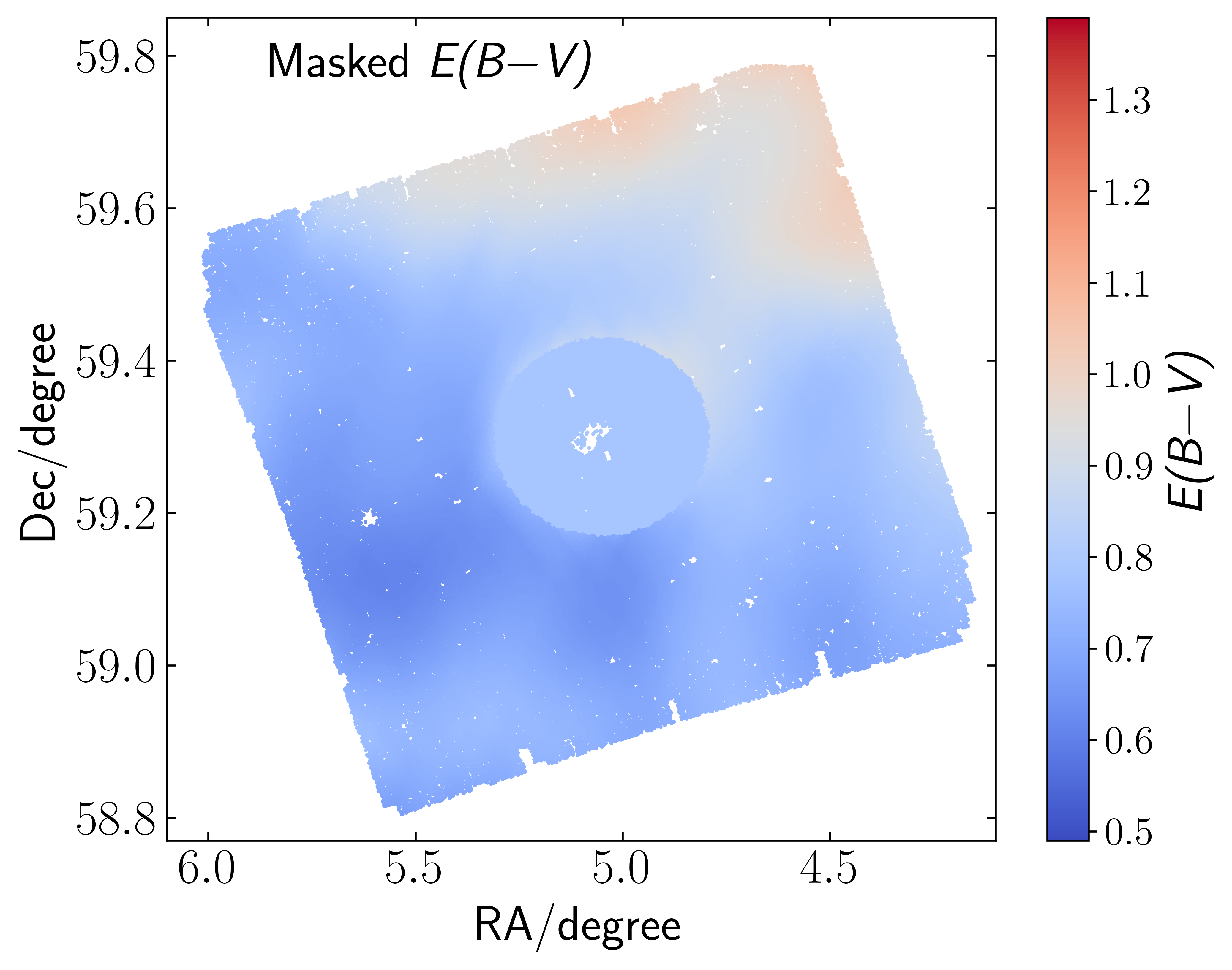}
\caption{\emph {Top:} Spatial map of \ebv\ in IC\,10 over the entire \Euclid FoV as derived from the \texttt{dustmaps} code assuming the \cite{schlegel98} 
dust maps. Foreground extinction increasing from south-east to north-west. The high \ebv \ values at the centre of the field are due to spurious contamination from IC\,10 in the dust maps. \emph{Bottom:} \ebv \ masked spatial map, where we assign  a constant reddening value equal to the average \ebv\ computed in an adjacent external annulus to a central 8\arcmin-radius circular region.} 
\label{fig:ebv}
\end{figure}

The reddening correction was performed in two steps, considering both far-infrared (FIR) dust maps and the \Euclid colours of resolved stars. Initially, we applied a spatially variable extinction correction to all sources based on the 
\citet{schlegel98} reddening maps, as described in \citetalias{ERONearbyGals}. Specifically, the foreground extinction at each source's position was estimated through the publicly
available Python package \texttt{dustmaps},\footnote{\texttt{dustmaps} is found at 
\url{https://dustmaps.readthedocs.io/en/latest/maps.html} and the dust maps themselves can be accessed and downloaded in the context of this package.} assuming the recalibration of \citet{schlafly11}. At each location, the code returns the corresponding \ebv\ value derived through linear interpolation on the dust maps. 
Figure~\ref{fig:ebv} (top) shows the \ebv\ map derived across the full \Euclid\ FoV. The field exhibits strong and spatially variable reddening, with \ebv\ increasing from $\sim$\,0.6\,mag to $\sim$\,1\,mag from south-east to north-west. The highest values near the galaxy centre [$E(B-V)\sim1.4$] are spurious, caused by IC\,10 contamination not properly removed from the \cite{schlegel98} dust maps. We therefore masked a central 8\arcmin\ radius region and assigned it a constant reddening of $E(B-V)=0.78$, corresponding to the 3\,$\sigma$-clipped mean \ebv\ in an adjacent $\sim$\,\mbox{\ang{;0.6;}}-wide annulus. The resulting masked map (Fig.~\ref{fig:ebv}, bottom) was used to apply a first-order extinction correction to individual sources in our photometric catalogue.
The \ebv\ values were converted into \av\ adopting $R_V=3.1$. 
Extinction values in the \Euclid bands were derived adopting the photometric bands from \citet{laureijs11}, the extinction curve from 
\cite{gordon23},\footnote{The G23 extinction curve also incorporates 
results by \citet{gordon09}, \citet{fitzpatrick19}, \citet{gordon21}, and \citet{decleir22}.} 
 and assuming a G2V stellar spectrum,  taken as a compromise between the blue and the red populations studied in this work. This provides relative ratios of $A_\lambda/A_V=0.726$, 0.375, 0.266, and 0.173, for \IE, \YE, \JE, and \HE, respectively.

Since this initial correction only accounts for foreground extinction and the \citet{schlegel98} maps lack the resolution to capture small-scale variations, we implemented a second, more refined procedure based on the colour of RGB stars.
Before this second step, foreground contaminants were removed from the catalogue by applying cuts in a colour-colour diagram as described in Sect.~\ref{sc:mw} and Appendix~\ref{sc:mw_appendix}, using foreground reddening-corrected magnitudes to achieve a cleaner separation between IC\,10 and Galactic stars.

\begin{figure}[]
\includegraphics[width=0.86\columnwidth,trim={-0.7cm 0 0 0 0}]{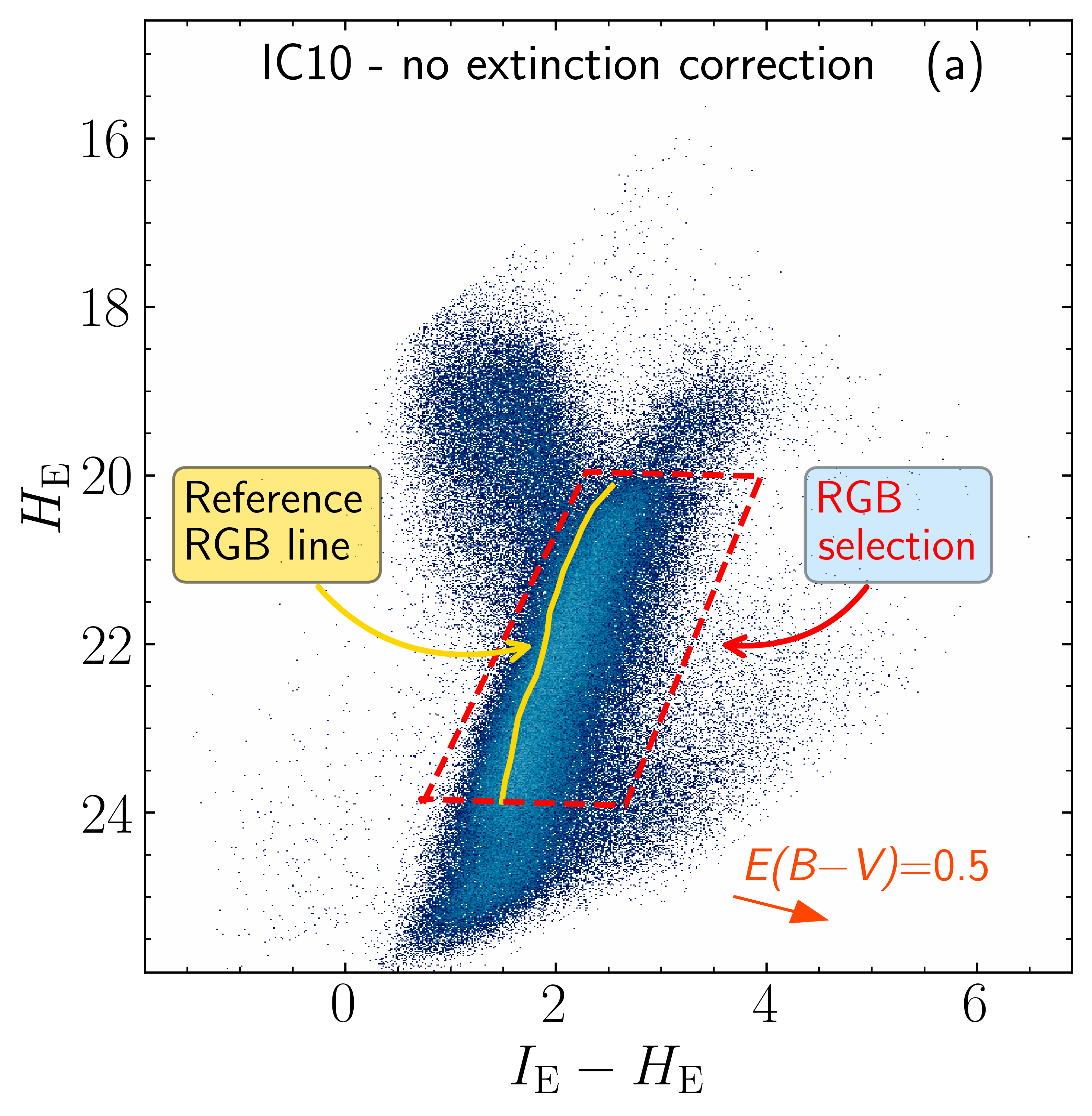} 
\includegraphics[width=1\columnwidth,trim={-0.3cm 0 0 0 0}]{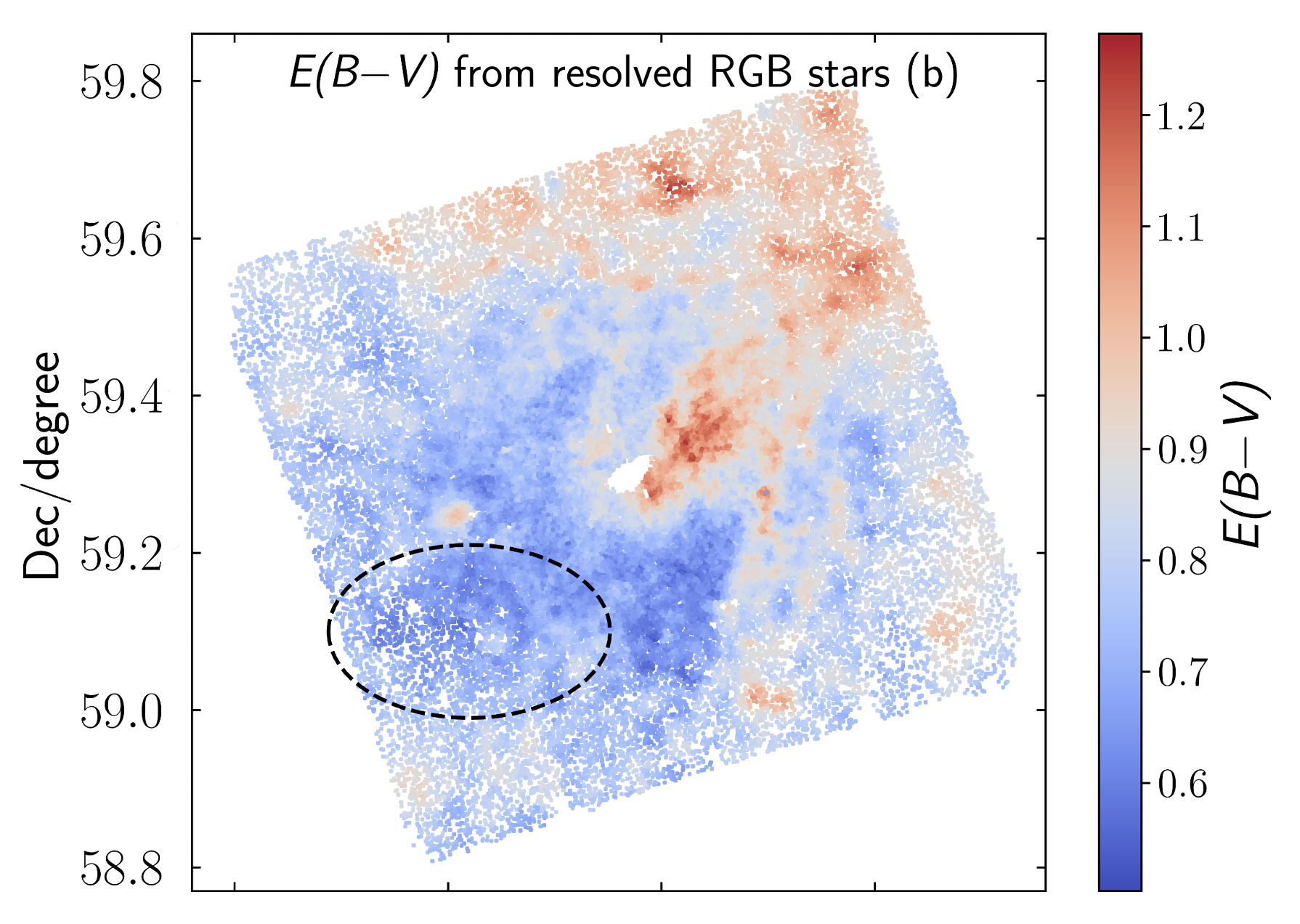} 
\includegraphics[width=0.98\columnwidth]{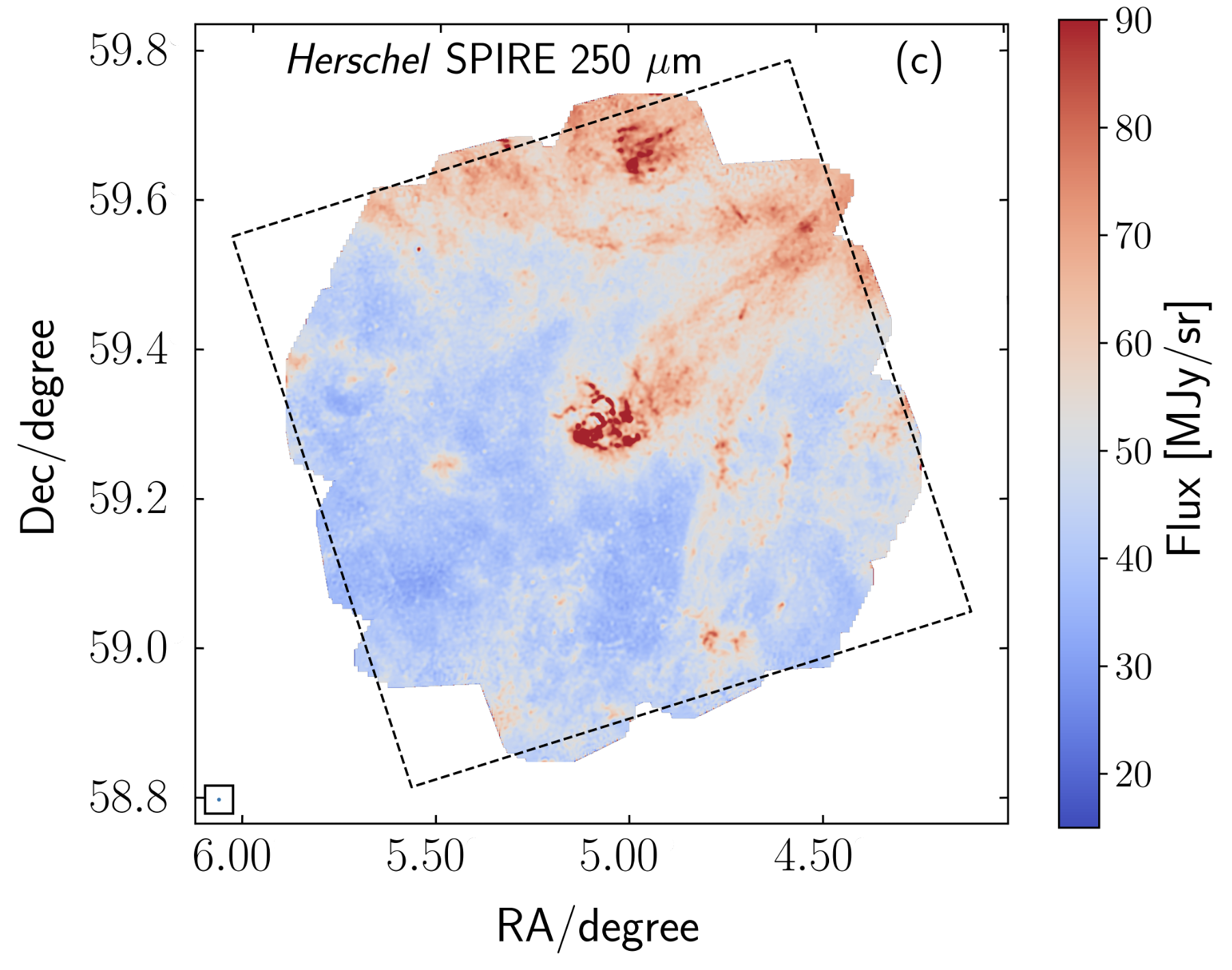}
\caption{\emph{Panel (a)}: \HE versus \IH CMD of IC\,10 prior to the reddening correction. The red polygon indicates our 
selection of RGB stars for the dust distribution, while the yellow curve is the reference RGB line with $E(B-V)\sim0.65$. The orange arrow is the reddening vector 
for $E(B-V)=0.5$. \emph{Panel (b)}: \ebv \  map derived from the RGB resolved stars, as described in Sect.~\ref{sc:ebv}. The dashed ellipse denotes the region used to infer the reference RGB line displayed in (a). \emph{Panel (c)}: FIR dust emission map from {\it Herschel} SPIRE/250 taken from \cite{Madden2013}. The \Euclid footprint is indicated for reference as the  dashed square}.
\label{fig:ebv_rgb}
\end{figure}

The cleaned \HE versus \IH CMD in Fig.~\ref{fig:ebv_rgb}a, with magnitudes not corrected for extinction,  shows a significant colour spread of $\Delta(I_\sfont{E}-H_\sfont{E})\sim1.4$. Although age and metallicity play a role, differential reddening is the main cause. In fact, the $E(B-V)$ variations shown in Fig.~$\ref{fig:ebv}$ translate into variations in $E(I_\sfont{E}-H_\sfont{E})$ of $\sim$\,1\,mag or more.  This exceeds, for instance, the $\sim$\,0.5\,mag shift in colour at the RGB tip predicted by stellar models \citep{Bressan2012, Marigo2017} when assuming a spread in metallicity from ${[\rm M/H]=-1.4}$ to the present-day value of about 
 $-0.4$ derived in \ion{H}{ii} regions by \cite{Magrini2009}.

We selected RGB stars within the polygon outlined in Fig.~\ref{fig:ebv_rgb}a and inferred a `reference' RGB line (yellow curve) from a sub-sample in a relatively low extinction region, with a median value from the Schlegel maps of $E(B-V)\sim0.65$ 
(dashed ellipse in Fig.~\ref{fig:ebv_rgb}b).
The reference RGB line presents some wiggles, likely due to residual foreground MW stars that had not been fully removed by our selections.
Starting from the reference RGB line, we generated a grid of RGB lines for extinctions in the range of $0<E(B-V)<1.6$,  in steps of $\Delta\,E(B-V)=0.01$, adopting the $A_\lambda/A_V$ values reported above.

Ultimately, our extinction map was derived following an approach similar to that described in \cite{Sarajedini2007}, \cite{Bernard2012}, and \cite{Milone2012}.
For each RGB star selected within the polygon in Fig.~\ref{fig:ebv_rgb}a, we identified the 50 nearest RGB neighbours in the image, from which we computed a median  \HE magnitude and  \IH colour. 
Then, the best-fit RGB line reproducing these median values provided the \ebv~value at the location of each star. This allows us to construct a high-resolution spatial map of extinction. 
 
This method relies on the assumption that the RGB population is uniform across the entire field of view (FoV), so that any displacement from the reference sequence can be attributed solely to reddening variations. This inevitably prevents us from probing possible age or metallicity gradients within the RGB population, but given the severe and highly inhomogeneous foreground extinction toward IC\,10, this assumption is unavoidable.

Our results are displayed in Fig.~\ref{fig:ebv_rgb}b. At the very centre of IC\,10, the hole in the extinction map is due to severe stellar crowding, which prevents us from resolving individual RGB stars. The poorer sampling in the outermost regions is caused by the decrease in star counts with increasing distance from the IC\,10 centre. The derived reddening values (internal plus foreground) are in the range 
$0.4\lesssim E(B-V)\lesssim1.4$, which is compatible with the range covered by the Schlegel et al. map. However, the spatial resolution of the \ebv \ map derived from resolved star counts is remarkably better than the latter, revealing the presence of filaments, knots, and spatial variations on scales of $\sim$\,1\arcmin. An approach based on resolved stars was also used by \cite{Sanna2008} for a small region of IC\,10 observed with the {\it Hubble} Space Telescope. 
Although a large fraction of their footprint falls in the central hole in our map, 
there is still some overlap between the two sets of photometry; for the region in common, our map provides a median reddening of $E(B-V)=0.8\pm0.1$, in excellent agreement with their value of $E(B-V)=0.78\pm0.06$ inferred from deep CMD analysis. 

Finally, in Fig.~\ref{fig:ebv_rgb}c we compare our results with the FIR dust emission map from {\it Herschel} SPIRE/250 taken from the 
Dwarf Galaxy Survey \citep{Madden2013}. The agreement between the two is remarkable, with the high extinction regions in our map corresponding to peaks in the {\it Herschel} emission map and, vice versa, low \ebv \ values being associated with depression values in the FIR emission. The excellent agreement between the two maps showcases the power of the resolved-star approach for deriving high-resolution reddening maps toward nearby galaxies affected by high extinction. In the end, we used the extinction map of Fig.~\ref{fig:ebv_rgb}b, to correct the entire (including also foreground stars) photometric catalogue.

\subsection{\label{sc:mw}Source contamination}

The dominant source of contamination in IC\,10 arises from Galactic foreground stars, mainly coming from the MW disc. To remove as many of these contaminating sources as possible, we followed the same procedure as in \citetalias{ERONearbyGals},
based both on {\it Gaia} proper motions (PMs) and parallaxes, and on selection of sources in a colour-colour diagram. We describe the procedure in details in Appendix~\ref{sc:mw_appendix}. 

While {\it Gaia} PMs enabled the identification of only a relatively small fraction of bright MW contaminants ($\sim$\num{7720} stars), the vast majority of foreground sources ($\sim$\num{99350} stars) was removed using the colour-colour diagram. This diagram also proved effective in rejecting faint, compact background galaxies ($\sim$\num{630}) that were not excluded by the differential aperture cuts applied in Sect.~\ref{sc:photometry}. Following the removal of contaminants, the
final IC\,10 photometric catalogue of reliable sources having photometry in all four \Euclid bands contained \num{227128} stars.

\section{\label{sc:cmd} Colour-magnitude diagrams}

\subsection{Stellar populations and evolutionary features}

The final calibrated, reddening-corrected, \Ho versus \IoHo and \Jo versus \YoHo CMDs of resolved stars in IC\,10 are shown in Fig.~\ref{fig:cmd}.
We show also the \Io versus \IoYo and \Io versus \IoHo CMDs in Fig.~\ref{fig:other_cmds} in Appendix~\ref{sc:other_cmds}.

The CMD in Fig.~\ref{fig:cmd}a is before removal of the contaminants described in Sect.~\ref{sc:mw}. 
Stars in IC\,10, mainly RGB and AGB stars, define a distinct sequence that extends from  
$I_\sfont{E,0}-H_\sfont{E,0}\sim3$, $H_\sfont{E,0}\sim18$ down to 
$I_\sfont{E,0}-H_\sfont{E,0}\sim-1$, $H_\sfont{E,0}\sim25.5$; on the other hand,  
Galactic disc stars appear organised along a different sequence that runs from 
$I_\sfont{E,0}-H_\sfont{E,0}\sim-1$, $H_\sfont{E,0}\sim17$ to 
$I_\sfont{E,0}-H_\sfont{E,0}\sim2$, $H_\sfont{E,0}\sim24$, and give the semblance of an X-feature on the CMD. Bright foreground contaminants identified through {\it Gaia} proper motions and parallaxes (see  Sect.~\ref{sc:mw}), over-plotted on the total CMD, amount to just  $\sim20\,\%$ of the total contamination from blue MW stars.

\begin{figure*}[h]
\includegraphics[height=0.35\textwidth,trim={0 0 0.cm 0}]{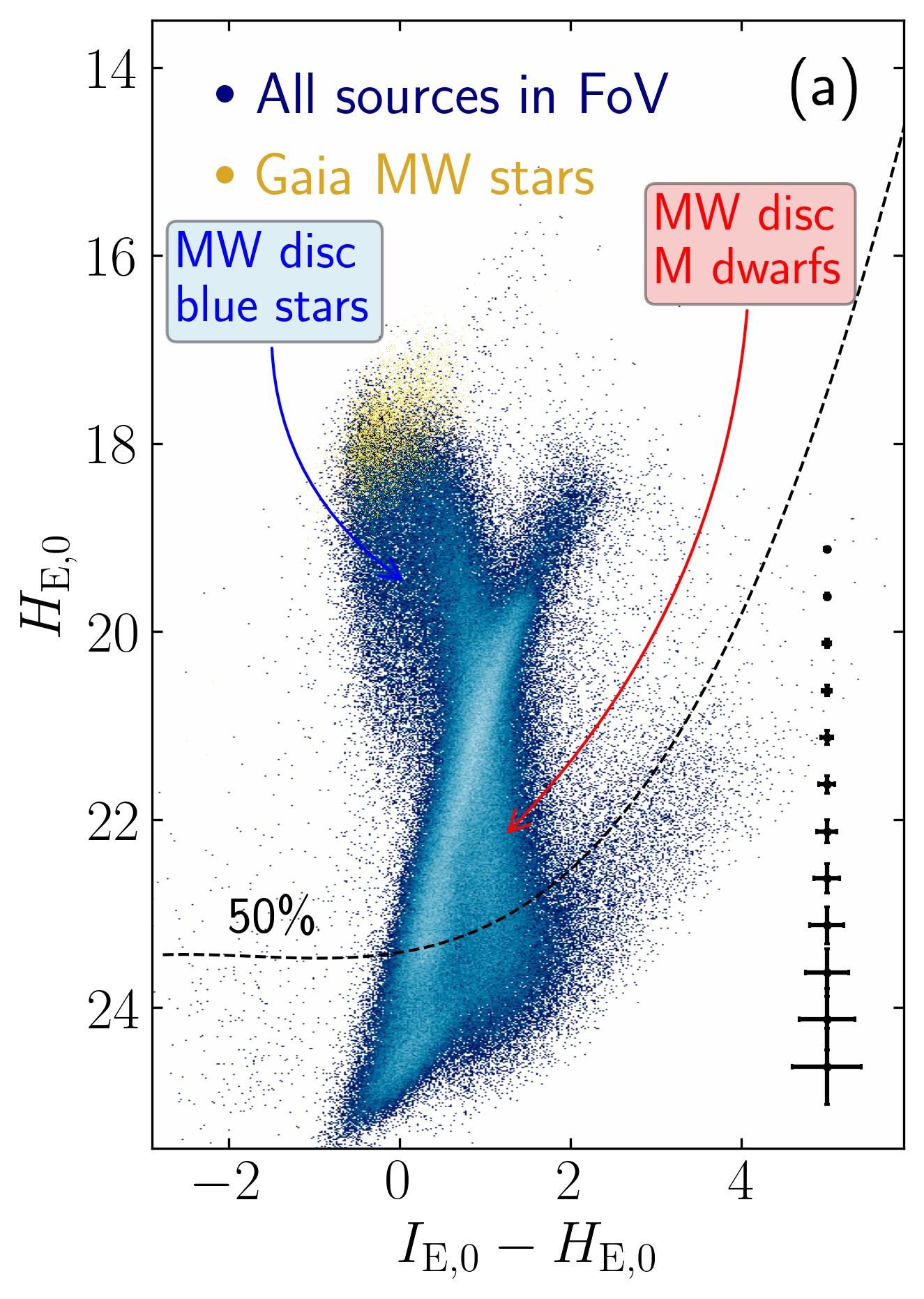} \hfill
\hspace*{0cm} \includegraphics[height=0.35\textwidth, clip, trim={1cm 0 0.1cm 0}]{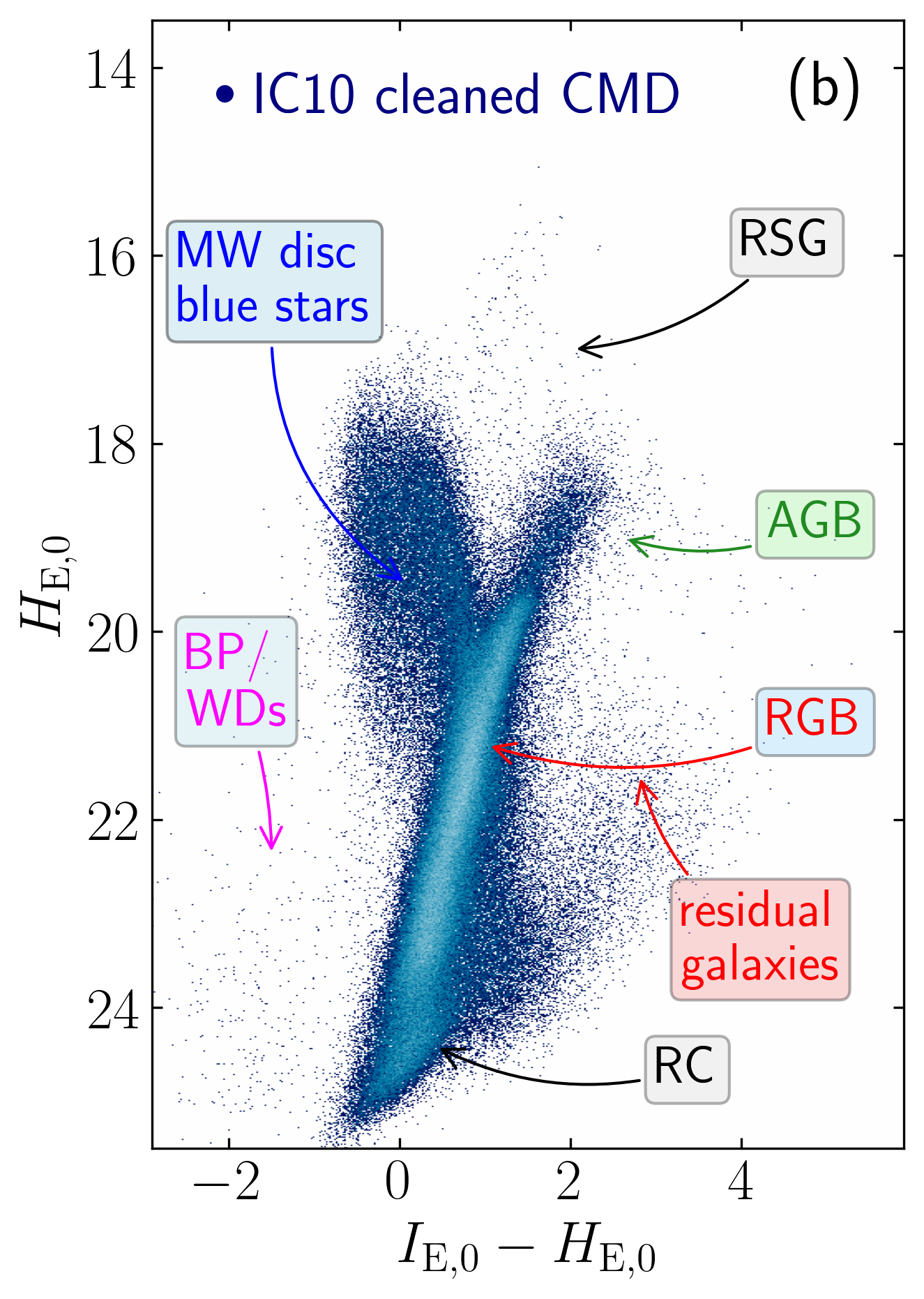} \hfill
\hspace*{0cm} \includegraphics[height=0.35\textwidth, clip, trim={1.cm 0 0.1cm 0}]{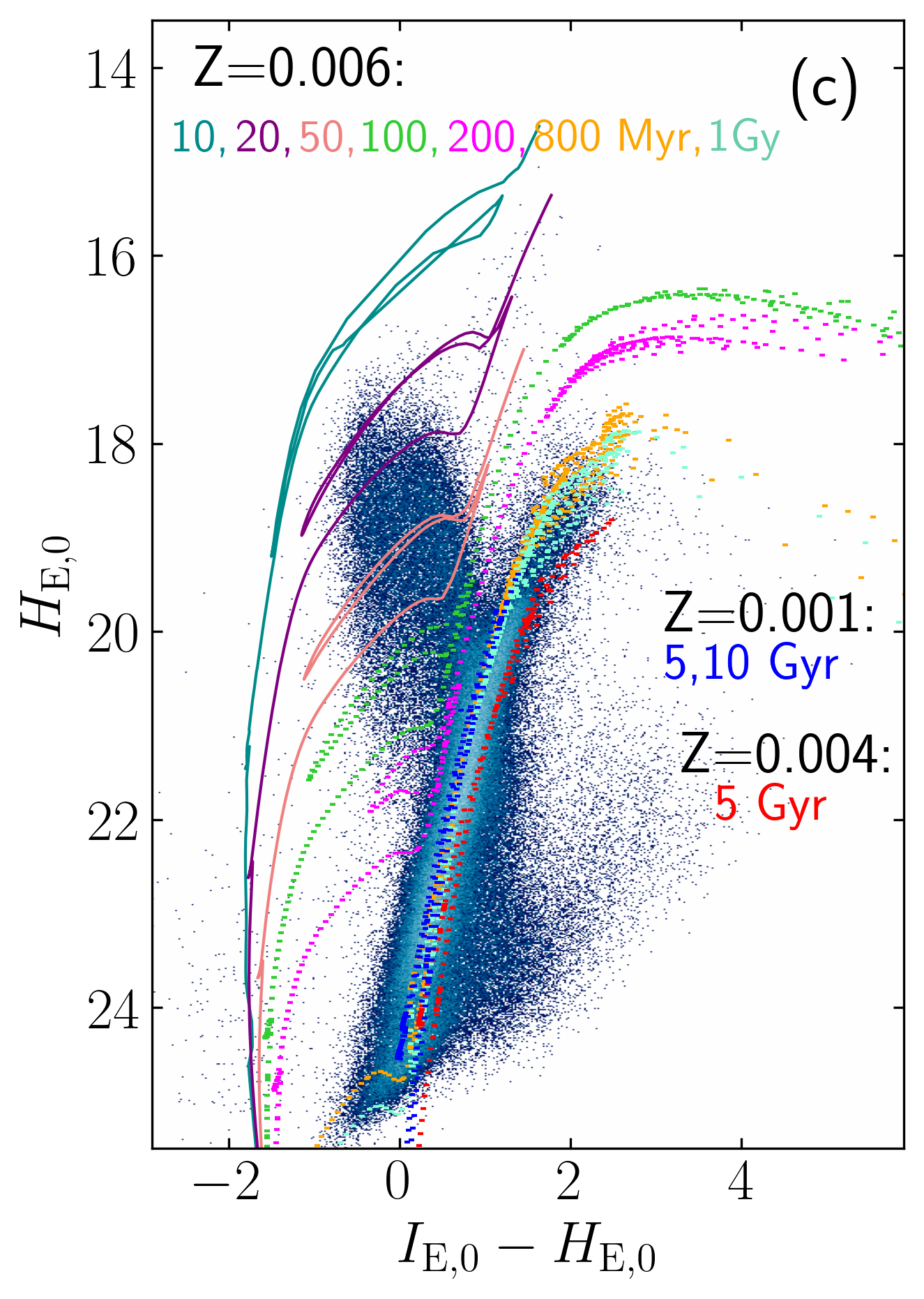} \hfill
\hspace*{0cm} \includegraphics[height=0.35\textwidth, clip, trim={0 0 0 0}]{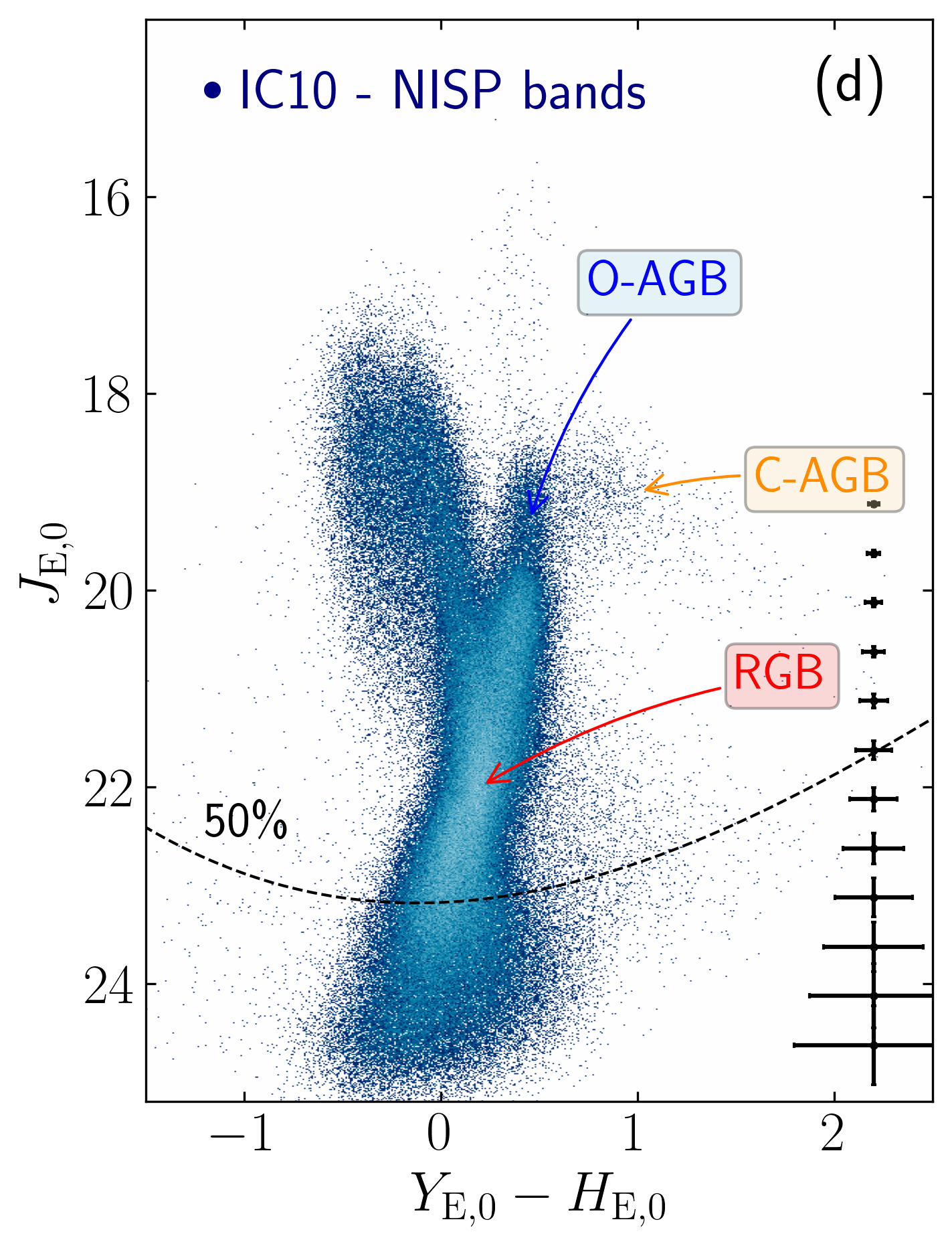} 
\caption{\emph{Panel (a)}: \Ho versus \IoHo reddening-corrected CMD of \num{336957} sources with four-band photometry in the entire \Euclid FoV after selection cuts (see Sect.~\ref{sc:photometry}). 
The CMD was corrected for spatially variable reddening as described in Sect.~\ref{sc:ebv}. Indicated is the contribution from MW disc stars (both blue stars, and red M dwarf stars); yellow dots denote the \num{7724} MW stars removed from {\it Gaia} DR3 proper motions and parallaxes.  On the right side, we indicate the photometric errors for $0\leq I_\sfont{E,0}-H_\sfont{E,0}\leq1$, while the black dashed curve denotes the average 50\% completeness level, both from the results of the artificial stars over the entire \Euclid field.
\emph{Panel (b)}: cleaned CMD (\num{227128} sources) after removal of Galactic M dwarfs and compact red galaxies as 
described in Sect.~\ref{sc:mw} and Appendix~\ref{sc:mw_appendix}. The main stellar evolutionary sequences are indicated: RSGs with ages from about 20\,Myr to 50\,Myr;  bright and red AGB stars with ages from $\sim0.1$ to a few Gyrs;  RGB stars, with ages older than 1--2 Gyr; at the faintest magnitudes is a hint of the RC,  with ages $>$1--2\,Gyr. The faint cloud of blue stars at $I_\sfont{E,0}-H_\sfont{E,0}\lesssim-1$, $H_\sfont{E,0}\gtrsim22$ is likely dominated by Galactic WDs. Faint red sources at $I_\sfont{E,0}-H_\sfont{E,0}\gtrsim2.5$, $H_\sfont{E,0}\gtrsim21$ are residual background galaxies.
\emph{Panel (c)}: same CMD as in \emph{(b)} with superimposed PARSEC stellar isochrones for 
a metallicity of $Z=0.006$ ($\sim$\,40\% solar) and ages in the range of 10\,Myr--1\,Gyr, and for lower metallicities of  $Z=0.004$ and 0.001 (26\%, and 6\% solar, respectively) and ages of 5 and 10\,Gyr. \emph{Panel (d)}: \Jo versus \YoHo CMD, where the oxygen-rich and the carbon-rich AGB stars (O-AGB and C-AGB) appear well separated and define vertical and horizontal sequences, respectively. Indicated is the average 50\% completeness level and the photometric errors at $-0.5\leq Y_\sfont{E,0}-H_\sfont{E,0}\leq0.5$.}
\label{fig:cmd}
\end{figure*}

The final cleaned CMD displayed in Fig.~\ref{fig:cmd}b still suffers from major contamination by MW foreground stars that are bluer than 
$I_\sfont{E,0}-H_\sfont{E,0}\sim1$ and brighter than $H_\sfont{E,0}\sim21$,  
 while redder and fainter Galactic M dwarf stars appear to have been largely removed. 
 In the clean diagram, the cloud of blue stars at 
 $I_\sfont{E,0}-H_\sfont{E,0}\lesssim-1$, $H_\sfont{E,0}\gtrsim22$, 
 is compatible both with blue plume (BP) stars in IC\,10, i.e. massive main sequence (MS) stars and post-MS stars in the hot core helium-burning phase, with ages $\la$\,200\,Myr (see also Fig.~\ref{fig:cmd}c for a direct comparison with stellar models) and with Galactic white dwarfs (WDs). However, as discussed 
in Appendix~\ref{sc:mw_appendix}, these sources appear distributed quite homogeneously over the \Euclid FoV, indicating the vast majority of them are Galactic WDs rather than BP stars in IC\,10.  

In the CMD, we further identified a vertical sequence of red supergiants (RSGs) at 
$I_\sfont{E,0}-H_\sfont{E,0}\sim1.5$, $15.5\la H_\sfont{E,0}\la 18$, 
with ages from $\sim\,20$\,Myr to $\sim$\,50\,Myr; 
 luminous (as bright as $H_\sfont{E,0}\sim18$) and red ($1\la I_\sfont{E,0}-H_\sfont{E,0}\la 5$) AGB stars with ages from $\sim$\,0.1\,Gyr to a few Gyrs; RGB stars with 
$0\la I_\sfont{E,0}-H_\sfont{E,0}\la 2$, $H_\sfont{E,0}\gtrsim19.7$
and ages older than 1--2\,Gyr (and potentially as old as $\sim$\,13\,Gyr). 
Also visible at 
$-1\la I_\sfont{E,0}-H_\sfont{E,0}\la 1$, $H_\sfont{E,0}\ga 24$, 
toward our detection limit, is a hint of the red clump (RC), populated by low-mass stars in the core-helium burning phase, with ages $>$1--2\,Gyr. Faint red sources at 
$I_\sfont{E,0}-H_\sfont{E,0}\gtrsim2.5$, $H_\sfont{E,0}\gtrsim 21$ are residual background galaxies, based both on direct visual inspection on the higher resolution VIS images, revealing that they are not point-like, and on their uniform spatial distribution. 

In Fig.~\ref{fig:cmd}c, the \Ho versus \IoHo CMD is compared with the PARSEC stellar isochrones 
\citep{Bressan2012,Chen2014,Chen2015,Tang2014,Marigo2017,Pastorelli2019,Pastorelli2020} in the \Euclid bands, 
shifted according to a distance modulus of $(m-M)_0=24.41\pm 0.05$, which corresponds to a distance of $D=(0.762\pm 0.02)$\,Mpc derived in Sect.~\ref{sc:trgb}.  
We display isochrones for stellar ages in the range of 10\,Myr--10\,Gyr. 
Metallicities as low as $Z=0.004$ and $Z=0.001$ ($\sim$\,26\% and 6\% Solar, 
respectively\footnote{We assumed a metallicity of $Z_{\odot}=0.0153$ and oxygen abundance of $\mathrm{12 + \logten(O/H)_{\odot}=8.76}$ for the Sun, from \cite{Caffau2011}.}), adopted for the 5--10\,Gyr old isochrones, seem to reasonably agree with the RGB colours. For the population younger than 1\,Gyr, a higher metallicity of  $Z=0.006$ appears compatible with the observed CMD (in particular, see the RSGs), and also in agreement, within the errors, with spectroscopic oxygen abundance estimates of $\mathrm{12+\logten(O/H)=8.30\pm0.2}$ \citep{Magrini2009} and $8.14\pm0.08$ \citep{pilyugin14}. The age-dependent metallicity range that we find provides the best match to the resolved stars is also in excellent agreement with that derived for the star cluster population by \cite{Howell2026}.

Finally, in Fig.~\ref{fig:cmd}d, we show the  \Jo versus \YoHo CMD of IC\,10. 
Besides the same evolutionary phases already described for the \Ho versus \IoHo diagram (with the exception of the RC,  which is fainter than the photometric depth in the NISP bands), we see  
that near-infrared (NIR) CMDs are more effective than optical ones in separating  oxygen-rich versus carbon-rich AGB stars 
\citep[see also, for NGC\,6822,][\citetalias{ERONearbyGals}]{Nally2024}, with the former delineating an almost vertical sequence with colours 
$0.35\la Y_\sfont{E,0}-H_\sfont{E,0}\la0.55$, 
and the latter forming a horizontal feature at  
$18.5\la J_\sfont{E,0}\la20$.  
This horizontal feature is considered to be an effective distance indicator in the NIR bands \citep[][and references therein]{Lee2025}. In the CMDs, we display the 50\% completeness curves and the photometric errors computed through the artificial star tests described in Appendix~\ref{sc:artificial}.

\subsection{Distance \label{sc:trgb}}

The large sample of RGB stars in IC\,10 allows for a robust estimate of the distance to the galaxy using one of the most widely used standard candles, the TRGB \citep[see, e.g.][and references therein]{lee93,salacas98,madore95,mbtip08,serenelli17,madore20,hoyt23}.
We can take advantage of the calibration in the \IE band recently derived by \citet[][hereafter \citetalias{bp24}]{bp24} using {\it Gaia} DR3 synthetic photometry \citep{XPSP23} of the Small Magellanic Cloud (SMC) and of the Large Magellanic Cloud (LMC). While the minimum dependency of the absolute magnitude of the TRGB on metallicity and colour is achieved in passbands sampling the wavelength range $\sim$800--1000\,nm (e.g. in the classical Johnson-Cousins $I$ band), the broad \IE band ($\simeq$500--930\,nm) is also well behaved in the metal-poor range typical of dwarf galaxies 
(\citetalias{bp24}). For instance, \citetalias{bp24} found that the LMC and the SMC differ in $M_{I_\sfont{E}}^{\sfont{TRGB}}$ by just $0.053\pm 0.072$\,mag, in good agreement with theoretical predictions.

\begin{figure}[]
\includegraphics[width=0.95\columnwidth]{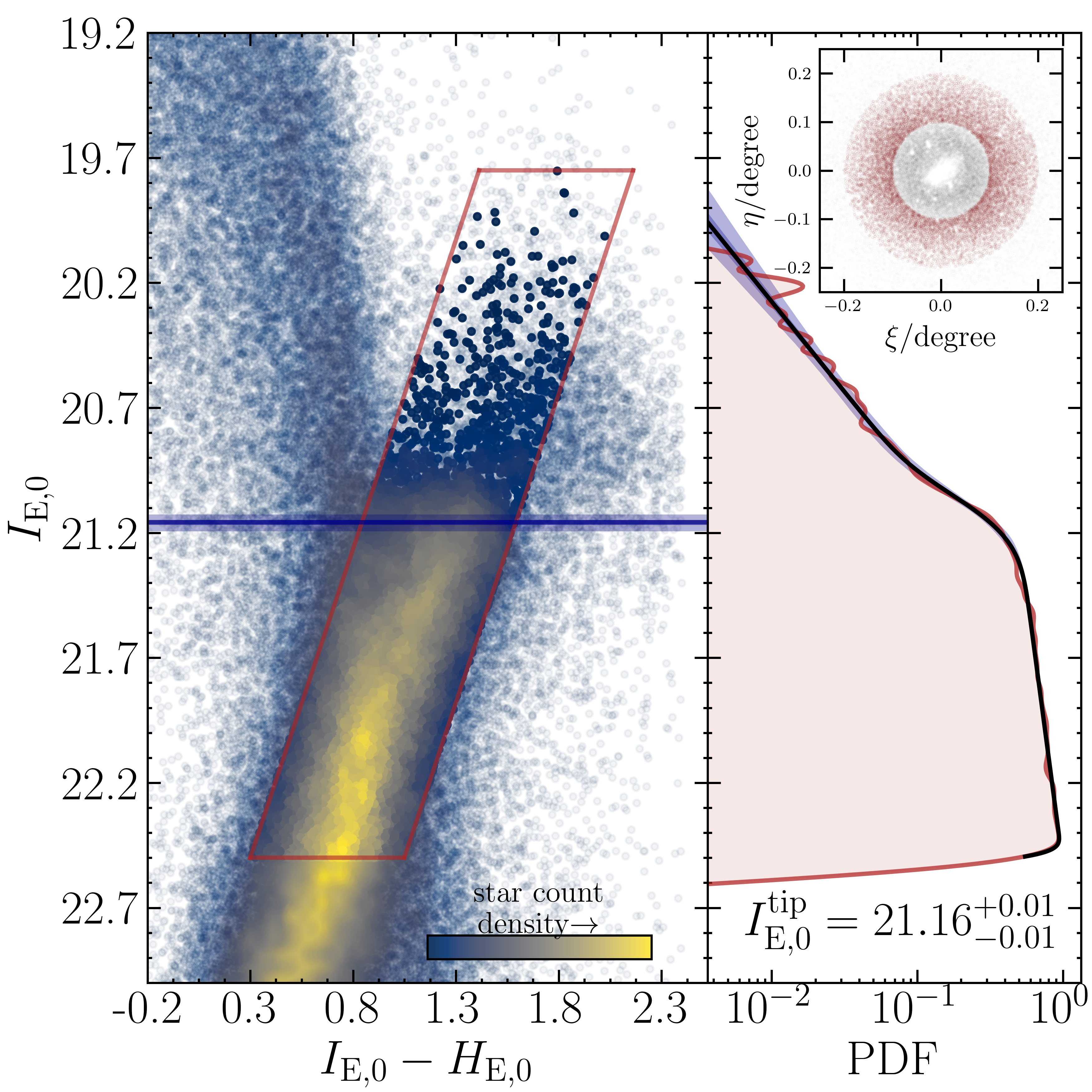}
\caption{Detection of the TRGB of IC\,10 in the annulus $\ang{0.1;;}\leq R\leq\ang{0.2;;}$. Left panel: CMD of the considered sample focused on the upper RGB. The red polygon depicts the stars selected for the TRGB detection algorithm, the horizontal line marks the position of the tip, while the purple shaded area encloses the $\pm\,3\,\sigma$ uncertainty range. Right panel: Probability density function (PDF) of the selected sample (red line) with the highest likelihood model superimposed (black line, with $\pm\,3\,\sigma$ purple shaded area).
}
\label{fig:tip}
\end{figure}

The location of the TRGB was measured with the Bayesian approach described in detail in \citetalias{bp24} and illustrated, for the present application, in Fig.~\ref{fig:tip}. Using the entire IC\,10 catalogue, we obtain $I_{\sfont{E,0}}^{\sfont{TRGB}}=21.12\pm 0.01$. However, our sample of RGB stars is so large that we can neglect the innermost regions of the galaxy where the incompleteness and the crowding are most severe (see Fig.~\ref{fig:cmd_grad} and Appendix~\ref{sc:artificial}). Thus, for a radial annulus at $\ang{0.1;;}\leq R\leq\ang{0.2;;}$, we obtain a more reliable detection of $I_{\sfont{E,0}}^{\sfont{TRGB}}=21.16\pm 0.01$ (Fig.~\ref{fig:tip}). 
From the dataset by \citet{Sanna2008}, we measure a mean colour of the TRGB 
$(V-I)_0=1.822^{+0.14}_{-0.12}$, compatible with that measured in the LMC by \citetalias{bp24}, $(V-I)_0=1.716^{+0.14}_{-0.15}$. Hence, we adopt as a reference $M_{I_\sfont{E}}^{\sfont{TRGB}}=3.247\pm 0.050$, as measured by \citetalias{bp24} for the LMC, obtaining $(m-M)_0=24.41\pm 0.05$, corresponding to $D=(762\pm 20)$\,kpc. This is well within the range of the estimates in the literature, listed by \citet{Gholami2025}, and, in particular, is in good agreement with  
the values of $(m-M)_0=24.51\pm 0.08$ \citep{Sanna2008}, $(m-M)_0=24.43\pm 0.03$ \citep{mcquinn17}, and $(m-M)_0=24.43$ (no error provided) from \citet{dellagli18}.

\section{\label{sc:spatial} Spatial distribution of stars}

In this section, we characterise the spatial distribution of stars in IC\,10 and infer the stellar radial profile from individual star counts. This approach is complementary to the method presented in \citetalias{ERONearbyGals} based on an analysis of the integrated light. 

\begin{figure*}[ht]
\centering
\hspace*{0.4cm}
\includegraphics[height=7.cm, trim={0cm 0cm 0cm 0cm}, clip]{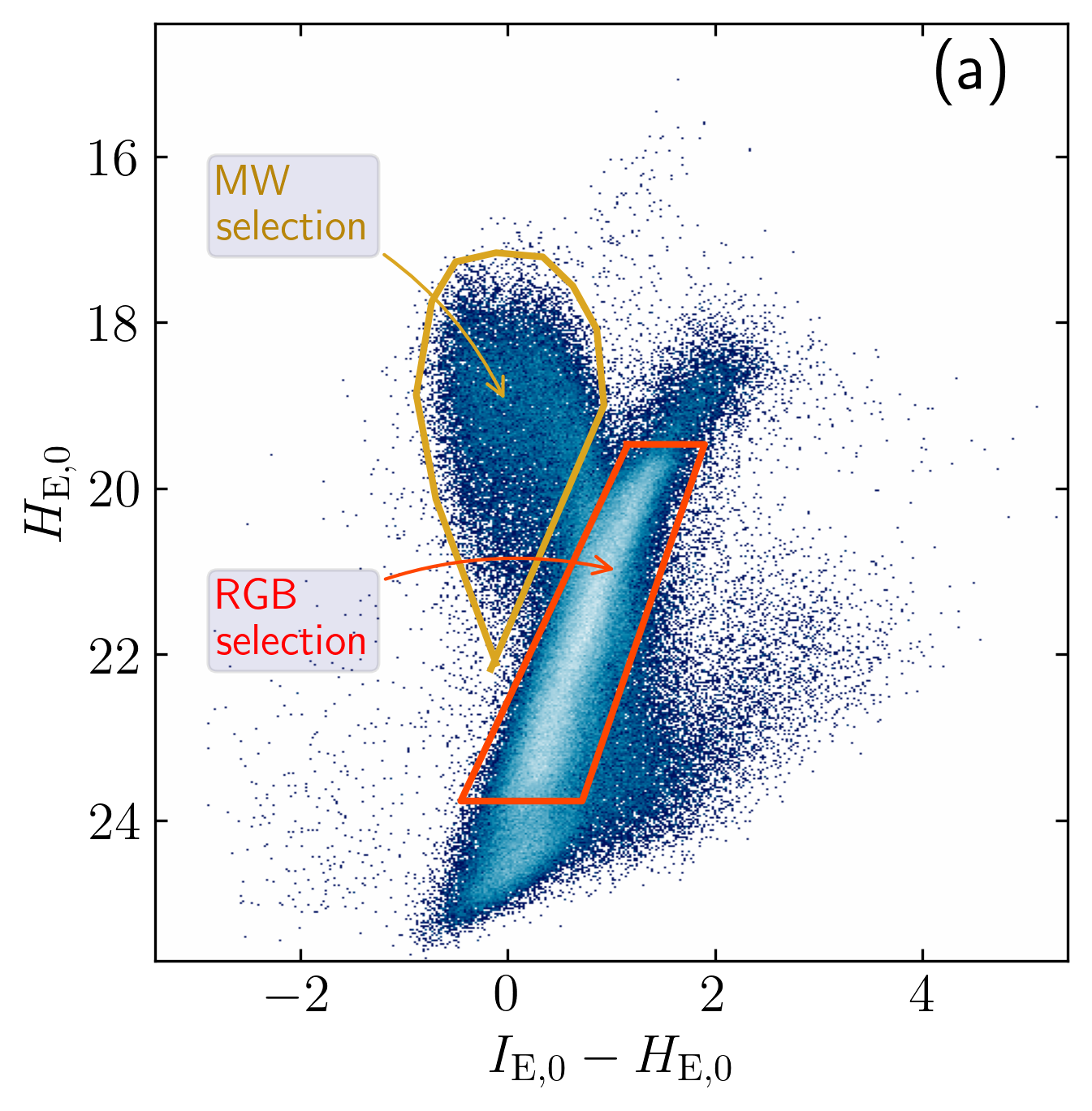}
\hfill
\centering
\hspace*{0.cm}
\includegraphics[height=7.5cm, trim={0cm 0cm 0cm 0cm}, clip]{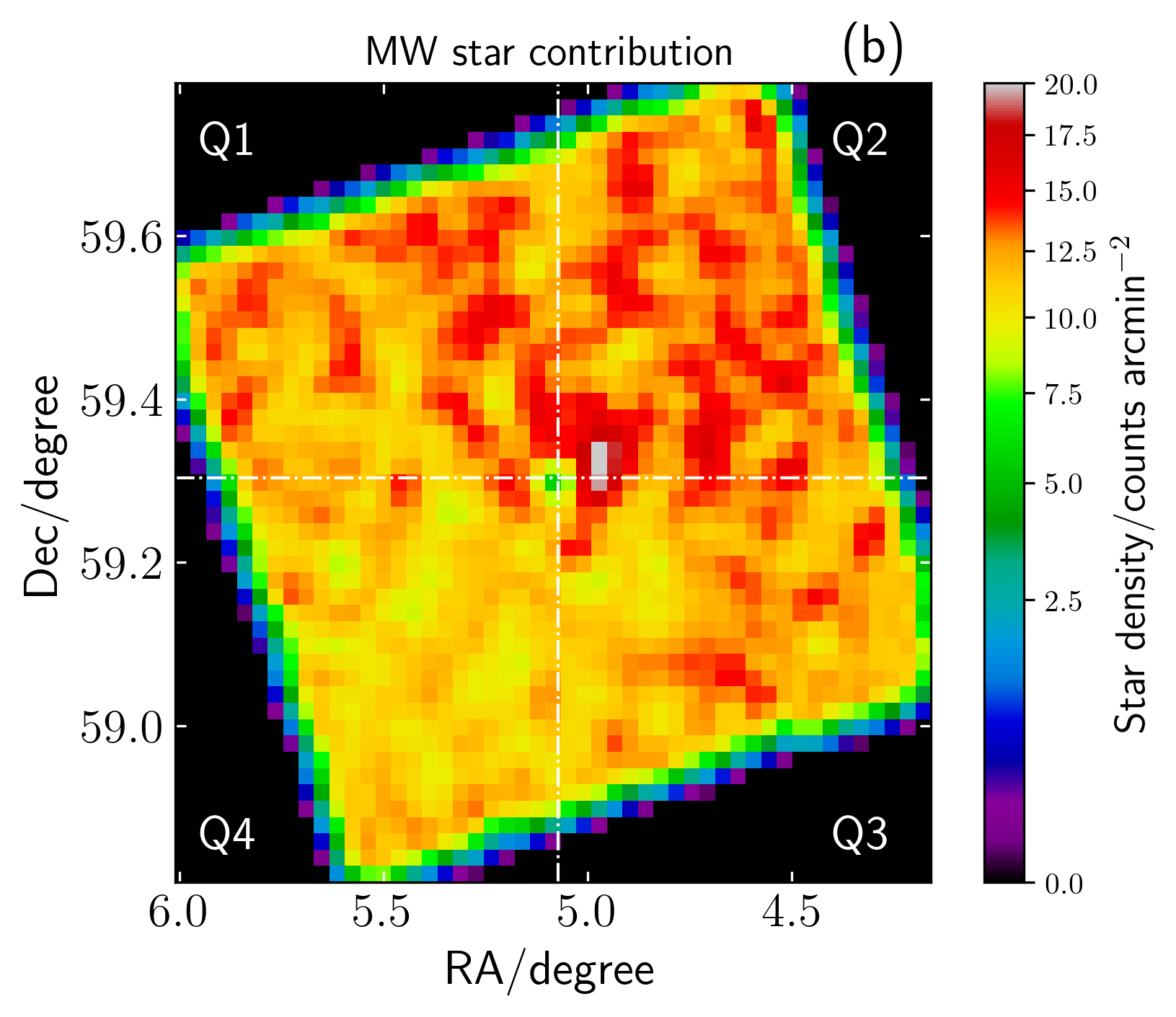}
\vspace{-0.cm}
\centering
\includegraphics[height=7.5cm, trim={0cm 0cm 0cm 0cm}, clip]{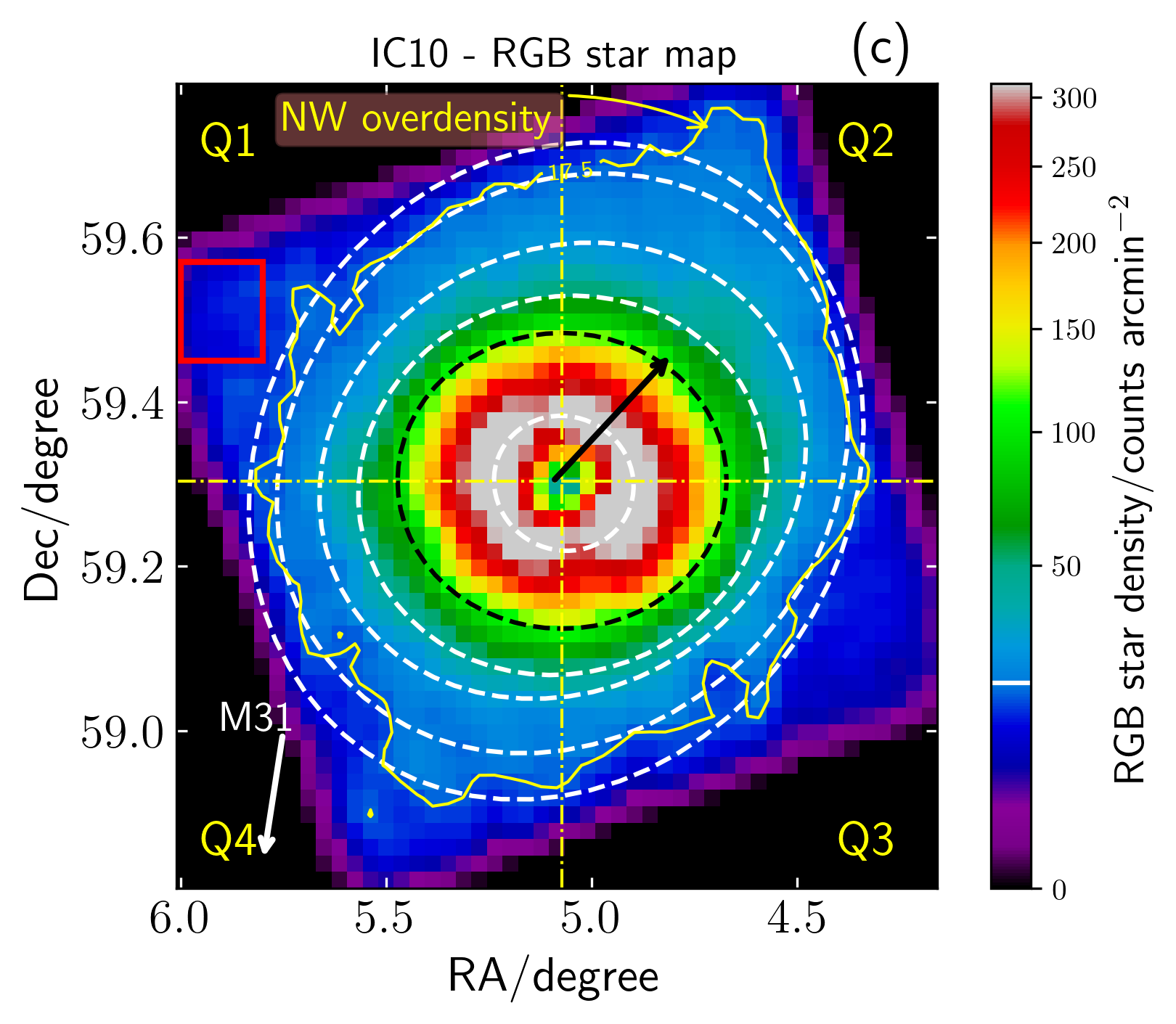}
\hfill
\centering
\includegraphics[height=7.5cm, trim={0cm 0cm 0cm 0cm}, clip]{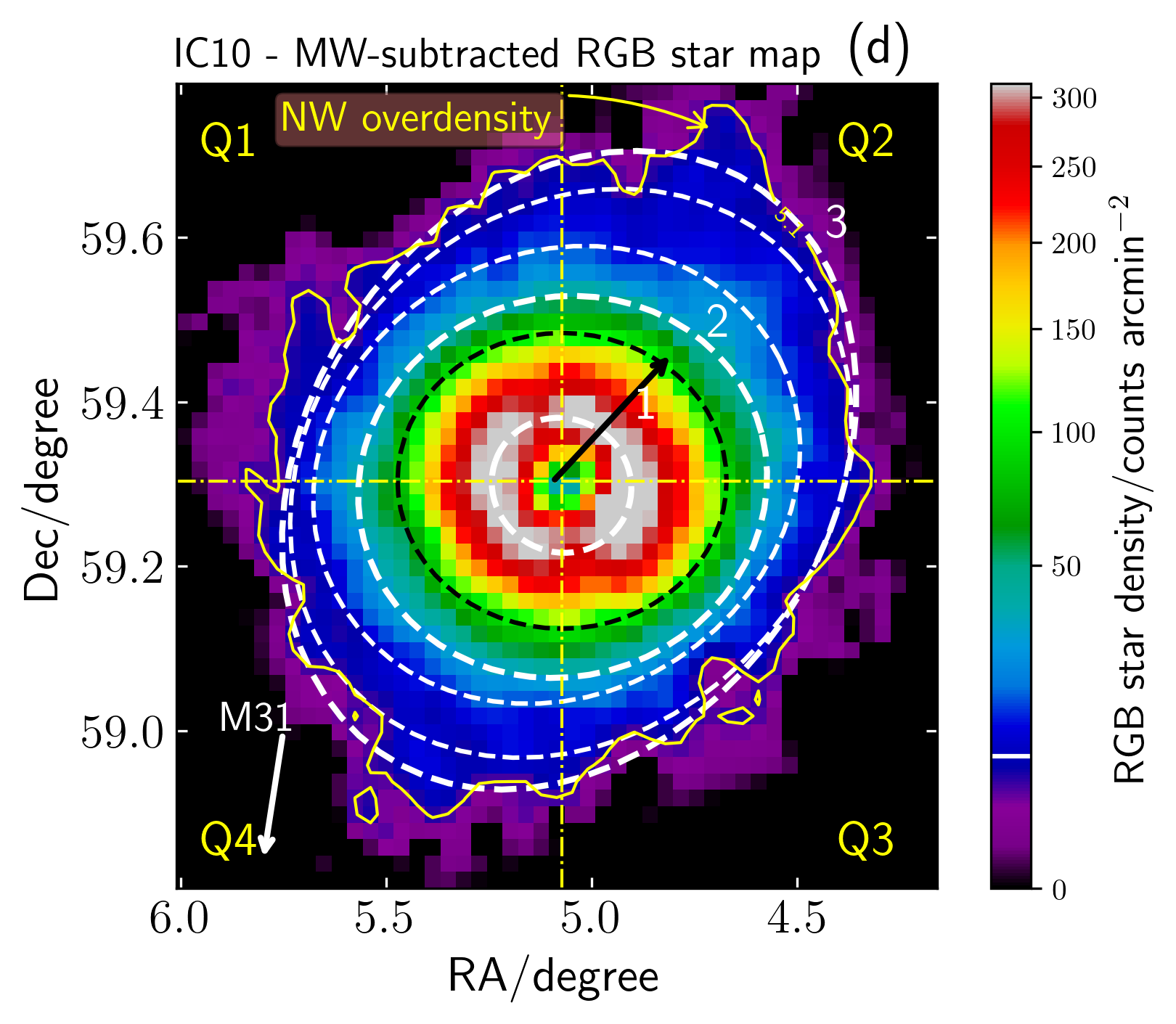}
\caption{ 
{\it Panel a:} Cleaned, reddening-corrected \Ho versus \IoHo CMD of IC\,10. The red and golden polygons denote our selection of RGB stars and bright MW foreground stars, respectively. 
{\it Panel b:} Spatial density map, in units of counts per arcmin$^{2}$, of the contribution from foreground MW stars to the RGB counts.  
{\it Panel c:} Spatial density map of sources within the red polygon of panel {\it(a)}, containing both RGB stars and MW stars. In the right-hand colour bar, the horizontal white segment denotes $3\,\sigma$ above the background computed from the counts in the region outlined by the red box. The black dashed ellipse with $e=0.12$ and $\mathrm{PA}=\ang{0;;}$ provides a good fit to the bulk of the old stellar population out to a radius $\lesssim\,20\arcmin$. The most external white dashed ellipses highlight the change in PA and ellipticity driven by a star count excess in the north-west direction, 
and by a less prominent excess in the south-east direction. The isodensity contour at $2\,\sigma$ above the background is in yellow.  
The RGB density drop in the most central galaxy region is due to high incompleteness. Q1 to Q4 denote the location of the four quadrants described in Sect.\,\ref{sc:profile}. 
The white arrow indicates the direction of M31,
while the black arrow is the direction of IC\,10's velocity with respect to M31 projected on the sky.
 {\it Panel d:} Same as in {\it (c)}, but after subtracting the contribution from MW stars represented in {\it (b)}. Ellipses 1--2--3 correspond to the regions selected for the CMD spatial analysis in Fig.~\ref{fig:cmd_grad}.}
\label{fig:rgb}
\end{figure*}

The RGB stars were selected from the cleaned, reddening-corrected CMDs presented in Sect.~\ref{sc:cmd}. 
In the extinction-corrected CMD of Fig.~\ref{fig:rgb}a, the RGB sequence is significantly tighter than in the 
uncorrected one, allowing us to adopt a more restrictive polygon compared to that displayed in Fig.~\ref{fig:ebv_rgb}a, and thus to achieve a cleaner selection of RGB stars down to $\sim4$ magnitudes below the RGB tip.
Figure~\ref{fig:rgb}c displays the resulting RGB star density map derived using 2\,arcmin-wide bins and applying a 2D Gaussian smoothing with a standard deviation of 0.2 times the bin size. 
In the colour bar, the white horizontal line indicates the level of $3\,\sigma$ above the average background of ${\rm 12\pm3\, counts~arcmin^{-2}}$, computed from an external 6\,arcmin size box to the north-east of the galaxy, whose position is indicated on the map. For comparison, when the same box is placed on the empty region to the south-west, the background results into ${\rm 11\pm2\, counts~arcmin^{-2}}$. 

The distribution of the old stars appears relatively round; with \texttt{PhotUtils} \citep{photutils}, an \texttt{Astropy} package for photometry, we fit elliptical isocontours to the RGB density map \citep[see also][]{Jedrzejewski87}. 
When averaging the derived ellipse parameters for radii less than $20\arcmin$ (black dashed ellipse in Fig.~\ref{fig:rgb}), we obtain a centre of 
$\alpha\sim$\mbox{\ang{5.068;;}}, $\delta\sim$\mbox{\ang{59.301;;}}, 
an ellipticity  $e=1-b/a=0.12\pm0.02$ and a position angle of 
$\mathrm{PA}=\mbox{\ang{0;;}}\pm\mbox{\ang{10;;}}$. 
This ellipticity is lower than that of $e=0.31\pm0.02$ derived with \texttt{AutoProf} \citep{autoprof} from the integrated-light fit, indicating that the old stellar population has a more circular morphology than the young one, which dominates the integrated light; this age-dependent ellipticity is also seen in the star cluster population \citep{Howell2026}. On the other hand, at galactocentric distances larger than $\sim$\,20\arcmin, the distribution of the old stars appears quite asymmetric, with a visible excess toward the north-west direction, where previous studies have identified the presence of a gaseous feature 
\citep{nidever13,namumba19}, and a less prominent excess toward the south-east direction, where a gaseous streamer was identified by \cite{nidever13}. 
As shown by the white dashed ellipses in Fig.~\ref{fig:rgb}c, this drives a noticeable change in the fitted isocontours' PA, accompanied by a slight shift of the centres toward the north-west direction, while the ellipticity remains similar. 
Interestingly, the north-west--south-east elongation derived for the outer stellar component is roughly aligned in the direction of M31 and with the motion of IC\,10  with respect to M31 computed from the PMs of \cite{Bennet2024} and \cite{Salomon2021}.

As discussed in Sect.~\ref{sc:cmd}, the final cleaned CMD we used to select RGB stars still suffers from a non-negligible contamination by MW stars. A non-uniform distribution of foreground contaminants across the imaged FoV could potentially bias our conclusions on the spatial distribution of RGB stars. Indeed, the reddening map in Fig.~\ref{fig:ebv_rgb} shows that the foreground extinction is highly variable, with the largest values coinciding with the north-west overdensity revealed by the RGB counts.

We therefore proceeded to subtract the contribution of foreground stars from our RGB counts. To this end, we first constructed a map of MW contaminants as follows. From the CMD displayed in Fig.~\ref{fig:rgb}a, we selected bright, blue foreground stars within the golden polygon. This CMD region contains a negligible contribution from IC\,10 stars and, therefore, it is effective in tracing the spatial variation of MW stars. The resulting map is then normalised to the mean stellar density computed in the red box region in Fig.~\ref{fig:rgb}c, which we assume to be free of RGB stars (see Sect.~\ref{sec:cmd_grad}). The normalised foreground contamination map is shown in  Fig.~\ref{fig:rgb}b. The average contamination amounts to ${\sim}\,12\pm1$ stars per arcmin$^{2}$, with a higher density in the north-west quadrant (Q2) than in the south-east one (Q4).

\begin{figure}
\includegraphics[width=0.95\columnwidth]{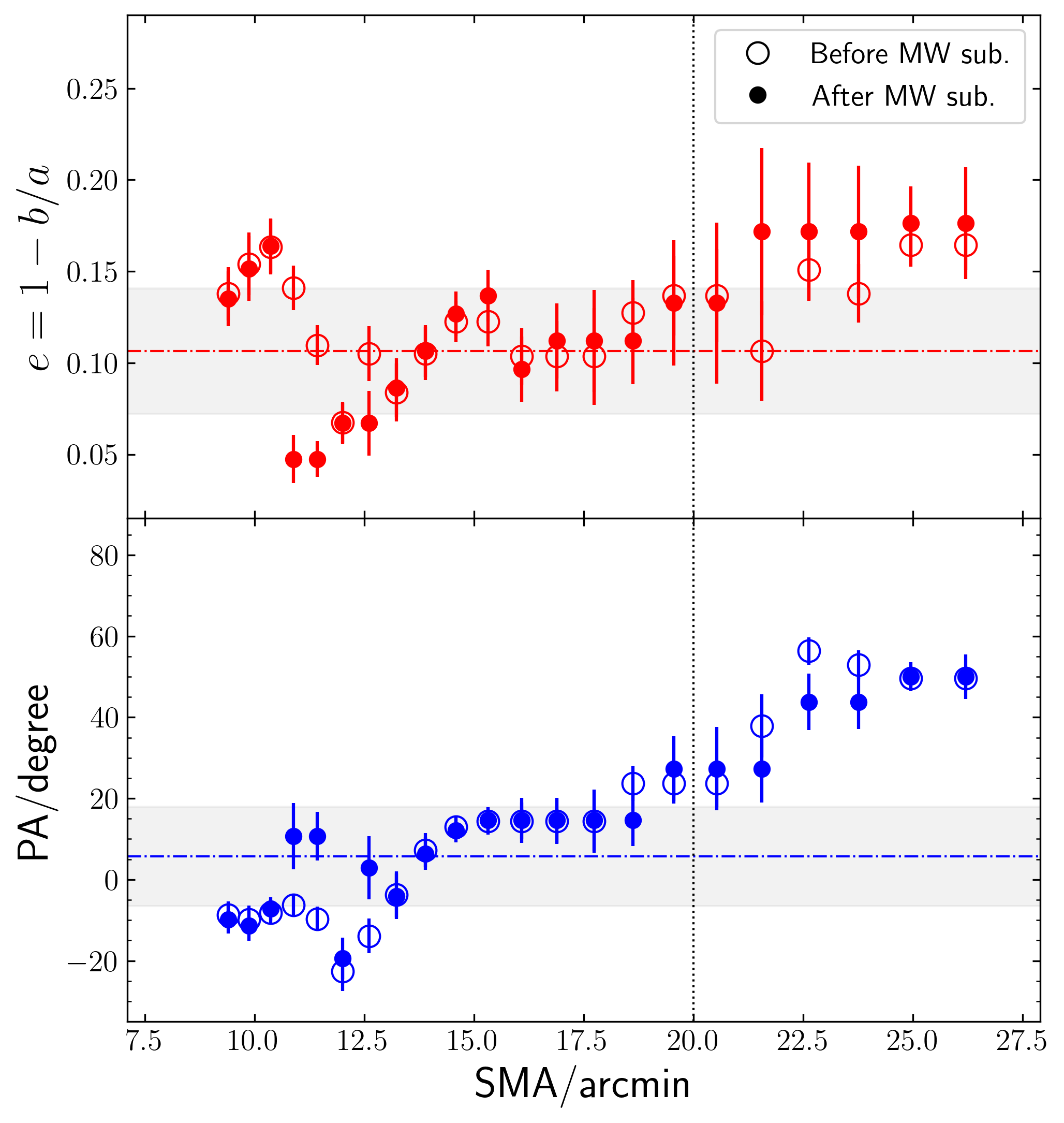}
\caption{
\emph{Top panel}: Ellipticity as a function of semi-major axis resulting from ellipse fitting to the RGB star density map before (open circles) and after
(filled circles) MW foreground subtraction, as in Fig.~\ref{fig:rgb}c and~\ref{fig:rgb}d, respectively. The horizontal line and the shaded area denote the average value of $e=0.12\pm0.02$ derived from the foreground-subtracted ellipses at SMA$\lesssim$\,20\arcmin. 
\emph{Bottom panel}: Same as top, but for the PA, measured from north to east counter-clockwise, with the minor axis oriented north-south.  The horizontal line and shaded area correspond to an average  value of $\mathrm{PA}=\mbox{\ang{0;;}}\pm\mbox{\ang{10;;}}$. }
\label{fig:ellipse_fit}
\end{figure}

The RGB star count map after subtraction of the variable foreground is displayed in Fig.~\ref{fig:rgb}d.
Here, the north-west overdensity (and more generally, the elongation from south-east to north-west) remains clearly visible, and is perhaps even more pronounced. Ellipse fitting shows no significant change in the innermost regions, while the outer ellipses become even more eccentric than in the uncorrected case.
Plots showing the trend of the ellipticity and PA as a function of semi-major axis, both before and after MW foreground subtraction,  are shown in Fig.~\ref{fig:ellipse_fit}.
This analysis thus confirms, and further strengthens, our conclusion that IC\,10 hosts an extended, elongated stellar component in its outskirts.

\subsection{CMD radial trends \label{sec:cmd_grad}}

\begin{figure*}[]
\includegraphics[width=1\textwidth]{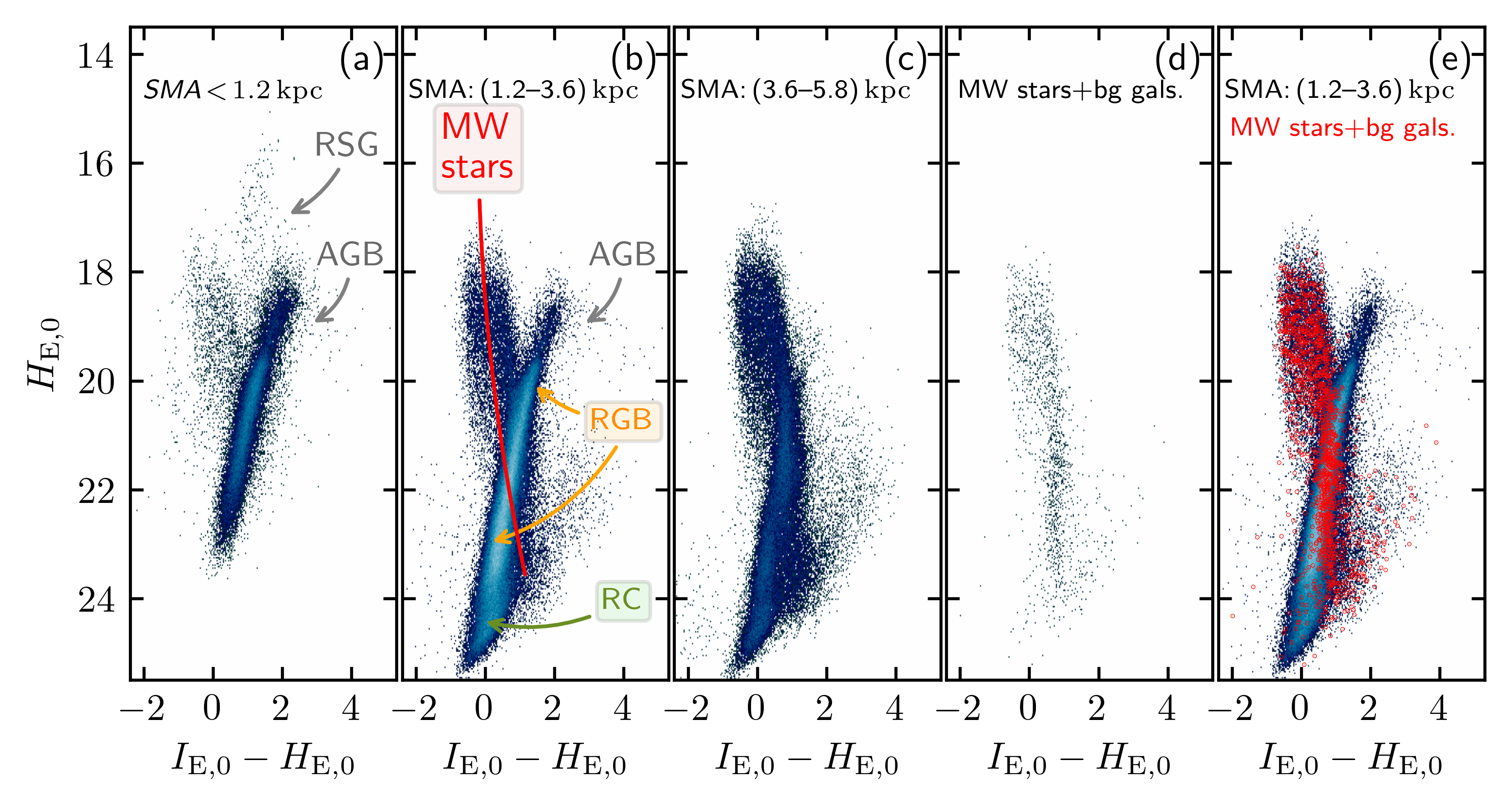}
\caption{\Ho versus \IoHo reddening-corrected CMD for four regions in IC\,10 at different galactocentric distances: 
one  within ellipse 1 in Fig.~\ref{fig:rgb}d, at semi-major axis 
SMA$\,<\,$1.2\,kpc (\emph{panel a}), showing both RSG stars and AGB stars, but only the brightest portion of the RGB due to severe incompleteness; one between ellipses 1 and 2, at 1.2\,kpc$\,<\,$SMA$\,<\,$3.6\,kpc (\emph{panel b}), 
showing a well defined RGB and RC, AGB stars, but no young RSG; a more external one 
between ellipses 2 and 3, at 
3.6\,kpc$\,<\,$SMA$\,<\,$5.8\,kpc (\emph{panel c}), with a persistent RGB but
negligible RSG and AGB stars; 
an external region (\emph{panel d}), corresponding to the red box in Fig.~\ref{fig:rgb}c, sampling MW stars and background galaxies, and appearing almost free of stars in IC\,10.
\emph{Panel (e)} displays the CMD of panel (b) with the stars in  panel (d) overplotted in red.  
The red curve in panel (b) describes the position of residual MW foreground stars not removed by our selections in Sect.~\ref{sc:mw}.
}
\label{fig:cmd_grad}
\end{figure*}

In this section, we investigate how different stellar populations are distributed within IC\,10 
through CMD spatial variations.  Figure~\ref{fig:cmd_grad} presents the CMDs for four representative regions which illustrate how the various stellar populations of IC\,10, as well as the foreground MW stars, are distributed across our FoV.

In Fig.~\ref{fig:cmd_grad}a, we see that young RSGs (age\,$\lesssim$\,50\,Myr) are confined within an elliptical region with semi-major axis (SMA) less than $\sim$\,1.2\,kpc, shown in Fig.~\ref{fig:rgb}d as ellipse 1. In this region the incompleteness is severe (see Appendix~\ref{sc:artificial}), and we only detect the brightest portion of
the RGB, together with AGB stars. At larger radii of 1.2\,kpc$\,<\,$SMA$\,<\,$3.6\,kpc, between ellipses 1 and 2, the CMD exhibits a well-defined RGB and RC due to stars older than $\sim$\,1--2\,Gyr, and a conspicuous population of intermediate-age AGB stars as old as $\sim$\,0.1--2\,Gyr (Fig.~\ref{fig:cmd_grad}, panel b). 
At larger distances of 3.6\,kpc$\,<\,$SMA$\,<\,$5.8\,kpc, between ellipses 2 and 3,  few AGB stars are observed, and the IC\,10 stellar population mostly consists of old RGB plus RC stars (Fig.~\ref{fig:cmd_grad}, panel c).  At these radii, the contribution of MW foreground stars compared to stars in IC\,10 is relatively high, and the brightest part of the RGB can hardly be recognised. Nonetheless, the depth of \Euclid photometry is such that the faintest portion of the RGB at $H_\sfont{E,0}\sim22$ and the RC are clearly detected. 
The radial distribution of the stellar populations is in agreement with the spatial maps presented by \citetalias{ERONearbyGals}, showing an increasing segregation of stars toward the galaxy centre with decreasing age. This distinction is also seen when examining the spatial distribution of IC\,10's star cluster population as a function of age \citep{Howell2026}. 

 Finally, panel (d) of Fig.~\ref{fig:cmd_grad} shows that the CMD extracted in the external region corresponding to the red box in Fig.~\ref{fig:rgb}c  
is dominated by MW foreground stars, with evidence for a marginal, if any, population of RGB/RC stars in IC\,10. This is best highlighted in Fig.~\ref{fig:cmd_grad}, panel (e), where these stars are plotted on top of the  1.2\,kpc$\,<\,$SMA$\,<\,$3.6\,kpc CMD. Despite the fact that bright RGB stars are absent in the outer region, we notice a few stars fainter than $H_\sfont{E,0}\sim22$ that could potentially be 
RGB/RC stars in IC\,10, although this population is highly subdominant compared to MW stars.

Although deeper data reaching fainter evolutionary phases (such as the old MS turnoff) would be required to confirm the presence of a very diffuse stellar component associated with IC\,10, our analysis indicates that, for the purposes of this study, it is reasonable to assume that this remote outer region is devoid of IC\,10 stars. It therefore traces the contribution of foreground MW stars and background galaxies to the IC\,10 point-source photometry.

\subsection{\label{sc:profile} Resolved star count profile}

\begin{figure}[]
\includegraphics[width=1\columnwidth]{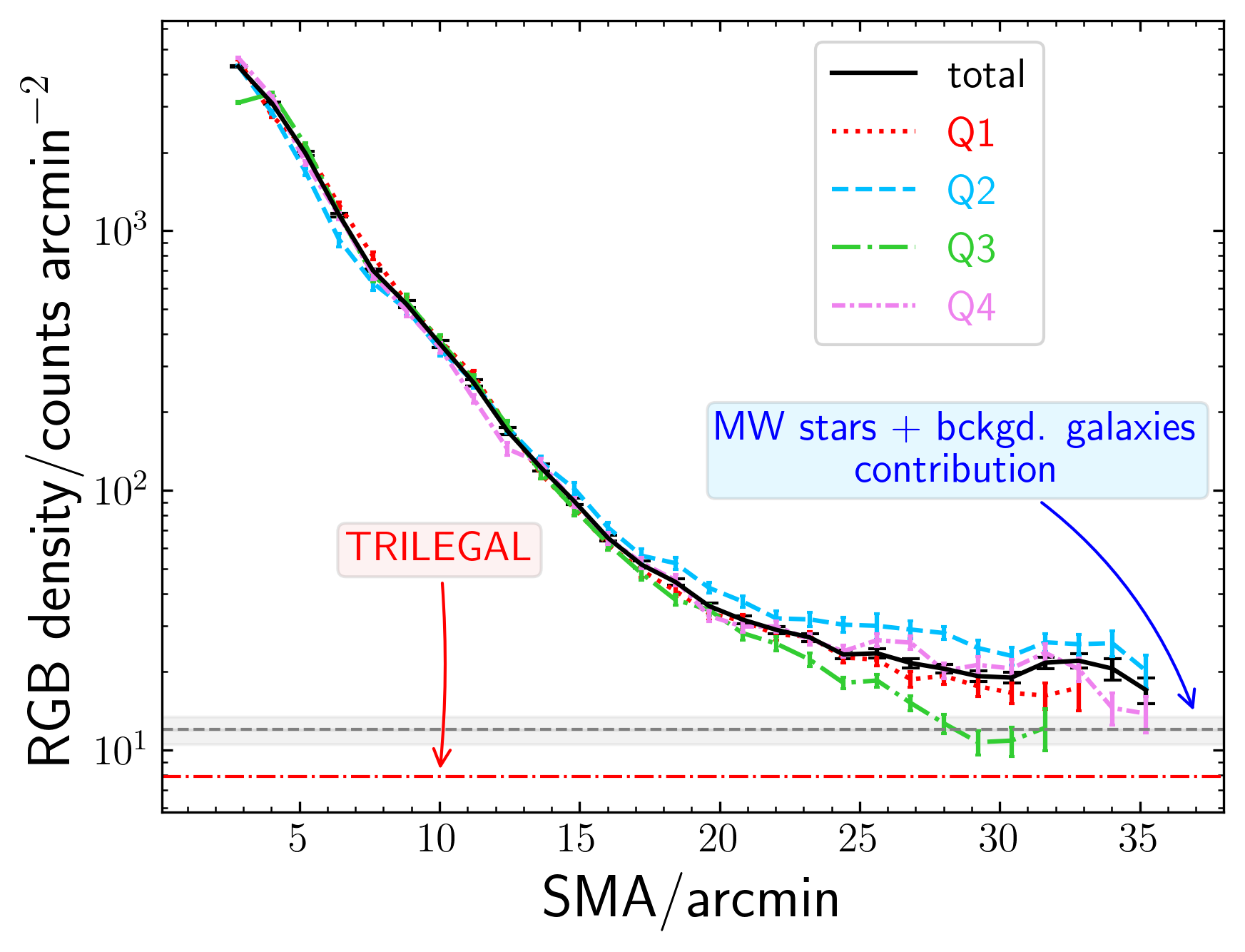}
\includegraphics[width=1\columnwidth]{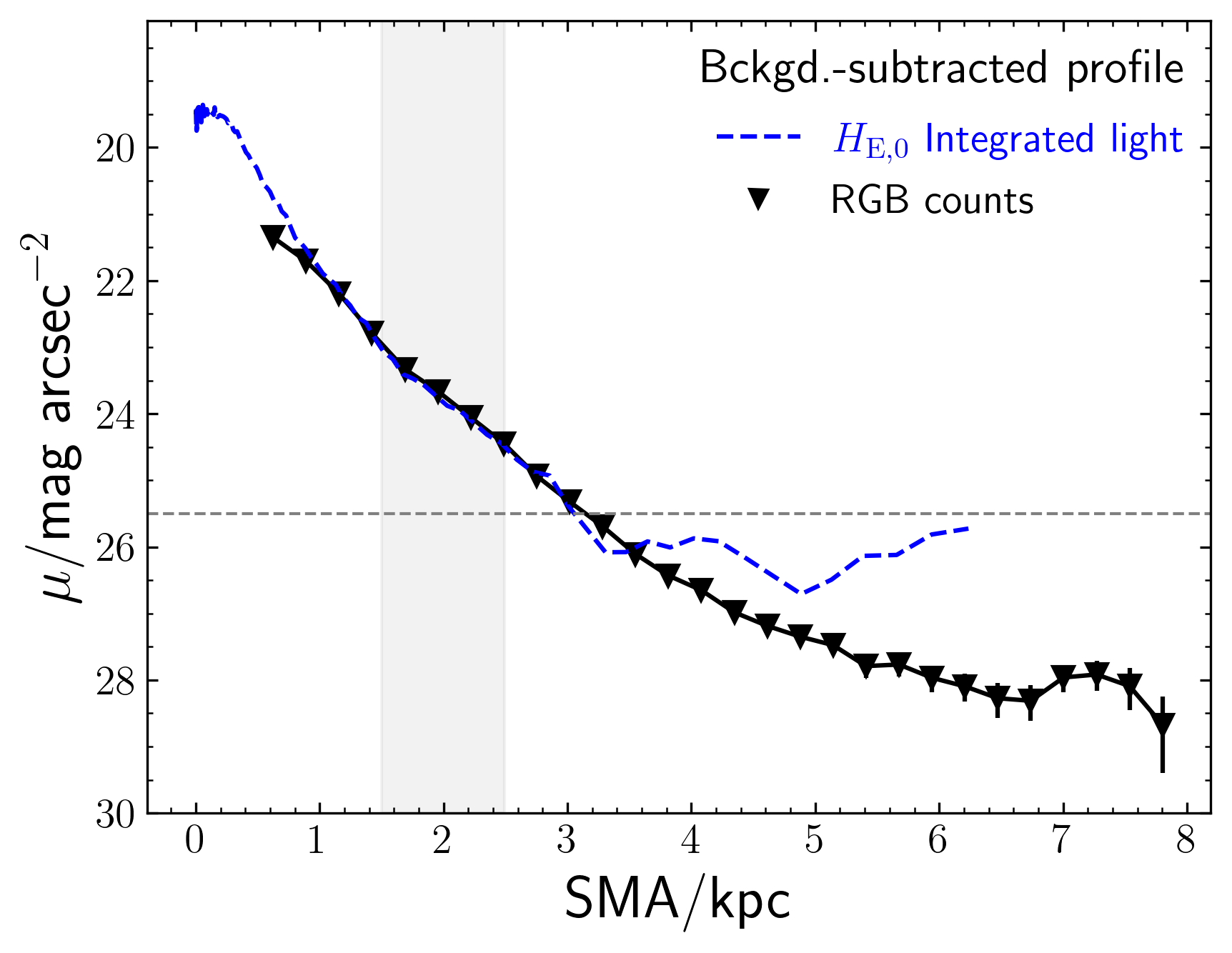}
\caption{\emph{Top panel}: RGB count density in units of number of counts per arcmin$^{2}$ as a function of semi-major axis. The solid black line is the total profile, while the coloured dashed/dotted lines correspond to the profiles in four different quadrants: Q1 (north-east), Q2 (north-west), Q3 (south-west), and Q4 (south-east). The horizontal dashed line is the average contribution from MW stars plus background galaxies directly inferred from our data (see Fig.~\ref{fig:rgb}), while the shaded area is the $\pm$\,1 standard deviation region.
The dotted-dashed line is instead the MW contribution 
predicted by TRILEGAL. \emph{Bottom panel}: 
RGB profile in mag\,arcsec$^{-2}$  (black triangle symbols) 
after subtraction of the foreground and background contaminants corresponding to the horizontal dashed line in the top panel.
The profile was normalised to the reddening-corrected, integrated-light profile from \texttt{Autoprof} (dashed blue line)
in the region indicated by the shaded area. 
The dashed horizontal line is the $1\,\sigma$ surface brightness limit from \texttt{Autoprof} in \citetalias{ERONearbyGals}.
}
\label{fig:profile}
\end{figure}

We inferred the resolved stellar profile of IC\,10 by computing the number density of RGB stars in 
concentric elliptical annuli with a fixed radial step of
0.25\,kpc, starting from an inner radius of $\sim$0.5\,kpc and extending outward to the maximum radius probed by our data.
For the annuli, we adopted the ellipse parameters of $e=0.12$ and $\mathrm{PA}=\ang{0;;}$ derived for the bulk of the stellar population in Fig.~\ref{fig:ellipse_fit}. 
RGB stars were selected from the reddening-corrected CMD, but completeness corrections, as derived in 
Appendix~\ref{sc:artificial}, were applied to the observed, uncorrected magnitudes.  
For each annulus, we binned the RGB stars in intervals of 0.5\,mag in the \IH colour and 0.5\,mag in \HE, computed the counts in each bin, and then divided them by the completeness of the corresponding colour-magnitude bin computed in the same annulus. The final counts were then computed as the sum of the corrected counts in all the bins.
Errors on the profile were derived accounting for both the Poissonian $\sqrt{N}$ error on the observed number of star counts (which is the dominant component) and 
the uncertainty on the completeness fraction, estimated as $\sqrt{N_{\rm out}}/N_{\rm inp}$, where $N_{\rm inp}$ and $N_{\rm out}$ are, for each colour-magnitude bin, 
the number of injected and recovered artificial stars, respectively. 

Given the severe incompleteness toward the inner galaxy regions, as evident in Fig.~\ref{fig:cmd_grad}a 
and Fig.~\ref{fig:spatial_compl},  we adopted for the most internal stars a brighter RGB selection than the one outlined by the polygon in the top panel of Fig.~\ref{fig:rgb} and considered only the portion of the RGB with 
$H_\sfont{E,0}\lesssim21.5$ \citep[see][for a similar approach]{Smercina23}.
Despite their different absolute values, the bright-limit and faint-limit  
RGB profiles share a common slope in a region at 1.9\,kpc$\,\lesssim\,$SMA$\,\lesssim\,$2.6\,kpc, which we adopted to normalise the bright profile to the faint one.  Eventually, the final merged profile results from the combination of the bright-limit one at SMA$\,\lesssim\,$1.9\,kpc and the faint-limit one at larger radii.
The profile was sampled out to a semi-major axis of SMA$\,\sim\,$35\arcmin, or $\sim$\,7.7\,kpc. 
At SMA$\,\gtrsim\,$24\arcmin, increasingly smaller fractions of the elliptical annuli are covered by the \Euclid FoV, implying poor statistics at the largest galactocentric distances.  

Besides the total profile, we also computed profiles for four different quadrants anchored on IC\,10's centre, as indicated in Figs.~\ref{fig:rgb}c and~\ref{fig:rgb}d: Q1 to the north-east, Q2, to the north-west, Q3 to the south-west, and Q4 to the south-east. The results are displayed in the top panel of Fig.~\ref{fig:profile}, where both the total and quadrant RGB count profiles are shown as a function of the semi-major axis. The quadrant profiles are seen to be in very good agreement with each other at SMA$\,\lesssim\,$20\arcmin. However, at larger radii, the Q1 and Q3 profiles fall below the Q2 and Q4 ones, 
in agreement with the north-west overdensity and the preferential distribution of RGB stars along the north-west--south-east direction in the map of Fig.~\ref{fig:rgb}d. 
It is difficult to explain these differences with the uncertainties associated with the star count profiles, as shown by the error bars in Fig.~\ref{fig:profile}. The horizontal line indicates the contribution from MW foreground stars and unresolved background galaxies, computed as the average density from the map in Fig.~\ref{fig:rgb}b, while the shaded area is its standard deviation. Ultimately, we subtracted this contribution from the total RGB profile.

The background-subtracted star count profile was normalised to surface brightness units by anchoring it to the \HE-band integrated-light profile inferred with \texttt{Autoprof}. 
This approach is commonly applied \citep[e.g.][]{barker12, Smercina23} and exploits the complementarity of the two methods: resolved star counts provide a robust tracer of the stellar density at large radii and low surface brightness levels, while the integrated-light profile offers a reliable normalisation in the inner regions, where crowding and incompleteness affect the resolved stellar component.
To this end, we ran \texttt{Autoprof} on the \HE image forcing the elliptical isophotes to have the same centre, ellipticity and PA as those adopted to derive the RGB star count profile. 
The \HE band was chosen as more sensitive than the bluer bands to the old stellar component \citepalias[see however][for a comparison with the profiles in the other \Euclid bands]{ERONearbyGals}.  We corrected the profile adopting $E(B-V)=0.8$, corresponding to the median reddening value over the \Euclid FoV. The normalisation of the resolved star-count profile to the integrated light one was performed in the region at 
1.5\,kpc$\,\lesssim\,$SMA$\,\lesssim\,$2.5\,kpc, where the CMD spatial analysis indicates negligible contribution from young stars \citep[see also][]{Sanna2010} and the 
slopes of the integrated-light and resolved RGB star profiles are in excellent agreement.  
Our final calibrated profile is shown in the bottom panel of Fig.~\ref{fig:profile}.

At SMA$\,\lesssim\,$1\,kpc, the integrated-light profile exhibits a rapid steepening, as expected from the increasingly large contribution of young, luminous stars,  as well as old, highly crowded stars. In the present work, we were not able to robustly 
trace the RGB profile because of the severe incompleteness toward the galaxy centre. At SMA$\,\gtrsim\,$3\,kpc the 
integrated-light profile flattens out, reaching the $1\,\sigma$ surface brightness limit from \citetalias{ERONearbyGals}, due to the impossibility of properly removing foreground star and background galaxy contaminants. On the other hand, the RGB star count profile keeps decreasing down to $\sim$\,29\,mag\,arcsec$^{-2}$ in \Ho\ (reddening-corrected). 
The small bump at SMA$\gtrsim$\,7\,kpc, already visible in the quadrant profiles of Q2 and Q4 prior to azimuthal averaging, is likely associated with the asymmetric outer stellar distribution along the north-west--south-east direction, including the prominent north-west overdensity.

We fit Sérsic and exponential profiles to the normalised RGB star profile (with no inclusion of the inner integrated-light profile) following the procedure by Sánchez-Alarcón et al. (in prep). 
In a nutshell, we used the Sérsic function in mag\,arcsec$^{-2}$, defined as
\begin{equation}
    \mu_{\rm{Sersic}} = \mu_{\rm{e}} + \frac{2.5 b_{n}}{\ln{10}}\left[(r/r_{\rm e})^{1/n} - 1\right]
    \label{eq:sersic},
\end{equation}
where the $\mu_{\rm{e}}$ is the surface brightness level at the effective radius $r_{\rm e}$, $n$ is the Sérsic index, 
and $b_{n}$ is related to $n$ as defined in \cite{Ciotti1999}.

For the exponential profile, we adopted the same functional form with $n=1$, which can be written as

\begin{equation}
    \mu_{\rm{exp}} = \mu_{\rm{0}} + \frac{2.5}{\ln{10}}\frac{r}{h} 
    \label{eq:sersic},
\end{equation}
where $\mu_{\rm{0}}$ is the surface brightness level at the center of the profile ($r=0$), and $h$ is the scale length. We first fit a single Sérsic profile to the RGB profile minimising the residuals. Then, we fit a two-component model, which results from the addition of a Sérsic and exponential profile. 

\begin{figure*}[]
\centering
\includegraphics[width=0.9\columnwidth]{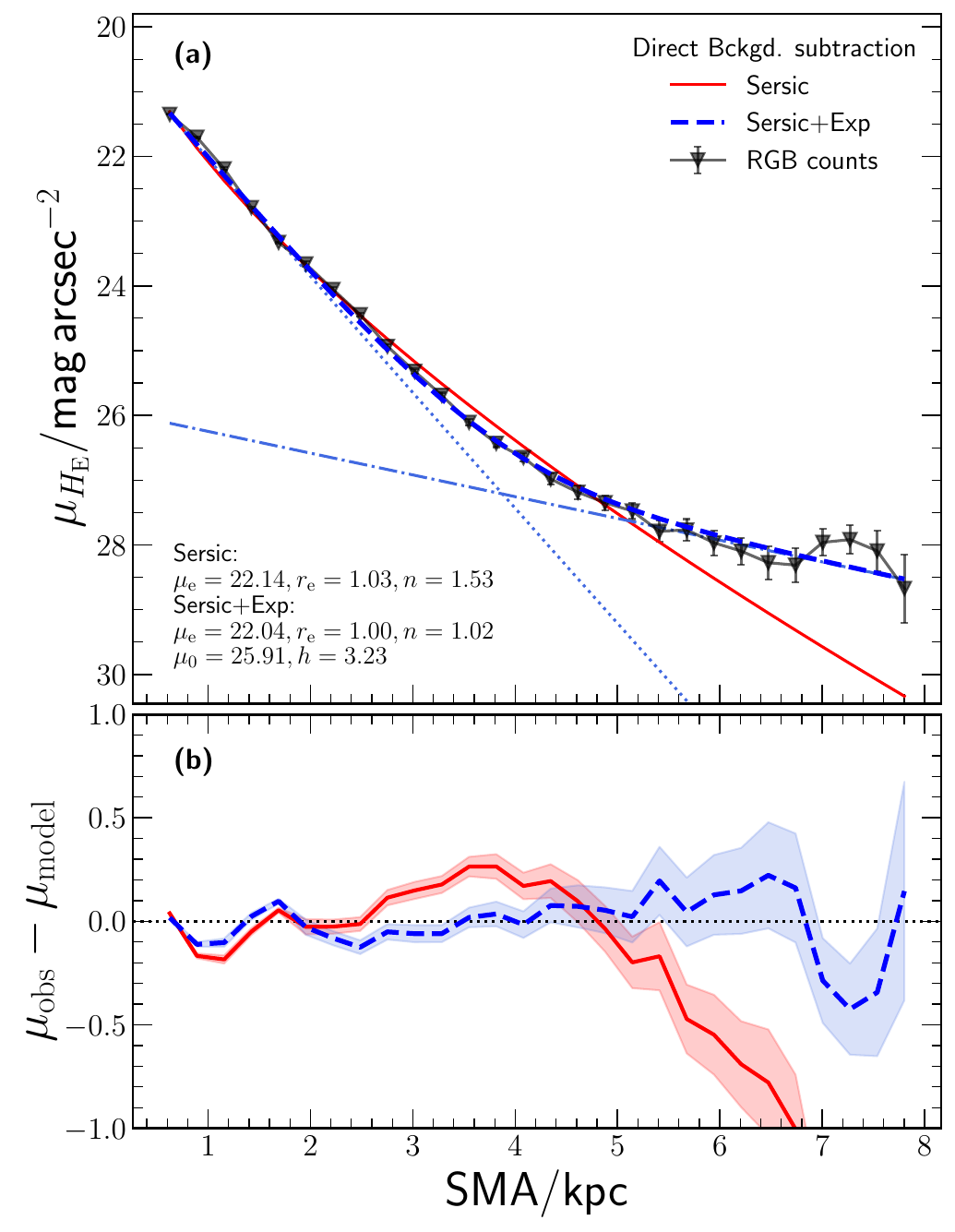}
\includegraphics[width=0.9\columnwidth]{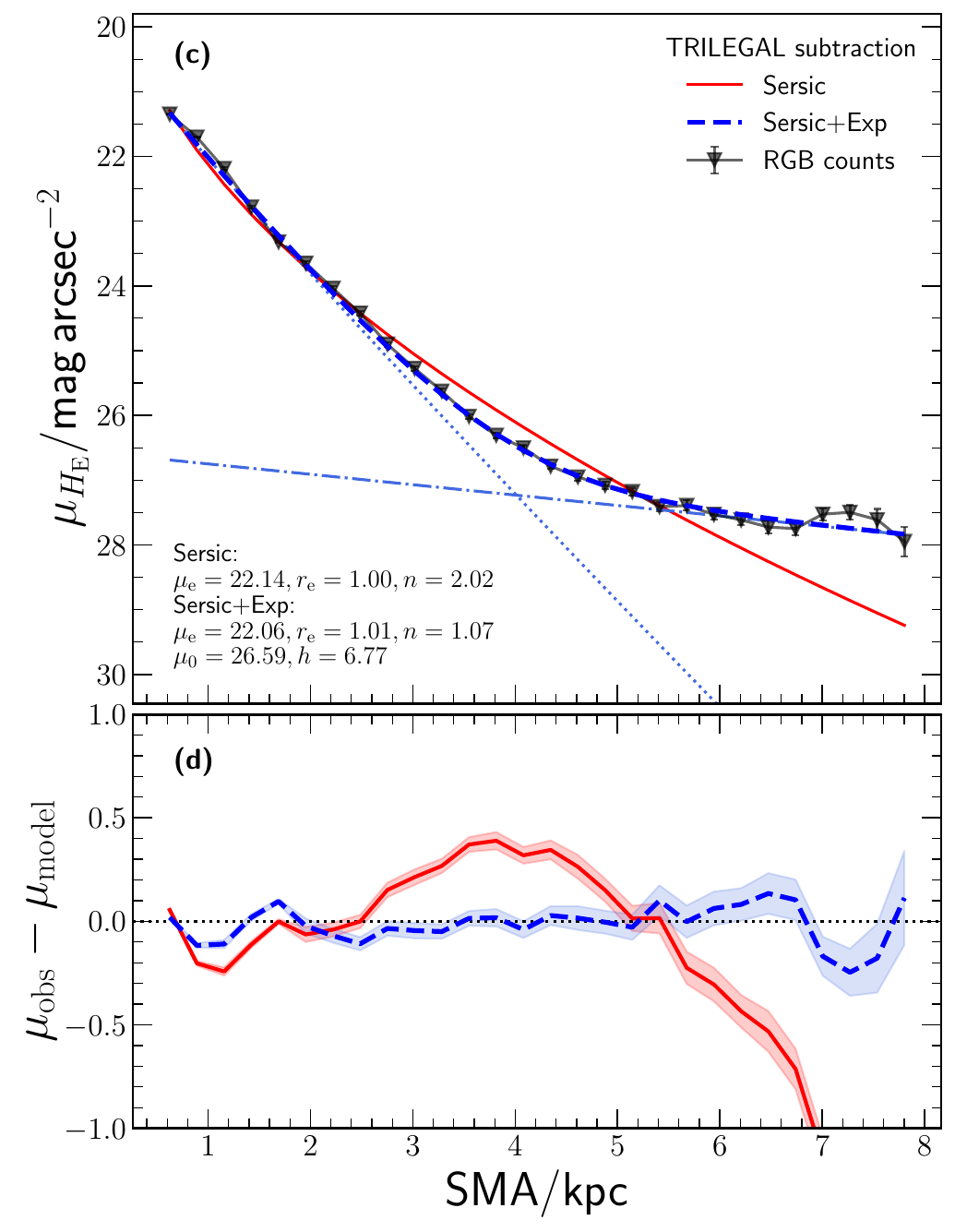}
\caption{Background-subtracted RGB profiles (black triangle symbols) normalised to the reddening-corrected \Ho\ integrated-light profile.
In \emph{panel (a)} the profile was obtained by subtracting the background (MW stars plus compact background 
galaxies) estimated from the star counts at large radii, while in \emph{panel (c)} the contribution from MW stars was inferred from TRILEGAL. In both panels, the solid red line is the best-fit Sérsic  model. 
The dashed blue line is instead the fitted two-component model consisting of a Sérsic  profile (dotted blue line) and an exponential disc (dot-dash blue line). \emph{Panels (b)} and \emph{(d)} display the residuals in 
surface brightness for the one and two-component best-fit models.
}
\label{fig:profile_fit}
\end{figure*}

For a single Sérsic component, we found best-fit parameters of $\mu_{\mathrm{e},\HE}=22.14$,  
$n=1.5$, and $r_{\rm e}=1.0\,$kpc. This best-fit model is shown with a red solid line in panel (a) of Fig.~\ref{fig:profile_fit}. 
Although the observed RGB counts are in good agreement with the fitted 
profile at SMA$\,\lesssim\,$5\,kpc, they appear to be systematically above the model at larger radii.
This is illustrated in panel (b) of Fig.~\ref{fig:profile_fit}, showing the surface brightness residuals from the observed profile and the best-fit Sérsic model.   
This indicates the presence of a second component that dominates in the outer regions.
To compensate for this extended emission, we fit a two-component model, with a Sérsic profile and an exponential disc (equivalent to a Sérsic with $n=1$). The two-component model, shown with the dashed blue line in Fig.~\ref{fig:profile_fit}, is able to reproduce both the internal and external regions at 
SMA$\,\gtrsim\,$5\,kpc, as evident from the residuals in panel (b). 
The parameters of the one- and two-component best-fit models are listed in Table~\ref{tab:tab1}. In addition to the parameters of the fit (i.e. the Sérsic index $n$, the effective radius $r_{\rm e}$, the reddening-corrected effective surface brightness $\mu_{\mathrm{e},\HE}$, the reddening-corrected central surface brightness $\mu_{0,\HE}$ and the scale~height $h$ of the exponential disc component), we provide the reddening-corrected 
\Ho-band magnitude in the individual components, the total \Ho magnitude from both components, the stellar mass of the individual components, and the total stellar mass. The \Ho-band magnitudes were obtained by integrating the profile to a radius of 50\,kpc, so that the outer wings of the profile are fully included; specifically, this radius was chosen such that the integrated magnitude changes by less than 0.1\%, well below the uncertainty.
The stellar mass in the various components was computed by adopting a stellar mass-to-light ratio in the {\it H} band of 
$\Upsilon_H\sim1.16$, 
derived from the PARSEC\footnote{These can be downloaded via the web page \url{https://stev.oapd.inaf.it/cgi-bin/cmd}} simple stellar populations (SSPs) for $Z=0.001$ and a canonical two-part-power law initial mass function from \cite{Kroupa2001,Kroupa2002} and \cite{Kroupa2013}, corrected for unresolved binaries; SSPs of different ages were weighted according to the star formation history inferred by \cite{weisz14} at look-back times   $\gtrsim$\,1\,Gyr, and then linearly combined. We adopted an {\it H}-band AB magnitude of $M_{H, \odot}=4.66$ for the Sun from \cite{Willmer2018}. To account for the uncertainty in mass due to the metallicity dependence, we computed the mass-to-light ratio for two extreme metallicity values of $Z=0.0004$ and $Z=0.004$, obtaining $\Upsilon_H\sim1.3$ and $\sim1.0$, respectively. This results in a stellar mass of $M_{\star}=6.7^{+0.9}_{-0.9}\times 10^8$\,\msun, which is a lower limit since it does not account for the amount of mass in stars younger than 1\,Gyr. 

\begin{table}[ht]
\centering
\small 
\setlength{\tabcolsep}{5pt} 
\renewcommand{\arraystretch}{1.25} 
\caption{Structural properties of the old stellar component in IC\,10.}
\label{tab:tab1}
\begin{tabular}{llcc}
\hline\hline
\multicolumn{4}{c}{\rule{0pt}{3.2ex}Single Sérsic  model} \\
\hline
Parameter & Unit & Direct bckgd.& TRILEGAL bckgd.\\
\hline
$\mu_{\mathrm{e},\HE}$ & mag arcsec$^{-2}$ & 22.14 & 22.14 \\
$r_\mathrm{e}$              & kpc               & 1.0 & 1.0 \\
$n$                 & ---               & 1.5  & 2.0 \\
\hline
\multicolumn{4}{c}{\rule{0pt}{3.2ex}Sérsic  + Exponential disc model} \\
\hline
Parameter & Unit & Direct bckgd.& TRILEGAL bckgd.\\
\hline
$\mu_{\mathrm{e},\HE}$ & mag arcsec$^{-2}$ & 22.04 & 22.06 \\
$r_\mathrm{e}$              & kpc               & 1.0 & 1.0 \\
$n$               & ---               & 1.0 & 1.1 \\
$\mu_{0,\HE}$ & mag arcsec$^{-2}$ & 25.91 & 26.59 \\
$h$                & kpc               & 3.2 & 6.8 \\
$H_{\sfont{E,}\mathrm{S\acute{e}rsic}}$  & mag   & 7.32 & 7.29 \\
$H_{\sfont{E,}\mathrm{exp}}$ & mag   & 9.35 & 8.43 \\
$H_{\sfont{E,}\mathrm{tot}}$ & mag   & 7.16 & 6.96 \\
$M_{\star,\mathrm{S\acute{e}rsic}}$   & $M_{\odot}$   & $5.8^{+0.7}_{-0.8}\times10^8$ & $6.0^{+0.8}_{-0.8}\times10^8$ \\
$M_{\star,\mathrm{exp}}$   & $M_{\odot}$   & $0.9^{+0.1}_{-0.1} \times 10^8$ & $2.1^{+0.3}_{-0.3}\times10^8$ \\
$M_{\star,\mathrm{tot}}$ & $M_{\odot}$  & $6.7^{+0.9}_{-0.9}\times 10^8$ & $8.1^{+1.0}_{-1.1}\times 10^8$ \\
\hline
\end{tabular}
\vspace{0.5em}
\begin{minipage}{\columnwidth}
\small
\textit{Notes.} All listed values are obtained assuming an average $E(B-V)=0.8$ correction and $D=0.762\,$Mpc (Sect.~\ref{sc:trgb}). 
The symbols $\mu_{\mathrm{e},\HE}$, $r_\mathrm{e}$, and $n$ correspond to the following Sérsic fit parameters: reddening-corrected surface brightness at $r_\mathrm{e}$, effective radius, and Sérsic index, respectively. 
The exponential disc parameters -- central surface brightness and scale height -- are $\mu_{0,\HE}$ and $h$. 
The quantities $H_{\sfont{E,}\mathrm{S\acute{e}rsic}}$, $H_{\sfont{E,}\mathrm{exp}}$, and $H_{\sfont{E,}\mathrm{tot}}$ 
are the apparent, reddening-corrected magnitudes in the \HE band for the Sérsic, disc, and total components. 
$M_\star$ represents the stellar mass in old ($\gtrsim$\,1\,Gyr) stars for all components, with the errors reflecting the uncertainty in metallicity. 
Values are provided for the two cases of the background directly estimated from the counts in an outer region of IC\,10 
(Direct bckgd.) and of the background inferred from the TRILEGAL Galactic model (TRILEGAL bckgd.)
\end{minipage}
\end{table}

At the largest sampled galactocentric distances, the most important source of uncertainty in the profile is the adopted contribution from MW foreground stars, vastly dominating over unresolved background galaxies at IC\,10's Galactic latitude. 
Thus, as a consistency check, we computed the number density of MW stars predicted by TRILEGAL \citep{trilegal05,trilegal12} that fall within the RGB selection polygon displayed in Fig.~\ref{fig:rgb}a. More specifically, TRILEGAL was run assuming no dust extinction, but then applying to the simulated CMD in the \Euclid filters an average extinction of $A_{I\sfont{E}}=1.7$ and $A_{H\sfont{E}}=0.4$, consistent with 
the average reddening value inferred in Sect.~\ref{sc:ebv}. This provides a density of ${\sim}\,8\pm1$ stars per arcmin$^{2}$, where the error accounts for the spread in reddening, to be compared with an average density of 
${\sim}\,12\pm1$ stars per arcmin$^{2}$ computed from the map in Fig.~\ref{fig:rgb}b.  Since the TRILEGAL computation does not account for the contribution of unresolved background galaxies, this value should be considered as a lower limit. From other ERO Showcase data with low contamination from the MW disc, 
we estimate that unresolved background galaxies contribute less than 10\% of the MW star contamination.

The profile and the fits resulting from the subtraction of the TRILEGAL 
background value are shown in panel (c) of Fig.~\ref{fig:profile_fit}, while the residuals are displayed in panel (d). A one-component Sérsic  fit results into $\mu_{\mathrm{e},\HE}=22.14$,  
$n=2.0$, and $r_{\rm e}=1.0\,$kpc which, as in the previous case, is unable to reproduce the counts at SMA$\,\gtrsim\,$5\,kpc, indicating the presence of an additional, more extended stellar component.
Thus, we fit a two-component model, with a Sérsic  and an exponential disc, finding again that this is able to reproduce the profile both in the inner and in the outer regions. The results of the fit are given in Table~\ref{tab:tab1}.

\section{\label{sc:conclusions} Discussion}

Using resolved stellar photometry in the \Euclid \IE, \YE, \JE, and \HE bands, we characterised the extended, low surface brightness,  old stellar component of IC\,10 out to a radius of $\sim$\,35\arcmin\ ($\sim$\,8\,kpc, considering our distance estimate 
of $\sim$\,0.76\,Mpc) from the galaxy centre, corresponding to about 5 times the $\mu_B=25$\,mag\,arcsec$^{-2}$ isophotal radius \citep{RC3}.
This was achieved by selecting RGB stars in the 
reddening-corrected \Ho versus \IoHo CMD, after subtracting the contribution of unresolved background galaxies and MW foreground stars. 
Thanks to the wide FoV and superb spatial resolution of \Euclid, this is the first time the stellar component of IC\,10  has been studied to such large radii and to (reddening-corrected) surface brightness levels as faint as $\mu_{\HE}\sim$\,29\,mag\,arcsec$^{-2}$.

A previous study by \cite{Demers2004} identified RGB stars out to $\sim$\,16\arcmin, while 
\cite{Tikhonov2009} reported a thick disc with dimensions 14\farcm5$\,\times\,$10\farcm5 (semi-major times semi-minor axis).  
From Suprime-Cam Subaru data, \cite{Sanna2010} detected  sizable samples of RGB stars out to radial distances of 
$18\arcmin$--$23\arcmin$ from the galactic centre and argued, based on predictions for the foreground contamination from Galactic models, that IC\,10  stars may extend as far as $\sim$34\arcmin--$42\arcmin$. Using optical and NIR wide-field imaging with MegaCam and the Wide Field Camera at the Canada France Hawaii Telescope (CFHT) and the United Kingdom Infra-Red Telescope (UKIRT), respectively,  \cite{gerbrandt15} traced the old stellar population of  IC\,10 out to a radial distance of $\sim$\,20\arcmin. However, all these studies achieved shallower surface brightness and point-source depths than our current analysis. 
While their CMDs were limited to the upper $\sim$1.5--3 magnitudes of the RGB, our reddening-corrected photometry probes down to $H_\sfont{E,0}\gtrsim24$, i.e. $\sim$\,4 magnitudes below the RGB tip. Indeed, our CMD spatial analysis (see Sect.~\ref{sec:cmd_grad} and Fig.~\ref{fig:cmd_grad}) highlights how critical CMD depth is for detecting low-density stellar components; beyond SMA$\,\gtrsim\,$3.6\,kpc ($\gtrsim$\,16\arcmin), nearly no RGB stars would be seen if we were limited to the top $\sim$\,2\,mag of the RGB.
Our results demonstrate that IC\,10's stellar component is significantly more extended than previously thought: RGB stars are still detected at the edge of the \Euclid FoV, suggesting the galaxy could extend nearly as far as the enormous $62\arcmin\times80\arcmin$ \ion{H}{i} envelope found by \citet{Huchtmeier1979}.
Other studies have shown that when traced down to very low surface brightness levels, the stellar component of galaxies can be as extended as the \ion{H}{i}  distribution \citep{Bellazzini2014,okamoto15,Annibali2020,Mancera2024}.

The RGB star count map reveals a relatively round and regular morphology within a radius of $\sim$\,20\arcmin, consistent with \citet{gerbrandt15}. 
In this region, the star count map is well fitted with isodensity contours characterised by ellipticity $e\sim0.12$ and position angle of $\mathrm{PA}\sim\ang{0;;}$. 
However, at larger radii, the isocontours begin to twist, driven by an excess of stars to the north-west and 
a less prominent excess to the south-east (see Sect.~\ref{sc:spatial}). 
Similar features have been observed in tidally disturbed dwarf galaxies \citep[e.g.][]{Okamoto23,Golini2024}.
Indeed, the north-west overdensity is in the same direction where \cite{nidever13} identified the \ang{1.3;;} long \ion{H}{i}  feature (corresponding to $\sim$\,17\,kpc at a distance of 0.76\,Mpc), culminating in a high-density clump later confirmed by \citet{namumba19}. As introduced in Sect.~\ref{sc:Intro}, this structure was interpreted as the result of a recent interaction or merger with a companion dwarf galaxy, a scenario that,  according to \cite{Ashley2014}, could also explain other peculiar morphological and kinematical properties of the gas in IC\,10:  
the southern plume, the three spurs, and the outer \ion{H}{i} region counter-rotating with respect to the inner gaseous disc  \citep{Shostak1989,wilcots98}.  We may now be detecting, for the first time, a stellar counterpart to this accretion event.

However another intriguing possibility is that the north-west--south-east elongated stellar component results from a tidal interaction with M31.  Using PMs, \cite{Bennet2024} 
derived the orbital history of IC\,10 and found that the galaxy underwent pericentric passage with M31 about 1\,Gyr ago and that it is still inside its virial radius.  The alignment of the elongated stellar feature that we infer through RGB counts with the orbit of IC\,10 around M31
\citep{nidever13,Bennet2024}, together with a marked flattening of the outer density profile (see discussion below), is suggestive of a tidal origin \citep[e.g.][]{Johnston1999,Munoz2008}. As our current \Euclid coverage does not extend to the full extent of either the stellar or the \ion{H}{i} component of IC\,10, 
data from a larger FoV than is currently available would be required to clarify the nature of 
the outer elongated stellar emission.

We derived the radial profile of the old ($\gtrsim$1–2\,Gyr) stellar component in IC\,10 from RGB star counts, corrected for incompleteness through artificial-star experiments.
The global contribution from compact background galaxies and, predominantly, foreground MW stars was subtracted using an external region assumed to be free of IC\,10 members, while accounting for spatial variations in MW star density across the field.
The RGB star density was then converted to surface brightness (mag\,arcsec$^{-2}$) by scaling to the \HE-band integrated-light profile, following a method widely used in studies of resolved stellar systems \citep[e.g.][]{Bellazzini2011,Rys2011,Bernard2012,Bellazzini2014,Kniazev2016,Annibali2020,Higgs2021}.
The reddening-corrected surface brightness profile reaches $\mu_{\HE}\sim$\,29\,mag\,arcsec$^{-2}$.

The dominant uncertainty arises from the correction for background and foreground contamination, which critically affects the profile in the galaxy outskirts, where low surface brightness values are highly sensitive to small subtraction errors.
Our CMD analysis indicates that a few RGB stars from IC\,10 are still present near the edge of the \Euclid FoV, implying that we may be slightly oversubtracting the background.
For comparison, the TRILEGAL model predicts a foreground density about 30\% lower than our empirical estimate, yielding a flatter outer profile.
Contamination from unresolved background galaxies is insufficient to explain this discrepancy, as their contribution is estimated to be $<10$\% of that from MW stars, based on other ERO datasets.
Given these uncertainties, the true IC\,10 profile likely lies between the two profiles obtained under different subtraction assumptions \citep[e.g. see ][for similar challenges]{Martin2008, McMonigal2014, Collins2021}.

The stellar density profile of dwarf galaxies retains the imprint of both internal and environmental evolutionary processes \citep[e.g.][]{Martin2025}. Departures from a single Sérsic component, due to an excess of stars at large radii, can signal the action of galactic tides from interactions with a more massive host, which are expected to redistribute stars and produce extended outer envelopes \citep{Penarrubia2009,Lokas2013}. Alternatively, diffuse outskirts may arise from dwarf-dwarf encounters or mergers \citep[e.g.][]{Deason2022}, or more generally from the gradual build-up of a stellar halo through the accretion of low-mass companions over cosmic time \citep{Cooper2025,Celiz2025}.

We fitted the surface brightness profile of IC\,10 with Sérsic  models. A single-component model with $n=1.5$\,--2 ( depending on the adopted method of subtraction) reproduces the inner profile. However, at radii larger than about 5\,kpc, we observe a significant excess with respect to the Sérsic  model. This indicates the  presence of a second component dominating the galaxy outskirts. Nevertheless, given that it is virtually impossible to trace its entire extension with the data in hand makes its modelling quite uncertain. With these limitations in mind, we adopted a two-component model comprising a Sérsic  plus an exponential profile. The inner Sérsic  component is well constrained, with $n\sim1.0$\,--1.1, $r_{\rm e}\sim1$\,kpc, and $\mu_{\mathrm{e},\HE}\sim22$\,--22.1\,mag\,arcsec$^{-2}$ (thus insensitive to which of the two profiles is adopted). Our Sérsic  index is in good agreement with that derived from RGB counts by \cite{McQuinn2017}, and somewhat larger than the value of $n=0.75\pm0.06$ previously inferred by \cite{gerbrandt15}, while our effective radius ($r_{\rm e}=1.0$\,kpc) is smaller than their $r_{\rm e}\sim1.2$\,kpc. 
On the other hand, the exponential component shows greater sensitivity to which of the two profiles is adopted, with central surface brightness values in the range of $\mu_{0,\HE}\sim25.9$\,--26.6\,mag\,arcsec$^{-2}$ and a scale height in the range of $h=3.2$\,--6.8\,kpc, where the larger value of $h$ is associated with the flatter profile. From the results of the fit, we infer a total reddening-corrected magnitude in the range of $H_{\sfont{E,}\mathrm{tot}}\sim7.0$\,--7.2, 
from which we derive a stellar mass of $M_{\star}=(6.7$--8.1)$\times10^8$\,\msun \ in stars older than $\sim$1\,Gyr. According to the star formation history derived by \cite{weisz14}, stars in this age range contribute at least 95\% of the total stellar mass in IC\,10. 
 Our mass estimate is larger than the value of ${\sim}\,4\times10^8$\,\msun \ inferred from integrated light by \cite{Lee2003} and \cite{nersesian19}.

Our structural parameters are consistent with scaling relations of effective radius and effective surface brightness versus magnitude for dwarf galaxies 
\citep[e.g.][]{McConnachie2012}.
The inner Sérsic  index $n\sim1$ agrees with values reported from previous studies of dwarf irregulars based on integrated-light photometry down to $\mu\sim26$\,mag\,arcsec$^{-2}$.
From {\it H}-band data, \cite{Kirby2008} derived 
$0.6\lesssim n\lesssim1.8$ for dwarf irregular galaxies within $\sim$\,10\,Mpc distance,  
while \cite{Young2014} found  $1\lesssim n\lesssim2$ for nearby dwarfs with  luminosity similar to  IC\,10; 
from $g$-band photometry, \cite{Poulain2021} found values of  $0.6\lesssim n\lesssim1.4$ for dwarf 
irregulars with $M_g<-14$. 
However, these comparisons are limited by the shallower depth of these studies, which typically miss the low-density outskirts  explored here.

Studies of the extended stellar component of dwarf galaxies down to $\mu_V\sim30$\,--31\,mag\,arcsec$^{-2}$  have been presented for some nearby objects based on a resolved star approach. \cite{Bellazzini2014} derived the RGB star count profiles of the dwarf irregulars Sextans A and B and found both galaxies to extend for $\sim$\,4\,kpc from their centres along their major axes.
Both profiles show significant changes of slope that can not be reproduced with 
single Sérsic  models, with Sextans A exhibiting an external flattening at galactocentric distances larger than $\sim$\,2.5\,kpc likely due to the presence of a tidal tail. On the other hand, other dwarfs are well fitted by single Sérsic  models, such as UGC\,4879 \citep[$n=1.3$,][]{Bellazzini2011},  NGC\,6822 \citep[$n=1$,][]{Zhang2021}, Sagittarius \citep[$n=1$,][]{Beccari2014}, and Phoenix \citep[$n=0.8$,][] {Battaglia2012}. \cite{Higgs2021} presented an extensive study of structural parameters from RGB count profiles in 12 nearby dwarfs (IC\,10 is not included in their sample). They found the majority of dwarfs to be more suitably described by single component Sérsic  models with 
$n\sim0.4$\,--1.3, with the exception of UGC\,4879, DDO\,210, and WLM for which the preference between one or two components is less clear. Their study shows no extended stellar substructure, which could be signs of a recent merger or accretion of satellites, to be present in any of the studied galaxies. It remains an open question whether IC\,10 represents a rare case of an exceptionally extended stellar envelope, due to a recent merger history with a smaller companion or to its interaction with M31, 
 or whether such features are common in dwarfs but have largely gone undetected due to the lack of imaging that combines both sufficient depth and wide spatial coverage.

Assuming that the extended stellar population of IC\,10 traces its hierarchical assembly history, 
the inferred properties provide an opportunity to confront with the predictions from $N$-body simulations of dwarf galaxies in a $\Lambda$CDM cosmological framework. 
According to simulations, the typical stellar mass fraction accreted by an IC\,10 mass-like galaxy is below 
10\%, although this fraction can be as high as 100\% in some cases \citep[e.g.][]{Deason2022,Tau2025,Cooper2025}.  The accreted stars are expected to be mostly centrally concentrated,
with only a small fraction of them being observable as an extended stellar halo beyond the transition radius, defined as the radius beyond which the ex situ component dominates over the in situ one \citep{Cooper2025,Celiz2025}. 
In the \cite{Cooper2025} simulations, the transition radius for an IC\,10-mass galaxy occurs at values below 10\,kpc, consistent with the radius of $\sim$\,5\,kpc where our profile starts to deviate from the one-component Sérsic  model. Assuming that the second, more extended component in our fit is entirely due to accreted stars, we would end up with an ex situ fraction of $\sim$\,13--26\%, 
(where the large range reflects the uncertainty in the background and foreground contamination to the star count profile), larger than what typically predicted by the simulations. On the other hand, \cite{Tau2025} argue that, for galaxies with the mass of IC\,10, 
the outer stellar haloes are largely dominated by in situ material formed in the inner regions of the galaxies and subsequently ejected into the outskirts during interactions and
merger events with satellite galaxies. In this scenario, the stellar mass accreted by IC\,10 could be lower than what is inferred 
from the fit to the extended component.

\section{\label{sc:summary} Summary}

Using the \Euclid VIS and NISP photometry of resolved stars, we characterised the extended old 
(age $\gtrsim$\,1--2\,Gyr) stellar component of IC\,10 traced by RGB stars out to a SMA of 
about 35\arcmin\ ($\sim$\,8\,kpc), reaching an intrinsic (reddening-corrected) 
surface brightness limit of $\mu_{\HE}\sim29$\,mag\,arcsec$^{-2}$. 

\begin{itemize}

\item While earlier studies predicted that the old stellar population of IC\,10 could extend to very large distances, this work provides the first robust evidence that it does so, reaching such large radii and faint surface brightness levels.

\item Thanks to the large sample of well measured RGB stars, we inferred a robust distance estimate through the RGB tip method. We obtained a distance 
modulus of $(m-M)_0=24.41\pm 0.05$, corresponding to $D=(762\pm 20)$\,kpc, in good agreement with the recent measures by \citet{Sanna2008}, \citet{mcquinn17}, and \citet{dellagli18}.

\item The old stellar component shows a regular, mildly elliptical morphology ($e=0.12$) within 
SMA$\,\sim\,$20\arcmin\ .  At larger radii, asymmetries emerge, particularly a stellar excess to the north-west and a more subtle feature towards the south-east direction. Interestingly, the north-west  excess lies in the same direction as a previously identified $\sim$\,17\,kpc-long \ion{H}{i} structure interpreted as evidence of a past merger, raising the possibility that we are detecting a stellar counterpart to this feature. This possibility ought to be investigated through further study. Another possibility is that the extended features we see are due to the tidal interaction of IC\,10 with M31. 

\item From RGB star counts, we derived the radial surface brightness profile of the old stellar component in IC\,10. The main source of uncertainty is the correction for contamination from background galaxies and foreground stars, which becomes especially critical in the galaxy’s outskirts where the stellar density is low.  
It is difficult to accurately estimate the contamination level, as our CMD analysis reveals that RGB stars could still be present at the edge of the \Euclid\ field, indicating that we have not reached the full extent of the galaxy.

\item Despite the uncertainty in the star counts at large galactocentric distances, the profile presents a clear excess in the outskirts 
compared to a single-component Sérsic  model. 
In fact, the surface brightness profile of IC\,10 is best described by a two-component model: an inner Sérsic  profile (with $n\sim1.0$--$1.1$, $r_{\rm e}\sim1$\,kpc) and an outer component (tentatively fit with an exponential disc) that becomes dominant beyond $\sim$\,5\,kpc, for which the fit parameters are more uncertain. This second,  extended stellar component might trace 
the ex situ material acquired during the past merging history of IC\,10 with smaller satellites or in situ material stripped from its tidal interaction with M31.  

\item From the parameters of our profile fit, we estimated a stellar mass of $M_{\star}=(6.7$--$8.1)\times10^8$\, \msun \ in stars older than $\sim$\,1\,Gyr,  which is higher than previous estimates for IC\,10. 

The extended stellar component accounts for approximately 13--26\% of the total mass in old stars, a higher fraction than typically predicted by cosmological simulations for the ex situ component, although such simulations also allow for accreted fractions of up to 100\% in some cases.

\end{itemize}

Our \Euclid data provide unprecedented depth and spatial resolution over a wide FoV, far surpassing all previous studies of IC\,10. They showcase the transformative impact of deep, wide-field, space-based imaging for studies of nearby galaxies, which will fully flourish in the coming years as subsequent \Euclid data releases enable homogeneous analyses across large galaxy samples and as the Roman Space Telescope begins its scientific operations.

\begin{acknowledgements}
We thank N. Sanna for providing the HST photometric catalog of IC\,10. 
We are grateful to the anonymous referee for their useful comments and suggestions that helped to improve the paper. 
FA acknowledges funding from the Italian INAF MINI GRANT RSN2:C33C24001360005 and 
project PRIN MUR 2022 (code 2022ARWP9C) ``Early Formation and Evolution of Bulge and HalO (EFEBHO)'', 
PI: Marconi, M., funded by European Union--Next Generation EU. AMNF is supported by UK Research and Innovation (UKRI) under the UK government’s Horizon Europe funding guarantee [grant number EP/Z534353/1] and by the UK Science and Technology Facilities Council [grant number ST/Y001281/1].
GB acknowledges support from the Agencia Estatal de Investigación del
Ministerio de Ciencia, Innovación
y Universidades (MCIU/AEI) under grant EN LA FRONTERA DE LA ARQUEOLOGÍA
GALÁCTICA: EVOLUCIÓN DE LA MATERIA LUMINOSA Y OSCURA DE LA VÍA LÁCTEA Y
LAS GALAXIAS ENANAS DEL GRUPO LOCAL EN LA ERA DE GAIA. (FOGALERA) and
the European Regional Development Fund (ERDF) with reference
PID2023-150319NB-C21 / 10.13039/501100011033. 
This research has made use of the NASA/IPAC Extragalactic Database (NED), which is funded 
by the National Aeronautics and Space Administration and operated by the
California Institute of Technology. 
This research has made use of \texttt{AutoProf},
a package for galaxy image photometry \citep{autoprof}, and on Astropy
(http://www.astropy.org), a community-developed core Python package
and an ecosystem of tools and resources for astronomy \citep{astropy_13,astropy_18,astropy_22}.
This research has also made use of \texttt{Photutils}, 
an \texttt{Astropy} package for detection and photometry of astronomical 
sources \citep{photutils}. 
This work has made use of the Early Release Observations (ERO) 
data \citep{EROcite} from the Euclid mission of the European Space Agency
(ESA). 

\AckEC

\end{acknowledgements}

%
%
\bibliography{Euclid}

@ARTICLE{Q1-TP004,
       author = {{Euclid Collaboration: Romelli}, E. and {K\"ummel}, M. and {Dole}, H. and others},
        title = "{Euclid Quick Data Release (Q1). From images to multiwavelength catalogues: the Euclid MERge Processing Function}",
      journal = {A\&A, in press (Euclid Q1 SI), \url{https://doi.org/10.1051/0004-6361/202554586}},
         year = 2025,
        month = mar,
          eid = {arXiv:2503.15305},
        pages = {arXiv:2503.15305},
archivePrefix = {arXiv},
       eprint = {2503.15305},
 primaryClass = {astro-ph.CO},
       adsurl = {https://ui.adsabs.harvard.edu/abs/2025arXiv250315305E}
}

@ARTICLE{EuclidSkyOverview,
author = {{Euclid Collaboration: Mellier}, Y. and {Abdurro'uf} and {Acevedo~Barroso}, J.A. and others},
	title = {Euclid - I. Overview of the Euclid mission},
	DOI= "10.1051/0004-6361/202450810",
	url= "https://doi.org/10.1051/0004-6361/202450810",
	journal = {A\&A},
	year = 2025,
	volume = 697,
	pages = "A1",
}

@ARTICLE{EuclidSkyVIS,
author = {{Euclid Collaboration: Cropper}, M. and {Al-Bahlawan}, A. and {Amiaux}, J. and others},
	title = {Euclid - II. The VIS instrument},
	DOI= "10.1051/0004-6361/202450996",
	url= "https://doi.org/10.1051/0004-6361/202450996",
	journal = {A\&A},
	year = 2025,
	volume = 697,
	pages = "A2",
}

@ARTICLE{EuclidSkyNISP,
author = {{Euclid Collaboration: Jahnke}, K. and {Gillard}, W. and {Schirmer}, M. and others},
	title = {Euclid - III. The NISP Instrument},
	DOI= "10.1051/0004-6361/202450786",
	url= "https://doi.org/10.1051/0004-6361/202450786",
	journal = {A\&A},
	year = 2025,
	volume = 697,
	pages = "A3",
}

@ARTICLE{EROData,
author = {{Cuillandre}, J.-C. and {Bertin}, E. and {Bolzonella}, M. and others},
	title = {Euclid: Early Release Observations – Programme overview and pipeline for compact- and diffuse-emission photometry},
	DOI= "10.1051/0004-6361/202450803",
	url= "https://doi.org/10.1051/0004-6361/202450803",
	journal = {A\&A},
	year = 2025,
	volume = 697,
	pages = "A6",
}

@ARTICLE{EROGalGCs,
        author = {{Massari}, D. and {Dalessandro}, E. and {Erkal}, D. and others},
	title = {Euclid: Early Release Observations – Unveiling the morphology of two Milky Way globular clusters out to their periphery},
	DOI= "10.1051/0004-6361/202449696",
	url= "https://doi.org/10.1051/0004-6361/202449696",
	journal = {A\&A},
	year = 2025,
	volume = 697,
	pages = "A8",
}

@ARTICLE{ERONearbyGals,
author = {{Hunt}, L.K. and {Annibali}, F. and {Cuillandre}, J.-C. and others},
	title = {Euclid: Early Release Observations – Deep anatomy of nearby galaxies },
	DOI= "10.1051/0004-6361/202450781",
	url= "https://doi.org/10.1051/0004-6361/202450781",
	journal = {A\&A},
	year = 2025,
	volume = 697,
	pages = "A9",
}

@misc{EROcite,
author = "{Euclid Early Release Observations}",
howpublished = "\url{https://doi.org/10.57780/esa-qmocze3}",
year = 2024
}

@ARTICLE{Scaramella-EP1,
       author = {{Euclid Collaboration: Scaramella}, R. and {Amiaux}, J. and {Mellier}, Y. and others},
        title = "{Euclid preparation. I. The Euclid Wide Survey}",
      journal = {\aap},
         year = 2022,
        month = jun,
       volume = {662},
          eid = {A112},
        pages = {A112},
          doi = {10.1051/0004-6361/202141938},
archivePrefix = {arXiv},
       eprint = {2108.01201},
 primaryClass = {astro-ph.CO},
       adsurl = {https://ui.adsabs.harvard.edu/abs/2022A&A...662A.112E}
}

@ARTICLE{Howell2026,
       author = {{Howell}, J.~M. and {Ferguson}, A.~M.~N. and {Larsen}, S.~S. and {Lan{\c{c}}on}, A. and {Annibali}, F. and {Cuillandre}, J.-C. and {Hunt}, L.~K. and {Mart{\'\i}nez-Delgado}, D. and {Massari}, D. and {Saifollahi}, T. and {Voggel}, K. and {Altieri}, B. and {Andreon}, S. and {Auricchio}, N. and {Baccigalupi}, C. and {Baldi}, M. and {Bardelli}, S. and {Biviano}, A. and {Branchini}, E. and {Brescia}, M. and {Brinchmann}, J. and {Camera}, S. and {Ca{\~n}as-Herrera}, G. and {Candini}, G.~P. and {Capobianco}, V. and {Carbone}, C. and {Carretero}, J. and {Castellano}, M. and {Castignani}, G. and {Cavuoti}, S. and {Cimatti}, A. and {Colodro-Conde}, C. and {Congedo}, G. and {Conselice}, C.~J. and {Conversi}, L. and {Copin}, Y. and {Courbin}, F. and {Courtois}, H.~M. and {Cropper}, M. and {Da Silva}, A. and {Degaudenzi}, H. and {De Lucia}, G. and {Dubath}, F. and {Duncan}, C.~A.~J. and {Dupac}, X. and {Dusini}, S. and {Escoffier}, S. and {Farina}, M. and {Farinelli}, R. and {Faustini}, F. and {Ferriol}, S. and {Finelli}, F. and {Frailis}, M. and {Franceschi}, E. and {Fumana}, M. and {Galeotta}, S. and {George}, K. and {Gillis}, B. and {Giocoli}, C. and {Gracia-Carpio}, J. and {Grazian}, A. and {Grupp}, F. and {Haugan}, S.~V.~H. and {Hoekstra}, H. and {Holmes}, W. and {Hormuth}, F. and {Hornstrup}, A. and {Jahnke}, K. and {Jhabvala}, M. and {Keih{\"a}nen}, E. and {Kermiche}, S. and {Kubik}, B. and {K{\"u}mmel}, M. and {Kunz}, M. and {Kurki-Suonio}, H. and {Le Brun}, A.~M.~C. and {Le Mignant}, D. and {Ligori}, S. and {Lilje}, P.~B. and {Lindholm}, V. and {Lloro}, I. and {Mainetti}, G. and {Maino}, D. and {Maiorano}, E. and {Mansutti}, O. and {Marggraf}, O. and {Martinelli}, M. and {Martinet}, N. and {Marulli}, F. and {Massey}, R.~J. and {Medinaceli}, E. and {Mei}, S. and {Melchior}, M. and {Mellier}, Y. and {Meneghetti}, M. and {Merlin}, E. and {Meylan}, G. and {Mora}, A. and {Moresco}, M. and {Moscardini}, L. and {Nakajima}, R. and {Neissner}, C. and {Niemi}, S.-M. and {Padilla}, C. and {Paltani}, S. and {Pasian}, F. and {Pedersen}, K. and {Percival}, W.~J. and {Pettorino}, V. and {Pires}, S. and {Polenta}, G. and {Poncet}, M. and {Popa}, L.~A. and {Raison}, F. and {Renzi}, A. and {Rhodes}, J. and {Riccio}, G. and {Romelli}, E. and {Roncarelli}, M. and {Saglia}, R. and {Sakr}, Z. and {Sapone}, D. and {Sartoris}, B. and {Schirmer}, M. and {Schneider}, P. and {Schrabback}, T. and {Secroun}, A. and {Seidel}, G. and {Serrano}, S. and {Simon}, P. and {Sirignano}, C. and {Sirri}, G. and {Skottfelt}, J. and {Stanco}, L. and {Steinwagner}, J. and {Tallada-Cresp{\'\i}}, P. and {Taylor}, A.~N. and {Teplitz}, H.~I. and {Tereno}, I. and {Toft}, S. and {Toledo-Moreo}, R. and {Torradeflot}, F. and {Tutusaus}, I. and {Valenziano}, L. and {Valiviita}, J. and {Vassallo}, T. and {Wang}, Y. and {Weller}, J. and {Zamorani}, G. and {Zinchenko}, I.~A. and {Mart{\'\i}n-Fleitas}, J. and {Scottez}, V.},
        title = "{Euclid: Early Release Observations ─ The star cluster systems of the Local Group dwarf galaxies IC 10 and NGC 6822}",
      journal = {\aap},
         year = 2026,
        month = feb,
       volume = {706},
          eid = {A185},
        pages = {A185},
          doi = {10.1051/0004-6361/202556953},
archivePrefix = {arXiv},
       eprint = {2509.10440},
 primaryClass = {astro-ph.GA},
       adsurl = {https://ui.adsabs.harvard.edu/abs/2026A&A...706A.185H}
}

@ARTICLE{Libralato24,
       author = {{Libralato}, M. and {Bedin}, L.~R. and {Griggio}, M. and others},
        title = "{Euclid: High-precision imaging astrometry and photometry from Early Release Observations: I. Internal kinematics of NGC6397 by combining Euclid and Gaia data}",
      journal = {\aap},
         year = 2024,
        month = dec,
       volume = {692},
          eid = {A96},
        pages = {A96},
          doi = {10.1051/0004-6361/202452295},
archivePrefix = {arXiv},
       eprint = {2411.02487},
 primaryClass = {astro-ph.SR},
       adsurl = {https://ui.adsabs.harvard.edu/abs/2024A&A...692A..96L}
}

@ARTICLE{Laureijs11,
       author = {{Laureijs}, R. and {Amiaux}, J. and {Arduini}, S. and {Augu{\`e}res}, J. -L. and {Brinchmann}, J. and {Cole}, R. and {Cropper}, M. and {Dabin}, C. and {Duvet}, L. and {Ealet}, A. and {Garilli}, B. and {Gondoin}, P. and {Guzzo}, L. and {Hoar}, J. and {Hoekstra}, H. and {Holmes}, R. and {Kitching}, T. and {Maciaszek}, T. and {Mellier}, Y. and {Pasian}, F. and {Percival}, W. and {Rhodes}, J. and {Saavedra Criado}, G. and {Sauvage}, M. and {Scaramella}, R. and {Valenziano}, L. and {Warren}, S. and {Bender}, R. and {Castander}, F. and {Cimatti}, A. and {Le F{\`e}vre}, O. and {Kurki-Suonio}, H. and {Levi}, M. and {Lilje}, P. and {Meylan}, G. and {Nichol}, R. and {Pedersen}, K. and {Popa}, V. and {Rebolo Lopez}, R. and {Rix}, H. -W. and {Rottgering}, H. and {Zeilinger}, W. and {Grupp}, F. and {Hudelot}, P. and {Massey}, R. and {Meneghetti}, M. and {Miller}, L. and {Paltani}, S. and {Paulin-Henriksson}, S. and {Pires}, S. and {Saxton}, C. and {Schrabback}, T. and {Seidel}, G. and {Walsh}, J. and {Aghanim}, N. and {Amendola}, L. and {Bartlett}, J. and {Baccigalupi}, C. and {Beaulieu}, J. -P. and {Benabed}, K. and {Cuby}, J. -G. and {Elbaz}, D. and {Fosalba}, P. and {Gavazzi}, G. and {Helmi}, A. and {Hook}, I. and {Irwin}, M. and {Kneib}, J. -P. and {Kunz}, M. and {Mannucci}, F. and {Moscardini}, L. and {Tao}, C. and {Teyssier}, R. and {Weller}, J. and {Zamorani}, G. and {Zapatero Osorio}, M.~R. and {Boulade}, O. and {Foumond}, J.~J. and {Di Giorgio}, A. and {Guttridge}, P. and {James}, A. and {Kemp}, M. and {Martignac}, J. and {Spencer}, A. and {Walton}, D. and {Bl{\"u}mchen}, T. and {Bonoli}, C. and {Bortoletto}, F. and {Cerna}, C. and {Corcione}, L. and {Fabron}, C. and {Jahnke}, K. and {Ligori}, S. and {Madrid}, F. and {Martin}, L. and {Morgante}, G. and {Pamplona}, T. and {Prieto}, E. and {Riva}, M. and {Toledo}, R. and {Trifoglio}, M. and {Zerbi}, F. and {Abdalla}, F. and {Douspis}, M. and {Grenet}, C. and {Borgani}, S. and {Bouwens}, R. and {Courbin}, F. and {Delouis}, J. -M. and {Dubath}, P. and {Fontana}, A. and {Frailis}, M. and {Grazian}, A. and {Koppenh{\"o}fer}, J. and {Mansutti}, O. and {Melchior}, M. and {Mignoli}, M. and {Mohr}, J. and {Neissner}, C. and {Noddle}, K. and {Poncet}, M. and {Scodeggio}, M. and {Serrano}, S. and {Shane}, N. and {Starck}, J. -L. and {Surace}, C. and {Taylor}, A. and {Verdoes-Kleijn}, G. and {Vuerli}, C. and {Williams}, O.~R. and {Zacchei}, A. and {Altieri}, B. and {Escudero Sanz}, I. and {Kohley}, R. and {Oosterbroek}, T. and {Astier}, P. and {Bacon}, D. and {Bardelli}, S. and {Baugh}, C. and {Bellagamba}, F. and {Benoist}, C. and {Bianchi}, D. and {Biviano}, A. and {Branchini}, E. and {Carbone}, C. and {Cardone}, V. and {Clements}, D. and {Colombi}, S. and {Conselice}, C. and {Cresci}, G. and {Deacon}, N. and {Dunlop}, J. and {Fedeli}, C. and {Fontanot}, F. and {Franzetti}, P. and {Giocoli}, C. and {Garcia-Bellido}, J. and {Gow}, J. and {Heavens}, A. and {Hewett}, P. and {Heymans}, C. and {Holland}, A. and {Huang}, Z. and {Ilbert}, O. and {Joachimi}, B. and {Jennins}, E. and {Kerins}, E. and {Kiessling}, A. and {Kirk}, D. and {Kotak}, R. and {Krause}, O. and {Lahav}, O. and {van Leeuwen}, F. and {Lesgourgues}, J. and {Lombardi}, M. and {Magliocchetti}, M. and {Maguire}, K. and {Majerotto}, E. and {Maoli}, R. and {Marulli}, F. and {Maurogordato}, S. and {McCracken}, H. and {McLure}, R. and {Melchiorri}, A. and {Merson}, A. and {Moresco}, M. and {Nonino}, M. and {Norberg}, P. and {Peacock}, J. and {Pello}, R. and {Penny}, M. and {Pettorino}, V. and {Di Porto}, C. and {Pozzetti}, L. and {Quercellini}, C. and {Radovich}, M. and {Rassat}, A. and {Roche}, N. and {Ronayette}, S. and {Rossetti}, E. and {Sartoris}, B. and {Schneider}, P. and {Semboloni}, E. and {Serjeant}, S. and {Simpson}, F. and {Skordis}, C. and {Smadja}, G. and {Smartt}, S. and {Spano}, P. and {Spiro}, S. and {Sullivan}, M. and {Tilquin}, A. and {Trotta}, R. and {Verde}, L. and {Wang}, Y. and {Williger}, G. and {Zhao}, G. and {Zoubian}, J. and {Zucca}, E.},
        title = "{Euclid Definition Study Report}",
      journal = {ESA/SRE(2011)12},
         year = 2011,
        month = oct,
          eid = {arXiv:1110.3193},
        pages = {arXiv:1110.3193},
          doi = {10.48550/arXiv.1110.3193},
archivePrefix = {arXiv},
       eprint = {1110.3193},
 primaryClass = {astro-ph.CO},
       adsurl = {https://ui.adsabs.harvard.edu/abs/2011arXiv1110.3193L}
}

@ARTICLE{XPSP23,
       author = {{Gaia Collaboration: Montegriffo}, P. and {Bellazzini}, M. and {De Angeli}, F. and {Andrae}, R. and {Barstow}, M.~A. and {Bossini}, D. and {Bragaglia}, A. and {Burgess}, P.~W. and {Cacciari}, C. and {Carrasco}, J.~M. and {Chornay}, N. and {Delchambre}, L. and {Evans}, D.~W. and {Fouesneau}, M. and {Fr{\'e}mat}, Y. and {Garabato}, D. and {Jordi}, C. and {Manteiga}, M. and {Massari}, D. and {Palaversa}, L. and {Pancino}, E. and {Riello}, M. and {Ruz Mieres}, D. and {Sanna}, N. and {Santove{\~n}a}, R. and {Sordo}, R. and {Vallenari}, A. and {Walton}, N.~A. and {Brown}, A.~G.~A. and {Prusti}, T. and {de Bruijne}, J.~H.~J. and {Arenou}, F. and {Babusiaux}, C. and {Biermann}, M. and {Creevey}, O.~L. and {Ducourant}, C. and {Eyer}, L. and {Guerra}, R. and {Hutton}, A. and {Klioner}, S.~A. and {Lammers}, U.~L. and {Lindegren}, L. and {Luri}, X. and {Mignard}, F. and {Panem}, C. and {Pourbaix}, D. and {Randich}, S. and {Sartoretti}, P. and {Soubiran}, C. and {Tanga}, P. and {Bailer-Jones}, C.~A.~L. and {Bastian}, U. and {Drimmel}, R. and {Jansen}, F. and {Katz}, D. and {Lattanzi}, M.~G. and {van Leeuwen}, F. and {Bakker}, J. and {Casta{\~n}eda}, J. and {Fabricius}, C. and {Galluccio}, L. and {Guerrier}, A. and {Heiter}, U. and {Masana}, E. and {Messineo}, R. and {Mowlavi}, N. and {Nicolas}, C. and {Nienartowicz}, K. and {Pailler}, F. and {Panuzzo}, P. and {Riclet}, F. and {Roux}, W. and {Seabroke}, G.~M. and {Th{\'e}venin}, F. and {Gracia-Abril}, G. and {Portell}, J. and {Teyssier}, D. and {Altmann}, M. and {Audard}, M. and {Bellas-Velidis}, I. and {Benson}, K. and {Berthier}, J. and {Blomme}, R. and {Busonero}, D. and {Busso}, G. and {C{\'a}novas}, H. and {Carry}, B. and {Cellino}, A. and {Cheek}, N. and {Clementini}, G. and {Damerdji}, Y. and {Davidson}, M. and {de Teodoro}, P. and {Nu{\~n}ez Campos}, M. and {Dell'Oro}, A. and {Esquej}, P. and {Fern{\'a}ndez-Hern{\'a}ndez}, J. and {Fraile}, E. and {Garc{\'\i}a-Lario}, P. and {Gosset}, E. and {Haigron}, R. and {Halbwachs}, J.-L. and {Hambly}, N.~C. and {Harrison}, D.~L. and {Hern{\'a}ndez}, J. and {Hestroffer}, D. and {Hodgkin}, S.~T. and {Holl}, B. and {Jan{\ss}en}, K. and {Jevardat de Fombelle}, G. and {Jordan}, S. and {Krone-Martins}, A. and {Lanzafame}, A.~C. and {L{\"o}ffler}, W. and {Marchal}, O. and {Marrese}, P.~M. and {Moitinho}, A. and {Muinonen}, K. and {Osborne}, P. and {Pauwels}, T. and {Recio-Blanco}, A. and {Reyl{\'e}}, C. and {Rimoldini}, L. and {Roegiers}, T. and {Rybizki}, J. and {Sarro}, L.~M. and {Siopis}, C. and {Smith}, M. and {Sozzetti}, A. and {Utrilla}, E. and {van Leeuwen}, M. and {Abbas}, U. and {{\'A}brah{\'a}m}, P. and {Abreu Aramburu}, A. and {Aerts}, C. and {Aguado}, J.~J. and {Ajaj}, M. and {Aldea-Montero}, F. and {Altavilla}, G. and {{\'A}lvarez}, M.~A. and {Alves}, J. and {Anderson}, R.~I. and {Anglada Varela}, E. and {Antoja}, T. and {Baines}, D. and {Baker}, S.~G. and {Balaguer-N{\'u}{\~n}ez}, L. and {Balbinot}, E. and {Balog}, Z. and {Barache}, C. and {Barbato}, D. and {Barros}, M. and {Bartolom{\'e}}, S. and {Bassilana}, J.-L. and {Bauchet}, N. and {Becciani}, U. and {Berihuete}, A. and {Bernet}, M. and {Bertone}, S. and {Bianchi}, L. and {Binnenfeld}, A. and {Blanco-Cuaresma}, S. and {Boch}, T. and {Bombrun}, A. and {Bouquillon}, S. and {Bramante}, L. and {Breedt}, E. and {Bressan}, A. and {Brouillet}, N. and {Brugaletta}, E. and {Bucciarelli}, B. and {Burlacu}, A. and {Butkevich}, A.~G. and {Buzzi}, R. and {Caffau}, E. and {Cancelliere}, R. and {Cantat-Gaudin}, T. and {Carballo}, R. and {Carlucci}, T. and {Carnerero}, M.~I. and {Casamiquela}, L. and {Castellani}, M. and {Castro-Ginard}, A. and {Chaoul}, L. and {Charlot}, P. and {Chemin}, L. and {Chiaramida}, V. and {Chiavassa}, A. and {Comoretto}, G. and {Contursi}, G. and {Cooper}, W.~J. and {Cornez}, T. and {Cowell}, S. and {Crifo}, F. and {Cropper}, M. and {Crosta}, M. and {Crowley}, C. and {Dafonte}, C. and {Dapergolas}, A.},
        title = "{Gaia Data Release 3. The Galaxy in your preferred colours: Synthetic photometry from Gaia low-resolution spectra}",
      journal = {\aap},
         year = 2023,
        month = jun,
       volume = {674},
          eid = {A33},
        pages = {A33},
          doi = {10.1051/0004-6361/202243709},
archivePrefix = {arXiv},
       eprint = {2206.06215},
 primaryClass = {astro-ph.SR},
       adsurl = {https://ui.adsabs.harvard.edu/abs/2023A&A...674A..33G}
}

@ARTICLE{dellagli18,
       author = {{Dell'Agli}, F. and {Di Criscienzo}, M. and {Ventura}, P. and {Limongi}, M. and {Garc{\'\i}a-Hern{\'a}ndez}, D.~A. and {Marini}, E. and {Rossi}, C.},
        title = "{Evolved stars in the Local Group galaxies - II. AGB, RSG stars, and dust production in IC10}",
      journal = {\mnras},
         year = 2018,
        month = oct,
       volume = {479},
       number = {4},
        pages = {5035-5048},
          doi = {10.1093/mnras/sty1614},
archivePrefix = {arXiv},
       eprint = {1806.04160},
 primaryClass = {astro-ph.SR},
       adsurl = {https://ui.adsabs.harvard.edu/abs/2018MNRAS.479.5035D}
}

@ARTICLE{bp24,
       author = {{Bellazzini}, M. and {Pascale}, R.},
        title = "{The red giant branch tip in the SDSS, PS1, JWST, NGRST, and Euclid photometric systems: Calibration in optical passbands using Gaia DR3 synthetic photometry}",
      journal = {\aap},
         year = 2024,
        month = nov,
       volume = {691},
          eid = {A42},
        pages = {A42},
          doi = {10.1051/0004-6361/202449575},
archivePrefix = {arXiv},
       eprint = {2406.04781},
 primaryClass = {astro-ph.GA},
       adsurl = {https://ui.adsabs.harvard.edu/abs/2024A&A...691A..42B}
}

@ARTICLE{hoyt23,
       author = {{Hoyt}, Taylor J.},
        title = "{Sub-per-cent determination of the brightness at the tip of the red giant branch in the Magellanic Clouds}",
      journal = {Nature Astronomy},
         year = 2023,
        month = may,
       volume = {7},
        pages = {590-601},
          doi = {10.1038/s41550-023-01913-1},
       adsurl = {https://ui.adsabs.harvard.edu/abs/2023NatAs...7..590H}
}

@ARTICLE{serenelli17,
       author = {{Serenelli}, A. and {Weiss}, A. and {Cassisi}, S. and {Salaris}, M. and {Pietrinferni}, A.},
        title = "{The brightness of the red giant branch tip. Theoretical framework, a set of reference models, and predicted observables}",
      journal = {\aap},
         year = 2017,
        month = oct,
       volume = {606},
          eid = {A33},
        pages = {A33},
          doi = {10.1051/0004-6361/201731004},
archivePrefix = {arXiv},
       eprint = {1706.09910},
 primaryClass = {astro-ph.SR},
       adsurl = {https://ui.adsabs.harvard.edu/abs/2017A&A...606A..33S}

}

@ARTICLE{mbtip08,
       author = {{Bellazzini}, M.},
        title = "{The Tip of the Red Giant Branch}",
      journal = {\memsai},
         year = 2008,
        month = jan,
       volume = {79},
        pages = {440},
          doi = {10.48550/arXiv.0711.2016},
archivePrefix = {arXiv},
       eprint = {0711.2016},
 primaryClass = {astro-ph},
       adsurl = {https://ui.adsabs.harvard.edu/abs/2008MmSAI..79..440B}
}

@ARTICLE{madore95,
       author = {{Madore}, Barry F. and {Freedman}, Wendy L.},
        title = "{The Tip of the Red Giant Branch as a Distance Indicator for Resolved Galaxies.II.Computer Simulations}",
      journal = {\aj},
         year = 1995,
        month = apr,
       volume = {109},
        pages = {1645},
          doi = {10.1086/117391},
       adsurl = {https://ui.adsabs.harvard.edu/abs/1995AJ....109.1645M}
}

@ARTICLE{madore20,
       author = {{Madore}, Barry F. and {Freedman}, Wendy L.},
        title = "{Mathematical Underpinnings of the Multiwavelength Structure of the Tip of the Red Giant Branch}",
      journal = {\aj},
         year = 2020,
        month = oct,
       volume = {160},
       number = {4},
          eid = {170},
        pages = {170},
          doi = {10.3847/1538-3881/abab9a},
archivePrefix = {arXiv},
       eprint = {2008.00341},
 primaryClass = {astro-ph.GA},
       adsurl = {https://ui.adsabs.harvard.edu/abs/2020AJ....160..170M}
}

@ARTICLE{lee93,
       author = {{Lee}, Myung Gyoon and {Freedman}, Wendy L. and {Madore}, Barry F.},
        title = "{The Tip of the Red Giant Branch as a Distance Indicator for Resolved Galaxies}",
      journal = {\apj},
         year = 1993,
        month = nov,
       volume = {417},
        pages = {553},
          doi = {10.1086/173334},
       adsurl = {https://ui.adsabs.harvard.edu/abs/1993ApJ...417..553L}
}

@ARTICLE{salacas98,
       author = {{Salaris}, Maurizio and {Cassisi}, Santi},
        title = "{A new analysis of the red giant branch `tip' distance scale and the value of the Hubble constant}",
      journal = {\mnras},
         year = 1998,
        month = jul,
       volume = {298},
       number = {1},
        pages = {166-178},
          doi = {10.1046/j.1365-8711.1998.01598.x},
archivePrefix = {arXiv},
       eprint = {astro-ph/9803103},
 primaryClass = {astro-ph},
       adsurl = {https://ui.adsabs.harvard.edu/abs/1998MNRAS.298..166S}
}

@ARTICLE{gordon09,
       author = {{Gordon}, Karl D. and {Cartledge}, Stefan and {Clayton}, Geoffrey C.},
        title = "{FUSE Measurements of Far-Ultraviolet Extinction. III. The Dependence on R(V) and Discrete Feature Limits from 75 Galactic Sightlines}",
      journal = {\apj},
         year = 2009,
        month = nov,
       volume = {705},
       number = {2},
        pages = {1320-1335},
          doi = {10.1088/0004-637X/705/2/1320},
archivePrefix = {arXiv},
       eprint = {0909.3087},
 primaryClass = {astro-ph.GA},
       adsurl = {https://ui.adsabs.harvard.edu/abs/2009ApJ...705.1320G}
}

@ARTICLE{gordon21,
       author = {{Gordon}, Karl D. and {Misselt}, Karl A. and {Bouwman}, Jeroen and {Clayton}, Geoffrey C. and {Decleir}, Marjorie and {Hines}, Dean C. and {Pendleton}, Yvonne and {Rieke}, George and {Smith}, J.~D.~T. and {Whittet}, D.~C.~B.},
        title = "{Milky Way Mid-Infrared Spitzer Spectroscopic Extinction Curves: Continuum and Silicate Features}",
      journal = {\apj},
         year = 2021,
        month = jul,
       volume = {916},
       number = {1},
          eid = {33},
        pages = {33},
          doi = {10.3847/1538-4357/ac00b7},
archivePrefix = {arXiv},
       eprint = {2105.05087},
 primaryClass = {astro-ph.GA},
       adsurl = {https://ui.adsabs.harvard.edu/abs/2021ApJ...916...33G}
}

@ARTICLE{decleir22,
       author = {{Decleir}, Marjorie and {Gordon}, Karl D. and {Andrews}, Jennifer E. and {Clayton}, Geoffrey C. and {Cushing}, Michael C. and {Misselt}, Karl A. and {Pendleton}, Yvonne and {Rayner}, John and {Vacca}, William D. and {Whittet}, D.~C.~B.},
        title = "{SpeX Near-infrared Spectroscopic Extinction Curves in the Milky Way}",
      journal = {\apj},
         year = 2022,
        month = may,
       volume = {930},
       number = {1},
          eid = {15},
        pages = {15},
          doi = {10.3847/1538-4357/ac5dbe},
archivePrefix = {arXiv},
       eprint = {2204.13716},
 primaryClass = {astro-ph.GA},
       adsurl = {https://ui.adsabs.harvard.edu/abs/2022ApJ...930...15D}
}

@ARTICLE{fitzpatrick19,
       author = {{Fitzpatrick}, E.~L. and {Massa}, Derck and {Gordon}, Karl D. and {Bohlin}, Ralph and {Clayton}, Geoffrey C.},
        title = "{An Analysis of the Shapes of Interstellar Extinction Curves. VII. Milky Way Spectrophotometric Optical-through-ultraviolet Extinction and Its R-dependence}",
      journal = {\apj},
         year = 2019,
        month = dec,
       volume = {886},
       number = {2},
          eid = {108},
        pages = {108},
          doi = {10.3847/1538-4357/ab4c3a},
archivePrefix = {arXiv},
       eprint = {1910.08852},
 primaryClass = {astro-ph.GA},
       adsurl = {https://ui.adsabs.harvard.edu/abs/2019ApJ...886..108F}
}

@ARTICLE{trilegal05,
       author = {{Girardi}, L. and {Groenewegen}, M.~A.~T. and {Hatziminaoglou}, E. and {da Costa}, L.},
        title = "{Star counts in the Galaxy. Simulating from very deep to very shallow photometric surveys with the TRILEGAL code}",
      journal = {\aap},
         year = 2005,
        month = jun,
       volume = {436},
       number = {3},
        pages = {895-915},
          doi = {10.1051/0004-6361:20042352},
archivePrefix = {arXiv},
       eprint = {astro-ph/0504047},
 primaryClass = {astro-ph},
       adsurl = {https://ui.adsabs.harvard.edu/abs/2005A&A...436..895G}
}

@INPROCEEDINGS{trilegal12,
       author = {{Girardi}, L{\'e}o and {Barbieri}, Mauro and {Groenewegen}, Martin A.~T. and {Marigo}, Paola and {Bressan}, Alessandro and {Rocha-Pinto}, Helio J. and {Santiago}, Bas{\'\i}lio X. and {Camargo}, Julio I.~B. and {da Costa}, Luiz N.},
        title = "{TRILEGAL, a TRIdimensional modeL of thE GALaxy: Status and Future}",
    booktitle = {Red Giants as Probes of the Structure and Evolution of the Milky Way},
         year = 2012,
       series = {Astrophysics and Space Science Proceedings},
       volume = {26},
        month = jan,
        pages = {165},
          doi = {10.1007/978-3-642-18418-5_17},
       adsurl = {https://ui.adsabs.harvard.edu/abs/2012ASSP...26..165G}
}

@ARTICLE{Sanna2008,
       author = {{Sanna}, N. and {Bono}, G. and {Stetson}, P.~B. and {Monelli}, M. and {Pietrinferni}, A. and {Drozdovsky}, I. and {Caputo}, F. and {Cassisi}, S. and {Gennaro}, M. and {Prada Moroni}, P.~G. and {Buonanno}, R. and {Corsi}, C.~E. and {Degl'Innocenti}, S. and {Ferraro}, I. and {Iannicola}, G. and {Nonino}, M. and {Pulone}, L. and {Romaniello}, M. and {Walker}, A.~R.},
        title = "{On the Distance and Reddening of the Starburst Galaxy IC 10}",
      journal = {\apjl},
         year = 2008,
        month = dec,
       volume = {688},
       number = {2},
        pages = {L69},
          doi = {10.1086/595551},
archivePrefix = {arXiv},
       eprint = {0810.1210},
 primaryClass = {astro-ph},
       adsurl = {https://ui.adsabs.harvard.edu/abs/2008ApJ...688L..69S}
}

@ARTICLE{Sanna2010,
       author = {{Sanna}, N. and {Bono}, G. and {Stetson}, P.~B. and {Ferraro}, I. and {Monelli}, M. and {Nonino}, M. and {Prada Moroni}, P.~G. and {Bresolin}, R. and {Buonanno}, R. and {Caputo}, F. and {Cignoni}, M. and {Degl'Innocenti}, S. and {Iannicola}, G. and {Matsunaga}, N. and {Pietrinferni}, A. and {Romaniello}, M. and {Storm}, J. and {Walker}, A.~R.},
        title = "{On the Radial Extent of the Dwarf Irregular Galaxy IC10}",
      journal = {\apjl},
         year = 2010,
        month = oct,
       volume = {722},
       number = {2},
        pages = {L244-L249},
          doi = {10.1088/2041-8205/722/2/L244},
archivePrefix = {arXiv},
       eprint = {1009.3917},
 primaryClass = {astro-ph.SR},
       adsurl = {https://ui.adsabs.harvard.edu/abs/2010ApJ...722L.244S}
}

@ARTICLE{gordon23,
       author = {{Gordon}, Karl D. and {Clayton}, Geoffrey C. and {Decleir}, Marjorie and {Fitzpatrick}, E.~L. and {Massa}, Derck and {Misselt}, Karl A. and {Tollerud}, Erik J.},
        title = "{One Relation for All Wavelengths: The Far-ultraviolet to Mid-infrared Milky Way Spectroscopic R(V)-dependent Dust Extinction Relationship}",
      journal = {\apj},
         year = 2023,
        month = jun,
       volume = {950},
       number = {2},
          eid = {86},
        pages = {86},
          doi = {10.3847/1538-4357/accb59},
archivePrefix = {arXiv},
       eprint = {2304.01991},
 primaryClass = {astro-ph.GA},
       adsurl = {https://ui.adsabs.harvard.edu/abs/2023ApJ...950...86G}
}

@ARTICLE{McConnachie2021,
       author = {{McConnachie}, Alan W. and {Higgs}, Clare R. and {Thomas}, Guillaume F. and {Venn}, Kim A. and {C{\^o}t{\'e}}, Patrick and {Battaglia}, Giuseppina and {Lewis}, Geraint F.},
        title = "{Solo dwarfs - III. Exploring the orbital origins of isolated Local Group galaxies with Gaia Data Release 2}",
      journal = {\mnras},
         year = 2021,
        month = feb,
       volume = {501},
       number = {2},
        pages = {2363-2377},
          doi = {10.1093/mnras/staa3740},
archivePrefix = {arXiv},
       eprint = {2012.01586},
 primaryClass = {astro-ph.GA},
       adsurl = {https://ui.adsabs.harvard.edu/abs/2021MNRAS.501.2363M}
}

@ARTICLE{Madden2013,
       author = {{Madden}, S.~C. and {R{\'e}my-Ruyer}, A. and {Galametz}, M. and {Cormier}, D. and {Lebouteiller}, V. and {Galliano}, F. and {Hony}, S. and {Bendo}, G.~J. and {Smith}, M.~W.~L. and {Pohlen}, M. and {Roussel}, H. and {Sauvage}, M. and {Wu}, R. and {Sturm}, E. and {Poglitsch}, A. and {Contursi}, A. and {Doublier}, V. and {Baes}, M. and {Barlow}, M.~J. and {Boselli}, A. and {Boquien}, M. and {Carlson}, L.~R. and {Ciesla}, L. and {Cooray}, A. and {Cortese}, L. and {de Looze}, I. and {Irwin}, J.~A. and {Isaak}, K. and {Kamenetzky}, J. and {Karczewski}, O. {\L}. and {Lu}, N. and {MacHattie}, J.~A. and {O'Halloran}, B. and {Parkin}, T.~J. and {Rangwala}, N. and {Schirm}, M.~R.~P. and {Schulz}, B. and {Spinoglio}, L. and {Vaccari}, M. and {Wilson}, C.~D. and {Wozniak}, H.},
        title = "{An Overview of the Dwarf Galaxy Survey}",
      journal = {\pasp},
         year = 2013,
        month = jun,
       volume = {125},
       number = {928},
        pages = {600},
          doi = {10.1086/671138},
archivePrefix = {arXiv},
       eprint = {1305.2628},
 primaryClass = {astro-ph.GA},
       adsurl = {https://ui.adsabs.harvard.edu/abs/2013PASP..125..600M}
}

@ARTICLE{Caffau2011,
       author = {{Caffau}, E. and {Ludwig}, H. -G. and {Steffen}, M. and {Freytag}, B. and {Bonifacio}, P.},
        title = "{Solar Chemical Abundances Determined with a CO5BOLD 3D Model Atmosphere}",
      journal = {\solphys},
         year = 2011,
        month = feb,
       volume = {268},
       number = {2},
        pages = {255-269},
          doi = {10.1007/s11207-010-9541-4},
archivePrefix = {arXiv},
       eprint = {1003.1190},
 primaryClass = {astro-ph.SR},
       adsurl = {https://ui.adsabs.harvard.edu/abs/2011SoPh..268..255C}
}

@ARTICLE{Magrini2009,
       author = {{Magrini}, Laura and {Gon{\c{c}}alves}, Denise R.},
        title = "{IC10: the history of the nearest starburst galaxy through its Planetary Nebula and HII region populations}",
      journal = {\mnras},
         year = 2009,
        month = sep,
       volume = {398},
       number = {1},
        pages = {280-292},
          doi = {10.1111/j.1365-2966.2009.15124.x},
archivePrefix = {arXiv},
       eprint = {0905.3630},
 primaryClass = {astro-ph.CO},
       adsurl = {https://ui.adsabs.harvard.edu/abs/2009MNRAS.398..280M}
}

@ARTICLE{Bressan2012,
       author = {{Bressan}, Alessandro and {Marigo}, Paola and {Girardi}, L{\'e}o. and {Salasnich}, Bernardo and {Dal Cero}, Claudia and {Rubele}, Stefano and {Nanni}, Ambra},
        title = "{PARSEC: stellar tracks and isochrones with the PAdova and TRieste Stellar Evolution Code}",
      journal = {\mnras},
         year = 2012,
        month = nov,
       volume = {427},
       number = {1},
        pages = {127-145},
          doi = {10.1111/j.1365-2966.2012.21948.x},
archivePrefix = {arXiv},
       eprint = {1208.4498},
 primaryClass = {astro-ph.SR},
       adsurl = {https://ui.adsabs.harvard.edu/abs/2012MNRAS.427..127B}
}

@software{photutils,
       author = {{Bradley}, Larry and {Sip{\H{o}}cz}, Brigitta and {Robitaille}, Thomas and {Tollerud}, Erik and {Vin{\'\i}cius}, Z{\'e} and {Deil}, Christoph and {Barbary}, Kyle and {Wilson}, Tom J and {Busko}, Ivo and {Donath}, Axel and {G{\"u}nther}, Hans Moritz and {Cara}, Mihai and {Lim}, P.~L. and {Me{\ss}linger}, Sebastian and {Conseil}, Simon and {Burnett}, Zach and {Bostroem}, Azalee and {Droettboom}, Michael and {Bray}, E.~M. and {Andersen Bratholm}, Lars and {Jamieson}, William and {Ginsburg}, Adam and {Barentsen}, Geert and {Craig}, Matt and {Morris}, Brett M. and {Perrin}, Marshall and {Rathi}, Shivangee and {Pascual}, Sergio and {Perren}, Gabriel and {Georgiev}, Iskren Y.},
        title = "{astropy/photutils: 1.10.0}",
         year = 2023,
        month = nov,
          eid = {10.5281/zenodo.1035865},
          doi = {10.5281/zenodo.1035865},
      version = {1.10.0},
    publisher = {Zenodo},
       adsurl = {https://ui.adsabs.harvard.edu/abs/2023zndo...1035865B}
}

@ARTICLE{Smercina23,
       author = {{Smercina}, Adam and {Dalcanton}, Julianne J. and {Williams}, Benjamin F. and {Durbin}, Meredith J. and {Lazzarini}, Margaret and {Bell}, Eric F. and {Choi}, Yumi and {Dolphin}, Andrew and {Gilbert}, Karoline and {Guhathakurta}, Puragra and {Koch}, Eric W. and {Quirk}, Amanda C.~N. and {Rix}, Hans-Walter and {Rosolowsky}, Erik and {Seth}, Anil and {Skillman}, Evan and {Weisz}, Daniel R.},
        title = "{The Panchromatic Hubble Andromeda Treasury: Triangulum Extended Region (PHATTER). V. The Structure of M33 in Resolved Stellar Populations}",
      journal = {\apj},
         year = 2023,
        month = nov,
       volume = {957},
       number = {1},
          eid = {3},
        pages = {3},
          doi = {10.3847/1538-4357/acf3e8},
archivePrefix = {arXiv},
       eprint = {2308.11618},
 primaryClass = {astro-ph.GA},
       adsurl = {https://ui.adsabs.harvard.edu/abs/2023ApJ...957....3S}
}

@ARTICLE{autoprof,
       author = {{Stone}, Connor J. and {Arora}, Nikhil and {Courteau}, St{\'e}phane and {Cuillandre}, Jean-Charles},
        title = "{AutoProf - I. An automated non-parametric light profile pipeline for modern galaxy surveys}",
      journal = {\mnras},
         year = 2021,
        month = dec,
       volume = {508},
       number = {2},
        pages = {1870-1887},
          doi = {10.1093/mnras/stab2709},
archivePrefix = {arXiv},
       eprint = {2106.13809},
 primaryClass = {astro-ph.GA},
       adsurl = {https://ui.adsabs.harvard.edu/abs/2021MNRAS.508.1870S}
}

@ARTICLE{Gaia2016,
       author = {{Gaia Collaboration: Prusti}, T. and {de Bruijne}, J.~H.~J. and {Brown}, A.~G.~A. and {Vallenari}, A. and {Babusiaux}, C. and {Bailer-Jones}, C.~A.~L. and {Bastian}, U. and {Biermann}, M. and {Evans}, D.~W. and {Eyer}, L. and {Jansen}, F. and {Jordi}, C. and {Klioner}, S.~A. and {Lammers}, U. and {Lindegren}, L. and {Luri}, X. and {Mignard}, F. and {Milligan}, D.~J. and {Panem}, C. and {Poinsignon}, V. and {Pourbaix}, D. and {Randich}, S. and {Sarri}, G. and {Sartoretti}, P. and {Siddiqui}, H.~I. and {Soubiran}, C. and {Valette}, V. and {van Leeuwen}, F. and {Walton}, N.~A. and {Aerts}, C. and {Arenou}, F. and {Cropper}, M. and {Drimmel}, R. and {H{\o}g}, E. and {Katz}, D. and {Lattanzi}, M.~G. and {O'Mullane}, W. and {Grebel}, E.~K. and {Holland}, A.~D. and {Huc}, C. and {Passot}, X. and {Bramante}, L. and {Cacciari}, C. and {Casta{\~n}eda}, J. and {Chaoul}, L. and {Cheek}, N. and {De Angeli}, F. and {Fabricius}, C. and {Guerra}, R. and {Hern{\'a}ndez}, J. and {Jean-Antoine-Piccolo}, A. and {Masana}, E. and {Messineo}, R. and {Mowlavi}, N. and {Nienartowicz}, K. and {Ord{\'o}{\~n}ez-Blanco}, D. and {Panuzzo}, P. and {Portell}, J. and {Richards}, P.~J. and {Riello}, M. and {Seabroke}, G.~M. and {Tanga}, P. and {Th{\'e}venin}, F. and {Torra}, J. and {Els}, S.~G. and {Gracia-Abril}, G. and {Comoretto}, G. and {Garcia-Reinaldos}, M. and {Lock}, T. and {Mercier}, E. and {Altmann}, M. and {Andrae}, R. and {Astraatmadja}, T.~L. and {Bellas-Velidis}, I. and {Benson}, K. and {Berthier}, J. and {Blomme}, R. and {Busso}, G. and {Carry}, B. and {Cellino}, A. and {Clementini}, G. and {Cowell}, S. and {Creevey}, O. and {Cuypers}, J. and {Davidson}, M. and {De Ridder}, J. and {de Torres}, A. and {Delchambre}, L. and {Dell'Oro}, A. and {Ducourant}, C. and {Fr{\'e}mat}, Y. and {Garc{\'\i}a-Torres}, M. and {Gosset}, E. and {Halbwachs}, J. -L. and {Hambly}, N.~C. and {Harrison}, D.~L. and {Hauser}, M. and {Hestroffer}, D. and {Hodgkin}, S.~T. and {Huckle}, H.~E. and {Hutton}, A. and {Jasniewicz}, G. and {Jordan}, S. and {Kontizas}, M. and {Korn}, A.~J. and {Lanzafame}, A.~C. and {Manteiga}, M. and {Moitinho}, A. and {Muinonen}, K. and {Osinde}, J. and {Pancino}, E. and {Pauwels}, T. and {Petit}, J. -M. and {Recio-Blanco}, A. and {Robin}, A.~C. and {Sarro}, L.~M. and {Siopis}, C. and {Smith}, M. and {Smith}, K.~W. and {Sozzetti}, A. and {Thuillot}, W. and {van Reeven}, W. and {Viala}, Y. and {Abbas}, U. and {Abreu Aramburu}, A. and {Accart}, S. and {Aguado}, J.~J. and {Allan}, P.~M. and {Allasia}, W. and {Altavilla}, G. and {{\'A}lvarez}, M.~A. and {Alves}, J. and {Anderson}, R.~I. and {Andrei}, A.~H. and {Anglada Varela}, E. and {Antiche}, E. and {Antoja}, T. and {Ant{\'o}n}, S. and {Arcay}, B. and {Atzei}, A. and {Ayache}, L. and {Bach}, N. and {Baker}, S.~G. and {Balaguer-N{\'u}{\~n}ez}, L. and {Barache}, C. and {Barata}, C. and {Barbier}, A. and {Barblan}, F. and {Baroni}, M. and {Barrado y Navascu{\'e}s}, D. and {Barros}, M. and {Barstow}, M.~A. and {Becciani}, U. and {Bellazzini}, M. and {Bellei}, G. and {Bello Garc{\'\i}a}, A. and {Belokurov}, V. and {Bendjoya}, P. and {Berihuete}, A. and {Bianchi}, L. and {Bienaym{\'e}}, O. and {Billebaud}, F. and {Blagorodnova}, N. and {Blanco-Cuaresma}, S. and {Boch}, T. and {Bombrun}, A. and {Borrachero}, R. and {Bouquillon}, S. and {Bourda}, G. and {Bouy}, H. and {Bragaglia}, A. and {Breddels}, M.~A. and {Brouillet}, N. and {Br{\"u}semeister}, T. and {Bucciarelli}, B. and {Budnik}, F. and {Burgess}, P. and {Burgon}, R. and {Burlacu}, A. and {Busonero}, D. and {Buzzi}, R. and {Caffau}, E. and {Cambras}, J. and {Campbell}, H. and {Cancelliere}, R. and {Cantat-Gaudin}, T. and {Carlucci}, T. and {Carrasco}, J.~M. and {Castellani}, M. and {Charlot}, P. and {Charnas}, J. and {Charvet}, P. and {Chassat}, F. and {Chiavassa}, A. and {Clotet}, M. and {Cocozza}, G. and {Collins}, R.~S. and {Collins}, P. and {Costigan}, G.},
        title = "{The Gaia mission}",
      journal = {\aap},
         year = 2016,
        month = nov,
       volume = {595},
          eid = {A1},
        pages = {A1},
          doi = {10.1051/0004-6361/201629272},
archivePrefix = {arXiv},
       eprint = {1609.04153},
 primaryClass = {astro-ph.IM},
       adsurl = {https://ui.adsabs.harvard.edu/abs/2016A&A...595A...1G}
}

@ARTICLE{Gaia2023,
       author = {{Gaia Collaboration: Vallenari}, A. and {Brown}, A.~G.~A. and {Prusti}, T. and {de Bruijne}, J.~H.~J. and {Arenou}, F. and {Babusiaux}, C. and {Biermann}, M. and {Creevey}, O.~L. and {Ducourant}, C. and {Evans}, D.~W. and {Eyer}, L. and {Guerra}, R. and {Hutton}, A. and {Jordi}, C. and {Klioner}, S.~A. and {Lammers}, U.~L. and {Lindegren}, L. and {Luri}, X. and {Mignard}, F. and {Panem}, C. and {Pourbaix}, D. and {Randich}, S. and {Sartoretti}, P. and {Soubiran}, C. and {Tanga}, P. and {Walton}, N.~A. and {Bailer-Jones}, C.~A.~L. and {Bastian}, U. and {Drimmel}, R. and {Jansen}, F. and {Katz}, D. and {Lattanzi}, M.~G. and {van Leeuwen}, F. and {Bakker}, J. and {Cacciari}, C. and {Casta{\~n}eda}, J. and {De Angeli}, F. and {Fabricius}, C. and {Fouesneau}, M. and {Fr{\'e}mat}, Y. and {Galluccio}, L. and {Guerrier}, A. and {Heiter}, U. and {Masana}, E. and {Messineo}, R. and {Mowlavi}, N. and {Nicolas}, C. and {Nienartowicz}, K. and {Pailler}, F. and {Panuzzo}, P. and {Riclet}, F. and {Roux}, W. and {Seabroke}, G.~M. and {Sordo}, R. and {Th{\'e}venin}, F. and {Gracia-Abril}, G. and {Portell}, J. and {Teyssier}, D. and {Altmann}, M. and {Andrae}, R. and {Audard}, M. and {Bellas-Velidis}, I. and {Benson}, K. and {Berthier}, J. and {Blomme}, R. and {Burgess}, P.~W. and {Busonero}, D. and {Busso}, G. and {C{\'a}novas}, H. and {Carry}, B. and {Cellino}, A. and {Cheek}, N. and {Clementini}, G. and {Damerdji}, Y. and {Davidson}, M. and {de Teodoro}, P. and {Nu{\~n}ez Campos}, M. and {Delchambre}, L. and {Dell'Oro}, A. and {Esquej}, P. and {Fern{\'a}ndez-Hern{\'a}ndez}, J. and {Fraile}, E. and {Garabato}, D. and {Garc{\'\i}a-Lario}, P. and {Gosset}, E. and {Haigron}, R. and {Halbwachs}, J. -L. and {Hambly}, N.~C. and {Harrison}, D.~L. and {Hern{\'a}ndez}, J. and {Hestroffer}, D. and {Hodgkin}, S.~T. and {Holl}, B. and {Jan{\ss}en}, K. and {Jevardat de Fombelle}, G. and {Jordan}, S. and {Krone-Martins}, A. and {Lanzafame}, A.~C. and {L{\"o}ffler}, W. and {Marchal}, O. and {Marrese}, P.~M. and {Moitinho}, A. and {Muinonen}, K. and {Osborne}, P. and {Pancino}, E. and {Pauwels}, T. and {Recio-Blanco}, A. and {Reyl{\'e}}, C. and {Riello}, M. and {Rimoldini}, L. and {Roegiers}, T. and {Rybizki}, J. and {Sarro}, L.~M. and {Siopis}, C. and {Smith}, M. and {Sozzetti}, A. and {Utrilla}, E. and {van Leeuwen}, M. and {Abbas}, U. and {{\'A}brah{\'a}m}, P. and {Abreu Aramburu}, A. and {Aerts}, C. and {Aguado}, J.~J. and {Ajaj}, M. and {Aldea-Montero}, F. and {Altavilla}, G. and {{\'A}lvarez}, M.~A. and {Alves}, J. and {Anders}, F. and {Anderson}, R.~I. and {Anglada Varela}, E. and {Antoja}, T. and {Baines}, D. and {Baker}, S.~G. and {Balaguer-N{\'u}{\~n}ez}, L. and {Balbinot}, E. and {Balog}, Z. and {Barache}, C. and {Barbato}, D. and {Barros}, M. and {Barstow}, M.~A. and {Bartolom{\'e}}, S. and {Bassilana}, J. -L. and {Bauchet}, N. and {Becciani}, U. and {Bellazzini}, M. and {Berihuete}, A. and {Bernet}, M. and {Bertone}, S. and {Bianchi}, L. and {Binnenfeld}, A. and {Blanco-Cuaresma}, S. and {Blazere}, A. and {Boch}, T. and {Bombrun}, A. and {Bossini}, D. and {Bouquillon}, S. and {Bragaglia}, A. and {Bramante}, L. and {Breedt}, E. and {Bressan}, A. and {Brouillet}, N. and {Brugaletta}, E. and {Bucciarelli}, B. and {Burlacu}, A. and {Butkevich}, A.~G. and {Buzzi}, R. and {Caffau}, E. and {Cancelliere}, R. and {Cantat-Gaudin}, T. and {Carballo}, R. and {Carlucci}, T. and {Carnerero}, M.~I. and {Carrasco}, J.~M. and {Casamiquela}, L. and {Castellani}, M. and {Castro-Ginard}, A. and {Chaoul}, L. and {Charlot}, P. and {Chemin}, L. and {Chiaramida}, V. and {Chiavassa}, A. and {Chornay}, N. and {Comoretto}, G. and {Contursi}, G. and {Cooper}, W.~J. and {Cornez}, T. and {Cowell}, S. and {Crifo}, F. and {Cropper}, M. and {Crosta}, M. and {Crowley}, C. and {Dafonte}, C. and {Dapergolas}, A. and {David}, M. and {David}, P. and {de Laverny}, P. and {De Luise}, F. and {De March}, R.},
        title = "{Gaia Data Release 3. Summary of the content and survey properties}",
      journal = {\aap},
         year = 2023,
        month = jun,
       volume = {674},
          eid = {A1},
        pages = {A1},
          doi = {10.1051/0004-6361/202243940},
archivePrefix = {arXiv},
       eprint = {2208.00211},
 primaryClass = {astro-ph.GA},
       adsurl = {https://ui.adsabs.harvard.edu/abs/2023A&A...674A...1G}
}

@ARTICLE{Milone2012,
       author = {{Milone}, A.~P. and {Piotto}, G. and {Bedin}, L.~R. and {Aparicio}, A. and {Anderson}, J. and {Sarajedini}, A. and {Marino}, A.~F. and {Moretti}, A. and {Davies}, M.~B. and {Chaboyer}, B. and {Dotter}, A. and {Hempel}, M. and {Mar{\'\i}n-Franch}, A. and {Majewski}, S. and {Paust}, N.~E.~Q. and {Reid}, I.~N. and {Rosenberg}, A. and {Siegel}, M.},
        title = "{The ACS survey of Galactic globular clusters. XII. Photometric binaries along the main sequence}",
      journal = {\aap},
         year = 2012,
        month = apr,
       volume = {540},
          eid = {A16},
        pages = {A16},
          doi = {10.1051/0004-6361/201016384},
archivePrefix = {arXiv},
       eprint = {1111.0552},
 primaryClass = {astro-ph.SR},
       adsurl = {https://ui.adsabs.harvard.edu/abs/2012A&A...540A..16M}
}

@BOOK{RC3,
       author = {{de Vaucouleurs}, Gerard and {de Vaucouleurs}, Antoinette and {Corwin}, Jr., Herold G. and {Buta}, Ronald J. and {Paturel}, Georges and {Fouque}, Pascal},
        title = "{Third Reference Catalogue of Bright Galaxies}",
         year = 1991,
        publisher = {Springer, New York},
       adsurl = {https://ui.adsabs.harvard.edu/abs/1991rc3..book.....D}
}

@ARTICLE{Tikhonov2009,
       author = {{Tikhonov}, N.~A. and {Galazutdinova}, O.~A.},
        title = "{Stellar population of the irregular galaxy IC 10}",
      journal = {Astronomy Letters},
         year = 2009,
        month = nov,
       volume = {35},
       number = {11},
        pages = {748-763},
          doi = {10.1134/S1063773709110036},
       adsurl = {https://ui.adsabs.harvard.edu/abs/2009AstL...35..748T}
}

@ARTICLE{Demers2004,
       author = {{Demers}, S. and {Battinelli}, P. and {Letarte}, B.},
        title = "{A Carbon star approach to IC 10: Distance and correct size}",
      journal = {\aap},
         year = 2004,
        month = sep,
       volume = {424},
        pages = {125-132},
          doi = {10.1051/0004-6361:20040552},
       adsurl = {https://ui.adsabs.harvard.edu/abs/2004A&A...424..125D}
}

@ARTICLE{Huchtmeier1979,
       author = {{Huchtmeier}, W.~K.},
        title = "{The giant H I-envelope of the irregular galaxy IC 10.}",
      journal = {\aap},
         year = 1979,
        month = may,
       volume = {75},
        pages = {170-175},
       adsurl = {https://ui.adsabs.harvard.edu/abs/1979A&A....75..170H}
}

@ARTICLE{Shostak1989,
       author = {{Shostak}, G.~S. and {Skillman}, E.~D.},
        title = "{Neutral hydrogen observations of the irregular galaxy IC 10.}",
      journal = {\aap},
         year = 1989,
        month = apr,
       volume = {214},
        pages = {33-42},
       adsurl = {https://ui.adsabs.harvard.edu/abs/1989A&A...214...33S}
}

@ARTICLE{Ashley2014,
       author = {{Ashley}, Trisha and {Elmegreen}, Bruce G. and {Johnson}, Megan and {Nidever}, David L. and {Simpson}, Caroline E. and {Pokhrel}, Nau Raj},
        title = "{The H I Chronicles of LITTLE THINGS BCDs II: The Origin of IC 10's H I Structure}",
      journal = {\aj},
         year = 2014,
        month = dec,
       volume = {148},
       number = {6},
          eid = {130},
        pages = {130},
          doi = {10.1088/0004-6256/148/6/130},
archivePrefix = {arXiv},
       eprint = {1409.5406},
 primaryClass = {astro-ph.GA},
       adsurl = {https://ui.adsabs.harvard.edu/abs/2014AJ....148..130A}
}

@ARTICLE{Higgs2021,
       author = {{Higgs}, C.~R. and {McConnachie}, A.~W. and {Annau}, N. and {Irwin}, M. and {Battaglia}, G. and {C{\^o}t{\'e}}, P. and {Lewis}, G.~F. and {Venn}, K.},
        title = "{Solo dwarfs II: the stellar structure of isolated Local Group dwarf galaxies}",
      journal = {\mnras},
         year = 2021,
        month = may,
       volume = {503},
       number = {1},
        pages = {176-199},
          doi = {10.1093/mnras/stab002},
archivePrefix = {arXiv},
       eprint = {2101.03189},
 primaryClass = {astro-ph.GA},
       adsurl = {https://ui.adsabs.harvard.edu/abs/2021MNRAS.503..176H}
}

@ARTICLE{Bellazzini2014,
       author = {{Bellazzini}, M. and {Beccari}, G. and {Fraternali}, F. and {Oosterloo}, T.~A. and {Sollima}, A. and {Testa}, V. and {Galleti}, S. and {Perina}, S. and {Faccini}, M. and {Cusano}, F.},
        title = "{The extended structure of the dwarf irregular galaxies Sextans A and Sextans B. Signatures of tidal distortion in the outskirts of the Local Group}",
      journal = {\aap},
         year = 2014,
        month = jun,
       volume = {566},
          eid = {A44},
        pages = {A44},
          doi = {10.1051/0004-6361/201423659},
archivePrefix = {arXiv},
       eprint = {1404.1697},
 primaryClass = {astro-ph.GA},
       adsurl = {https://ui.adsabs.harvard.edu/abs/2014A&A...566A..44B}
}

@ARTICLE{Kniazev2016,
       author = {{Kniazev}, Alexei Y. and {Brosch}, Noah and {Hoffman}, G. Lyle and {Grebel}, Eva K. and {Zucker}, Daniel B. and {Pustilnik}, Simon A.},
        title = "{The faint outer regions of the Pegasus dwarf irregular galaxy: a much larger and undisturbed galaxy}",
      journal = {\mnras},
         year = 2009,
        month = dec,
       volume = {400},
       number = {4},
        pages = {2054-2069},
          doi = {10.1111/j.1365-2966.2009.15601.x},
archivePrefix = {arXiv},
       eprint = {0908.3621},
 primaryClass = {astro-ph.GA},
       adsurl = {https://ui.adsabs.harvard.edu/abs/2009MNRAS.400.2054K}
}

@ARTICLE{Annibali2020,
       author = {{Annibali}, F. and {Beccari}, G. and {Bellazzini}, M. and {Tosi}, M. and {Cusano}, F. and {Paris}, D. and {Cignoni}, M. and {Ciotti}, L. and {Nipoti}, C. and {Sacchi}, E.},
        title = "{The Smallest Scale of Hierarchy survey (SSH) - I. Survey description}",
      journal = {\mnras},
         year = 2020,
        month = feb,
       volume = {491},
       number = {4},
        pages = {5101-5125},
          doi = {10.1093/mnras/stz3185},
archivePrefix = {arXiv},
       eprint = {1911.08543},
 primaryClass = {astro-ph.GA},
       adsurl = {https://ui.adsabs.harvard.edu/abs/2020MNRAS.491.5101A}
}

@ARTICLE{Rys2011,
       author = {{Ry{\'s}}, A. and {Grocholski}, A.~J. and {van der Marel}, R.~P. and {Aloisi}, A. and {Annibali}, F.},
        title = "{Hubble Space Telescope study of resolved red giant stars in the outer halos of nearby dwarf starburst galaxies}",
      journal = {\aap},
         year = 2011,
        month = jun,
       volume = {530},
          eid = {A23},
        pages = {A23},
          doi = {10.1051/0004-6361/201015881},
archivePrefix = {arXiv},
       eprint = {1104.0899},
 primaryClass = {astro-ph.CO},
       adsurl = {https://ui.adsabs.harvard.edu/abs/2011A&A...530A..23R}
}

@ARTICLE{Bellazzini2011,
       author = {{Bellazzini}, M. and {Beccari}, G. and {Oosterloo}, T.~A. and {Galleti}, S. and {Sollima}, A. and {Correnti}, M. and {Testa}, V. and {Mayer}, L. and {Cignoni}, M. and {Fraternali}, F. and {Gallozzi}, S.},
        title = "{An optical and H i study of the dwarf Local Group galaxy VV124 = UGC4879. A gas-poor dwarf with a stellar disk?}",
      journal = {\aap},
         year = 2011,
        month = mar,
       volume = {527},
          eid = {A58},
        pages = {A58},
          doi = {10.1051/0004-6361/201016159},
archivePrefix = {arXiv},
       eprint = {1012.3757},
 primaryClass = {astro-ph.CO},
       adsurl = {https://ui.adsabs.harvard.edu/abs/2011A&A...527A..58B}
}

@ARTICLE{Beccari2014,
       author = {{Beccari}, G. and {Bellazzini}, M. and {Fraternali}, F. and {Battaglia}, G. and {Perina}, S. and {Sollima}, A. and {Oosterloo}, T.~A. and {Testa}, V. and {Galleti}, S.},
        title = "{The extended structure of the dwarf irregular galaxy Sagittarius}",
      journal = {\aap},
         year = 2014,
        month = oct,
       volume = {570},
          eid = {A78},
        pages = {A78},
          doi = {10.1051/0004-6361/201424411},
archivePrefix = {arXiv},
       eprint = {1409.0715},
 primaryClass = {astro-ph.GA},
       adsurl = {https://ui.adsabs.harvard.edu/abs/2014A&A...570A..78B}
}

@ARTICLE{Young2014,
       author = {{Young}, T. and {Jerjen}, H. and {L{\'o}pez-S{\'a}nchez}, {\'A}. R. and {Koribalski}, B.~S.},
        title = "{Deep near-infrared surface photometry and properties of Local Volume dwarf irregular galaxies}",
      journal = {\mnras},
         year = 2014,
        month = nov,
       volume = {444},
       number = {4},
        pages = {3052-3077},
          doi = {10.1093/mnras/stu1646},
archivePrefix = {arXiv},
       eprint = {1408.2609},
 primaryClass = {astro-ph.GA},
       adsurl = {https://ui.adsabs.harvard.edu/abs/2014MNRAS.444.3052Y}
}

@ARTICLE{Poulain2021,
       author = {{Poulain}, M{\'e}lina and {Marleau}, Francine R. and {Habas}, Rebecca and {Duc}, Pierre-Alain and {S{\'a}nchez-Janssen}, Rub{\'e}n and {Durrell}, Patrick R. and {Paudel}, Sanjaya and {Ahad}, Syeda Lammim and {Chougule}, Abhishek and {M{\"u}ller}, Oliver and {Lim}, Sungsoon and {B{\'\i}lek}, Michal and {Fensch}, J{\'e}r{\'e}my},
        title = "{Structure and morphology of the MATLAS dwarf galaxies and their central nuclei}",
      journal = {\mnras},
         year = 2021,
        month = oct,
       volume = {506},
       number = {4},
        pages = {5494-5511},
          doi = {10.1093/mnras/stab2092},
archivePrefix = {arXiv},
       eprint = {2107.09379},
 primaryClass = {astro-ph.GA},
       adsurl = {https://ui.adsabs.harvard.edu/abs/2021MNRAS.506.5494P}
}

@ARTICLE{Kirby2008,
       author = {{Kirby}, Emma M. and {Jerjen}, Helmut and {Ryder}, Stuart D. and {Driver}, Simon P.},
        title = "{Deep Near-Infrared Surface Photometry of 57 Galaxies in the Local Sphere of Influence}",
      journal = {\aj},
         year = 2008,
        month = nov,
       volume = {136},
       number = {5},
        pages = {1866-1888},
          doi = {10.1088/0004-6256/136/5/1866},
archivePrefix = {arXiv},
       eprint = {0808.2529},
 primaryClass = {astro-ph},
       adsurl = {https://ui.adsabs.harvard.edu/abs/2008AJ....136.1866K}
}

@ARTICLE{McConnachie2012,
       author = {{McConnachie}, Alan W.},
        title = "{The Observed Properties of Dwarf Galaxies in and around the Local Group}",
      journal = {\aj},
         year = 2012,
        month = jul,
       volume = {144},
       number = {1},
          eid = {4},
        pages = {4},
          doi = {10.1088/0004-6256/144/1/4},
archivePrefix = {arXiv},
       eprint = {1204.1562},
 primaryClass = {astro-ph.CO},
       adsurl = {https://ui.adsabs.harvard.edu/abs/2012AJ....144....4M}
}

@ARTICLE{Battaglia2012,
       author = {{Battaglia}, G. and {Rejkuba}, M. and {Tolstoy}, E. and {Irwin}, M.~J. and {Beccari}, G.},
        title = "{A wide-area view of the Phoenix dwarf galaxy from Very Large Telescope/FORS imaging$^{{\ensuremath{\star}}}$}",
      journal = {\mnras},
         year = 2012,
        month = aug,
       volume = {424},
       number = {2},
        pages = {1113-1131},
          doi = {10.1111/j.1365-2966.2012.21286.x},
archivePrefix = {arXiv},
       eprint = {1205.2704},
 primaryClass = {astro-ph.GA},
       adsurl = {https://ui.adsabs.harvard.edu/abs/2012MNRAS.424.1113B}
}

@ARTICLE{McMonigal2014,
       author = {{McMonigal}, B. and {Bate}, N.~F. and {Lewis}, G.~F. and {Irwin}, M.~J. and {Battaglia}, G. and {Ibata}, R.~A. and {Martin}, N.~F. and {McConnachie}, A.~W. and {Guglielmo}, M. and {Conn}, A.~R.},
        title = "{Sailing under the Magellanic Clouds: a DECam view of the Carina dwarf}",
      journal = {\mnras},
         year = 2014,
        month = nov,
       volume = {444},
       number = {4},
        pages = {3139-3149},
          doi = {10.1093/mnras/stu1659},
archivePrefix = {arXiv},
       eprint = {1408.2907},
 primaryClass = {astro-ph.GA},
       adsurl = {https://ui.adsabs.harvard.edu/abs/2014MNRAS.444.3139M}
}

@ARTICLE{Collins2021,
       author = {{Collins}, Michelle L.~M. and {Read}, Justin I. and {Ibata}, Rodrigo A. and {Rich}, R. Michael and {Martin}, Nicolas F. and {Pe{\~n}arrubia}, Jorge and {Chapman}, Scott C. and {Tollerud}, Erik J. and {Weisz}, Daniel R.},
        title = "{Andromeda XXI - a dwarf galaxy in a low-density dark matter halo}",
      journal = {\mnras},
         year = 2021,
        month = aug,
       volume = {505},
       number = {4},
        pages = {5686-5701},
          doi = {10.1093/mnras/stab1624},
archivePrefix = {arXiv},
       eprint = {2102.11890},
 primaryClass = {astro-ph.GA},
       adsurl = {https://ui.adsabs.harvard.edu/abs/2021MNRAS.505.5686C}
}

@ARTICLE{Martin2008,
       author = {{Martin}, Nicolas F. and {de Jong}, Jelte T.~A. and {Rix}, Hans-Walter},
        title = "{A Comprehensive Maximum Likelihood Analysis of the Structural Properties of Faint Milky Way Satellites}",
      journal = {\apj},
         year = 2008,
        month = sep,
       volume = {684},
       number = {2},
        pages = {1075-1092},
          doi = {10.1086/590336},
archivePrefix = {arXiv},
       eprint = {0805.2945},
 primaryClass = {astro-ph},
       adsurl = {https://ui.adsabs.harvard.edu/abs/2008ApJ...684.1075M}
}

@ARTICLE{Bernard2012,
       author = {{Bernard}, Edouard J. and {Ferguson}, Annette M.~N. and {Barker}, Michael K. and {Irwin}, Michael J. and {Jablonka}, Pascale and {Arimoto}, Nobuo},
        title = "{A deep, wide-field study of Holmberg II with Suprime-Cam: evidence for ram pressure stripping}",
      journal = {\mnras},
         year = 2012,
        month = nov,
       volume = {426},
       number = {4},
        pages = {3490-3500},
          doi = {10.1111/j.1365-2966.2012.22025.x},
archivePrefix = {arXiv},
       eprint = {1208.4808},
 primaryClass = {astro-ph.GA},
       adsurl = {https://ui.adsabs.harvard.edu/abs/2012MNRAS.426.3490B}
}

@ARTICLE{Cooper2025,
       author = {{Cooper}, Andrew P. and {Frenk}, Carlos S. and {Hellwing}, Wojciech A. and {Bose}, Sownak},
        title = "{Simulations of the accreted stellar haloes of low-mass field galaxies}",
      journal = {\mnras},
         year = 2025,
        month = jul,
       volume = {540},
       number = {3},
        pages = {2049-2080},
          doi = {10.1093/mnras/staf833},
archivePrefix = {arXiv},
       eprint = {2501.13317},
 primaryClass = {astro-ph.GA},
       adsurl = {https://ui.adsabs.harvard.edu/abs/2025MNRAS.540.2049C}
}

@ARTICLE{Kang2019,
       author = {{Kang}, Hoyoung D. and {Ricotti}, Massimo},
        title = "{Ghostly haloes in dwarf galaxies: constraints on the star formation efficiency before reionization}",
      journal = {\mnras},
         year = 2019,
        month = sep,
       volume = {488},
       number = {2},
        pages = {2673-2688},
          doi = {10.1093/mnras/stz1886},
archivePrefix = {arXiv},
       eprint = {1810.01437},
 primaryClass = {astro-ph.GA},
       adsurl = {https://ui.adsabs.harvard.edu/abs/2019MNRAS.488.2673K}
}

@ARTICLE{Deason2022,
       author = {{Deason}, Alis J. and {Bose}, Sownak and {Fattahi}, Azadeh and {Amorisco}, Nicola C. and {Hellwing}, Wojciech and {Frenk}, Carlos S.},
        title = "{Dwarf stellar haloes: a powerful probe of small-scale galaxy formation and the nature of dark matter}",
      journal = {\mnras},
         year = 2022,
        month = apr,
       volume = {511},
       number = {3},
        pages = {4044-4059},
          doi = {10.1093/mnras/stab3524},
archivePrefix = {arXiv},
       eprint = {2110.05499},
 primaryClass = {astro-ph.GA},
       adsurl = {https://ui.adsabs.harvard.edu/abs/2022MNRAS.511.4044D}
}

@ARTICLE{Tau2025,
       author = {{Tau}, Elisa A. and {Monachesi}, Antonela and {Gomez}, Facundo A. and {Grand}, Robert J.~J. and {Pakmor}, R{\"u}diger and {van de Voort}, Freeke and {Gonzalez-Jara}, Jenny and {Tissera}, Patricia B. and {Marinacci}, Federico and {Bieri}, Rebekka},
        title = "{The role of accreted and in situ populations in shaping the stellar halos of low-mass galaxies}",
      journal = {\aap},
         year = 2025,
        month = jul,
       volume = {699},
          eid = {A93},
        pages = {A93},
          doi = {10.1051/0004-6361/202453488},
archivePrefix = {arXiv},
       eprint = {2412.13807},
 primaryClass = {astro-ph.GA},
       adsurl = {https://ui.adsabs.harvard.edu/abs/2025A&A...699A..93T}
}

@ARTICLE{Lee2003,
       author = {{Lee}, Henry and {McCall}, Marshall L. and {Kingsburgh}, Robin L. and {Ross}, Robert and {Stevenson}, Chris C.},
        title = "{Uncovering Additional Clues to Galaxy Evolution. I. Dwarf Irregular Galaxies in the Field}",
      journal = {\aj},
         year = 2003,
        month = jan,
       volume = {125},
       number = {1},
        pages = {146-165},
          doi = {10.1086/345384},
archivePrefix = {arXiv},
       eprint = {astro-ph/0210098},
 primaryClass = {astro-ph},
       adsurl = {https://ui.adsabs.harvard.edu/abs/2003AJ....125..146L}
}

@ARTICLE{White1978,
       author = {{White}, S.~D.~M. and {Rees}, M.~J.},
        title = "{Core condensation in heavy halos: a two-stage theory for galaxy formation and clustering.}",
      journal = {\mnras},
         year = 1978,
        month = may,
       volume = {183},
        pages = {341-358},
          doi = {10.1093/mnras/183.3.341},
       adsurl = {https://ui.adsabs.harvard.edu/abs/1978MNRAS.183..341W}
}

@ARTICLE{Springel2006,
       author = {{Springel}, Volker and {Frenk}, Carlos S. and {White}, Simon D.~M.},
        title = "{The large-scale structure of the Universe}",
      journal = {\nat},
         year = 2006,
        month = apr,
       volume = {440},
       number = {7088},
        pages = {1137-1144},
          doi = {10.1038/nature04805},
archivePrefix = {arXiv},
       eprint = {astro-ph/0604561},
 primaryClass = {astro-ph},
       adsurl = {https://ui.adsabs.harvard.edu/abs/2006Natur.440.1137S}
}

@ARTICLE{Peebles1982,
       author = {{Peebles}, P.~J.~E.},
        title = "{Large-scale background temperature and mass fluctuations due to scale-invariant primeval perturbations}",
      journal = {\apjl},
         year = 1982,
        month = dec,
       volume = {263},
        pages = {L1-L5},
          doi = {10.1086/183911},
       adsurl = {https://ui.adsabs.harvard.edu/abs/1982ApJ...263L...1P}
}

@ARTICLE{Diemand2008,
       author = {{Diemand}, J. and {Kuhlen}, M. and {Madau}, P. and {Zemp}, M. and {Moore}, B. and {Potter}, D. and {Stadel}, J.},
        title = "{Clumps and streams in the local dark matter distribution}",
      journal = {\nat},
         year = 2008,
        month = aug,
       volume = {454},
       number = {7205},
        pages = {735-738},
          doi = {10.1038/nature07153},
archivePrefix = {arXiv},
       eprint = {0805.1244},
 primaryClass = {astro-ph},
       adsurl = {https://ui.adsabs.harvard.edu/abs/2008Natur.454..735D}
}

@ARTICLE{Bullock2005,
       author = {{Bullock}, James S. and {Johnston}, Kathryn V.},
        title = "{Tracing Galaxy Formation with Stellar Halos. I. Methods}",
      journal = {\apj},
         year = 2005,
        month = dec,
       volume = {635},
       number = {2},
        pages = {931-949},
          doi = {10.1086/497422},
archivePrefix = {arXiv},
       eprint = {astro-ph/0506467},
 primaryClass = {astro-ph},
       adsurl = {https://ui.adsabs.harvard.edu/abs/2005ApJ...635..931B}
}

@ARTICLE{Johnston2008,
       author = {{Johnston}, Kathryn V. and {Bullock}, James S. and {Sharma}, Sanjib and {Font}, Andreea and {Robertson}, Brant E. and {Leitner}, Samuel N.},
        title = "{Tracing Galaxy Formation with Stellar Halos. II. Relating Substructure in Phase and Abundance Space to Accretion Histories}",
      journal = {\apj},
         year = 2008,
        month = dec,
       volume = {689},
       number = {2},
        pages = {936-957},
          doi = {10.1086/592228},
archivePrefix = {arXiv},
       eprint = {0807.3911},
 primaryClass = {astro-ph},
       adsurl = {https://ui.adsabs.harvard.edu/abs/2008ApJ...689..936J}
}

@ARTICLE{Fakhouri2010, 
       author = {{Fakhouri}, Onsi and {Ma}, Chung-Pei and {Boylan-Kolchin}, Michael},
        title = "{The merger rates and mass assembly histories of dark matter haloes in the two Millennium simulations}",
      journal = {\mnras},
         year = 2010,
        month = aug,
       volume = {406},
       number = {4},
        pages = {2267-2278},
          doi = {10.1111/j.1365-2966.2010.16859.x},
archivePrefix = {arXiv},
       eprint = {1001.2304},
 primaryClass = {astro-ph.CO},
       adsurl = {https://ui.adsabs.harvard.edu/abs/2010MNRAS.406.2267F}
}

@ARTICLE{DeLucia2008,
       author = {{De Lucia}, Gabriella and {Helmi}, Amina},
        title = "{The Galaxy and its stellar halo: insights on their formation from a hybrid cosmological approach}",
      journal = {\mnras},
         year = 2008,
        month = nov,
       volume = {391},
       number = {1},
        pages = {14-31},
          doi = {10.1111/j.1365-2966.2008.13862.x},
archivePrefix = {arXiv},
       eprint = {0804.2465},
 primaryClass = {astro-ph},
       adsurl = {https://ui.adsabs.harvard.edu/abs/2008MNRAS.391...14D}
}

@ARTICLE{Ibata2001a,
       author = {{Ibata}, Rodrigo and {Irwin}, Michael and {Lewis}, Geraint F. and {Stolte}, Andrea},
        title = "{Galactic Halo Substructure in the Sloan Digital Sky Survey: The Ancient Tidal Stream from the Sagittarius Dwarf Galaxy}",
      journal = {\apjl},
         year = 2001,
        month = feb,
       volume = {547},
       number = {2},
        pages = {L133-L136},
          doi = {10.1086/318894},
archivePrefix = {arXiv},
       eprint = {astro-ph/0004255},
 primaryClass = {astro-ph},
       adsurl = {https://ui.adsabs.harvard.edu/abs/2001ApJ...547L.133I}
}

@ARTICLE{Ibata2001b,
       author = {{Ibata}, Rodrigo and {Irwin}, Michael and {Lewis}, Geraint and {Ferguson}, Annette M.~N. and {Tanvir}, Nial},
        title = "{A giant stream of metal-rich stars in the halo of the galaxy M31}",
      journal = {\nat},
         year = 2001,
        month = jul,
       volume = {412},
       number = {6842},
        pages = {49-52},
          doi = {10.1038/35083506},
archivePrefix = {arXiv},
       eprint = {astro-ph/0107090},
 primaryClass = {astro-ph},
       adsurl = {https://ui.adsabs.harvard.edu/abs/2001Natur.412...49I}
}

@ARTICLE{Belokurov2006,
       author = {{Belokurov}, V. and {Zucker}, D.~B. and {Evans}, N.~W. and {Gilmore}, G. and {Vidrih}, S. and {Bramich}, D.~M. and {Newberg}, H.~J. and {Wyse}, R.~F.~G. and {Irwin}, M.~J. and {Fellhauer}, M. and {Hewett}, P.~C. and {Walton}, N.~A. and {Wilkinson}, M.~I. and {Cole}, N. and {Yanny}, B. and {Rockosi}, C.~M. and {Beers}, T.~C. and {Bell}, E.~F. and {Brinkmann}, J. and {Ivezi{\'c}}, {\v{Z}}. and {Lupton}, R.},
        title = "{The Field of Streams: Sagittarius and Its Siblings}",
      journal = {\apjl},
         year = 2006,
        month = may,
       volume = {642},
       number = {2},
        pages = {L137-L140},
          doi = {10.1086/504797},
archivePrefix = {arXiv},
       eprint = {astro-ph/0605025},
 primaryClass = {astro-ph},
       adsurl = {https://ui.adsabs.harvard.edu/abs/2006ApJ...642L.137B}
}

@ARTICLE{Crnojevic2016,
       author = {{Crnojevi{\'c}}, D. and {Sand}, D.~J. and {Spekkens}, K. and {Caldwell}, N. and {Guhathakurta}, P. and {McLeod}, B. and {Seth}, A. and {Simon}, J.~D. and {Strader}, J. and {Toloba}, E.},
        title = "{The Extended Halo of Centaurus A: Uncovering Satellites, Streams, and Substructures}",
      journal = {\apj},
         year = 2016,
        month = may,
       volume = {823},
       number = {1},
          eid = {19},
        pages = {19},
          doi = {10.3847/0004-637X/823/1/19},
archivePrefix = {arXiv},
       eprint = {1512.05366},
 primaryClass = {astro-ph.GA},
       adsurl = {https://ui.adsabs.harvard.edu/abs/2016ApJ...823...19C}
}

@ARTICLE{Ferguson2002,
       author = {{Ferguson}, Annette M.~N. and {Irwin}, Michael J. and {Ibata}, Rodrigo A. and {Lewis}, Geraint F. and {Tanvir}, Nial R.},
        title = "{Evidence for Stellar Substructure in the Halo and Outer Disk of M31}",
      journal = {\aj},
         year = 2002,
        month = sep,
       volume = {124},
       number = {3},
        pages = {1452-1463},
          doi = {10.1086/342019},
archivePrefix = {arXiv},
       eprint = {astro-ph/0205530},
 primaryClass = {astro-ph},
       adsurl = {https://ui.adsabs.harvard.edu/abs/2002AJ....124.1452F}
}

@ARTICLE{Wheeler2015,
       author = {{Wheeler}, Coral and {O{\~n}orbe}, Jose and {Bullock}, James S. and {Boylan-Kolchin}, Michael and {Elbert}, Oliver D. and {Garrison-Kimmel}, Shea and {Hopkins}, Philip F. and {Kere{\v{s}}}, Du{\v{s}}an},
        title = "{Sweating the small stuff: simulating dwarf galaxies, ultra-faint dwarf galaxies, and their own tiny satellites}",
      journal = {\mnras},
         year = 2015,
        month = oct,
       volume = {453},
       number = {2},
        pages = {1305-1316},
          doi = {10.1093/mnras/stv1691},
archivePrefix = {arXiv},
       eprint = {1504.02466},
 primaryClass = {astro-ph.GA},
       adsurl = {https://ui.adsabs.harvard.edu/abs/2015MNRAS.453.1305W}
}

@ARTICLE{Sacchi2024,
       author = {{Sacchi}, Elena and {Bellazzini}, Michele and {Annibali}, Francesca and {Tosi}, Monica and {Beccari}, Giacomo and {Cannon}, John M. and {Hunter}, Laura C. and {Paris}, Diego and {Roychowdhury}, Sambit and {Schisgal}, Lila and {van Zee}, Liese and {Cignoni}, Michele and {Cusano}, Felice and {de Jong}, Roelof S. and {Hunt}, Leslie and {Pascale}, Raffaele},
        title = "{The Smallest Scale of Hierarchy Survey (SSH): III. Dwarf-dwarf satellite merging phenomena in the low-mass regime}",
      journal = {\aap},
         year = 2024,
        month = nov,
       volume = {691},
          eid = {A65},
        pages = {A65},
          doi = {10.1051/0004-6361/202450106},
archivePrefix = {arXiv},
       eprint = {2406.01683},
 primaryClass = {astro-ph.GA},
       adsurl = {https://ui.adsabs.harvard.edu/abs/2024A&A...691A..65S}
}

@ARTICLE{Annibali2016,
       author = {{Annibali}, Francesca and {Nipoti}, Carlo and {Ciotti}, Luca and {Tosi}, Monica and {Aloisi}, Alessandra and {Bellazzini}, Michele and {Cignoni}, Michele and {Cusano}, Felice and {Paris}, Diego and {Sacchi}, Elena},
        title = "{DDO 68: A Flea with Smaller Fleas that on Him Prey}",
      journal = {\apjl},
         year = 2016,
        month = aug,
       volume = {826},
       number = {2},
          eid = {L27},
        pages = {L27},
          doi = {10.3847/2041-8205/826/2/L27},
archivePrefix = {arXiv},
       eprint = {1607.02628},
 primaryClass = {astro-ph.GA},
       adsurl = {https://ui.adsabs.harvard.edu/abs/2016ApJ...826L..27A}
}

@ARTICLE{Kado2020,
       author = {{Kado-Fong}, Erin and {Greene}, Jenny E. and {Greco}, Johnny P. and {Beaton}, Rachael and {Goulding}, Andy D. and {Johnson}, Sean D. and {Komiyama}, Yutaka},
        title = "{Star Formation in Isolated Dwarf Galaxies Hosting Tidal Debris: Extending the Dwarf-Dwarf Merger Sequence}",
      journal = {\aj},
         year = 2020,
        month = mar,
       volume = {159},
       number = {3},
          eid = {103},
        pages = {103},
          doi = {10.3847/1538-3881/ab6ef3},
archivePrefix = {arXiv},
       eprint = {1911.08497},
 primaryClass = {astro-ph.GA},
       adsurl = {https://ui.adsabs.harvard.edu/abs/2020AJ....159..103K}
}

@ARTICLE{Koposov2018,
       author = {{Koposov}, Sergey E. and {Walker}, Matthew G. and {Belokurov}, Vasily and {Casey}, Andrew R. and {Geringer-Sameth}, Alex and {Mackey}, Dougal and {Da Costa}, Gary and {Erkal}, Denis and {Jethwa}, Prashin and {Mateo}, Mario and {Olszewski}, Edward W. and {Bailey}, John I.},
        title = "{Snake in the Clouds: a new nearby dwarf galaxy in the Magellanic bridge*}",
      journal = {\mnras},
         year = 2018,
        month = oct,
       volume = {479},
       number = {4},
        pages = {5343-5361},
          doi = {10.1093/mnras/sty1772},
archivePrefix = {arXiv},
       eprint = {1804.06430},
 primaryClass = {astro-ph.GA},
       adsurl = {https://ui.adsabs.harvard.edu/abs/2018MNRAS.479.5343K}
}

@ARTICLE{Paudel2018,
       author = {{Paudel}, Sanjaya and {Smith}, Rory and {Yoon}, Suk Jin and {Calder{\'o}n-Castillo}, Paula and {Duc}, Pierre-Alain},
        title = "{A Catalog of Merging Dwarf Galaxies in the Local Universe}",
      journal = {\apjs},
         year = 2018,
        month = aug,
       volume = {237},
       number = {2},
          eid = {36},
        pages = {36},
          doi = {10.3847/1538-4365/aad555},
archivePrefix = {arXiv},
       eprint = {1807.07195},
 primaryClass = {astro-ph.GA},
       adsurl = {https://ui.adsabs.harvard.edu/abs/2018ApJS..237...36P}
}

@ARTICLE{Belokurov2016,
       author = {{Belokurov}, Vasily and {Koposov}, Sergey E.},
        title = "{Stellar streams around the Magellanic Clouds}",
      journal = {\mnras},
         year = 2016,
        month = feb,
       volume = {456},
       number = {1},
        pages = {602-616},
          doi = {10.1093/mnras/stv2688},
archivePrefix = {arXiv},
       eprint = {1511.03667},
 primaryClass = {astro-ph.GA},
       adsurl = {https://ui.adsabs.harvard.edu/abs/2016MNRAS.456..602B}
}

@ARTICLE{Cooper2010,
       author = {{Cooper}, A.~P. and {Cole}, S. and {Frenk}, C.~S. and {White}, S.~D.~M. and {Helly}, J. and {Benson}, A.~J. and {De Lucia}, G. and {Helmi}, A. and {Jenkins}, A. and {Navarro}, J.~F. and {Springel}, V. and {Wang}, J.},
        title = "{Galactic stellar haloes in the CDM model}",
      journal = {\mnras},
         year = 2010,
        month = aug,
       volume = {406},
       number = {2},
        pages = {744-766},
          doi = {10.1111/j.1365-2966.2010.16740.x},
archivePrefix = {arXiv},
       eprint = {0910.3211},
 primaryClass = {astro-ph.GA},
       adsurl = {https://ui.adsabs.harvard.edu/abs/2010MNRAS.406..744C}
}

@ARTICLE{Bekki2001,
       author = {{Bekki}, Kenji and {Chiba}, Masashi},
        title = "{Formation of the Galactic Stellar Halo. I. Structure and Kinematics}",
      journal = {\apj},
         year = 2001,
        month = sep,
       volume = {558},
       number = {2},
        pages = {666-686},
          doi = {10.1086/322300},
archivePrefix = {arXiv},
       eprint = {astro-ph/0106523},
 primaryClass = {astro-ph},
       adsurl = {https://ui.adsabs.harvard.edu/abs/2001ApJ...558..666B}
}

@ARTICLE{Sestito2023,
       author = {{Sestito}, Federico and {Zaremba}, Daria and {Venn}, Kim A. and {D'Aoust}, Lina and {Hayes}, Christian and {Jensen}, Jaclyn and {Navarro}, Julio F. and {Jablonka}, Pascale and {Fern{\'a}ndez-Alvar}, Emma and {Glover}, Jennifer and {McConnachie}, Alan W. and {Chen{\'e}}, Andr{\'e}-Nicolas},
        title = "{The extended 'stellar halo' of the Ursa Minor dwarf galaxy}",
      journal = {\mnras},
         year = 2023,
        month = oct,
       volume = {525},
       number = {2},
        pages = {2875-2890},
          doi = {10.1093/mnras/stad2427},
archivePrefix = {arXiv},
       eprint = {2301.13214},
 primaryClass = {astro-ph.GA},
       adsurl = {https://ui.adsabs.harvard.edu/abs/2023MNRAS.525.2875S}
}

@ARTICLE{Kado-Fong2022,
       author = {{Kado-Fong}, Erin and {Sanderson}, Robyn E. and {Greene}, Jenny E. and {Cunningham}, Emily C. and {Wheeler}, Coral and {Chan}, T.~K. and {El-Badry}, Kareem and {Hopkins}, Philip F. and {Wetzel}, Andrew and {Boylan-Kolchin}, Michael and {Faucher-Gigu{\`e}re}, Claude-Andr{\'e} and {Huang}, Song and {Quataert}, Eliot and {Starkenburg}, Tjitske},
        title = "{The In Situ Origins of Dwarf Stellar Outskirts in FIRE-2}",
      journal = {\apj},
         year = 2022,
        month = jun,
       volume = {931},
       number = {2},
          eid = {152},
        pages = {152},
          doi = {10.3847/1538-4357/ac6c88},
archivePrefix = {arXiv},
       eprint = {2109.05034},
 primaryClass = {astro-ph.GA},
       adsurl = {https://ui.adsabs.harvard.edu/abs/2022ApJ...931..152K}
}

@ARTICLE{Sales2022,
       author = {{Sales}, Laura V. and {Wetzel}, Andrew and {Fattahi}, Azadeh},
        title = "{Baryonic solutions and challenges for cosmological models of dwarf galaxies}",
      journal = {Nature Astronomy},
         year = 2022,
        month = jun,
       volume = {6},
        pages = {897-910},
          doi = {10.1038/s41550-022-01689-w},
archivePrefix = {arXiv},
       eprint = {2206.05295},
 primaryClass = {astro-ph.GA},
       adsurl = {https://ui.adsabs.harvard.edu/abs/2022NatAs...6..897S}
}

@ARTICLE{Chiti2021,
       author = {{Chiti}, Anirudh and {Frebel}, Anna and {Simon}, Joshua D. and {Erkal}, Denis and {Chang}, Laura J. and {Necib}, Lina and {Ji}, Alexander P. and {Jerjen}, Helmut and {Kim}, Dongwon and {Norris}, John E.},
        title = "{An extended halo around an ancient dwarf galaxy}",
      journal = {Nature Astronomy},
         year = 2021,
        month = apr,
       volume = {5},
        pages = {392-400},
          doi = {10.1038/s41550-020-01285-w},
archivePrefix = {arXiv},
       eprint = {2012.02309},
 primaryClass = {astro-ph.GA},
       adsurl = {https://ui.adsabs.harvard.edu/abs/2021NatAs...5..392C}
}

@ARTICLE{Tarumi2021,
       author = {{Tarumi}, Yuta and {Yoshida}, Naoki and {Frebel}, Anna},
        title = "{Formation of an Extended Stellar Halo around an Ultra-faint Dwarf Galaxy Following One of the Earliest Mergers from Galactic Building Blocks}",
      journal = {\apjl},
         year = 2021,
        month = jun,
       volume = {914},
       number = {1},
          eid = {L10},
        pages = {L10},
          doi = {10.3847/2041-8213/ac024e},
archivePrefix = {arXiv},
       eprint = {2103.01962},
 primaryClass = {astro-ph.GA},
       adsurl = {https://ui.adsabs.harvard.edu/abs/2021ApJ...914L..10T}
}

@ARTICLE{Martin2021,
       author = {{Martin}, G. and {Jackson}, R.~A. and {Kaviraj}, S. and {Choi}, H. and {Devriendt}, J.~E.~G. and {Dubois}, Y. and {Kimm}, T. and {Kraljic}, K. and {Peirani}, S. and {Pichon}, C. and {Volonteri}, M. and {Yi}, S.~K.},
        title = "{The role of mergers and interactions in driving the evolution of dwarf galaxies over cosmic time}",
      journal = {\mnras},
         year = 2021,
        month = jan,
       volume = {500},
       number = {4},
        pages = {4937-4957},
          doi = {10.1093/mnras/staa3443},
archivePrefix = {arXiv},
       eprint = {2007.07913},
 primaryClass = {astro-ph.GA},
       adsurl = {https://ui.adsabs.harvard.edu/abs/2021MNRAS.500.4937M}
}

@ARTICLE{Pillepich2015,
       author = {{Pillepich}, Annalisa and {Madau}, Piero and {Mayer}, Lucio},
        title = "{Building Late-type Spiral Galaxies by In-situ and Ex-situ Star Formation}",
      journal = {\apj},
         year = 2015,
        month = feb,
       volume = {799},
       number = {2},
          eid = {184},
        pages = {184},
          doi = {10.1088/0004-637X/799/2/184},
archivePrefix = {arXiv},
       eprint = {1407.7855},
 primaryClass = {astro-ph.GA},
       adsurl = {https://ui.adsabs.harvard.edu/abs/2015ApJ...799..184P}
}

@ARTICLE{Jensen2024,
       author = {{Jensen}, Jaclyn and {Hayes}, Christian R. and {Sestito}, Federico and {McConnachie}, Alan W. and {Waller}, Fletcher and {Smith}, Simon E.~T. and {Navarro}, Julio and {Venn}, Kim A.},
        title = "{Small-scale stellar haloes: detecting low surface brightness features in the outskirts of Milky Way dwarf satellites}",
      journal = {\mnras},
         year = 2024,
        month = jan,
       volume = {527},
       number = {2},
        pages = {4209-4233},
          doi = {10.1093/mnras/stad3322},
archivePrefix = {arXiv},
       eprint = {2308.07394},
 primaryClass = {astro-ph.GA},
       adsurl = {https://ui.adsabs.harvard.edu/abs/2024MNRAS.527.4209J}
}

@ARTICLE{Willmer2018,
       author = {{Willmer}, Christopher N.~A.},
        title = "{The Absolute Magnitude of the Sun in Several Filters}",
      journal = {\apjs},
         year = 2018,
        month = jun,
       volume = {236},
       number = {2},
          eid = {47},
        pages = {47},
          doi = {10.3847/1538-4365/aabfdf},
archivePrefix = {arXiv},
       eprint = {1804.07788},
 primaryClass = {astro-ph.SR},
       adsurl = {https://ui.adsabs.harvard.edu/abs/2018ApJS..236...47W}
}

@ARTICLE{Roman2023,
       author = {{Rom{\'a}n}, J. and {S{\'a}nchez-Alarc{\'o}n}, P.~M. and {Knapen}, J.~H. and {Peletier}, R.},
        title = "{Evidence for globular cluster collapse after a dwarf-dwarf merger: A potential nuclear star cluster in formation}",
      journal = {\aap},
         year = 2023,
        month = mar,
       volume = {671},
          eid = {L7},
        pages = {L7},
          doi = {10.1051/0004-6361/202345928},
archivePrefix = {arXiv},
       eprint = {2302.08516},
 primaryClass = {astro-ph.GA},
       adsurl = {https://ui.adsabs.harvard.edu/abs/2023A&A...671L...7R}
}

@ARTICLE{Correnti2025,
       author = {{Correnti}, Matteo and {Annibali}, Francesca and {Bellazzini}, Michele and {Marinelli}, Mariarosa and {Aloisi}, Alessandra and {Cignoni}, Michele and {Tosi}, Monica and {Pascale}, Raffaele and {Cannon}, John M. and {Schisgal}, Lila and {Hunt}, Leslie K. and {Sacchi}, Elena and {Sohn}, Sangmo Tony},
        title = "{DDO 68-C: HST Confirms Yet Another Companion of the Isolated Dwarf Galaxy DDO 68}",
      journal = {\apj},
         year = 2025,
        month = mar,
       volume = {982},
       number = {1},
          eid = {31},
        pages = {31},
          doi = {10.3847/1538-4357/adb7e6},
archivePrefix = {arXiv},
       eprint = {2502.18171},
 primaryClass = {astro-ph.GA},
       adsurl = {https://ui.adsabs.harvard.edu/abs/2025ApJ...982...31C}
}

@ARTICLE{Massey1995,
       author = {{Massey}, Philip and {Armandroff}, Taft E.},
        title = "{The Massive Star Content, Reddening, and Distance of the Nearby Irregular Galaxy IC 10}",
      journal = {\aj},
         year = 1995,
        month = jun,
       volume = {109},
        pages = {2470},
          doi = {10.1086/117465},
       adsurl = {https://ui.adsabs.harvard.edu/abs/1995AJ....109.2470M}
}

@ARTICLE{Gholami2025,
       author = {{Gholami}, Mahtab and {Javadi}, Atefeh and {Mahani}, Hamidreza and {van Loon}, Jacco Th. and {Khosroshahi}, Habib and {Saremi}, Elham and {McDonald}, Iain and {Eftekhari}, Samaneh and {Ren}, Yi and {Altafi}, Hamed},
        title = "{The Isaac Newton Telescope Monitoring Survey of Local Group Dwarf Galaxies. VII. Long-period Variable Stars in the Nearest Starburst Dwarf Galaxy, IC 10}",
      journal = {\aj},
         year = 2025,
        month = jul,
       volume = {170},
       number = {1},
          eid = {54},
        pages = {54},
          doi = {10.3847/1538-3881/adda2b},
archivePrefix = {arXiv},
       eprint = {2505.03972},
 primaryClass = {astro-ph.GA},
       adsurl = {https://ui.adsabs.harvard.edu/abs/2025AJ....170...54G}
}

@ARTICLE{Tehrani2017,
       author = {{Tehrani}, Katie and {Crowther}, Paul A. and {Archer}, I.},
        title = "{Revealing the nebular properties and Wolf-Rayet population of IC10 with Gemini/GMOS}",
      journal = {\mnras},
         year = 2017,
        month = dec,
       volume = {472},
       number = {4},
        pages = {4618-4633},
          doi = {10.1093/mnras/stx2124},
archivePrefix = {arXiv},
       eprint = {1708.03634},
 primaryClass = {astro-ph.GA},
       adsurl = {https://ui.adsabs.harvard.edu/abs/2017MNRAS.472.4618T}
}

@ARTICLE{Crowther2009,
       author = {{Crowther}, P.~A. and {Bibby}, J.~L.},
        title = "{On the massive star content of the nearby dwarf irregular Wolf-Rayet galaxy IC 4662}",
      journal = {\aap},
         year = 2009,
        month = may,
       volume = {499},
       number = {2},
        pages = {455-464},
          doi = {10.1051/0004-6361/200911758},
archivePrefix = {arXiv},
       eprint = {0903.2288},
 primaryClass = {astro-ph.CO},
       adsurl = {https://ui.adsabs.harvard.edu/abs/2009A&A...499..455C}
}

@ARTICLE{Koch2025TheLGLBS,
       author = {{Koch}, Eric W. and {Leroy}, Adam K. and {Rosolowsky}, Erik W. and {Chomiuk}, Laura and {Dalcanton}, Julianne J. and {Pingel}, Nickolas M. and {Sarbadhicary}, Sumit K. and {Stanimirovi{\'c}}, Sne{\v{z}}ana and {Walter}, Fabian and {Archer}, Haylee N. and {Bolatto}, Alberto D. and {Busch}, Michael P. and {Chen}, Hongxing and {Chown}, Ryan and {Corbould}, Harrisen and {Cronin}, Serena A. and {Darling}, Jeremy and {Do}, Thomas and {Meyer}, Jennifer Donovan and {Eibensteiner}, Cosima and {Hunter}, Deidre and {Indebetouw}, R{\'e}my and {Jagannathan}, Preshanth and {Kepley}, Amanda A. and {Kim}, Chang-Goo and {Kim}, Shin-Jeong and {Kovacs}, Timea O. and {Marvil}, Joshua and {Murphy}, Eric J. and {Murray}, Claire E. and {Ott}, J{\"u}rgen and {Pisano}, D.~J. and {Putman}, Mary and {Rybarczyk}, Daniel R. and {Roman-Duval}, Julia and {Sandstrom}, Karin and {Schinnerer}, Eva and {Skillman}, Evan D. and {Smercina}, Adam and {Stelea}, Ioana and {Strader}, Jay and {Sun}, Jiayi and {Tallapaneni}, Devisree and {Tarantino}, Elizabeth and {Villanueva}, Vicente and {Weisz}, Daniel R. and {Williams}, Thomas G. and {Wong}, Tony},
        title = "{The Karl G. Jansky Very Large Array Local Group L-Band Survey (LGLBS)}",
      journal = {\apjs},
         year = 2025,
        month = aug,
       volume = {279},
       number = {2},
          eid = {35},
        pages = {35},
          doi = {10.3847/1538-4365/ade0ad},
archivePrefix = {arXiv},
       eprint = {2506.11792},
 primaryClass = {astro-ph.GA},
       adsurl = {https://ui.adsabs.harvard.edu/abs/2025ApJS..279...35K}
}

@ARTICLE{Mancera2024,
       author = {{Mancera Pi{\~n}a}, Pavel E. and {Golini}, Giulia and {Trujillo}, Ignacio and {Montes}, Mireia},
        title = "{Exploring the nature of dark matter with the extreme galaxy AGC 114905}",
      journal = {\aap},
         year = 2024,
        month = sep,
       volume = {689},
          eid = {A344},
        pages = {A344},
          doi = {10.1051/0004-6361/202450230},
archivePrefix = {arXiv},
       eprint = {2404.06537},
 primaryClass = {astro-ph.GA},
       adsurl = {https://ui.adsabs.harvard.edu/abs/2024A&A...689A.344M}
}

@ARTICLE{Peng2011,
       author = {{Peng}, Eric W. and {Ferguson}, Henry C. and {Goudfrooij}, Paul and {Hammer}, Derek and {Lucey}, John R. and {Marzke}, Ronald O. and {Puzia}, Thomas H. and {Carter}, David and {Balcells}, Marc and {Bridges}, Terry and {Chiboucas}, Kristin and {del Burgo}, Carlos and {Graham}, Alister W. and {Guzm{\'a}n}, Rafael and {Hudson}, Michael J. and {Matkovi{\'c}}, Ana and {Merritt}, David and {Miller}, Bryan W. and {Mouhcine}, Mustapha and {Phillipps}, Steven and {Sharples}, Ray and {Smith}, Russell J. and {Tully}, Brent and {Verdoes Kleijn}, Gijs},
        title = "{The HST/ACS Coma Cluster Survey. IV. Intergalactic Globular Clusters and the Massive Globular Cluster System at the Core of the Coma Galaxy Cluster}",
      journal = {\apj},
         year = 2011,
        month = mar,
       volume = {730},
       number = {1},
          eid = {23},
        pages = {23},
          doi = {10.1088/0004-637X/730/1/23},
archivePrefix = {arXiv},
       eprint = {1101.1000},
 primaryClass = {astro-ph.GA},
       adsurl = {https://ui.adsabs.harvard.edu/abs/2011ApJ...730...23P}
}

@ARTICLE{Lee2025,
       author = {{Lee}, Abigail J. and {Freedman}, Wendy L. and {Madore}, Barry F. and {Jang}, In Sung and {Owens}, Kayla A. and {Hoyt}, Taylor J.},
        title = "{The Chicago{\textendash}Carnegie Hubble Program: The JWST J-region Asymptotic Giant Branch Extragalactic Distance Scale}",
      journal = {\apj},
         year = 2025,
        month = jun,
       volume = {985},
       number = {2},
          eid = {182},
        pages = {182},
          doi = {10.3847/1538-4357/adc8a1},
archivePrefix = {arXiv},
       eprint = {2408.03474},
 primaryClass = {astro-ph.GA},
       adsurl = {https://ui.adsabs.harvard.edu/abs/2025ApJ...985..182L}
}

@ARTICLE{Nally2024,
       author = {{Nally}, Conor and {Jones}, Olivia C. and {Lenki{\'c}}, Laura and {Habel}, Nolan and {Hirschauer}, Alec S. and {Meixner}, Margaret and {Kavanagh}, P.~J. and {Boyer}, Martha L. and {Ferguson}, Annette M.~N. and {Sargent}, B.~A. and {Nayak}, Omnarayani and {Temim}, Tea},
        title = "{JWST MIRI and NIRCam unveil previously unseen infrared stellar populations in NGC 6822}",
      journal = {\mnras},
         year = 2024,
        month = jun,
       volume = {531},
       number = {1},
        pages = {183-198},
          doi = {10.1093/mnras/stae1163},
archivePrefix = {arXiv},
       eprint = {2309.13521},
 primaryClass = {astro-ph.GA},
       adsurl = {https://ui.adsabs.harvard.edu/abs/2024MNRAS.531..183N}
}

@ARTICLE{Richer01,
       author = {{Richer}, M.~G. and {Bullejos}, A. and {Borissova}, J. and {McCall}, M.~L. and {Lee}, H. and {Kurtev}, R. and {Georgiev}, L. and {Kingsburgh}, R.~L. and {Ross}, R. and {Rosado}, M.},
        title = "{IC 10: More evidence that it is a blue compact dwarf}",
      journal = {\aap},
         year = 2001,
        month = apr,
       volume = {370},
        pages = {34-42},
          doi = {10.1051/0004-6361:20010206},
archivePrefix = {arXiv},
       eprint = {astro-ph/0103066},
 primaryClass = {astro-ph},
       adsurl = {https://ui.adsabs.harvard.edu/abs/2001A&A...370...34R}
}

@ARTICLE{Binder2025The10,
       author = {{Binder}, Breanna A. and {Lazarus}, Rosalie and {Thoresen}, Mina and {Laycock}, Silas and {Bhattacharya}, Sayantan},
        title = "{The X-Ray Variability and Luminosity Function of High-mass X-Ray Binaries in the Dwarf Starburst Galaxy IC 10}",
      journal = {\apj},
         year = 2025,
        month = sep,
       volume = {991},
       number = {1},
          eid = {27},
        pages = {27},
          doi = {10.3847/1538-4357/adf7a7},
archivePrefix = {arXiv},
       eprint = {2508.02876},
 primaryClass = {astro-ph.HE},
       adsurl = {https://ui.adsabs.harvard.edu/abs/2025ApJ...991...27B}
}

@ARTICLE{Jacobs2009,
       author = {{Jacobs}, Bradley A. and {Rizzi}, Luca and {Tully}, R. Brent and {Shaya}, Edward J. and {Makarov}, Dmitry I. and {Makarova}, Lidia},
        title = "{The Extragalactic Distance Database: Color-Magnitude Diagrams}",
      journal = {\aj},
         year = 2009,
        month = aug,
       volume = {138},
       number = {2},
        pages = {332-337},
          doi = {10.1088/0004-6256/138/2/332},
archivePrefix = {arXiv},
       eprint = {0902.3675},
 primaryClass = {astro-ph.CO},
       adsurl = {https://ui.adsabs.harvard.edu/abs/2009AJ....138..332J}
}

@ARTICLE{Hunter2001,
       author = {{Hunter}, Deidre A.},
        title = "{The Stellar Population and Star Clusters in the Unusual Local Group Galaxy IC 10}",
      journal = {\apj},
         year = 2001,
        month = sep,
       volume = {559},
       number = {1},
        pages = {225-242},
          doi = {10.1086/322399},
archivePrefix = {arXiv},
       eprint = {astro-ph/0105456},
 primaryClass = {astro-ph},
       adsurl = {https://ui.adsabs.harvard.edu/abs/2001ApJ...559..225H}
}

@ARTICLE{Pace2025,
       author = {{Pace}, Andrew B},
        title = "{The Local Volume Database: a library of the observed properties of nearby dwarf galaxies and star clusters}",
      journal = {The Open Journal of Astrophysics},
         year = 2025,
        month = sep,
       volume = {8},
          eid = {142},
        pages = {142},
          doi = {10.33232/001c.144859},
archivePrefix = {arXiv},
       eprint = {2411.07424},
 primaryClass = {astro-ph.GA},
       adsurl = {https://ui.adsabs.harvard.edu/abs/2025OJAp....8E.142P}
}

@ARTICLE{astropy_13,
       author = {{Astropy Collaboration: Robitaille}, Thomas P. and {Tollerud}, Erik J. and {Greenfield}, Perry and {Droettboom}, Michael and {Bray}, Erik and {Aldcroft}, Tom and {Davis}, Matt and {Ginsburg}, Adam and {Price-Whelan}, Adrian M. and {Kerzendorf}, Wolfgang E. and {Conley}, Alexander and {Crighton}, Neil and {Barbary}, Kyle and {Muna}, Demitri and {Ferguson}, Henry and {Grollier}, Fr{\'e}d{\'e}ric and {Parikh}, Madhura M. and {Nair}, Prasanth H. and {Unther}, Hans M. and {Deil}, Christoph and {Woillez}, Julien and {Conseil}, Simon and {Kramer}, Roban and {Turner}, James E.~H. and {Singer}, Leo and {Fox}, Ryan and {Weaver}, Benjamin A. and {Zabalza}, Victor and {Edwards}, Zachary I. and {Azalee Bostroem}, K. and {Burke}, D.~J. and {Casey}, Andrew R. and {Crawford}, Steven M. and {Dencheva}, Nadia and {Ely}, Justin and {Jenness}, Tim and {Labrie}, Kathleen and {Lim}, Pey Lian and {Pierfederici}, Francesco and {Pontzen}, Andrew and {Ptak}, Andy and {Refsdal}, Brian and {Servillat}, Mathieu and {Streicher}, Ole},
        title = "{Astropy: A community Python package for astronomy}",
      journal = {\aap},
         year = 2013,
        month = oct,
       volume = {558},
          eid = {A33},
        pages = {A33},
          doi = {10.1051/0004-6361/201322068},
archivePrefix = {arXiv},
       eprint = {1307.6212},
 primaryClass = {astro-ph.IM},
       adsurl = {https://ui.adsabs.harvard.edu/abs/2013A&A...558A..33A}
}

@ARTICLE{astropy_18,
       author = {{Astropy Collaboration: Price-Whelan}, A.~M. and {Sip{\H{o}}cz}, B.~M. and {G{\"u}nther}, H.~M. and {Lim}, P.~L. and {Crawford}, S.~M. and {Conseil}, S. and {Shupe}, D.~L. and {Craig}, M.~W. and {Dencheva}, N. and {Ginsburg}, A. and {VanderPlas}, J.~T. and {Bradley}, L.~D. and {P{\'e}rez-Su{\'a}rez}, D. and {de Val-Borro}, M. and {Aldcroft}, T.~L. and {Cruz}, K.~L. and {Robitaille}, T.~P. and {Tollerud}, E.~J. and {Ardelean}, C. and {Babej}, T. and {Bach}, Y.~P. and {Bachetti}, M. and {Bakanov}, A.~V. and {Bamford}, S.~P. and {Barentsen}, G. and {Barmby}, P. and {Baumbach}, A. and {Berry}, K.~L. and {Biscani}, F. and {Boquien}, M. and {Bostroem}, K.~A. and {Bouma}, L.~G. and {Brammer}, G.~B. and {Bray}, E.~M. and {Breytenbach}, H. and {Buddelmeijer}, H. and {Burke}, D.~J. and {Calderone}, G. and {Cano Rodr{\'\i}guez}, J.~L. and {Cara}, M. and {Cardoso}, J.~V.~M. and {Cheedella}, S. and {Copin}, Y. and {Corrales}, L. and {Crichton}, D. and {D'Avella}, D. and {Deil}, C. and {Depagne}, {\'E}. and {Dietrich}, J.~P. and {Donath}, A. and {Droettboom}, M. and {Earl}, N. and {Erben}, T. and {Fabbro}, S. and {Ferreira}, L.~A. and {Finethy}, T. and {Fox}, R.~T. and {Garrison}, L.~H. and {Gibbons}, S.~L.~J. and {Goldstein}, D.~A. and {Gommers}, R. and {Greco}, J.~P. and {Greenfield}, P. and {Groener}, A.~M. and {Grollier}, F. and {Hagen}, A. and {Hirst}, P. and {Homeier}, D. and {Horton}, A.~J. and {Hosseinzadeh}, G. and {Hu}, L. and {Hunkeler}, J.~S. and {Ivezi{\'c}}, {\v{Z}}. and {Jain}, A. and {Jenness}, T. and {Kanarek}, G. and {Kendrew}, S. and {Kern}, N.~S. and {Kerzendorf}, W.~E. and {Khvalko}, A. and {King}, J. and {Kirkby}, D. and {Kulkarni}, A.~M. and {Kumar}, A. and {Lee}, A. and {Lenz}, D. and {Littlefair}, S.~P. and {Ma}, Z. and {Macleod}, D.~M. and {Mastropietro}, M. and {McCully}, C. and {Montagnac}, S. and {Morris}, B.~M. and {Mueller}, M. and {Mumford}, S.~J. and {Muna}, D. and {Murphy}, N.~A. and {Nelson}, S. and {Nguyen}, G.~H. and {Ninan}, J.~P. and {N{\"o}the}, M. and {Ogaz}, S. and {Oh}, S. and {Parejko}, J.~K. and {Parley}, N. and {Pascual}, S. and {Patil}, R. and {Patil}, A.~A. and {Plunkett}, A.~L. and {Prochaska}, J.~X. and {Rastogi}, T. and {Reddy Janga}, V. and {Sabater}, J. and {Sakurikar}, P. and {Seifert}, M. and {Sherbert}, L.~E. and {Sherwood-Taylor}, H. and {Shih}, A.~Y. and {Sick}, J. and {Silbiger}, M.~T. and {Singanamalla}, S. and {Singer}, L.~P. and {Sladen}, P.~H. and {Sooley}, K.~A. and {Sornarajah}, S. and {Streicher}, O. and {Teuben}, P. and {Thomas}, S.~W. and {Tremblay}, G.~R. and {Turner}, J.~E.~H. and {Terr{\'o}n}, V. and {van Kerkwijk}, M.~H. and {de la Vega}, A. and {Watkins}, L.~L. and {Weaver}, B.~A. and {Whitmore}, J.~B. and {Woillez}, J. and {Zabalza}, V. and {Astropy Contributors}},
        title = "{The Astropy Project: Building an Open-science Project and Status of the v2.0 Core Package}",
      journal = {\aj},
         year = 2018,
        month = sep,
       volume = {156},
       number = {3},
          eid = {123},
        pages = {123},
          doi = {10.3847/1538-3881/aabc4f},
archivePrefix = {arXiv},
       eprint = {1801.02634},
 primaryClass = {astro-ph.IM},
       adsurl = {https://ui.adsabs.harvard.edu/abs/2018AJ....156..123A}
}

@ARTICLE{astropy_22,
       author = {{Astropy Collaboration: Price-Whelan}, Adrian M. and {Lim}, Pey Lian and {Earl}, Nicholas and {Starkman}, Nathaniel and {Bradley}, Larry and {Shupe}, David L. and {Patil}, Aarya A. and {Corrales}, Lia and {Brasseur}, C.~E. and {N{\"o}the}, Maximilian and {Donath}, Axel and {Tollerud}, Erik and {Morris}, Brett M. and {Ginsburg}, Adam and {Vaher}, Eero and {Weaver}, Benjamin A. and {Tocknell}, James and {Jamieson}, William and {van Kerkwijk}, Marten H. and {Robitaille}, Thomas P. and {Merry}, Bruce and {Bachetti}, Matteo and {G{\"u}nther}, H. Moritz and {Aldcroft}, Thomas L. and {Alvarado-Montes}, Jaime A. and {Archibald}, Anne M. and {B{\'o}di}, Attila and {Bapat}, Shreyas and {Barentsen}, Geert and {Baz{\'a}n}, Juanjo and {Biswas}, Manish and {Boquien}, M{\'e}d{\'e}ric and {Burke}, D.~J. and {Cara}, Daria and {Cara}, Mihai and {Conroy}, Kyle E. and {Conseil}, Simon and {Craig}, Matthew W. and {Cross}, Robert M. and {Cruz}, Kelle L. and {D'Eugenio}, Francesco and {Dencheva}, Nadia and {Devillepoix}, Hadrien A.~R. and {Dietrich}, J{\"o}rg P. and {Eigenbrot}, Arthur Davis and {Erben}, Thomas and {Ferreira}, Leonardo and {Foreman-Mackey}, Daniel and {Fox}, Ryan and {Freij}, Nabil and {Garg}, Suyog and {Geda}, Robel and {Glattly}, Lauren and {Gondhalekar}, Yash and {Gordon}, Karl D. and {Grant}, David and {Greenfield}, Perry and {Groener}, Austen M. and {Guest}, Steve and {Gurovich}, Sebastian and {Handberg}, Rasmus and {Hart}, Akeem and {Hatfield-Dodds}, Zac and {Homeier}, Derek and {Hosseinzadeh}, Griffin and {Jenness}, Tim and {Jones}, Craig K. and {Joseph}, Prajwel and {Kalmbach}, J. Bryce and {Karamehmetoglu}, Emir and {Ka{\l}uszy{\'n}ski}, Miko{\l}aj and {Kelley}, Michael S.~P. and {Kern}, Nicholas and {Kerzendorf}, Wolfgang E. and {Koch}, Eric W. and {Kulumani}, Shankar and {Lee}, Antony and {Ly}, Chun and {Ma}, Zhiyuan and {MacBride}, Conor and {Maljaars}, Jakob M. and {Muna}, Demitri and {Murphy}, N.~A. and {Norman}, Henrik and {O'Steen}, Richard and {Oman}, Kyle A. and {Pacifici}, Camilla and {Pascual}, Sergio and {Pascual-Granado}, J. and {Patil}, Rohit R. and {Perren}, Gabriel I. and {Pickering}, Timothy E. and {Rastogi}, Tanuj and {Roulston}, Benjamin R. and {Ryan}, Daniel F. and {Rykoff}, Eli S. and {Sabater}, Jose and {Sakurikar}, Parikshit and {Salgado}, Jes{\'u}s and {Sanghi}, Aniket and {Saunders}, Nicholas and {Savchenko}, Volodymyr and {Schwardt}, Ludwig and {Seifert-Eckert}, Michael and {Shih}, Albert Y. and {Jain}, Anany Shrey and {Shukla}, Gyanendra and {Sick}, Jonathan and {Simpson}, Chris and {Singanamalla}, Sudheesh and {Singer}, Leo P. and {Singhal}, Jaladh and {Sinha}, Manodeep and {Sip{\H{o}}cz}, Brigitta M. and {Spitler}, Lee R. and {Stansby}, David and {Streicher}, Ole and {{\v{S}}umak}, Jani and {Swinbank}, John D. and {Taranu}, Dan S. and {Tewary}, Nikita and {Tremblay}, Grant R. and {de Val-Borro}, Miguel and {Van Kooten}, Samuel J. and {Vasovi{\'c}}, Zlatan and {Verma}, Shresth and {de Miranda Cardoso}, Jos{\'e} Vin{\'\i}cius and {Williams}, Peter K.~G. and {Wilson}, Tom J. and {Winkel}, Benjamin and {Wood-Vasey}, W.~M. and {Xue}, Rui and {Yoachim}, Peter and {Zhang}, Chen and {Zonca}, Andrea and {Astropy Project Contributors}},
        title = "{The Astropy Project: Sustaining and Growing a Community-oriented Open-source Project and the Latest Major Release (v5.0) of the Core Package}",
      journal = {\apj},
         year = 2022,
        month = aug,
       volume = {935},
       number = {2},
          eid = {167},
        pages = {167},
          doi = {10.3847/1538-4357/ac7c74},
archivePrefix = {arXiv},
       eprint = {2206.14220},
 primaryClass = {astro-ph.IM},
       adsurl = {https://ui.adsabs.harvard.edu/abs/2022ApJ...935..167A}
}

@ARTICLE{Sarajedini2007,
       author = {{Sarajedini}, Ata and {Bedin}, Luigi R. and {Chaboyer}, Brian and {Dotter}, Aaron and {Siegel}, Michael and {Anderson}, Jay and {Aparicio}, Antonio and {King}, Ivan and {Majewski}, Steven and {Mar{\'\i}n-Franch}, A. and {Piotto}, Giampaolo and {Reid}, I. Neill and {Rosenberg}, Alfred},
        title = "{The ACS Survey of Galactic Globular Clusters. I. Overview and Clusters without Previous Hubble Space Telescope Photometry}",
      journal = {\aj},
         year = 2007,
        month = apr,
       volume = {133},
       number = {4},
        pages = {1658-1672},
          doi = {10.1086/511979},
archivePrefix = {arXiv},
       eprint = {astro-ph/0612598},
 primaryClass = {astro-ph},
       adsurl = {https://ui.adsabs.harvard.edu/abs/2007AJ....133.1658S}
}

@ARTICLE{Zhang2021,
       author = {{Zhang}, Shumeng and {Mackey}, Dougal and {Da Costa}, Gary S.},
        title = "{A panoramic view of the Local Group dwarf galaxy NGC 6822}",
      journal = {\mnras},
         year = 2021,
        month = dec,
       volume = {508},
       number = {2},
        pages = {2098-2113},
          doi = {10.1093/mnras/stab2642},
archivePrefix = {arXiv},
       eprint = {2108.04431},
 primaryClass = {astro-ph.GA},
       adsurl = {https://ui.adsabs.harvard.edu/abs/2021MNRAS.508.2098Z}
}

@ARTICLE{Bennet2024,
       author = {{Bennet}, Paul and {Patel}, Ekta and {Sohn}, Sangmo Tony and {del Pino Molina}, Andr{\'e}s and {van der Marel}, Roeland P. and {Libralato}, Mattia and {Watkins}, Laura L. and {Aparicio}, Antonio and {Besla}, Gurtina and {Gallart}, Carme and {Fardal}, Mark A. and {Monelli}, Matteo and {Sacchi}, Elena and {Tollerud}, Erik and {Weisz}, Daniel R.},
        title = "{Proper Motions and Orbits of Distant Local Group Dwarf Galaxies from a Combination of Gaia and Hubble Data}",
      journal = {\apj},
         year = 2024,
        month = aug,
       volume = {971},
       number = {1},
          eid = {98},
        pages = {98},
          doi = {10.3847/1538-4357/ad5349},
archivePrefix = {arXiv},
       eprint = {2312.09276},
 primaryClass = {astro-ph.GA},
       adsurl = {https://ui.adsabs.harvard.edu/abs/2024ApJ...971...98B}
}

@ARTICLE{Brunthaler2007,
       author = {{Brunthaler}, A. and {Reid}, M.~J. and {Falcke}, H. and {Henkel}, C. and {Menten}, K.~M.},
        title = "{The proper motion of the Local Group galaxy IC 10}",
      journal = {\aap},
         year = 2007,
        month = jan,
       volume = {462},
       number = {1},
        pages = {101-106},
          doi = {10.1051/0004-6361:20066430},
archivePrefix = {arXiv},
       eprint = {astro-ph/0610774},
 primaryClass = {astro-ph},
       adsurl = {https://ui.adsabs.harvard.edu/abs/2007A&A...462..101B}
}

@ARTICLE{Munoz2008,
       author = {{Mu{\~n}oz}, Ricardo R. and {Majewski}, Steven R. and {Johnston}, Kathryn V.},
        title = "{Modeling the Structure and Dynamics of Dwarf Spheroidal Galaxies with Dark Matter and Tides}",
      journal = {\apj},
         year = 2008,
        month = may,
       volume = {679},
       number = {1},
        pages = {346-372},
          doi = {10.1086/587125},
archivePrefix = {arXiv},
       eprint = {0712.4312},
 primaryClass = {astro-ph},
       adsurl = {https://ui.adsabs.harvard.edu/abs/2008ApJ...679..346M}
}

@ARTICLE{Johnston1999,
       author = {{Johnston}, K.~V. and {Majewski}, S.~R. and {Siegel}, M.~H. and {Reid}, I.~N. and {Kunkel}, W.~E.},
        title = "{Constraining the History of the Sagittarius Dwarf Galaxy Using Observations of Its Tidal Debris}",
      journal = {\aj},
         year = 1999,
        month = oct,
       volume = {118},
       number = {4},
        pages = {1719-1726},
          doi = {10.1086/301037},
archivePrefix = {arXiv},
       eprint = {astro-ph/9906457},
 primaryClass = {astro-ph},
       adsurl = {https://ui.adsabs.harvard.edu/abs/1999AJ....118.1719J}
}

@ARTICLE{Kroupa2001,
       author = {{Kroupa}, Pavel},
        title = "{On the variation of the initial mass function}",
      journal = {\mnras},
         year = 2001,
        month = apr,
       volume = {322},
       number = {2},
        pages = {231-246},
          doi = {10.1046/j.1365-8711.2001.04022.x},
archivePrefix = {arXiv},
       eprint = {astro-ph/0009005},
 primaryClass = {astro-ph},
       adsurl = {https://ui.adsabs.harvard.edu/abs/2001MNRAS.322..231K}
}

@ARTICLE{Kroupa2002,
       author = {{Kroupa}, Pavel},
        title = "{The Initial Mass Function of Stars: Evidence for Uniformity in Variable Systems}",
      journal = {Science},
         year = 2002,
        month = jan,
       volume = {295},
       number = {5552},
        pages = {82-91},
          doi = {10.1126/science.1067524},
archivePrefix = {arXiv},
       eprint = {astro-ph/0201098},
 primaryClass = {astro-ph},
       adsurl = {https://ui.adsabs.harvard.edu/abs/2002Sci...295...82K}
}

@INCOLLECTION{Kroupa2013,
       author = {{Kroupa}, Pavel and {Weidner}, Carsten and {Pflamm-Altenburg}, Jan and {Thies}, Ingo and {Dabringhausen}, J{\"o}rg and {Marks}, Michael and {Maschberger}, Thomas},
        title = "{The Stellar and Sub-Stellar Initial Mass Function of Simple and Composite Populations}",
    booktitle = {Planets, Stars and Stellar Systems. Volume 5: Galactic Structure and Stellar Populations},
         year = 2013,
       editor = {{Oswalt}, Terry D. and {Gilmore}, Gerard},
       volume = {5},
        pages = {115},
          doi = {10.1007/978-94-007-5612-0_4},
       adsurl = {https://ui.adsabs.harvard.edu/abs/2013pss5.book..115K}
}

@ARTICLE{Tang2014,
       author = {{Tang}, Jing and {Bressan}, Alessandro and {Rosenfield}, Philip and {Slemer}, Alessandra and {Marigo}, Paola and {Girardi}, L{\'e}o and {Bianchi}, Luciana},
        title = "{New PARSEC evolutionary tracks of massive stars at low metallicity: testing canonical stellar evolution in nearby star-forming dwarf galaxies}",
      journal = {\mnras},
         year = 2014,
        month = dec,
       volume = {445},
       number = {4},
        pages = {4287-4305},
          doi = {10.1093/mnras/stu2029},
archivePrefix = {arXiv},
       eprint = {1410.1745},
 primaryClass = {astro-ph.SR},
       adsurl = {https://ui.adsabs.harvard.edu/abs/2014MNRAS.445.4287T}
}

@ARTICLE{Chen2014,
       author = {{Chen}, Yang and {Girardi}, L{\'e}o and {Bressan}, Alessandro and {Marigo}, Paola and {Barbieri}, Mauro and {Kong}, Xu},
        title = "{Improving PARSEC models for very low mass stars}",
      journal = {\mnras},
         year = 2014,
        month = nov,
       volume = {444},
       number = {3},
        pages = {2525-2543},
          doi = {10.1093/mnras/stu1605},
archivePrefix = {arXiv},
       eprint = {1409.0322},
 primaryClass = {astro-ph.SR},
       adsurl = {https://ui.adsabs.harvard.edu/abs/2014MNRAS.444.2525C}
}

@ARTICLE{Chen2015,
       author = {{Chen}, Yang and {Bressan}, Alessandro and {Girardi}, L{\'e}o and {Marigo}, Paola and {Kong}, Xu and {Lanza}, Antonio},
        title = "{PARSEC evolutionary tracks of massive stars up to 350 M$_{{\ensuremath{\odot}}}$ at metallicities 0.0001 {\ensuremath{\leq}} Z {\ensuremath{\leq}} 0.04}",
      journal = {\mnras},
         year = 2015,
        month = sep,
       volume = {452},
       number = {1},
        pages = {1068-1080},
          doi = {10.1093/mnras/stv1281},
archivePrefix = {arXiv},
       eprint = {1506.01681},
 primaryClass = {astro-ph.SR},
       adsurl = {https://ui.adsabs.harvard.edu/abs/2015MNRAS.452.1068C}
}

@ARTICLE{Pastorelli2020,
       author = {{Pastorelli}, Giada and {Marigo}, Paola and {Girardi}, L{\'e}o and {Aringer}, Bernhard and {Chen}, Yang and {Rubele}, Stefano and {Trabucchi}, Michele and {Bladh}, Sara and {Boyer}, Martha L. and {Bressan}, Alessandro and {Dalcanton}, Julianne J. and {Groenewegen}, Martin A.~T. and {Lebzelter}, Thomas and {Mowlavi}, Nami and {Chubb}, Katy L. and {Cioni}, Maria-Rosa L. and {de Grijs}, Richard and {Ivanov}, Valentin D. and {Nanni}, Ambra and {van Loon}, Jacco Th and {Zaggia}, Simone},
        title = "{Constraining the thermally pulsing asymptotic giant branch phase with resolved stellar populations in the Large Magellanic Cloud}",
      journal = {\mnras},
         year = 2020,
        month = nov,
       volume = {498},
       number = {3},
        pages = {3283-3301},
          doi = {10.1093/mnras/staa2565},
archivePrefix = {arXiv},
       eprint = {2008.08595},
 primaryClass = {astro-ph.SR},
       adsurl = {https://ui.adsabs.harvard.edu/abs/2020MNRAS.498.3283P}
}

@ARTICLE{Pastorelli2019,
       author = {{Pastorelli}, Giada and {Marigo}, Paola and {Girardi}, L{\'e}o and {Chen}, Yang and {Rubele}, Stefano and {Trabucchi}, Michele and {Aringer}, Bernhard and {Bladh}, Sara and {Bressan}, Alessandro and {Montalb{\'a}n}, Josefina and {Boyer}, Martha L. and {Dalcanton}, Julianne J. and {Eriksson}, Kjell and {Groenewegen}, Martin A.~T. and {H{\"o}fner}, Susanne and {Lebzelter}, Thomas and {Nanni}, Ambra and {Rosenfield}, Philip and {Wood}, Peter R. and {Cioni}, Maria-Rosa L.},
        title = "{Constraining the thermally pulsing asymptotic giant branch phase with resolved stellar populations in the Small Magellanic Cloud}",
      journal = {\mnras},
         year = 2019,
        month = jun,
       volume = {485},
       number = {4},
        pages = {5666-5692},
          doi = {10.1093/mnras/stz725},
archivePrefix = {arXiv},
       eprint = {1903.04499},
 primaryClass = {astro-ph.SR},
       adsurl = {https://ui.adsabs.harvard.edu/abs/2019MNRAS.485.5666P}
}

@ARTICLE{Moffat,
       author = {{Moffat}, A.~F.~J.},
        title = "{A Theoretical Investigation of Focal Stellar Images in the Photographic Emulsion and Application to Photographic Photometry}",
      journal = {\aap},
         year = 1969,
        month = dec,
       volume = {3},
        pages = {455},
       adsurl = {https://ui.adsabs.harvard.edu/abs/1969A&A.....3..455M}
}

@ARTICLE{McQuinn2017,
       author = {{McQuinn}, Kristen B.~W. and {Boyer}, Martha L. and {Mitchell}, Mallory B. and {Skillman}, Evan D. and {Gehrz}, R.~D. and {Groenewegen}, Martin A.~T. and {McDonald}, Iain and {Sloan}, G.~C. and {van Loon}, Jacco Th. and {Whitelock}, Patricia A. and {Zijlstra}, Albert A.},
        title = "{DUSTiNGS. III. Distribution of Intermediate-age and Old Stellar Populations in Disks and Outer Extremities of Dwarf Galaxies}",
      journal = {\apj},
         year = 2017,
        month = jan,
       volume = {834},
       number = {1},
          eid = {78},
        pages = {78},
          doi = {10.3847/1538-4357/834/1/78},
archivePrefix = {arXiv},
       eprint = {1611.05044},
 primaryClass = {astro-ph.GA},
       adsurl = {https://ui.adsabs.harvard.edu/abs/2017ApJ...834...78M}
}

@ARTICLE{Golini2024,
       author = {{Golini}, Giulia and {Montes}, Mireia and {Carrasco}, Eleazar R. and {Rom{\'a}n}, Javier and {Trujillo}, Ignacio},
        title = "{Ultra-deep imaging of NGC 1052-DF2 and NGC 1052-DF4 to unravel their origins}",
      journal = {\aap},
         year = 2024,
        month = apr,
       volume = {684},
          eid = {A99},
        pages = {A99},
          doi = {10.1051/0004-6361/202348300},
archivePrefix = {arXiv},
       eprint = {2402.04304},
 primaryClass = {astro-ph.GA},
       adsurl = {https://ui.adsabs.harvard.edu/abs/2024A&A...684A..99G}
}

@ARTICLE{Bedin2025,
       author = {{Bedin}, Luigi ''Rolly''},
        title = "{The case for an astrometric mission extension of Euclid: Extending Gaia by six magnitudes with Euclid covering one third of the sky}",
      journal = {\aap},
         year = 2025,
        month = dec,
       volume = {704},
          eid = {A193},
        pages = {A193},
          doi = {10.1051/0004-6361/202557407},
archivePrefix = {arXiv},
       eprint = {2510.23694},
 primaryClass = {astro-ph.IM},
       adsurl = {https://ui.adsabs.harvard.edu/abs/2025A&A...704A.193B}
}

@ARTICLE{Lokas2013,
       author = {{{\L}okas}, Ewa L. and {Gajda}, Grzegorz and {Kazantzidis}, Stelios},
        title = "{Tidal tails of dwarf galaxies on different orbits around the Milky Way}",
      journal = {\mnras},
         year = 2013,
        month = jul,
       volume = {433},
       number = {1},
        pages = {878-888},
          doi = {10.1093/mnras/stt774},
archivePrefix = {arXiv},
       eprint = {1302.2744},
 primaryClass = {astro-ph.GA},
       adsurl = {https://ui.adsabs.harvard.edu/abs/2013MNRAS.433..878L}
}

@ARTICLE{Penarrubia2009,
       author = {{Pe{\~n}arrubia}, Jorge and {Navarro}, Julio F. and {McConnachie}, Alan W. and {Martin}, Nicolas F.},
        title = "{The Signature of Galactic Tides in Local Group Dwarf Spheroidals}",
      journal = {\apj},
         year = 2009,
        month = jun,
       volume = {698},
       number = {1},
        pages = {222-232},
          doi = {10.1088/0004-637X/698/1/222},
archivePrefix = {arXiv},
       eprint = {0811.1579},
 primaryClass = {astro-ph},
       adsurl = {https://ui.adsabs.harvard.edu/abs/2009ApJ...698..222P}
}

@ARTICLE{Celiz2025,
       author = {{Celiz}, Bruno M. and {Navarro}, Julio F. and {Abadi}, Mario G.},
        title = "{Accreted stars and stellar haloes of simulated galaxies in TNG50}",
      journal = {\aap},
         year = 2025,
        month = nov,
       volume = {704},
          eid = {A57},
        pages = {A57},
          doi = {10.1051/0004-6361/202556633},
archivePrefix = {arXiv},
       eprint = {2510.18971},
 primaryClass = {astro-ph.GA},
       adsurl = {https://ui.adsabs.harvard.edu/abs/2025A&A...704A..57C}
}

@ARTICLE{Martin2025,
       author = {{Martin}, G. and {Watkins}, A.~E. and {Dubois}, Y. and {Devriendt}, J. and {Kaviraj}, S. and {Kim}, D. and {Kraljic}, K. and {Lazar}, I. and {Pearce}, F.~R. and {Peirani}, S. and {Pichon}, C. and {Slyz}, A. and {Yi}, S.~K.},
        title = "{Cosmic reflections I: the structural diversity of simulated and observed low-mass galaxy analogues}",
      journal = {\mnras},
         year = 2025,
        month = aug,
       volume = {541},
       number = {2},
        pages = {1831-1850},
          doi = {10.1093/mnras/staf1092},
archivePrefix = {arXiv},
       eprint = {2505.04509},
 primaryClass = {astro-ph.GA},
       adsurl = {https://ui.adsabs.harvard.edu/abs/2025MNRAS.541.1831M}
}

@ARTICLE{Salomon2021,
       author = {{Salomon}, J.-B. and {Ibata}, R. and {Reyl{\'e}}, C. and {Famaey}, B. and {Libeskind}, N.~I. and {McConnachie}, A.~W. and {Hoffman}, Y.},
        title = "{The proper motion of Andromeda from Gaia EDR3: confirming a nearly radial orbit}",
      journal = {\mnras},
         year = 2021,
        month = oct,
       volume = {507},
       number = {2},
        pages = {2592-2601},
          doi = {10.1093/mnras/stab2253},
archivePrefix = {arXiv},
       eprint = {2012.09204},
 primaryClass = {astro-ph.GA},
       adsurl = {https://ui.adsabs.harvard.edu/abs/2021MNRAS.507.2592S}
}

@ARTICLE{barker12,
       author = {{Barker}, Michael K. and {Ferguson}, Annette M.~N. and {Irwin}, M.~J. and {Arimoto}, N. and {Jablonka}, P.},
        title = "{Quantifying the faint structure of galaxies: the late-type spiral NGC 2403}",
      journal = {\mnras},
         year = 2012,
        month = jan,
       volume = {419},
       number = {2},
        pages = {1489-1506},
          doi = {10.1111/j.1365-2966.2011.19814.x},
archivePrefix = {arXiv},
       eprint = {1109.2625},
 primaryClass = {astro-ph.CO},
       adsurl = {https://ui.adsabs.harvard.edu/abs/2012MNRAS.419.1489B}
}

@ARTICLE{carlin19,
       author = {{Carlin}, Jeffrey L. and {Garling}, Christopher T. and {Peter}, Annika H.~G. and {Crnojevi{\'c}}, Denija and {Forbes}, Duncan A. and {Hargis}, Jonathan R. and {Mutlu-Pakdil}, Bur{\c{c}}in and {Pucha}, Ragadeepika and {Romanowsky}, Aaron J. and {Sand}, David J. and {Spekkens}, Kristine and {Strader}, Jay and {Willman}, Beth},
        title = "{Tidal Destruction in a Low-mass Galaxy Environment: The Discovery of Tidal Tails around DDO 44}",
      journal = {\apj},
         year = 2019,
        month = dec,
       volume = {886},
       number = {2},
          eid = {109},
        pages = {109},
          doi = {10.3847/1538-4357/ab4c32},
archivePrefix = {arXiv},
       eprint = {1906.08260},
 primaryClass = {astro-ph.GA},
       adsurl = {https://ui.adsabs.harvard.edu/abs/2019ApJ...886..109C}
}

@ARTICLE{Ciotti1999,
       author = {{Ciotti}, L. and {Bertin}, G.},
        title = "{Analytical properties of the R$^{1/m}$ law}",
      journal = {\aap},
         year = 1999,
        month = dec,
       volume = {352},
        pages = {447-451},
          doi = {10.48550/arXiv.astro-ph/9911078},
archivePrefix = {arXiv},
       eprint = {astro-ph/9911078},
 primaryClass = {astro-ph},
       adsurl = {https://ui.adsabs.harvard.edu/abs/1999A&A...352..447C}
}

@ARTICLE{gerbrandt15,
       author = {{Gerbrandt}, Stephanie A.~N. and {McConnachie}, Alan W. and {Irwin}, Mike},
        title = "{The red extended structure of IC 10, the nearest blue compact galaxy}",
      journal = {\mnras},
         year = 2015,
        month = nov,
       volume = {454},
       number = {1},
        pages = {1000-1011},
          doi = {10.1093/mnras/stv2029},
archivePrefix = {arXiv},
       eprint = {1509.05436},
 primaryClass = {astro-ph.GA},
       adsurl = {https://ui.adsabs.harvard.edu/abs/2015MNRAS.454.1000G}
}

@ARTICLE{Jedrzejewski87,
       author = {{Jedrzejewski}, Robert I.},
        title = "{CCD surface photometry of elliptical galaxies - I. Observations, reduction and results.}",
      journal = {\mnras},
         year = 1987,
        month = jun,
       volume = {226},
        pages = {747-768},
          doi = {10.1093/mnras/226.4.747},
       adsurl = {https://ui.adsabs.harvard.edu/abs/1987MNRAS.226..747J}
}

@ARTICLE{Marigo2017,
       author = {{Marigo}, Paola and {Girardi}, L{\'e}o and {Bressan}, Alessandro and {Rosenfield}, Philip and {Aringer}, Bernhard and {Chen}, Yang and {Dussin}, Marco and {Nanni}, Ambra and {Pastorelli}, Giada and {Rodrigues}, Tha{\'\i}se S. and {Trabucchi}, Michele and {Bladh}, Sara and {Dalcanton}, Julianne and {Groenewegen}, Martin A.~T. and {Montalb{\'a}n}, Josefina and {Wood}, Peter R.},
        title = "{A New Generation of PARSEC-COLIBRI Stellar Isochrones Including the TP-AGB Phase}",
      journal = {\apj},
         year = 2017,
        month = jan,
       volume = {835},
       number = {1},
          eid = {77},
        pages = {77},
          doi = {10.3847/1538-4357/835/1/77},
archivePrefix = {arXiv},
       eprint = {1701.08510},
 primaryClass = {astro-ph.SR},
       adsurl = {https://ui.adsabs.harvard.edu/abs/2017ApJ...835...77M}
}

@ARTICLE{martinez08,
       author = {{Mart{\'\i}nez-Delgado}, David and {Pe{\~n}arrubia}, Jorge and {Gabany}, R. Jay and {Trujillo}, Ignacio and {Majewski}, Steven R. and {Pohlen}, M.},
        title = "{The Ghost of a Dwarf Galaxy: Fossils of the Hierarchical Formation of the Nearby Spiral Galaxy NGC 5907}",
      journal = {\apj},
         year = 2008,
        month = dec,
       volume = {689},
       number = {1},
        pages = {184-193},
          doi = {10.1086/592555},
archivePrefix = {arXiv},
       eprint = {0805.1137},
 primaryClass = {astro-ph},
       adsurl = {https://ui.adsabs.harvard.edu/abs/2008ApJ...689..184M}
}

@ARTICLE{martinez10,
       author = {{Mart{\'\i}nez-Delgado}, David and {Gabany}, R. Jay and {Crawford}, Ken and {Zibetti}, Stefano and {Majewski}, Steven R. and {Rix}, Hans-Walter and {Fliri}, J{\"u}rgen and {Carballo-Bello}, Julio A. and {Bardalez-Gagliuffi}, Daniella C. and {Pe{\~n}arrubia}, Jorge and {Chonis}, Taylor S. and {Madore}, Barry and {Trujillo}, Ignacio and {Schirmer}, Mischa and {McDavid}, David A.},
        title = "{Stellar Tidal Streams in Spiral Galaxies of the Local Volume: A Pilot Survey with Modest Aperture Telescopes}",
      journal = {\aj},
         year = 2010,
        month = oct,
       volume = {140},
       number = {4},
        pages = {962-967},
          doi = {10.1088/0004-6256/140/4/962},
archivePrefix = {arXiv},
       eprint = {1003.4860},
 primaryClass = {astro-ph.CO},
       adsurl = {https://ui.adsabs.harvard.edu/abs/2010AJ....140..962M}
}

@ARTICLE{martinez12,
       author = {{Mart{\'\i}nez-Delgado}, David and {Romanowsky}, Aaron J. and {Gabany}, R. Jay and {Annibali}, Francesca and {Arnold}, Jacob A. and {Fliri}, J{\"u}rgen and {Zibetti}, Stefano and {van der Marel}, Roeland P. and {Rix}, Hans-Walter and {Chonis}, Taylor S. and {Carballo-Bello}, Julio A. and {Aloisi}, Alessandra and {Macci{\`o}}, Andrea V. and {Gallego-Laborda}, J. and {Brodie}, Jean P. and {Merrifield}, Michael R.},
        title = "{Dwarfs Gobbling Dwarfs: A Stellar Tidal Stream around NGC 4449 and Hierarchical Galaxy Formation on Small Scales}",
      journal = {\apjl},
         year = 2012,
        month = apr,
       volume = {748},
       number = {2},
          eid = {L24},
        pages = {L24},
          doi = {10.1088/2041-8205/748/2/L24},
archivePrefix = {arXiv},
       eprint = {1112.2154},
 primaryClass = {astro-ph.CO},
       adsurl = {https://ui.adsabs.harvard.edu/abs/2012ApJ...748L..24M}
}

@ARTICLE{mcquinn17,
       author = {{McQuinn}, Kristen. B.~W. and {Skillman}, Evan D. and {Dolphin}, Andrew E. and {Berg}, Danielle and {Kennicutt}, Robert},
        title = "{Accurate Distances to Important Spiral Galaxies: M63, M74, NGC 1291, NGC 4559, NGC 4625, and NGC 5398}",
      journal = {\aj},
         year = 2017,
        month = aug,
       volume = {154},
       number = {2},
          eid = {51},
        pages = {51},
          doi = {10.3847/1538-3881/aa7aad},
archivePrefix = {arXiv},
       eprint = {1706.06586},
 primaryClass = {astro-ph.GA},
       adsurl = {https://ui.adsabs.harvard.edu/abs/2017AJ....154...51M}
}

@ARTICLE{namumba19,
       author = {{Namumba}, B. and {Carignan}, C. and {Foster}, T. and {Deg}, N.},
        title = "{H I observations of IC 10 with the DRAO synthesis telescope}",
      journal = {\mnras},
         year = 2019,
        month = dec,
       volume = {490},
       number = {3},
        pages = {3365-3377},
          doi = {10.1093/mnras/stz2737},
archivePrefix = {arXiv},
       eprint = {1908.02198},
 primaryClass = {astro-ph.GA},
       adsurl = {https://ui.adsabs.harvard.edu/abs/2019MNRAS.490.3365N}
}

@ARTICLE{nersesian19,
       author = {{Nersesian}, A. and {Xilouris}, E.~M. and {Bianchi}, S. and {Galliano}, F. and {Jones}, A.~P. and {Baes}, M. and {Casasola}, V. and {Cassar{\`a}}, L.~P. and {Clark}, C.~J.~R. and {Davies}, J.~I. and {Decleir}, M. and {Dobbels}, W. and {De Looze}, I. and {De Vis}, P. and {Fritz}, J. and {Galametz}, M. and {Madden}, S.~C. and {Mosenkov}, A.~V. and {Tr{\v{c}}ka}, A. and {Verstocken}, S. and {Viaene}, S. and {Lianou}, S.},
        title = "{Old and young stellar populations in DustPedia galaxies and their role in dust heating}",
      journal = {\aap},
         year = 2019,
        month = apr,
       volume = {624},
          eid = {A80},
        pages = {A80},
          doi = {10.1051/0004-6361/201935118},
archivePrefix = {arXiv},
       eprint = {1903.05933},
 primaryClass = {astro-ph.GA},
       adsurl = {https://ui.adsabs.harvard.edu/abs/2019A&A...624A..80N}
}

@ARTICLE{nidever13,
       author = {{Nidever}, David L. and {Ashley}, Trisha and {Slater}, Colin T. and {Ott}, J{\"u}rgen and {Johnson}, Megan and {Bell}, Eric F. and {Stanimirovi{\'c}}, Sne{\v{z}}ana and {Putman}, Mary and {Majewski}, Steven R. and {Simpson}, Caroline E. and {J{\"u}tte}, Eva and {Oosterloo}, Tom A. and {Butler Burton}, W.},
        title = "{Evidence for an Interaction in the Nearest Starbursting Dwarf Irregular Galaxy IC 10}",
      journal = {\apjl},
         year = 2013,
        month = dec,
       volume = {779},
       number = {2},
          eid = {L15},
        pages = {L15},
          doi = {10.1088/2041-8205/779/2/L15},
archivePrefix = {arXiv},
       eprint = {1310.7573},
 primaryClass = {astro-ph.GA},
       adsurl = {https://ui.adsabs.harvard.edu/abs/2013ApJ...779L..15N}
}

@ARTICLE{okamoto15,
       author = {{Okamoto}, Sakurako and {Arimoto}, Nobuo and {Ferguson}, Annette M.~N. and {Bernard}, Edouard J. and {Irwin}, Mike J. and {Yamada}, Yoshihiko and {Utsumi}, Yousuke},
        title = "{A Hyper Suprime-Cam View of the Interacting Galaxies of the M81 Group}",
      journal = {\apjl},
         year = 2015,
        month = aug,
       volume = {809},
       number = {1},
          eid = {L1},
        pages = {L1},
          doi = {10.1088/2041-8205/809/1/L1},
archivePrefix = {arXiv},
       eprint = {1507.04889},
 primaryClass = {astro-ph.GA},
       adsurl = {https://ui.adsabs.harvard.edu/abs/2015ApJ...809L...1O}
}

@ARTICLE{pilyugin14,
       author = {{Pilyugin}, L.~S. and {Grebel}, E.~K. and {Kniazev}, A.~Y.},
        title = "{The Abundance Properties of Nearby Late-type Galaxies. I. The Data}",
      journal = {\aj},
         year = 2014,
        month = jun,
       volume = {147},
       number = {6},
          eid = {131},
        pages = {131},
          doi = {10.1088/0004-6256/147/6/131},
archivePrefix = {arXiv},
       eprint = {1403.5461},
 primaryClass = {astro-ph.GA},
       adsurl = {https://ui.adsabs.harvard.edu/abs/2014AJ....147..131P}
}

@ARTICLE{schlafly11,
       author = {{Schlafly}, Edward F. and {Finkbeiner}, Douglas P.},
        title = "{Measuring Reddening with Sloan Digital Sky Survey Stellar Spectra and Recalibrating SFD}",
      journal = {\apj},
         year = 2011,
        month = aug,
       volume = {737},
       number = {2},
          eid = {103},
        pages = {103},
          doi = {10.1088/0004-637X/737/2/103},
archivePrefix = {arXiv},
       eprint = {1012.4804},
 primaryClass = {astro-ph.GA},
       adsurl = {https://ui.adsabs.harvard.edu/abs/2011ApJ...737..103S}
}

@ARTICLE{schlegel98,
       author = {{Schlegel}, David J. and {Finkbeiner}, Douglas P. and {Davis}, Marc},
        title = "{Maps of Dust Infrared Emission for Use in Estimation of Reddening and Cosmic Microwave Background Radiation Foregrounds}",
      journal = {\apj},
         year = 1998,
        month = jun,
       volume = {500},
       number = {2},
        pages = {525-553},
          doi = {10.1086/305772},
archivePrefix = {arXiv},
       eprint = {astro-ph/9710327},
 primaryClass = {astro-ph},
       adsurl = {https://ui.adsabs.harvard.edu/abs/1998ApJ...500..525S}
}

@ARTICLE{Stetson1987,
       author = {{Stetson}, Peter B.},
        title = "{DAOPHOT: A Computer Program for Crowded-Field Stellar Photometry}",
      journal = {\pasp},
         year = 1987,
        month = mar,
       volume = {99},
        pages = {191},
          doi = {10.1086/131977},
       adsurl = {https://ui.adsabs.harvard.edu/abs/1987PASP...99..191S}
}

@ARTICLE{wilcots98,
       author = {{Wilcots}, Eric M. and {Miller}, Bryan W.},
        title = "{The Kinematics and Distribution of H I in IC 10}",
      journal = {\aj},
         year = 1998,
        month = nov,
       volume = {116},
       number = {5},
        pages = {2363-2394},
          doi = {10.1086/300595},
       adsurl = {https://ui.adsabs.harvard.edu/abs/1998AJ....116.2363W}
}

@ARTICLE{weisz14,
       author = {{Weisz}, Daniel R. and {Dolphin}, Andrew E. and {Skillman}, Evan D. and {Holtzman}, Jon and {Gilbert}, Karoline M. and {Dalcanton}, Julianne J. and {Williams}, Benjamin F.},
        title = "{The Star Formation Histories of Local Group Dwarf Galaxies. I. Hubble Space Telescope/Wide Field Planetary Camera 2 Observations}",
      journal = {\apj},
         year = 2014,
        month = jul,
       volume = {789},
       number = {2},
          eid = {147},
        pages = {147},
          doi = {10.1088/0004-637X/789/2/147},
archivePrefix = {arXiv},
       eprint = {1404.7144},
 primaryClass = {astro-ph.GA},
       adsurl = {https://ui.adsabs.harvard.edu/abs/2014ApJ...789..147W}
}

@ARTICLE{Okamoto23,
       author = {{Okamoto}, Sakurako and {Arimoto}, Nobuo and {Ferguson}, Annette M.~N. and {Irwin}, Mike J. and {{\v{Z}}emaitis}, Rokas},
        title = "{The Progenitor of the Peculiar Galaxy NGC 3077}",
      journal = {\apj},
         year = 2023,
        month = jul,
       volume = {952},
       number = {1},
          eid = {77},
        pages = {77},
          doi = {10.3847/1538-4357/acdad1},
archivePrefix = {arXiv},
       eprint = {2306.04102},
 primaryClass = {astro-ph.GA},
       adsurl = {https://ui.adsabs.harvard.edu/abs/2023ApJ...952...77O}
}

%
%

\begin{appendix}

\nolinenumbers

%
%
%
%
\section{\label{sc:mw_appendix} Removal of MW contaminants from the IC\,10 catalogue}

In this section, we describe the details of the procedure we adopted to remove MW contaminants (and residual background galaxies) from the IC\,10 photometric catalogue.

The first step consists of cross-correlating the source catalogue with the {\it Gaia} data release 3 
\citep[DR3,][]{Gaia2016,Gaia2023} and looking for objects with a non-zero proper motion PM or with a parallax (Plx) 
larger than zero. 
For the match with the {\it Gaia} catalogue, we adopted a 1\arcsecond\ maximum tolerance in distance, but explored also lower matching 
radii finding the results to differ by just $\sim$\,3\% for a tolerance radius as low as 0\farcs1.
Since IC\,10 has a {\it Gaia} PM compatible with zero within the errors \citep[e.g.][]{McConnachie2021,Bennet2024}, likely MW members were identified as those sources having a measured proper motion 
PM larger than $3\,\sigma_{\mathrm{PM}}$ or a parallax Plx larger than $3\,\sigma_{\mathrm{Plx}}$, where $\sigma_{\mathrm{PM}}$ and $\sigma_{\mathrm{Plx}}$ are the {\it Gaia} errors on PM and Plx, respectively. 
With this approach, we effectively removed \num{7724} bright foreground stars with 
$I_\sfont{E,0}\la19$ (reddening-corrected magnitudes) from our photometric catalogue. 
Second-epoch \Euclid observations would be highly effective to remove most contaminants via proper motion membership \citep{Bedin2025}.

\begin{figure}[htbp]
\includegraphics[width=0.49\textwidth]{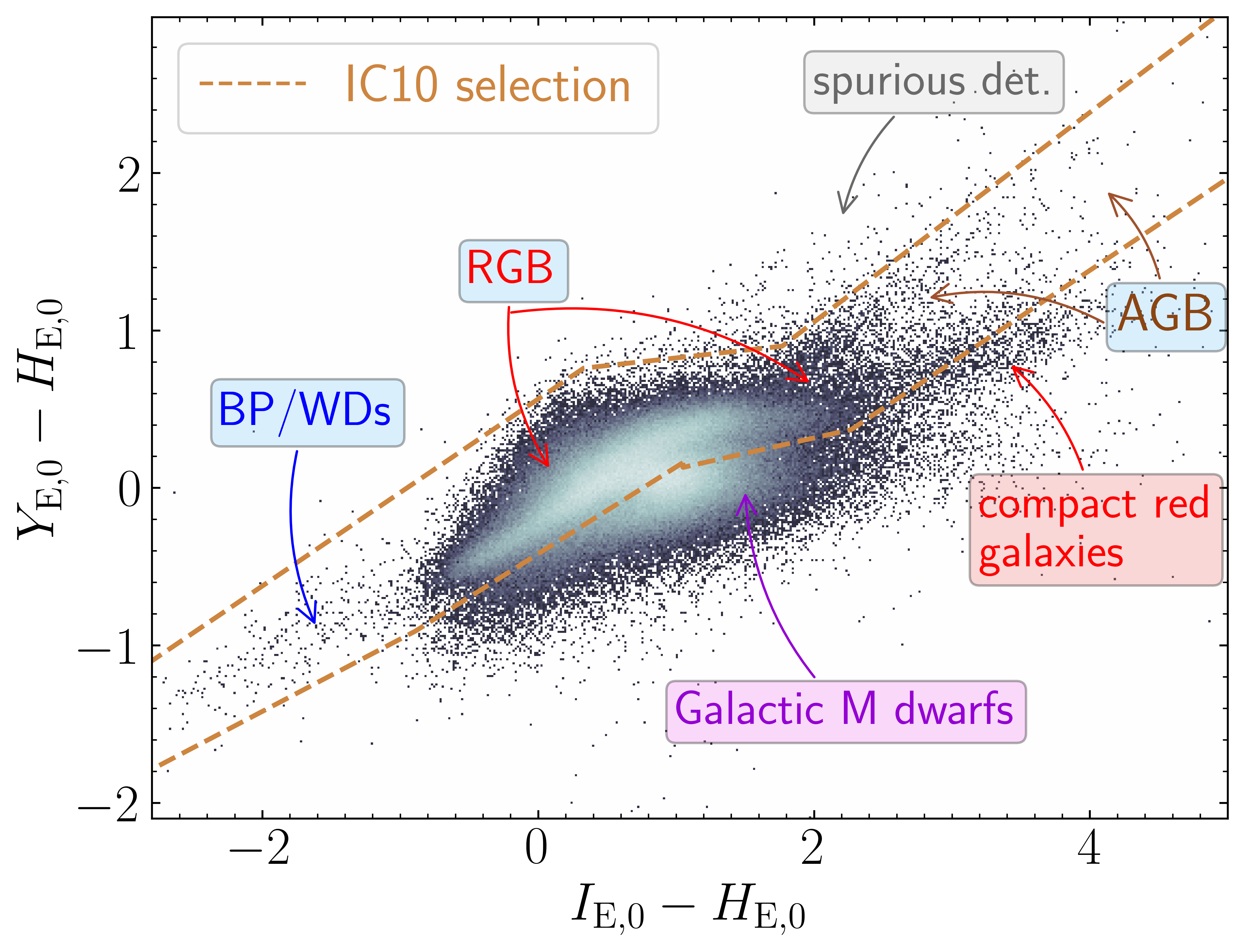}
\caption{$Y_{\protect\sfont{E,0}}-H_{\protect\sfont{E,0}}$ versus $I_{\protect\sfont{E,0}}-H_{\protect\sfont{E,0}}$
diagram for sources in the IC\,10 photometric catalogue, after the removal of bright 
$I_{\protect\sfont{E,0}}<19$ MW stars in {\it Gaia} DR3 and after reddening correction from the map of Fig.~\ref{fig:ebv_rgb}b. 
The locations of Galactic M dwarf stars, RGB and AGB stars in IC\,10, background compact red galaxies, 
and residual spurious detection are indicated. The bluest objects could be either young BP stars in IC\,10 or Galactic WDs.
The dashed orange polygon indicates our selection, aimed at retaining the largest possible number of stars belonging to IC\,10 while removing Galactic M dwarf contaminants, compact red galaxies, and residual spurious detections (see text for details).}
\label{fig:color_color_sel}
\end{figure}

\begin{figure*}[htpb]
\centering
\includegraphics[width=0.85\textwidth]{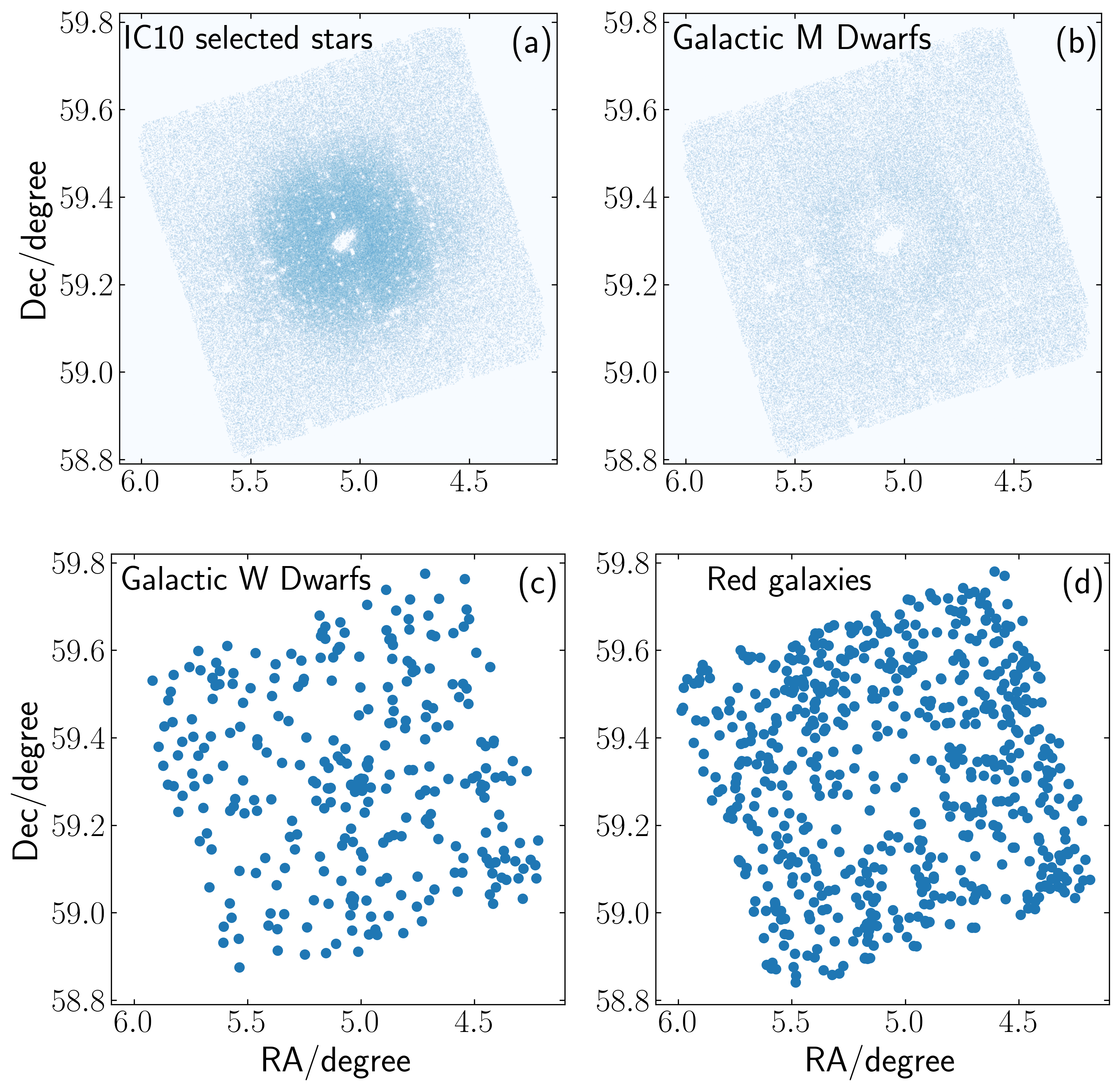} \hfill
\caption{Spatial distribution of stars in  IC\,10 after removal of different contaminants ({\emph {panel a}}) versus Galactic M dwarfs ({\emph {panel\,b}}), Galactic white dwarfs ({\emph {panel c}})
and compact red galaxies ({\emph {panel d}}).}
\label{fig:spat_conts}
\end{figure*}

The second step exploits the combination of VIS and NISP colours to remove Galactic M dwarf stars. 
The applied selection is illustrated in Fig.~\ref{fig:color_color_sel}, where we plot the reddening-corrected \YoHo colour versus the \IoHo colour for the
entire catalogue after removal of {\it Gaia} sources. At 
$I_\sfont{E,0}-H_\sfont{E,0}\lesssim1$, giant and dwarf stars populate the same diagram locus, since their colours are degenerate for early spectral types; as a consequence, it is not possible to disentangle bluer dwarf stars from giant stars in IC\,10 (mainly RGB and AGB stars).
On the other hand, for $I_\sfont{E,0}-H_\sfont{E,0}\gtrsim1$, the colours of dwarfs and giants start to diverge, with Galactic M dwarfs forming a \YoHo relatively bluer sequence than giant stars in IC\,10, as showcased in Fig.10 of \citetalias{ERONearbyGals} based on a comparison of the TRILEGAL Galactic model \citep{trilegal05,trilegal12} with stellar isochrones in the \YoHo versus the \IoHo  diagram.  
By removing sources at $I_\sfont{E,0}-H_\sfont{E,0}\gtrsim1$, $Y_\sfont{E,0}-H_\sfont{E,0}\lesssim0$,
as outlined in Fig.~\ref{fig:color_color_sel}, we eliminate the vast majority of Galactic M dwarf interlopers. The use of reddening-corrected  magnitudes guarantees the cleanest possible separation between the giant and dwarf sequences.

The selection described above also allows the removal of residual compact red galaxies that survived the initial cuts based on photometry at different apertures, as illustrated in  Fig.~\ref{fig:ap_selection}. These red galaxies form a separate sequence with $\Yo-\Ho$ colours intermediate between that defined by Galactic M dwarfs and that populated by AGB stars in IC\,10. A visual inspection of these objects in the VIS image shows in fact that they are associated with extended emission, with asymmetric or irregular morphology, confirming their extragalactic nature. 
This selection also removes spurious detections with \YoHo colours redder than the AGB and the thermal pulsing AGB sequence,  typically located at the edge of detectors, as illustrated in Fig.~\ref{fig:color_color_sel}. 

Figure~\ref{fig:color_color_sel} also shows a very blue sequence of objects, with 
$I_\sfont{E,0}-H_\sfont{E,0}\lesssim-1$ and $Y_\sfont{E,0}-H_\sfont{E,0}\lesssim-0.5$.
Based on their position in the colour-colour diagram and in the CMDs presented in Sect.~\ref{sc:cmd}, it is not possible to determine whether these sources are young luminous `blue plume' (BP) stars in IC\,10 or Galactic white dwarfs (WDs), and thus we retain them in our catalogue. 
Nonetheless, their quite uniform spatial distribution across the entire \Euclid field, shown in Fig.~\ref{fig:spat_conts}, suggests that they are mainly Galactic WDs, with a marginal contribution from young BP stars in IC\,10. Despite the very well-known presence of recent star formation in IC\,10 \citep{Massey1995,Crowther2009,Tehrani2017,Gholami2025}, luminous blue stars are mainly concentrated in the innermost crowded regions and are affected by severe blending; thus they have likely been removed from our catalogue by selection criteria aimed at rejecting extended sources. 
In Fig.~\ref{fig:spat_conts} we also display the spatial distributions for the selected Galactic M dwarfs and compact red galaxies to show that they are distributed quite uniformly over the \Euclid FoV as we would expect from foreground and background contaminants.

\section{\label{sc:other_cmds} Additional colour-magnitude diagrams}

\begin{figure}
\includegraphics[width=0.85\columnwidth]{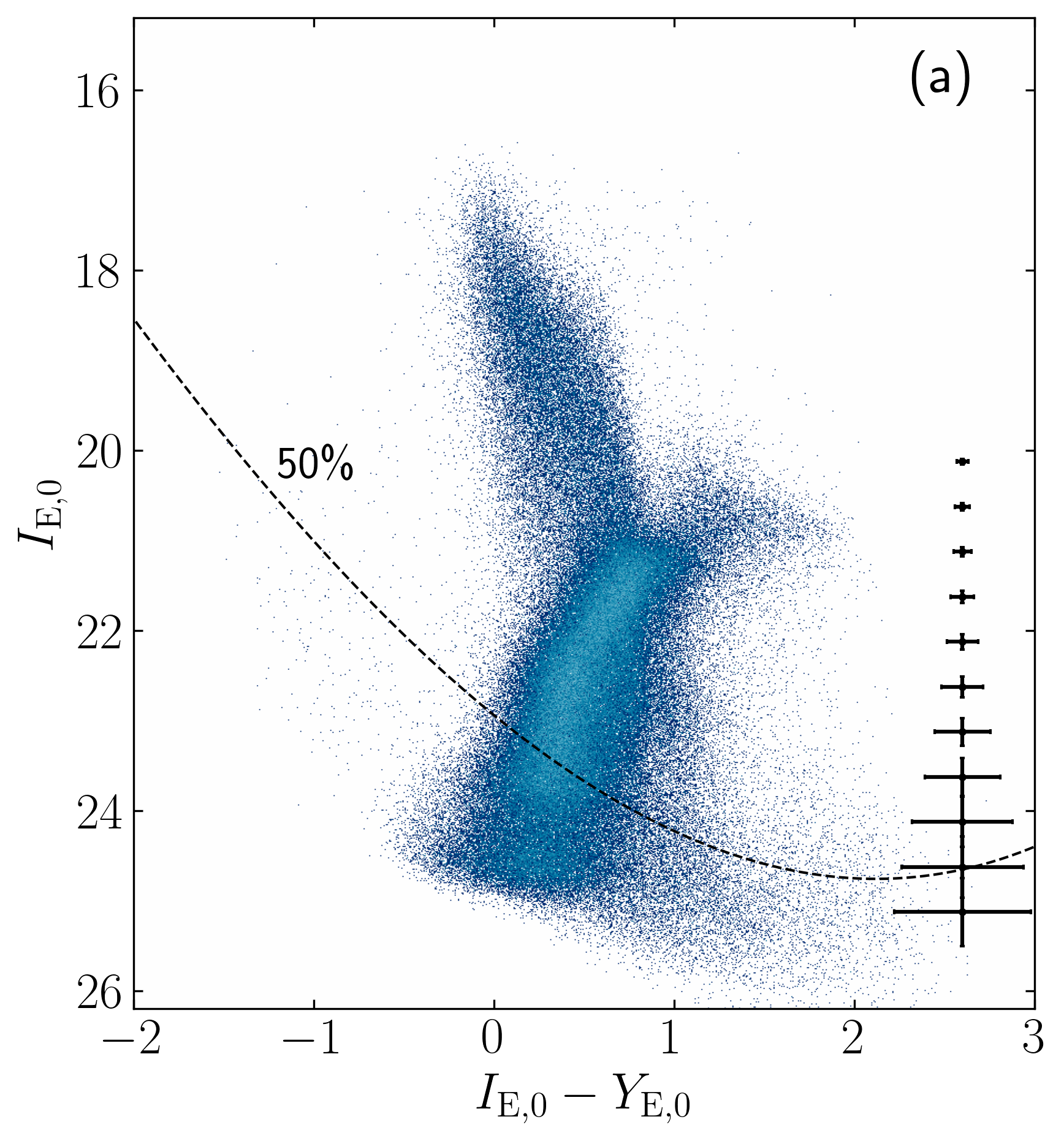}
\includegraphics[width=0.85\columnwidth]{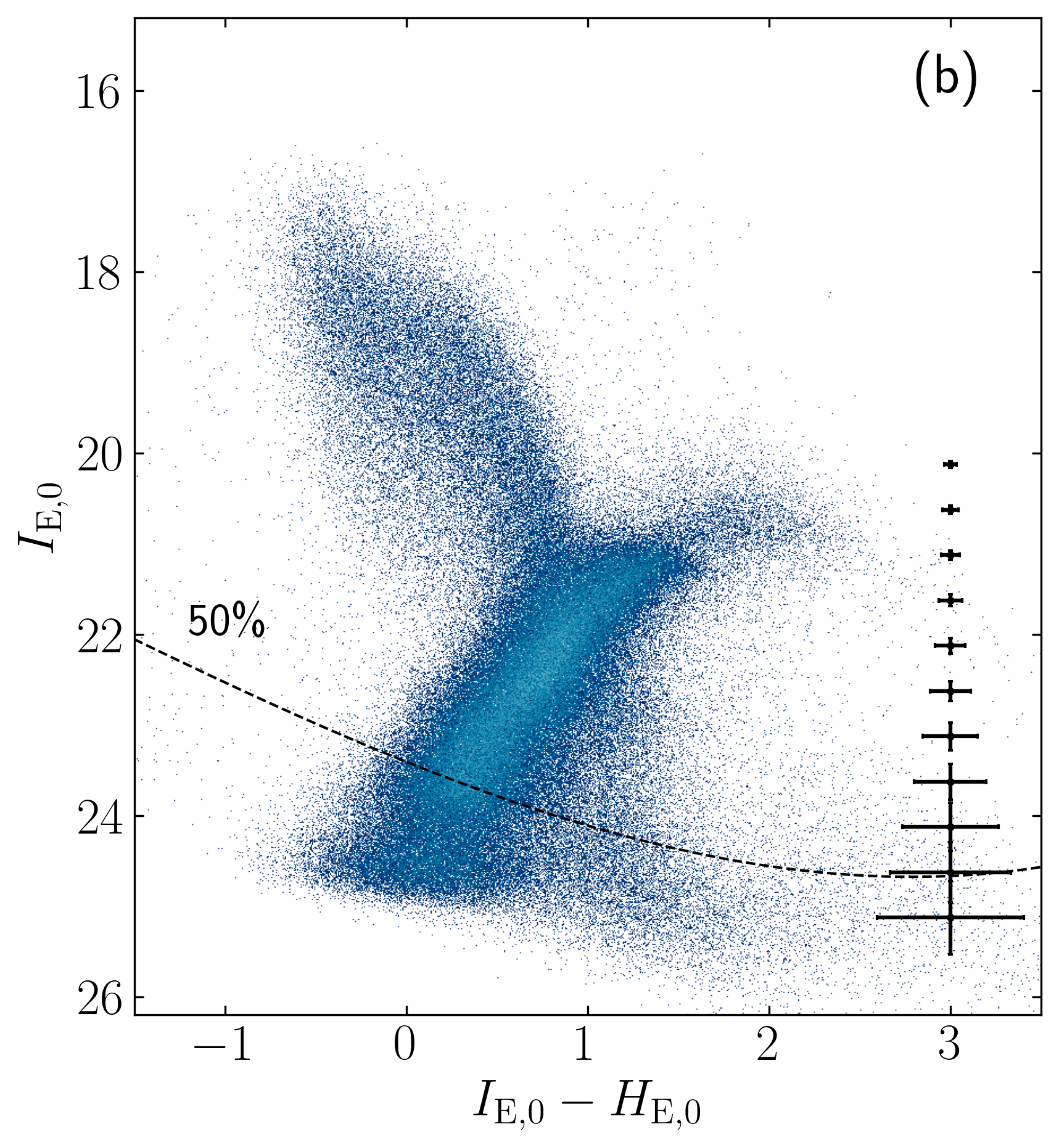}
\caption{\Io versus \IoYo ({\emph {panel a}}) and \Io versus \IoHo ({\emph {panel b}}) reddening-corrected CMDs after removal of foreground contaminants, as described in Sect.~\ref{sc:ebv_fg}.
The black dashed curves denote the average 50\% completeness levels, corrected for the median reddening in  IC\,10.}
\label{fig:other_cmds}
\end{figure}

In this section we present, for completeness, the \Io versus \IoYo (Fig.~\ref{fig:other_cmds}a) and \Io versus \IoHo CMDs (Fig.~\ref{fig:other_cmds}b) of the stars in IC\,10. As in Fig.~\ref{fig:cmd}, the CMDs were corrected for spatially variable reddening as described in Sect.~\ref{sc:ebv}, and cleaned for background galaxies and foreground stars as detailed in Sect.~\ref{sc:mw}. The displayed CMDs contain 
a total of \num{227 128} sources. The black dashed curves denote the 50\% completeness levels averaged over the entire \Euclid FoV. The curves were obtained from the results of the artificial star tests presented in Sect.~\ref{sc:artificial} and were corrected adopting a median reddening value of $E(B-V)=0.8$ for comparison with the displayed CMDs. 

\section{\label{sc:artificial}Artificial star tests and completeness}

To evaluate the completeness of our photometry at different magnitudes, we perform artificial star tests; such an analysis also quantifies the observational effects associated with the data reduction process, such as the accuracy of the photometric measurements and 
the probability of blending in crowded regions. 
The tests were performed by adding artificial stars to the \IE, \YE, \JE, and \HE images, and then reducing the frames following exactly the same procedure as for the real data. 

To place artificial stars into the images, we first created PSF models for each of the 4 \Euclid bands selecting the most isolated, bright, and unsaturated stars within the frames, and then running the PSF routine in \texttt{DAOPHOT}. 
To speed up the computation and better control the PSF, we divided each image into 3 $\times$ 3 smaller subsections, and modelled for each band the nine PSFs with an analytic Moffat function \citep{Moffat} plus additive corrections derived from the residuals of the fit to the bright selected stars. The additive corrections include only 
first-order derivatives of the PSF with respect to the $X$ and $Y$ positions in the sub-image, which is a reasonable choice given the smooth behaviour of the PSF over the \Euclid FoV \citepalias{EROData}.

To assign colours and magnitudes to the artificial stars, we exploited the deeper CMD of NGC\,6822 (containing about \num{200000} stars), the nearest dwarf irregular (at a distance of $\sim$\,0.5\,Mpc) observed in the context of the Showcase Galaxies ERO \citepalias{ERONearbyGals}. 
The CMD of NGC\,6822 was scaled to IC\,10's distance and reddening, providing a sample of stars with colours compatible with those observed in IC\,10 but extending to fainter magnitudes, as needed to properly evaluate the incompleteness at the faintest limits. Artificial stars  with \IE, \YE, \JE, and \HE magnitudes were extracted from this CMD and placed homogeneously on the images. To avoid artificial stars interfering with one another, thus creating spurious crowding, we added only \num{20000} stars at a time for each of the nine image subsections and repeated the process 10 times (20 times for the innermost, crowded section), for a total of 2 million artificial stars simulated in each band. 
Each star was placed in the IC\,10 image at the same exact position in both the VIS and NISP frames to properly evaluate our ability of recovering stars also as a function of their colours. 

Then, the photometric reduction of the VIS and NISP frames containing both real and artificial stars, the match of the catalogues in the four \IE, \YE, \JE, and \HE bands, and the selections of sources based on the \texttt{DAOPHOT} parameters were performed exactly as for the original images. In the end, the final catalogue with photometry in all four bands was cross-matched with the list of injected artificial stars. Once the 10 runs (or 20 for the innermost region) were completed for each of the nine image subsections, the outputs were joined together into a final master catalogue totalling 2 million simulated artificial stars. 

For each magnitude bin, the completeness was computed as the ratio of the number of remaining artificial stars over the number of added ones. 
The difference between the output and input magnitudes for the recovered stars provides a realistic measurement of the photometric uncertainties, with the exception of uncertainties related to the exact shape of the PSF.

\begin{figure}
\includegraphics[width=0.49\textwidth]{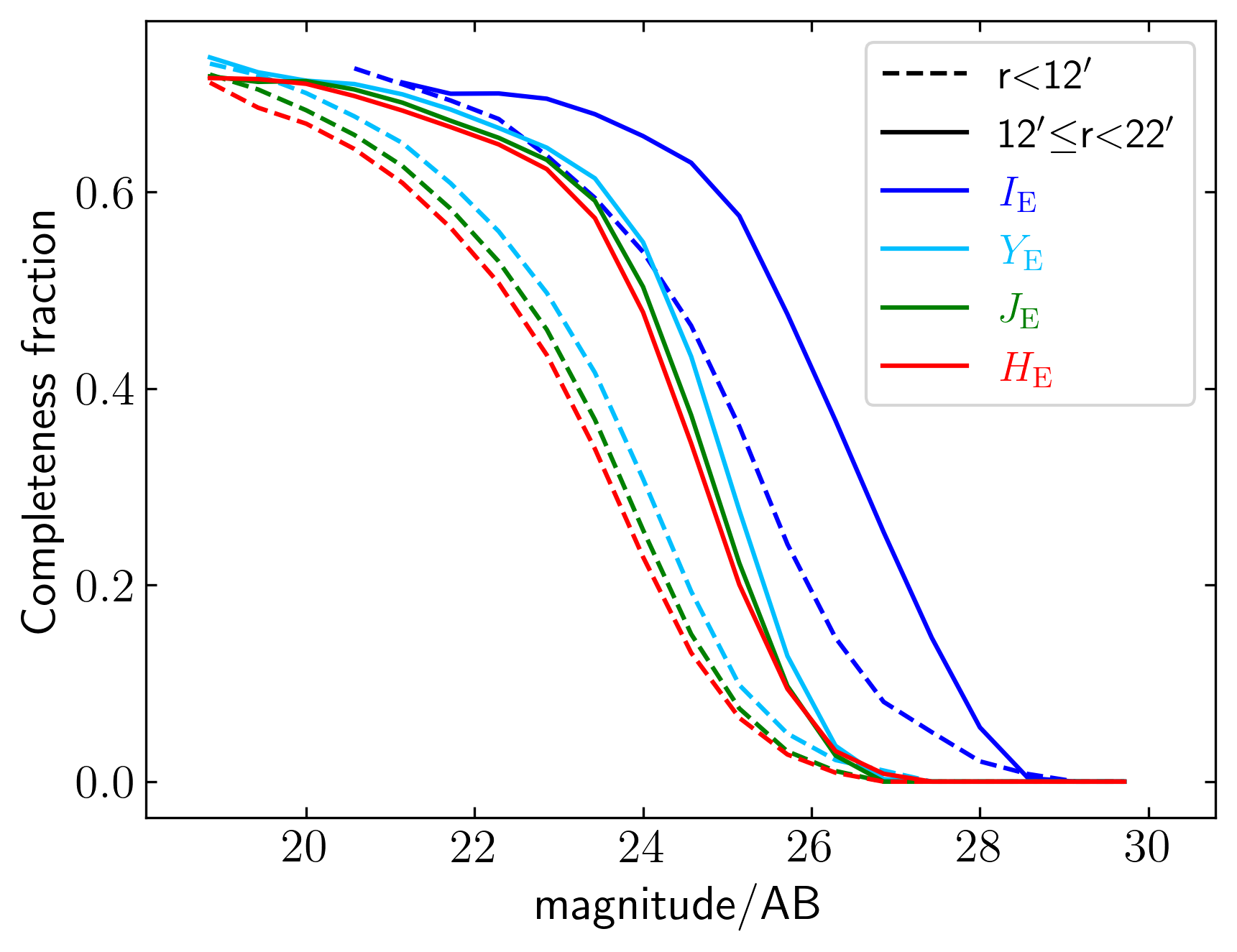}
\caption{Completeness fraction derived from artificial star tests as a function of magnitude in the four \Euclid bands: \IE (blue), \YE (cyan), \JE (green), and \HE (red). The dashed lines denote the completeness for a circular region within 12\arcmin\ from the centre, while the solid lines are for 
an annulus at $12\arcmin \leq r < 22\arcmin$. The magnitudes shown here are not corrected for reddening.} 
\label{fig:completeness}
\end{figure}

\begin{figure*}[]
\includegraphics[width=0.49\textwidth,clip, trim={0 0 0 0}]{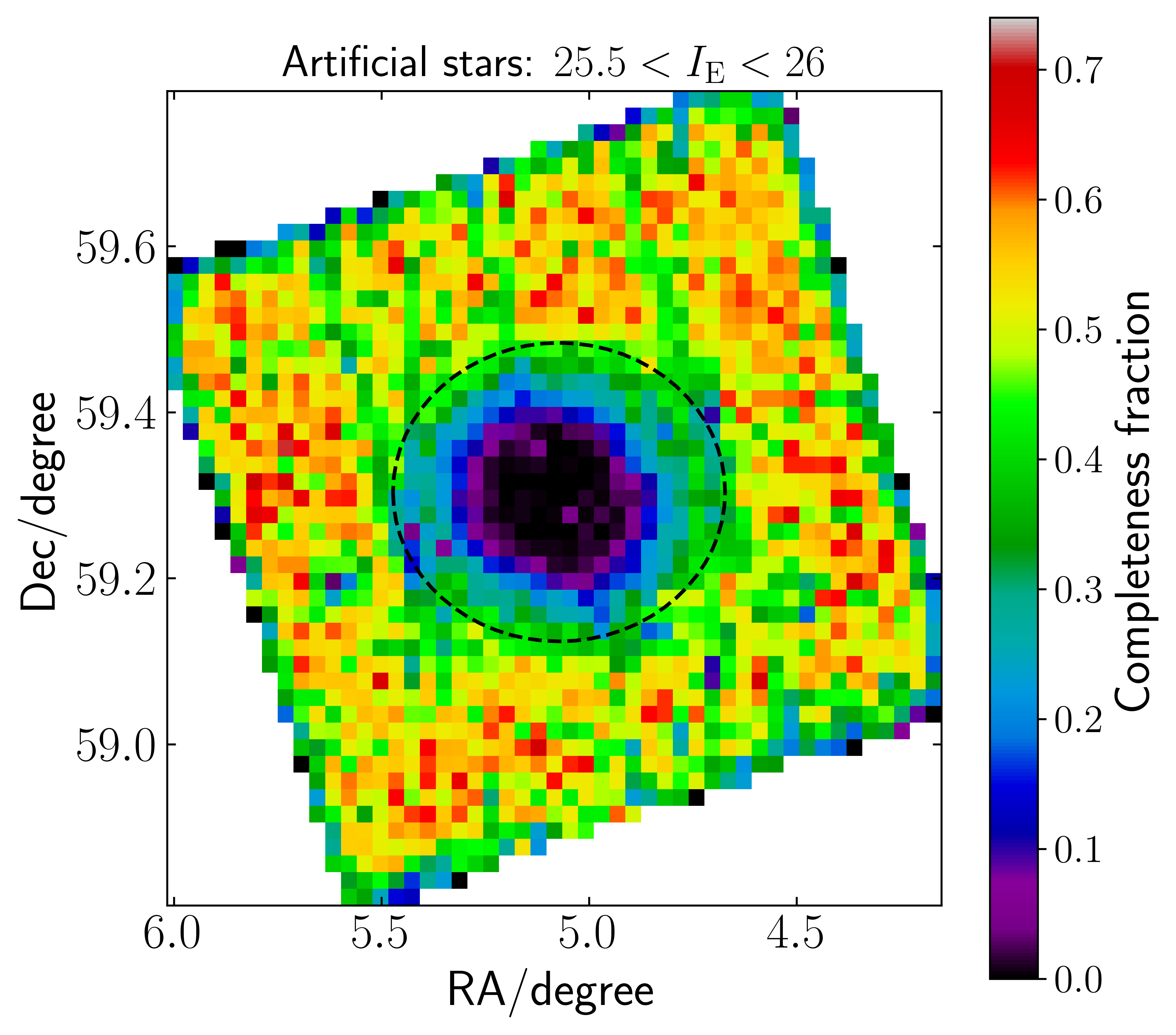} \hfill
\includegraphics[width=0.49\textwidth,clip, trim={0 0 0 0}]{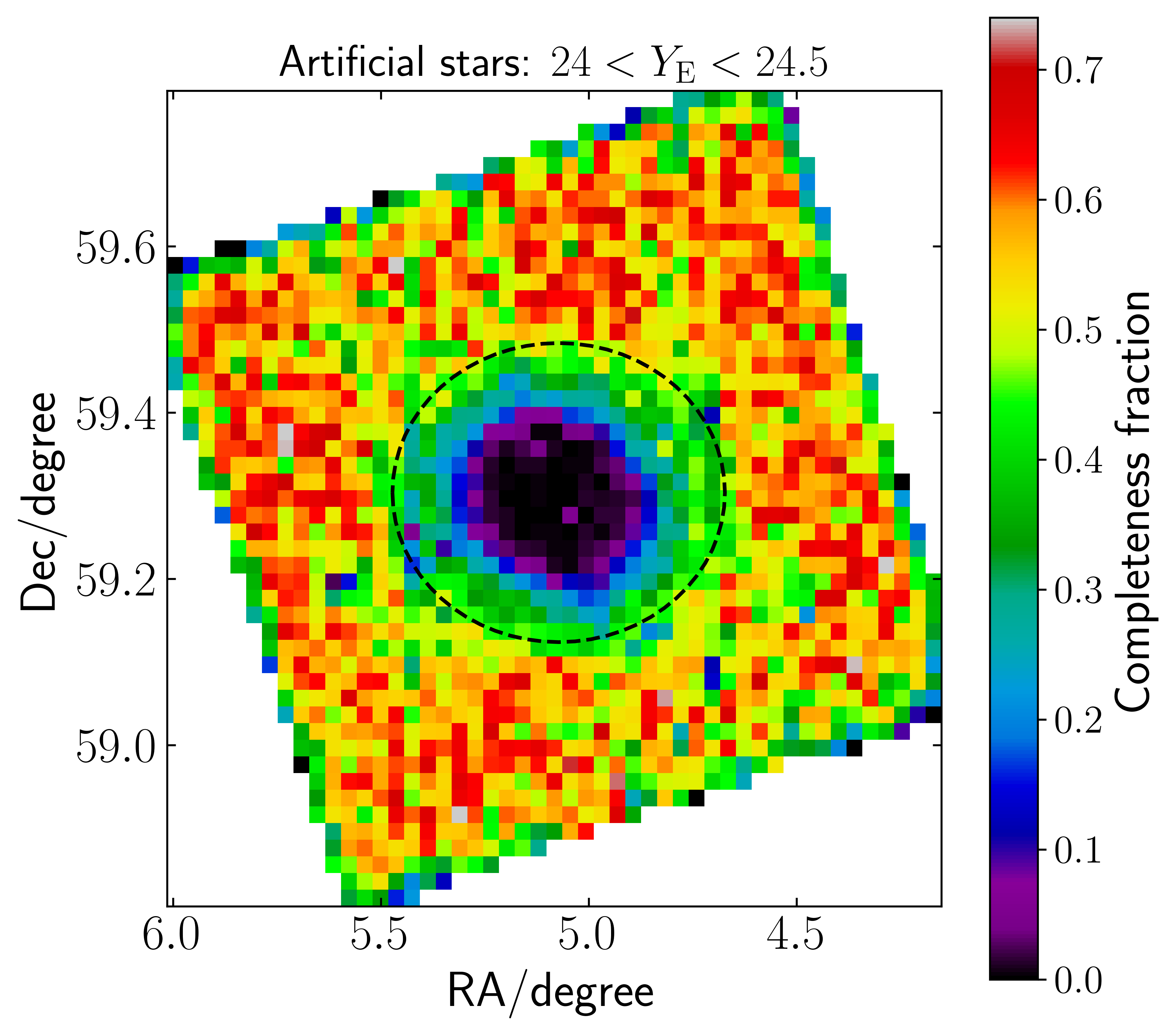} \hfill
\includegraphics[width=0.49\textwidth,clip, trim={0 0 0 0}]{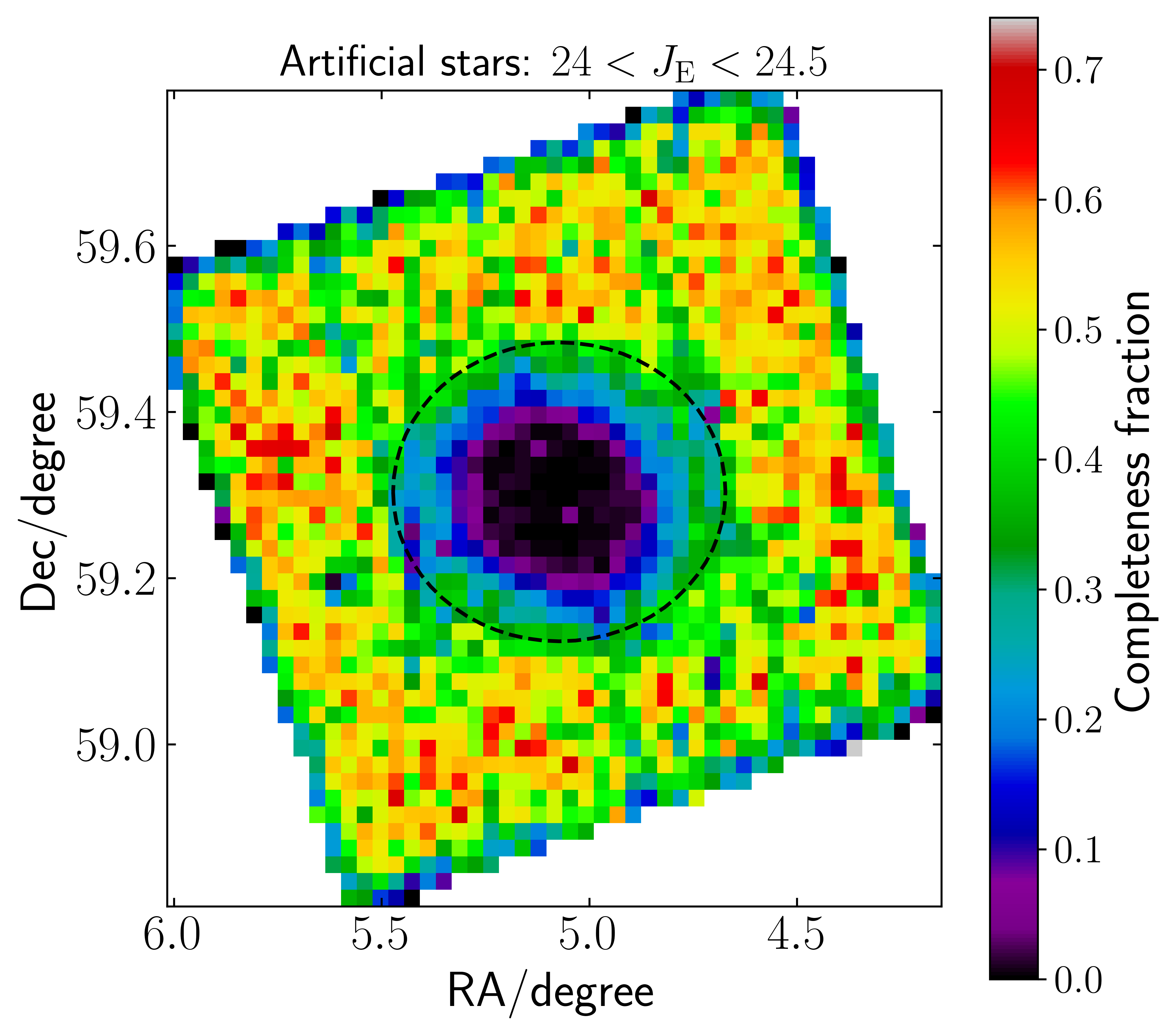} \hfill
\includegraphics[width=0.49\textwidth,clip, trim={0 0 0 0}]{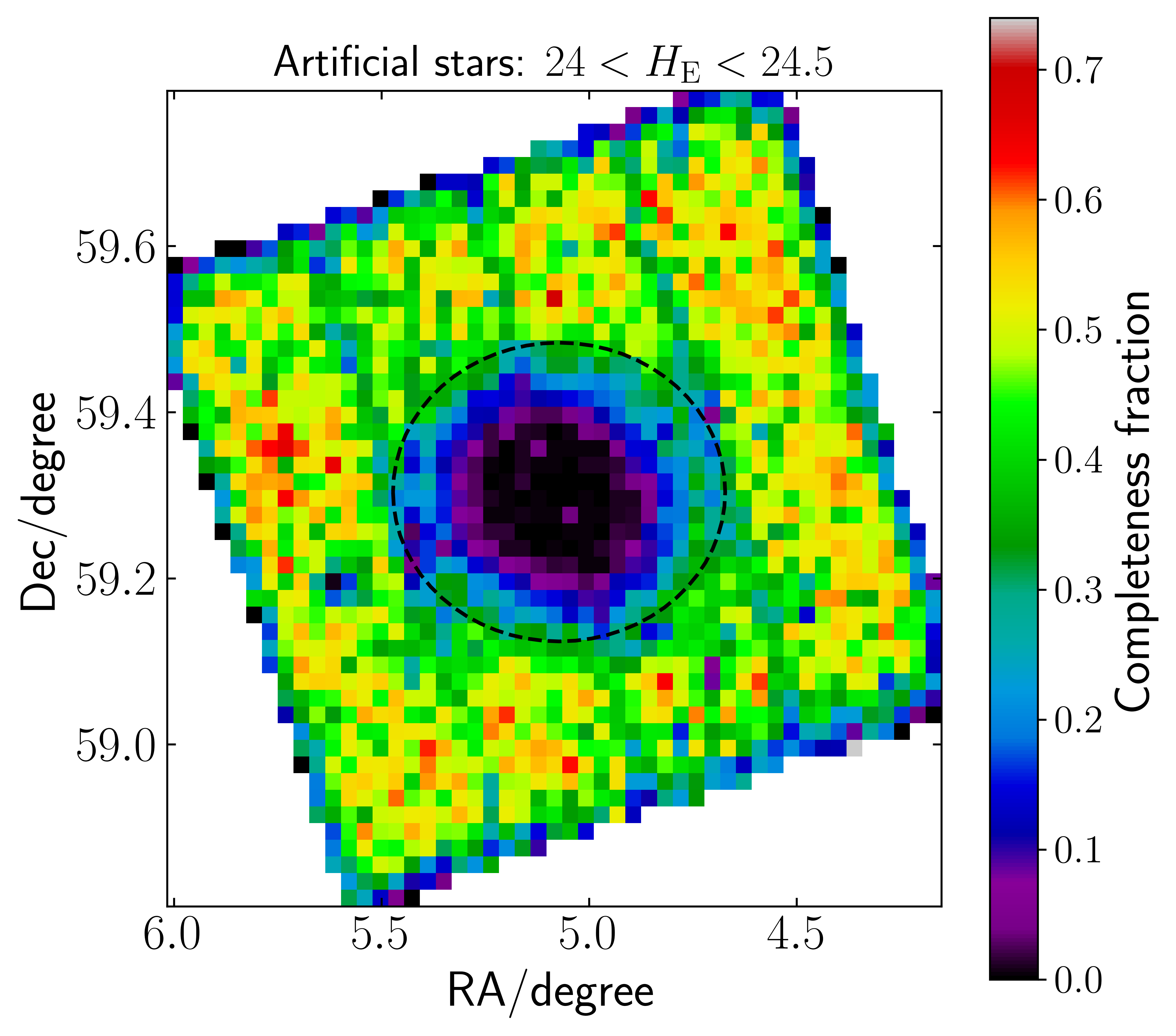}
\caption{
Spatial distribution of completeness in different \Euclid bands, as derived from artificial star experiments. The maps show the fraction of injected artificial stars that were successfully recovered by the photometric reduction process, after applying all selection criteria. Completeness is computed over the magnitude ranges $25.5$--$26$ for the \IE band and $24$--$24.5$ for the \YE, \JE, and \HE bands. The same black dashed ellipse as in Fig.~\ref{fig:rgb} is overplotted for direct comparison with the RGB star density map. Note that no reddening correction was applied to the artificial star magnitudes prior to producing these maps.}
\label{fig:spatial_compl}
\end{figure*}

\begin{figure*}[]
\includegraphics[width=0.49\textwidth,clip, trim={0 0 0 0}]{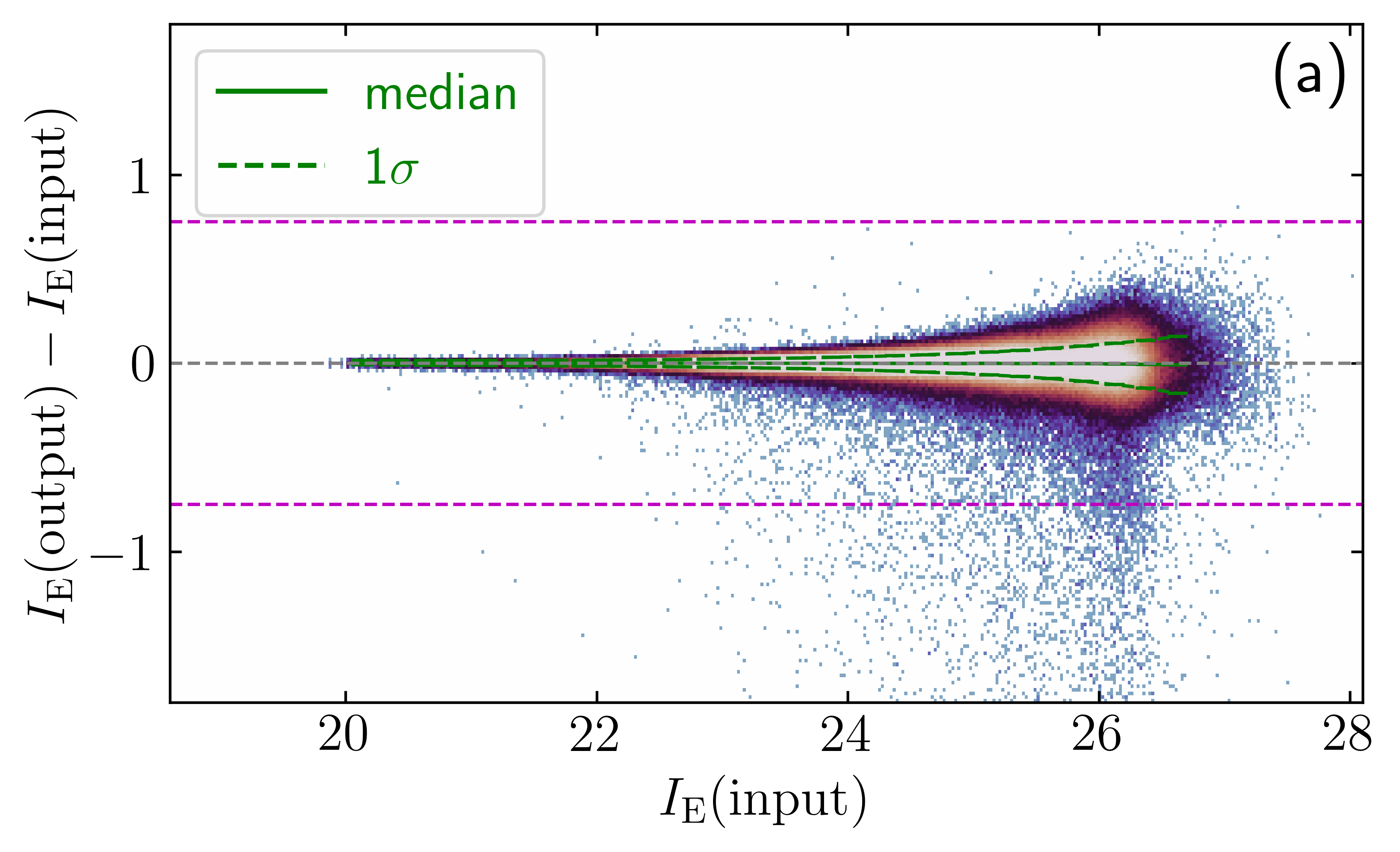} \hfill
\includegraphics[width=0.49\textwidth,clip, trim={0 0 0 0}]{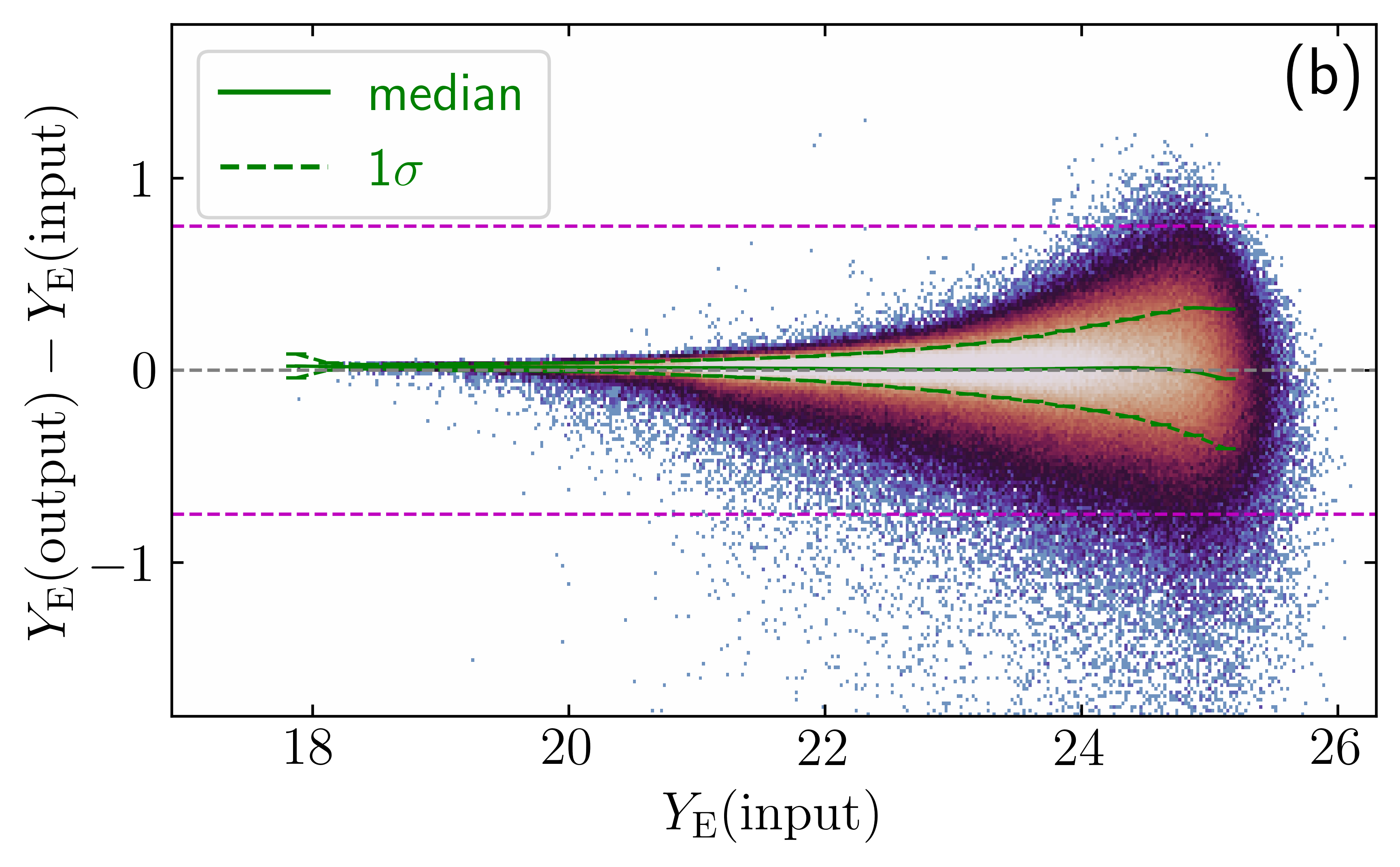} \hfill
\includegraphics[width=0.49\textwidth,clip, trim={0 0 0 0}]{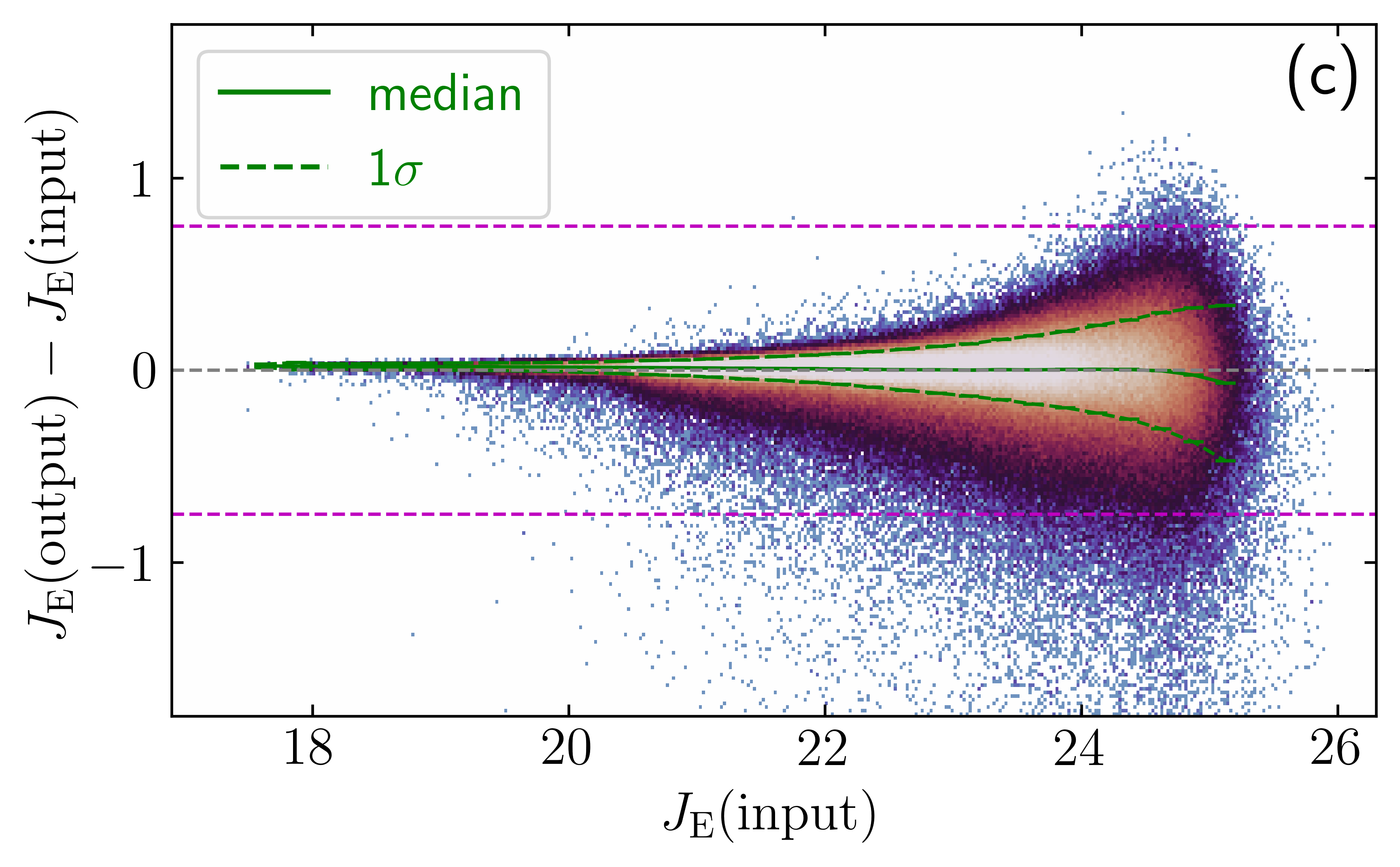} \hfill
\includegraphics[width=0.49\textwidth,clip, trim={0 0 0 0}]{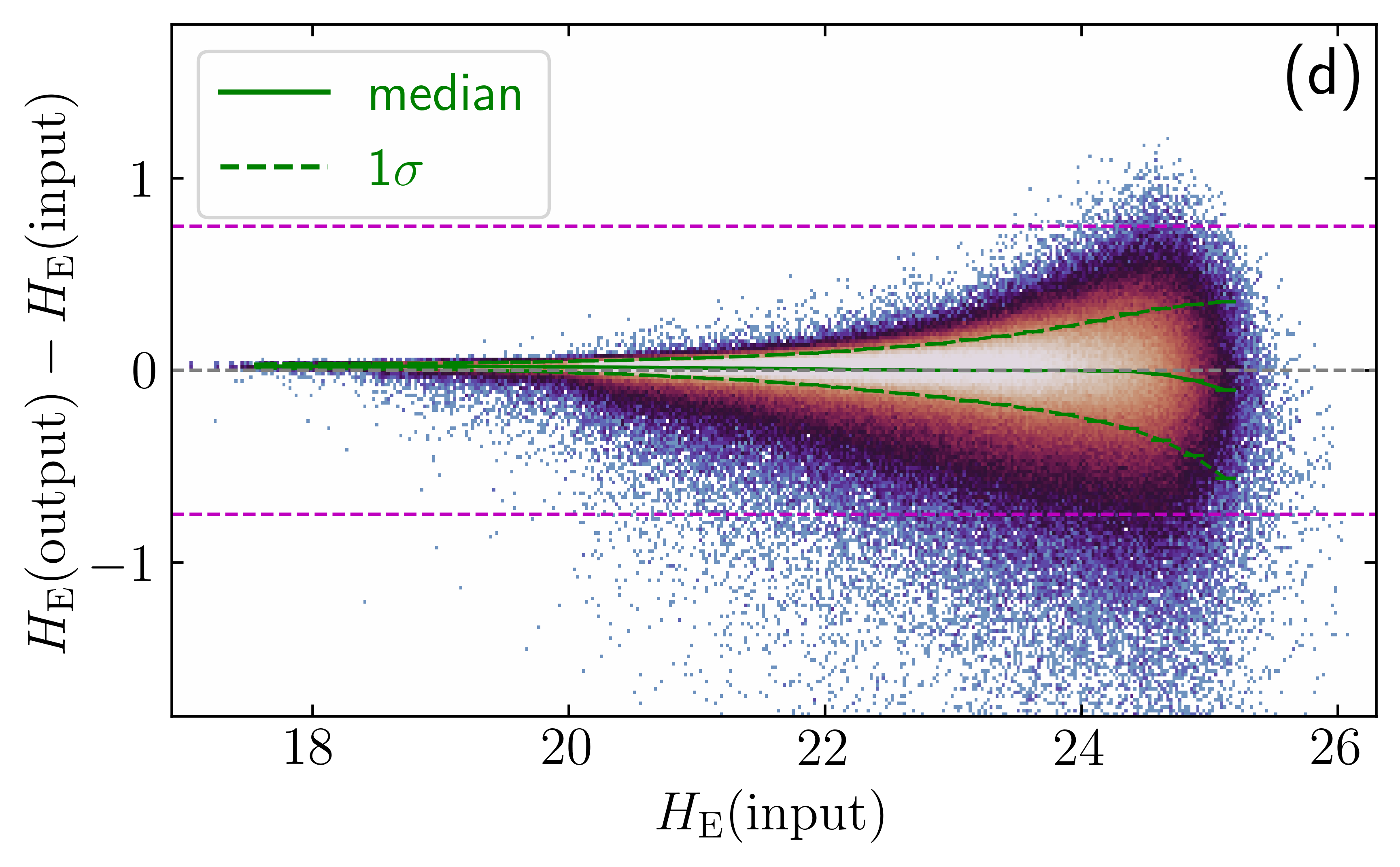}
\caption{Distribution of the output minus input magnitudes ($\Delta m$) for the artificial stars in different bands: \IE (\emph{panel a}), \YE (\emph{panel b}), \JE (\emph{panel c}), 
and \HE (\emph{panel d}). The green continuous line and the dashed lines denote the median and the 
$\pm$\,1\,$\sigma$ regions of the distributions, respectively, while the magenta dashed horizontal lines correspond to values of $\Delta m\pm 0.75$. In all bands, the distributions appear asymmetric, with a larger number of stars exhibiting negative $\Delta m$ values due to blending with neighbouring sources. For example, a star blended with another of equal flux will appear brighter by $\Delta m = -0.75$.}
\label{fig:delta_mag}
\end{figure*}

The results of our artificial star experiments are illustrated in Figs.~\ref{fig:completeness}, \ref{fig:spatial_compl}, and \ref{fig:delta_mag}. 
Figure~\ref{fig:completeness} shows the completeness fraction as a function of (reddening-uncorrected) magnitude for different bands. We display the completeness both for an inner region at galactocentric distance of $r\lesssim12\arcmin$\ and for an outer annulus at $12\arcmin\lesssim r\lesssim 22\arcmin$. As expected, completeness is lower in the centre than in the external region  
due to increased crowding: we infer 50\% completeness at 
$I_\sfont{E}\sim24.3$ in the inner region while the same level is reached at 
$I_\sfont{E}\sim25.5$ in the external region.  These values are similar to the completeness values obtained by \cite{EROGalGCs} for the \Euclid ERO data of NGC\,6254, a Milky Way globular cluster. 
A maximum completeness below 80\% is a consequence of stars having to spatially match in all four \Euclid bands to appear in the final photometric catalogue, coupled with the presence of severe Galactic foreground star contamination.
Artificial stars injected on top of bright saturated 
foreground stars tend to be lost due to the compromised photometry, explaining why even at the brightest 
magnitudes we never reach 100\% completeness.

Figure~\ref{fig:spatial_compl} illustrates the spatial variation of the completeness across the entire \Euclid FoV. These maps were constructed for specific magnitude intervals tailored to each band: $25.5$--$26$\,mag for the \IE band, and $24$--$24.5$\,mag for the \YE, \JE, and \HE bands. The maps reveal a clear spatial trend, with completeness significantly decreasing toward the central crowded regions of IC\,10. Conversely, in the outer, less crowded regions, the completeness remains relatively high, enabling a more reliable analysis of the faint, extended stellar populations of the galaxy.

Figure~\ref{fig:delta_mag} shows the distribution of the difference between output and input magnitudes as a function of the input magnitude for the recovered artificial stars across the different bands.
In all bands, the distributions are not symmetric around zero, but become more and more skewed toward negative values at the faintest magnitudes, as also denoted by the behaviour of the median line. This trend is due to the increasing importance of blending at fainter magnitudes, which causes the flux of stars to be overestimated. The artificial star tests also indicate that the photometric accuracy is better in \IE than in \YE, \JE,  and \HE, with $\sigma_{\IE}\sim0.1$ at 26\,mag compared to $\sigma_{\mathrm{NISP}}\sim0.2$ at 24\,mag in the NISP bands.
This effect is likely due to the undersampling of the \Euclid PSF, which is worse for the NISP bands than  for VIS \citep{EuclidSkyOverview,Libralato24}.

\end{appendix}

\end{document}